\documentclass[11pt,a4paper]{article}
\usepackage{jstyle}
\usepackage{tikz-cd}
\usetikzlibrary {decorations.pathmorphing, decorations.pathreplacing, decorations.shapes}
\usepackage{pdflscape}   
\usepackage{enumitem}
\usepackage{tabularray}

\usepackage{genyoungtabtikz}
\Yboxdim{8pt}
\Ylinethick{1pt}
\newcommand{\hyperboloid}[1]{%
    \begin{tikzpicture}[scale=#1,baseline=(a.center)]
    \node (a) at (0,0) {};
    \draw[fill=MidnightBlue!80, opacity=0.8]
        plot[smooth] coordinates
        {(-0.745,1) (-0.3,0) (-0.75,-1)}
        -- (-0.75,-1) arc (360:180:-0.75 and 0.2)
        -- plot[smooth] coordinates
        {(0.75,-1) (0.3,0) (0.745,1)};
    \draw[fill=MidnightBlue!20] (0,1) ellipse (0.74 and 0.15);
    \end{tikzpicture}
}
\newcommand{\coneUp}[1]{%
    \begin{tikzpicture}[scale=#1,baseline=(a.center)]
    \node (a) at (0,0.5) {};
    \draw[fill=BrickRed!80, opacity=0.8]
        (-0.75,1) -- (0,0) -- (0.75,1);
    \draw[fill=BrickRed!20] (0,1) ellipse (0.74 and 0.2);
    \end{tikzpicture}
}
\newcommand{\coneDown}[1]{%
    \begin{tikzpicture}[scale=#1,baseline=(a.center)]
    \node (a) at (0,-0.5) {};
    \draw[fill=BrickRed!80, opacity=0.8]
        (-0.75,-1) -- (0,0) -- (0.75,-1);
    \draw[fill=BrickRed!20] (0,-1) ellipse (0.74 and 0.2);
    \end{tikzpicture}
}
\newcommand{\bowlDown}[1]{%
    \begin{tikzpicture}[scale=#1,baseline=(a.center)]
    \node (a) at (0,0.5) {};
    \draw[fill=OliveGreen!80, opacity=0.5]
        plot[smooth] coordinates{(-0.75, 0) (0,1) (0.75,0)};
    \draw[fill=OliveGreen!20] (0,0) ellipse (0.75 and 0.2);
    \end{tikzpicture}
}
\newcommand{\bowlUp}[1]{%
    \begin{tikzpicture}[scale=#1,baseline=(a.center)]
    \node (a) at (0,-0.5) {};
    \draw[fill=OliveGreen!80, opacity=0.5]
        plot[smooth] coordinates{(-0.75, 0) (0,-1) (0.75,0)};
    \draw[fill=OliveGreen!20] (0,0) ellipse (0.75 and 0.2);
    \end{tikzpicture}
}
\newcommand{\C}{\mathbb{C}}
\newcommand{\N}{\mathbb{N}}
\newcommand{\R}{\mathbb{R}}
\newcommand{\Z}{\mathbb{Z}}
\newcommand{\rd}{\mathrm{d}}
\newcommand{\so}{\mathfrak{so}}
\newcommand{\uu}{\mathfrak{u}}
\newcommand{\su}{\mathfrak{su}}
\newcommand{\iso}{\mathfrak{iso}}
\renewcommand{\sp}{\mathfrak{sp}}
\newcommand{\heis}{\mathfrak{heis}}
\newcommand{\amb}[1]{\textsc{#1}}
\newcommand{\CasimirMass}{\mathcal{C}_M}
\newcommand{\CasimirSpin}{\mathcal{C}_S}
\newcommand{\Functions}{\mathscr{C}^\infty}
\newcommand{\Hilbert}{\mathcal{H}}

\newcommand{\Op}{\mathsf{Op}}
\newcommand{\1}{\mathbf 1}
\author[b]{Thomas BASILE}
\author[a]{\quad Euihun JOUNG}
\author[a]{\quad TaeHwan OH}

\affiliation[a]{%
    Department of Physics
    and Research Institute of Basic Science, \\
    Kyung Hee University, Seoul 02447, Korea}
\affiliation[b]{%
    Service de Physique de l’Univers,
    Champs et Gravitation,\\
    Universit\'e de Mons, 20 place du Parc,
    7000 Mons, Belgium
}

\emailAdd{thomas.basile@umons.ac.be}
\emailAdd{euihun.joung@khu.ac.kr}
\emailAdd{hadron@khu.ac.kr}

\title{\centering
Manifestly Covariant Worldline Actions \\ from Coadjoint Orbits \\
{\LARGE Part I: Generalities and Vectorial Descriptions}}

\abstract{
We derive manifestly covariant actions of spinning 
particles starting from coadjoint orbits of isometry groups,
by using Hamiltonian reductions. We show that the defining
conditions of  a classical Lie group can be treated
as Hamiltonian constraints which generate the coadjoint orbits
of another, \emph{dual}, Lie group. In case of (inhomogeneous) orthogonal
groups, the dual groups are (centrally-extended inhomogeneous) symplectic groups.
This defines a  
symplectic dual pair correspondence between the coadjoint orbits of the isometry group
and those of the dual Lie group,
whose quantum version is the reductive dual pair correspondence \`a la Howe.
We show explicitly how various particle species 
arise from the classification of coadjoint orbits of Poincar\'e and (A)dS symmetry.
In the Poincar\'e case, we recover the data 
of the Wigner classification, which includes
continuous spin particles, (spinning) tachyons 
and null particles with vanishing momenta, 
besides the usual massive and massless spinning particles.
In (A)dS case, our classification results
are not only consistent with the pattern of 
the corresponding unitary irreducible representations observed in the literature,
but also contain novel information.
In dS, we find the presence of partially massless spinning particles, but
continuous spin particles, spinning tachyons 
and null particles are absent.
The AdS case shows the largest diversity of particle species.
It has all particles species of Poincar\'e symmetry except for the null particle, but
allows in addition various exotic entities
such as 
one parameter extension of continuous particles
and conformal particles living on the boundary of AdS.
Notably, we also find a large class of particles 
living in ``bitemporal'' AdS space, including ones where mass and spin play an interchanged role.
We also discuss the relative inclusion structure
of the corresponding orbits.
}

\begin{document}

\maketitle

%%%%%%%%%%%%%%%%%%%%%%
\section{Introduction}
%%%%%%%%%%%%%%%%%%%%%%

A coadjoint orbit of a Lie group is equipped with
a symplectic structure \cite{Kirillov2004},
and therefore can be viewed as the phase space
of a classical mechanical system. When the Lie group
is the isometry group of a spacetime, and it is large enough
--- typically the relativistic ones (Poincar\'e and (A)dS group)
or their non-relativistic  counterparts (such as the Galilean group) ---
even the dynamics (that is, the time evolution) of the system
can be ascribed by the symmetry, making it integrable.
Such mechanical systems can be interpreted as particles
moving in the spacetime having this isometry group.
Therefore, actions for relativistic particles 
can be derived
from coadjoint orbits of their isometry group,
and there have been many works in this direction, which
we shall summarise shortly in one of the following paragraphs.
Typically, the resulting actions are not manifestly covariant
under the isometry group and heavily depend on the coordinate
system of the coadjoint orbits. Since the system is integrable,
too good coordinates, such as the action-angle variables,
would render the system essentially trivial, obscuring
the spacetime propagation.
Therefore, the art is in the choice of an appropriate
set of coordinates with which the mechanical system
can be interpreted as a dynamical worldline particle,
keeping both the spacetime motion as well as
the isometries explicit.
In this regard, the covariance of the system
under the isometry group is crucial. However,
this covariance will not be manifest unless we introduce
additional degrees of freedom together with constraints.

\medskip

Many relativistic spinning particle actions have been constructed
as spin generalisations of the relativistic scalar particle action,
without explicitly relying on coadjoint orbits.
Like the scalar case, such systems have Hamiltonian constraints
and involve additional variables to describe the spin degrees of freedom.
Since the spin degrees of freedom are discrete,\footnote{In the sense that, upon quantisation, they yield a \emph{finite-dimensional} Hilbert space.} the additional variables can be introduced
as fermionic ones and this leads to 
 supersymmetry.
One may also persist to use bosonic variables for the spin degrees of freedom.
Then, the classical system
has additional continuous degrees of freedom,
on top of the
position and momentum variables, 
rather than the desired
discrete ones.
These continuous
spin degrees of freedom
should be projected, afterwards, to discrete ones
in the course of a quantisation procedure.
The twistor formulations for spinning particles are also obtained
in a similar fashion, by employing an appropriate set of constraints.
Because these works do not make use of the coadjoint orbits,
or at least its role is implicit, one often needs a separate constraint
analysis to check whether the system indeed describes
the sought after spinning particles.

\medskip

In this work, we reconsider the worldline particle actions
from the vantage point of a manifestly covariant
description of coadjoint orbits of a classical Lie group.
Since the Poincar\'e, (A)dS as well as the Lie groups
behind twistor descriptions are all classical ones,
our approach is sufficiently general to cover particles 
in Minkowski and (A)dS spaces. Using the fact that
a classical Lie group is a subgroup of the matrix group
$GL(N,\R)$ subject to a certain set of defining conditions
compatible with the matrix product, we can describe
a coadjoint orbit of a classical Lie group $G$
as a reduced phase space lying inside a coadjoint orbit
of an embedding $GL(N,\R)$ group, where the Hamiltonian reduction
is induced by the Hamiltonian constraints stemming from
the defining conditions of the group $G$.
For a given $G$-coadjoint orbit, the resulting constraints
are given by components of the moment map
for another Lie group $\tilde G$, with certain constant shifts.
We will refer to this Lie group $\tilde G$ as the \emph{dual group}.
If there is no constant shift, all constraints are first class,
but for a non-vanishing shift,
they are a mixture of first and second class constraints.
The first class constraints generate a subgroup of $\tilde G$,
whereas the second class constraints
can be associated with a $\tilde G$-coadjoint orbit.
This establishes a correspondence between the set
of $G$-coadjoint orbits and a set of $\tilde G$-coadjoint orbits.
In physical terms, the information of particle species,
such as mass and spin, is originally encoded in the $G$-coadjoint orbit.
Then, our construction maps such information
to a $\tilde G$-coadjoint orbit through the constant shifts,
where the constants are given by  the particle labels.
As the information of the $G$-coadjoint orbit (the starting point)
is encoded in the particle action through the data of
the constant shift of the dual $\tilde G$-coadjoint orbit,
the action always enjoys a manifest $G$-symmetry.

In this setting, once the starting group $G$ is fixed,
the form of the particle action is essentially universal, and 
only the constant shift
differentiates particle species.\footnote{In this paper,
we often use a very loose terminology
and refer to a coadjoint orbit
of an isometry group as a \emph{particle} simply.}
Therefore, together with the general construction
of the above system, we devote a part of our work
to the classification of $G$-coadjoint orbits
as well as the identification
of the corresponding $\tilde G$-coadjoint orbits,
i.e. the identification of the corresponding constant shift.
In the case of Poincar\'e symmetry,
the classification of coadjoint orbits can be done
in a very analogous manner as in Wigner classification:
we classify the coadjoint orbits 
in terms of the representative coadjoint elements, like the way 
we choose the momentum in the rest frame for the representative 
momentum vector  of a massive particle in the Wigner classification.
This allows to identify the coadjoint orbits of massive,
massless, tachyonic spinning particles
and even those of the continuous spin particle
and the null particles with vanishing momentum.
The same classification scheme can be equally applied to (A)dS cases.
In dS, we find the presence of partially massless spinning
particles, but continuous spin particles, spinning tachyons and  
null particles are absent.
The AdS case shows the largest diversity of particle species. It has all particles species
of Poincar\'e symmetry except for the null particle, but allows in addition various exotic
entities such as 
particles with entangled mass and spin,
which contain a one parameter extension of continuous spin particle
as a subcase, and conformal particles
living on the boundary of AdS. Notably, we also find a large class of particles living in ``bitemporal'' AdS space,
defined by $X^2=+1$ in the ambient space
with the $(-,-,+,\ldots,+)$ metric.
This class includes ones where mass and spin play an interchanged role.
The classification can be
easily 
extended to mixed symmetry cases,
where we find 
various shortening conditions
consistent with the  pattern of the corresponding unitary irreducible representations
observed in the literature. 
In each of these cases,
we identify the
dual group $\tilde G$ and the dual coadjoint orbit 
from which 
the worldline particle action can be 
readily expressed.

\medskip

The general construction used in this paper
has a close relation to
the reductive dual pair correspondence,
about which the first two authors
of 
explicitly analysed in \cite{Basile:2020gqi}.
The relation works as follows.
After a part of the constraints simply removes non-dynamical
spectator variables, the effective embedding phase space
of our model
becomes a flat one, which is the minimal coadjoint orbit
of $Sp(2n,\R) \subset GL(N,\R)$. The pair of $G$-
and $\tilde G$-coadjoint orbits is an example of symplectic dual pair \cite{Weinstein1983},
and it ensures the one-to-one
correspondence between the coadjoint orbits of $G$ and $\tilde G$
when the group $G$ is reductive.
The reductive dual pairs ---  pairs of subgroups
$(G,\tilde G) \subset Sp(2n,\R)$ which are mutual stabilisers ---
ensure even the existence of a one-to-one correspondence
between the $G$-irreducible representations (irreps)
and the $\tilde G$-irreps which arise in the  restriction
of the metaplectic representation of $Sp(2n,\R)$
onto $G \times \tilde G$.
This correspondence
is known as  the
reductive dual pair correspondence or
simply Howe duality \cite{Howe1989i, Howe1989ii}. 
Since a suitable quantisation
--- such as the geometric quantisation --- of $G$-coadjoint orbits
and $\tilde G$-coadjoint orbits would result in $G$-irreps
and $\tilde G$-irreps, respectively, the current picture
can be viewed as the classical counterpart
of the reductive dual pair correspondence.

\medskip

Let us provide a brief overview of previous works
on particle actions.
As previously mentioned, one of the most common ways
of describing spinning particles
consists in introducing fermionic variables
to the phase space.\footnote{The first introduction
of Grassmannian variables in a classical mechanics
setting seems to go back to the paper \cite{Martin1959}.}
The latter are used to realise supersymmetry on the worldline,
with the number $\cN$ of supercharges corresponding to 
a particle of spin-$\tfrac\cN2$,
as shown in \cite{Gershun:1979fb, Howe:1988ft},
drawing on earlier work on massive superparticles
\cite{Casalbuoni:1976tz, Berezin:1976eg, Brink:1976sz, Barducci:1976qu}
(see also \cite{Galvao:1980cu, Siegel:1988ru, Marcus:1994mm, 
Gorbunov:1998id, Kuzenko:1995mg, Fedoruk:2006jm, Fedoruk:2006it, Bastianelli:2007pv, Bastianelli:2008nm, Bonezzi:2012cf, Bastianelli:2014lia}).
Another approach is to use (super)twistor variables in $d=3,\ 4,$
and $6$ dimensions to describe spinning massive
\cite{Bette:1984qt, Lyakhovich:1996we, Kuzenko:1994ju, Fedoruk:2014vqa, Mezincescu:2013nta, Routh:2015ifa, Mezincescu:2015apa}
and massless particles \cite{Shirafuji:1983zd,Bengtsson:1987ap,
Bengtsson:1987si,Townsend:1991sj,Howe:1992bv}
in flat spacetime,
as well as in AdS$_{d+1}$
\cite{Cederwall:2000km, Bars:2005ze, Arvanitakis:2016wdn, Arvanitakis:2016vnp, Arvanitakis:2017cpk}.
More recently, these techniques were also
used to obtain actions for continuous spin particles \cite{Buchbinder:2018soq,Buchbinder:2019gbq,Buchbinder:2019iwi,Buchbinder:2019sie,Buchbinder:2021bgv}.

The use of the symplectic structure
on a coadjoint orbit in describing
a particle dynamics
has also a long history
starting from the pioneering work of Souriau \cite{Souriau2012}.
In the formulation with twistor variables
of relativistic particles 
this was used starting from the early works \cite{Tod:1977vf,Balachandran:1983oit,Bette:1984qt, Howe:1992bv}
to a more recent one \cite{Fedoruk:2014vqa}.
In the formulation with spacetime variables,
this appeared in e.g.
\cite{Kuzenko:1994vh, Kuzenko:1995aq,   Lyakhovich:1998ij, Lyakhovich:1998ve,Andrzejewski:2020qxt}. 
See also \cite{Gorbunov:1999jg,Duval:2014ppa, Batlle:2017cfa} 
for other applications to particle dynamics.

A closely related set up to derive a particle action
starting from a Lie group is known as
the \emph{nonlinear realisation} method
which proved particularly useful to construct actions
for $p$-branes as well as non-relativistic particle actions,
see e.g. \cite{Ivanov:1999fwa, Ivanov:1999gy, Bellucci:1998mk, Bellucci:2000bd, Bellucci:2002ji, Gomis:2006xw,Gomis:2006wu,Gomis:2007zz}
and references therein. See also \cite{Gomis:2021irw}
for its use in a color-extension of spacetime symmetry,
\cite{Batlle:2023zhp} for particles in BMS space,
and \cite{Alekseev:1988vx, Alekseev:1988tj,Alekseev:1988ce,   Aratyn:1990dj, Alekseev:2015hda, Alekseev:2018pbv}
for discussions of the path integral quantisation of this kind of model.

The aforementioned references have covered a wide range of relativistic particle species,
but each paper was focused on a specific particle model.
In the end, many of particle species are left aside and 
the underlying systematics were missing.
Our work aims at generalising these efforts and
adding a clean systematics: worldline actions for any particle species
in any constant curvature spacetime are obtained in a universal fashion
from the first principle of the coadjoint orbit and particle correspondence.

Let us end this brief tour of the literature by mentioning
that worldline models can serve in various quantum field theory
contexts \cite{Schubert:1996jj, Bastianelli:2009mw},
for instance to compute heat kernel/effective action coefficients
\cite{Bastianelli:2008vh, Bonezzi:2017mwr, Bastianelli:2022pqq, Bastianelli:2023oca, Bastianelli:2023oyz} and 
scattering amplitudes \cite{Albonico:2022pmd}
or to probe properties of the gauge theory associated
with background fields \cite{Howe:1989vn, Casalbuoni:2014ofa, Casalbuoni:2014qia, Rempel:2016elp, Grigoriev:2021bes, Boffo:2023fsz}.
In the context of higher spin gravity, coadjoint orbits 
play an important role, in that several higher spin algebras
arise as the quantisation of particular orbits of $\so(2,d-1)$.
To be more precise, the simplest higher spin algebra
(sometimes referred to as the type-A algebra), 
is the symmetry algebra of the minimal representation
of $\so(2,d-1)$, representation which is obtained by
quantising its minimal nilpotent orbit
\cite{Eastwood:2002su,Joung:2014qya} (see also \cite{Michel2011}
and \cite{Joung:2015jza} for a discussion
of the partially-massless generalisation in relation
to the quantisation of coadjoint orbits). On top of that,
higher spin algebras are commonly realised using the dual
pair correspondence (also known as Howe duality
\cite{Howe1989i, Howe1989ii},
see e.g. \cite{Rowe:2011zz, Rowe:2012ym, Basile:2020gqi} for reviews)
previously mentioned, a classical counterpart of which
is recovered in this paper.

\medskip

The organization of the paper is as follows:
In Section \ref{sec:coadjoint},
we start by reviewing the basics of coadjoint orbits
of a Lie group and their symplectic structures.
We explain how one can associate a particle action
to each orbit, and discuss the conditions under which 
the path integral is well-defined. After detailing
simple examples of three-dimensional Lie groups,
we point out the issue of coordinate choice in this action
and argue for the necessity of manifest covariant description
of the actions using Hamiltonian constraints.
In Section \ref{sec: constraint},
we present several general results of
a constrained Hamiltonian system where the constraints
are given by constant shifts of the moment map associated
with the dual Lie algebra. In particular, we demonstrate
that the second and first class constraints correspond to
a coadjoint orbit of the dual group
and its stabiliser of the dual group.
In Section \ref{sec: cov action},
we explain how the coadjoint action
for a classical Lie group and its semi-direct product
with an Abelian ideal can be reformulated
as a constrained Hamiltonian system
by making use of the set-up explained
in Section \ref{sec: constraint}.
After briefly covering the general cases,
we provide more details on the orthogonal
and inhomogeneous orthogonal group cases,
relevant to the symmetries of spacetime.
In Section \ref{sec: Minkowski}, we apply 
the construction of worldline action for semi-direct
product groups detailed in the previous section
to the Poincar\'e case, and (re)derive the actions
for various particles in Minkowski spacetime.
 In Section \ref{sec: AdS}, we move to the (A)dS
case and derive various particle actions by using
the same method. On top of the usual massive
and massless particles, we spell out 
various other particle species.
 In Section \ref{sec: inclusion}, we discuss 
the inclusion structure of both nilpotent orbits
--- which is known to admit a convenient description
in terms of Young and Hasse diagrams --- and semisimple
ones, which seem to have received less attention.
In Section \ref{sec: conclusion}, we conclude
this paper with a short discussion of the remaining
questions that we intend to address in our follow-up
paper \cite{partII}.
Finally, this paper includes several appendices
containing additional details and material complementing
its bulk. In Appendix \ref{app:notations}, we summarise
the conventions and notations used. In Appendix \ref{app:trick},
we explain how one can convert the second
class constraints appearing in the Hamiltonian system
detailed in Section \ref{sec: constraint}
into first class ones. Appendix \ref{app:O(n)} contains
details on the classification of orbits of the orthogonal
groups $O(n)$. We collect in Appendix \ref{app:tables}
the data defining the coadjoint orbits and their duals
identified in this paper, and detail in Appendix \ref{app: so(2,2)}
the relation between coadjoint orbits of $SO^+(2,2)$ and of $SO^+(2,1)$.
Finally, we compare our classification
with the results of Metsaev \cite{Metsaev:2019opn}
in Appendix \ref{sec: Metsaev}.

%%%%%%%%%%%%%%%%%%%%%%%%%%%%%%%%%%%%%%%%
\section{Coadjoint orbits and particles}
\label{sec:coadjoint}
%%%%%%%%%%%%%%%%%%%%%%%%%%%%%%%%%%%%%%%%

In order to understand how a particle action can be
obtained from a coadjoint orbit of the associated symmetry group,
let us first consider the simple example of a relativistic scalar particle action,
\be
	S=\int {\rm d}t \left[p_\mu\,\dot x^\mu-e\,(p^2-m^2)\right],
	\label{massive scalar mink}
\ee
where $\dot x^\mu := \tfrac{{\rm d} x^\mu}{{\rm d} t}$, and the einbein $e$ plays the role of a Lagrange multiplier which
sets the mass-shell constraint $p^2=m^2$. By solving the latter
as $p_0=\pm \sqrt{p_a\,p^a+m^2}$ and fixing $x^0$ to $t$ using the reparametrisation symmetry, we find an equivalent action,
\be
	S=\int {\rm d}t \left[p_a\,\dot x^a\pm\sqrt{p_a\,p^a+m^2}\right].
\ee
Here, the sign $\pm$ distinguishes the positive energy and negative energy solutions
which can be mapped to each other by the time inversion  $t\to -t$.

The same action can be obtained from a coadjoint orbit of the Poincar\'e group, whose Lie algebra
$\mathfrak{iso}(1,d-1)$ is generated by $P_\mu$ and $J_{\mu\nu}$
(see Appendix \ref{app:notations} for our conventions
regarding their Lie bracket).
A vector $\phi$ in the coadjoint space $\mathfrak{iso}(1,d-1)^*$ has the
form  $\phi=p_\mu\,\cP^\mu+j_{\m\n}\,\cJ^{\m\n}$
where $\cP^\mu$ and $\cJ^{\m\n}$ are the dual basis vectors
satisfying $\la \cP^\mu, P_\nu\ra=\delta^\mu_\n$,
$\la \cJ^{\m\n},J_{\r\s}\ra=\delta^{[\m}_{\r}\,\delta^{\n]}_{\s}$
and $\la \cP^\mu,J_{\r\s}\ra=0=\la \cJ^{\m\n},P_\r\ra$.
The orbit corresponding to a massive scalar particle
is given by the representative vector $\phi=m\,\cP^0$
whose only non-vanishing component is $p_0=m>0$.
Under the coadjoint action of the Poincar\'e group on $\phi$, all $j_{\m\n}$ components remain  zero,
while $p_\mu$ forms an upper hyperboloid given by $p_\mu\,p^\m=m^2$
and $p_0>0$, the typical momentum orbit.
Note that this orbit is embedded in the $d(d+1)/2$ dimensional space $\mathfrak{iso}(1,d-1)^*$.

The action corresponding to the orbit $\cO_\phi$ is given by
(we shall review the details later),
\be
	S[g] = \int {\rm d}t\,\la\phi, g^{-1}\dot g\ra\,,
	\label{MC action}
\ee
where $g$ is a generic element of the Poincar\'e group.
Parameterising the element as\footnote{In this paper,
we use the convention where the Lie algebra generators
are anti-Hermitian.}
\be
	g = e^{x^\mu\,P_\mu}\,e^{v^a\,J_{a0}}\,e^{\theta^{ab}\,J_{ab}}\,,
	\label{g para poincare}
\ee
we find
\be
	\la \phi,g^{-1}\dot g\ra
	=\dot x^\mu \la e^{v^a\,J_{a0}}\,m\,\cP^0\,e^{-v^a\,J_{a0}}, 
    P_\mu\,\ra.
\ee
The boost parameters $v^a$ parameterise the momentum orbit as
\be
 	e^{v^a\,J_{a0}}\,m\,\cP^0\,e^{-v^a\,J_{a0}}
    = -\sqrt{m^2+p_a\,p^a}\,\cP^0+p_a\,\cP^a\,,
\ee
where $p_a=m\,\frac{\sinh(v)}{v}\,v_a$ and $v=\sqrt{v^a\,v_a}$.
So we can reformulate the right hand side
of the above equation as $p_\mu\,\cP^\mu$ by appending 
the constraints $p^2=m^2$ and $p_0>0$. In this way,
we recover the action \eqref{massive scalar mink}
of a massive scalar particle in Minkowski space.
The method of using the Maurer--Cartan one-form
$g^{-1}\,{\rm d}g$ has been well developed
under the name of \emph{nonlinear realisation}
and it has been shown that this method can be applied
to various  particles
(or even branes) with different symmetries,
whether relativistic, non relativistic
or conformal, see e.g. \cite{Brugues:2004an, Gomis:2004pw, Gomis:2006xw, Brugues:2006yd, Casalbuoni:2008iy, Gomis:2011dw, Bergshoeff:2014jla, Bergshoeff:2015wma, Barducci:2017mse, Gomis:2021irw}
and references therein.

The aim of the current paper is to generalise
the above procedure of obtaining the constrained
action \eqref{massive scalar mink} with manifest covariance
to spinning particles as well as more exotic types
of particles such as continuous spin particle. 
For that purpose, in the current section we consider
the generalisation of the unconstrained action \eqref{MC action}.
In the following, to be self-contained,
we begin with reviewing the classical result of
Kirillov--Kostant--Souriau that there exists a $G$-invariant
symplectic structure on any coadjoint $G$-orbit.
Then, we discuss several issues arising in interpreting
the coadjoint orbit action \eqref{MC action}
as a particle action. Let us stress that
in this paper, we will be using the term `particle'
loosely to refer to the different types of coadjoint orbits
that we will encounter. We have in mind that, 
when the quantisation of these coadjoint orbits
is possible, it will give rise to a unitary 
and irreducible representation of the isometry group.
Moreover, we will see that the parameters that label
the different coadjoint orbits correspond,
in the `quantisable' case, to usual physical parameters
such as the mass and spin of the particle.

%*******************************************%
\subsection{Coadjoint orbits: generalities}
\label{sec:review}
%*******************************************%

Let us begin with 
the introduction of a few mathematical notions
relevant to the study of coadjoint orbits.
For a general introduction to the subject,
one can consult e.g. \cite{Kirillov2004, Kirillov1999, Vogan2000}.

Recall that given a Lie group $G$  with Lie algebra
$\mathfrak{g}$, the coadjoint orbit $\cO^{G}_\phi$ of
an element $\phi\in\mathfrak{g}^*$ is the submanifold
of $\mathfrak{g}^*$ whose points are related to $\phi$
by the coadjoint action of $G$, i.e.
\begin{equation}
  \cO^{G}_\phi := \big\{{\rm Ad}^*_g\,\phi\,,\ g \in G \big\}
  \subset \mathfrak{g}^*\,,
  \label{eq:def_orbit}
\end{equation}
where ${\rm Ad}^*$ denotes the coadjoint action of $G$
on $\mathfrak{g}^*$, defined by
\begin{equation}
  \langle {\rm Ad}^*_g \varphi, \xi \rangle
  = \langle \varphi, {\rm Ad}_{g^{-1}}\xi \rangle\,,
  \label{eq:Ad*}
\end{equation}
for any $\varphi \in \mathfrak{g}^*$, $\xi \in \mathfrak{g}$
and $g \in G$. Here ${\rm Ad}$ is the adjoint action
of $G$ on its Lie algebra $\mathfrak{g}$,
and  $\langle  \cdot, \cdot \rangle$ denotes the pairing
between $\mathfrak{g}^*$ and $\mathfrak{g}$.
The element $\phi \in \mathfrak g^*$ above
simply serves as a reference point for the coadjoint 
orbit $\cO^G_\phi$, and can be used as a label 
for the latter. Of course, there is no privileged choice
for this reference point as it is a representative
of the equivalence class of element in $\mathfrak g^*$
under the coadjoint action of $G$. In the rest of the paper,
we will use the representative $\phi$ to designate
the corresponding coadjoint orbit.
Note that when the Lie group $G$ has disconnected parts,
related by finite subgroups, their coadjoint orbits 
may have also disconnected parts.

One can identify a coadjoint orbit with the quotient space,
\begin{equation}
    \cO^{G}_\phi\, \simeq \, G/G_\phi
    =\{\,[g]\,, \forall g\in G\,|\,[g\,h]=[g]\,,
    \forall h\in G_\phi\,\}\,,
    \label{eq:orbit_quotient}
\end{equation}
where $G_\phi$ is the subgroup of $G$ which leaves $\phi$
invariant under its coadjoint action,
\begin{equation}
  G_\phi = \big\{ g \in G\,\rvert\,
  {\rm Ad}^*_g \,\phi = \phi \big\}\,,
\end{equation}
and is called its stabiliser or isotropy subgroup.
Therefore, the coadjoint orbit $\cO^{G}_\phi$
can be viewed as the base space of the principal
$G_\phi$-bundle whose total space is $G$ and
with projection map $\pi_\phi$,
\begin{equation}
	\begin{split}
		\pi_\phi: &\ G\ \to\ \cO^{G}_\phi\,, \\
		&\ g\ \ \mapsto\ \pi_\phi(g)={\rm Ad}^*_g\,\phi\,.
	\end{split}
\end{equation}
Notice that the stabilisers of any two elements of
a coadjoint orbit are isomorphic.\footnote{Indeed, 
a simple computation shows that
$G_{{\rm Ad}^*_g\phi} = g^{-1} G_\phi g$
for any $g \in G$, and
$\mathfrak g_{{\rm Ad}_g^*\phi}={\rm Ad}_{g^{-1}} \mathfrak g_\phi$.
Note also that the projection $G \twoheadrightarrow \cO^G_\phi$
does not depend on a choice of representative of the orbit:
one can verify that $\pi_{{\rm Ad}^*_h\phi}(g) = \pi_\phi(gh)$,
which implies that different choices of orbit representatives
to define the projection explicitly lead to diffeomorphic
$G_\phi$-principal bundle structures.}

The tangent space
of the quotient manifold \eqref{eq:orbit_quotient}
at a point $\varphi\in \cO_\phi^G$ is therefore
given by the quotient of the corresponding Lie algebras,
\begin{equation}
    T_\varphi \cO_\phi^{G}\, \cong\, \mathfrak{g} / \mathfrak{g}_\varphi\,,
    \label{eq:tangent_orbit}
\end{equation}
with $\mathfrak{g}_\varphi$ the Lie algebra of $G_\varphi$,
which can be described as
\begin{equation}
  \mathfrak{g}_\varphi = \big\{ \xi \in \mathfrak{g} \,\rvert\,
  {\rm ad}^*_\xi \,\varphi = 0 \}\,,
     \label{stabiliser}
\end{equation}
where ${\rm ad}^*_\xi\, \varphi := -\varphi \circ {\rm ad}_\xi$
denotes the coadjoint action of a Lie algebra element
$\xi\in\mathfrak{g}$ on  $\varphi\in\mathfrak{g}^*$.
Consequently, any vector $V_\xi \in T_\varphi \cO_\phi^{G}$
can be generated by an element
$\xi \in \mathfrak{g} / \mathfrak{g}_\varphi$,
\begin{equation}
  V_\xi := {\rm ad}^*_\xi\,\varphi\,.
\end{equation}
The coadjoint orbits can be grouped into two categories:
\emph{semisimple} and \emph{nilpotent} coadjoint orbits.
If a coadjoint orbit $\cO^{G}_\phi$ satisfies 
$\mathfrak{g}_\phi\subset {\rm Ker}\,\phi$,
that is,
\begin{equation}
	\la \phi, \xi\ra=0\,, \qquad \forall \xi\in \mathfrak{g}_\phi,
\end{equation}
the orbit is \emph{nilpotent}, and if not, the orbit is 
\emph{semisimple} (see e.g. \cite[Sec. 1.3]{Collingwood1993}).
For a given Lie algebra $\mathfrak{g}$, there is a 
continuum of semisimple orbits, and they are labelled by
a set of continuous parameters. On the contrary,
there is only a finite discretum of nilpotent orbits,
and hence representative vectors of nilpotent orbits
do not contain any parameters which label the orbits.
In other words, coadjoint vectors with rescaled
parameters belong to the same nilpotent coadjoint orbit.
As we shall review shortly below, each coadjoint orbit
is an even dimensional subspace of $\mathfrak{g}^*$
equipped with a $G$-invariant symplectic form.

Various properties of a coadjoint orbit can be
captured by the quotient Lie algebra,
\begin{equation}
    \mathfrak{g}^{\rm Ab}_\phi
    := \mathfrak{g}_\phi/[\mathfrak{g}_\phi,\mathfrak{g}_\phi]\,,
\end{equation}
the Abelianisation of $\mathfrak{g}_\phi$,
since the derived algebra $[\mathfrak{g}_\phi,\mathfrak{g}_\phi]$
verifies
\begin{equation}
    [\mathfrak{g}_\phi,\mathfrak{g}_\phi]
    \cong \big\{\xi \in \mathfrak{g}_\phi \mid
    \langle \phi, \xi \rangle = 0\big\}
    \equiv {\rm Ker}\phi \cap \mathfrak{g}_\phi\,.
\end{equation}
For a nilpotent orbit, $\mathfrak{g}_\phi^{\rm Ab}=\emptyset$
by definition, whereas $\mathfrak{g}_\phi^{\rm Ab}$
is non-trivial for a semisimple orbit, and it is \emph{elliptic}
if $\mathfrak{g}_\phi^{\rm Ab}$ is compact.

Let us conclude this section by recalling that,
when the Lie algebra $\mathfrak g$ is endowed with
a symmetric bilinear form
\begin{equation}
    \kappa: \mathfrak g \times \mathfrak g \longrightarrow \mathbb R\,,
\end{equation}
which is \emph{Ad-invariant}, meaning it verifies
\begin{equation}
    \kappa({\rm Ad}_g \xi, {\rm Ad}_g \zeta) = \kappa(\xi, \zeta)\,,
    \qquad 
    \xi, \zeta \in \mathfrak g\,,
\end{equation}
for any Lie group element $g \in G$, then one can relate
coadjoint orbits to adjoint ones --- orbits of the Lie group $G$
on its Lie algebra $\mathfrak g$ defined by the adjoint action. 
Indeed, one can define the `musical morphism',
\begin{equation}
    \begin{aligned}
        \kappa^\flat:\ \mathfrak g & \longrightarrow\ \mathfrak g^*\\
        \xi & \longmapsto\ \kappa^\flat(\xi) := \kappa(\xi,-)\,,
    \end{aligned}
\end{equation}
which, by Ad-invariance of $\kappa$, implies
\begin{equation}
    \kappa^\flat\big(\cO^G_\xi\big) = \cO^G_{\kappa^\flat(\xi)}\,,
\end{equation}
where on the left-hand side, one has the adjoint orbit
of $\xi \in \mathfrak g$, and on the right hand side
the coadjoint orbit of $\kappa^\flat(\xi)$. On top of that,
if $\kappa$ is non-degenerate, i.e. 
\begin{equation}
    \kappa(\xi,\zeta) = 0
    \quad
    \forall \zeta \in \mathfrak g
    \qquad \Rightarrow \qquad 
    \xi = 0\,,
\end{equation}
the musical morphism $\kappa^\flat$ is an isomorphism, 
and therefore defines a diffeomorphism between the adjoint
orbit of any $\xi \in \mathfrak g$ and coadjoint orbit
of $\kappa^\flat(\xi) \in \mathfrak g^*$.

%*************************************************************%
\subsubsection*{Coadjoint orbits of real semisimple Lie groups}
%*************************************************************%

Since the Killing form of real semisimple Lie groups
is non-degenerate, one can equivalently study their coadjoint 
or adjoint orbits.
The representative element  of an adjoint orbit
admits a unique decomposition,
the Jordan decomposition, in terms of 
elliptic, hyperbolic and nilpotent elements.
An element $\xi \in \mathfrak{g}$ is called  nilpotent if the matrix
${\rm ad}_\xi$ is a nilpotent matrix. 
An element $\xi \in \mathfrak{g}$ is called semisimple
if the matrix $\mathrm{ad}_\xi$ is diagonalisable
over the complex numbers.
Semisimple elements are divided into
elliptic and hyperbolic ones 
depending on whether their non-zero eigenvalues
are all pure imaginary or not
(with anti-Hermitian convention for $\mathfrak g$).
Compact semisimple Lie groups have only semisimple
coadjoint orbits, which are in one-to-one correspondence
with orbits of the Weyl group in the Cartan subalgebra.
For classical Lie groups, that is real forms of $GL_N$,
$O_N$ or $Sp_{2N}$ which can be compact or non-compact,
the classification of adjoint orbits has been worked out
in \cite{Burgoyne1977, Djokovic1983}.

Nilpotent orbits are of particular interest,
both in mathematics and physics:
see e.g. \cite{Hanany:2016gbz, Cabrera:2017ucb, Hanany:2017ooe, Hanany:2018uzt, Hanany:2018xth} 
for recent progress 
on complex nilpotent orbits.
These orbits have been classified, and can be labeled by
\emph{signed Young diagrams} \cite{Springer1970} 
(see also \cite{Gerstenhaber1961} for the classification
of nilpotent orbits of the complex forms,
and \cite[Chap. 9]{Collingwood1993}
for a textbook account), which are simply Young diagrams
whose boxes are filled in with plus or minus signs,
in a way that encodes the real form of the Lie algebra
of interest. The basic idea of this classification
comes from the Jacobson--Morozov theorem which states
that any nilpotent element of a semisimple Lie algebra
$E \in \mathfrak g$ fit into a triple $\{H,E,F\}$
which span an $\mathfrak{sl}(2,\mathbb R)$ subalgebra
in $\mathfrak g$, as its raising operator. The fundamental 
representation $V$ of $\mathfrak g$ is completely reducible
under the action of this $\mathfrak{sl}(2,\mathbb R)$,
as a direct sum of highest weight modules. This collection
of highest weights allows one to associate a partition
of the dimension of $V$, i.e. a Young diagram with 
$\dim V$ boxes, to a given nilpotent orbit. Moreover, 
each box of these Young diagrams should be filled in 
with either a $+$ or a $-$ sign, in an alternating manner
in each row, according to rules
that depend on the particular real form $\mathfrak g$.
Two signed Young diagrams are equivalent if one can be
related to the other by a permutation of its rows.
The interested reader may find a detailed account
of this classification in \cite[Chap. 9]{Collingwood1993}.

An adjoint orbit is called \emph{regular} if its elements
are regular, which is to say that their centralisers
are of minimal dimension, namely the rank of the algebra
\cite[Chap. II.2]{Knapp2013}.
Consequently, these orbits are of maximal dimensions,
and can be described as surfaces in $\mathfrak{g}^*$
defined as the common level sets of the functions
dual to the Casimir operators.\footnote{In the sense
that the space of polynomial functions on $\mathfrak{g}^*$
is isomorphic to $S(\mathfrak{g})$, to which the Casimir
operators of $\mathfrak{g}$ belong.} Hence, their dimension
is $\dim\mathfrak{g}- {\rm rank}\,\mathfrak{g}$. 
Among regular orbits, there is a unique nilpotent orbit,
usually called the principal nilpotent orbit,
defined by the zero locus of the Casimir functions. 
The other nilpotent orbits have smaller dimensions, 
as they are defined by a larger number of polynomial equations.
The nilpotent orbit with minimum dimension,
apart from the trivial orbit $\{0\}$, is also unique
and called the minimal orbit.

%***********************************************%
\subsection{Kirillov--Kostant--Souriau symplectic
two-form and symplectic potential}
\label{sec:KKS}
%***********************************************%

Coadjoint orbits form an interesting class of symplectic manifolds,
as they are endowed with a symplectic form, called the
Kirillov--Kostant--Souriau (KKS) symplectic form.
Its value at any point
$\varphi \in \cO_\phi^G$ is defined by
\begin{equation}
  \omega_\varphi(V_{\xi_1}, V_{\xi_2})
  = \langle \varphi, [\xi_1,\xi_2] \rangle\,,
  \label{eq:KKS}
\end{equation}
with $\xi_1, \xi_2 \in \mathfrak{g}$.
The pullback $\O$ of the KKS symplectic form $\o$ on $G$ gives
\be
	\O_g(\xi_1,\xi_2) := (\pi_\phi^*\omega)_g(\xi_1,\xi_2)=\la \phi, [\Theta_g(\xi_1),\Theta_g(\xi_2)]\ra\,,
\ee
where $\Theta$ is the Maurer--Cartan form --- the left-invariant $\mathfrak{g}$-valued one-form on $G$, locally given by
\be
	\Theta_g=g^{-1}{\rm d}_G g\,,
\ee
where ${\rm d}_G$ is the de Rham differential on the group manifold $G$.
Since the Maurer--Cartan form satisfies the Maurer--Cartan equation,
\be
	{\rm d}_G\,\Theta+\tfrac12\,[\Theta, \Theta]=0\,,
\ee
the two-form $\O$ is exact\,:
\be
	\O = -{\rm d}_G\,\la \phi,\Theta\ra\,.
\ee
The orbit $\cO^G_\phi$ can be covered by several coordinate patches $U_i\subset \cO^G_\phi$
with local sections $\sigma_i: U_i \hookrightarrow G$.
We can pullback the two-form 
 $\O$ by $\sigma_i$  to obtain
 the symplectic two-form $\o$ and the corresponding symplectic potential $\theta_i$
 in each $U_i$\,:
\be
	\omega=-{\rm d}_\cO\,\theta_i\,,\qquad \theta_i=\la \phi,\s_i^*\Theta\ra
	=\la \phi, \s_i{}^{-1}\,{\rm d}_\cO\s_i\ra\,,
\ee
where $\rm d_\cO$ is  the de Rham differential on the coadjoint
orbit $\cO^G_\phi$. 
Note that the two-form $\o$ does not depend on the choice of sections but $\theta_i$ does:
two sections are related by
\be
	\s_j=\s_i\,\t_{ij}\,,
\ee
with the transition map $\t_{ij}\,: U_i\cap U_j \to G_\phi$,
and consequently the symplectic potentials are related by
\be
	\theta_j=\theta_i+\la \phi,\t_{ij}{}^{-1}\,{\rm d}_\cO\t_{ij}\ra\,.
\ee
The fact that the second term is closed,
due to ${\rm Ad}^*_{\t_{ij}}\phi=\phi$,
shows that the two-form $\o$ is gauge independent,
that is, independent of sections.

%************************************************%
\subsection{Worldline action and its quantisation}
\label{sec:worldline_action}
%************************************************%

The worldline action is given by the integral of $\theta_i$
on a path $\g$ lying in $U_i\subset\cO_\phi^G$,
or equivalently, the integral of $\Theta$
on the lifted path $\sigma_i(\gamma)$
lying in $\sigma_i(U_i) \subset G$,
\begin{equation}
	S_i[\gamma] = \int_{\gamma \subset \cO_\phi} \theta_i
    =\int_{\sigma_i(\g)\subset G} \langle\phi,\Theta\rangle\,.
\end{equation}
Note that this type of action has been considered
in various contexts: see e.g. 
\cite{Alekseev:1988vx, Barnich:2017jgw, Ciambelli:2022cfr, Barnich:2022bni}.
The action transforms as 
\begin{equation}
	S_j=S_i+\int_{\g}\la \phi,\t_{ij}{}^{-1}\,{\rm d}_\cO\t_{ij}\ra\,.
\end{equation}
We consider only local change of section,
that is to say, the transition function $\t_{ij}$
becomes identity at the end points
of the path $\g$.\footnote{Transformations
of the end points may involve issues
of large gauge transformations. See \cite{Alekseev:2018pbv}
for related discussions.}
In the case the transition map $\t_{ij}$ is connected
to identity, the difference of the action vanishes.
In the other case, it gives a non-trivial contribution.
When we quantize the system through
the path integral,\footnote{The issues of quantisation,
including the path integral measure,
will be addressed in the forthcoming paper \cite{partII}.}
\begin{equation}
	Z = \int \cD\g\,\exp\left(\tfrac{i}{\hbar}\,S[\g]\right)\,,
\end{equation}
we can also ask the invariance of $Z$ under a change
of section by a transition map $\t_{ij}$
which may not be connected to identity.
The difference of the action by such a $\t_{ij}$
belongs to the first de Rham cohomology group $H^1(G^{\rm Ab}_\phi)$, the Lie group associated with
$\mathfrak{g}_\phi^{\rm Ab}$.
Since $\la \phi, [\xi,\zeta]\ra=0$
for any $\xi, \zeta\in \mathfrak{g}_\phi$,
it is sufficient to consider $\mathfrak{g}_\phi^{\rm Ab}$
instead of $\mathfrak{g}_\phi$.
The group $G^{\rm Ab}_\phi$ is Abelian, and
we can parameterise an element $\tau_{ij} \in G^{\rm Ab}_\phi$ as
\begin{equation}
	\t_{ij}=e^{\z^1_{ij}\,J_1}\,e^{\z^2_{ij}\,J_2}\cdots e^{\z^{k}_{ij}\,J_{k}}\,,
	\qquad k={\rm dim}G^{\rm Ab}_\phi\,,
\end{equation}
where $J_a$'s are the generators of  the Lie algebra $\mathfrak{g}^{\rm Ab}_\phi$.
We require that the $\t_{ij}$ transformation leaves the path integral invariant:
\begin{equation}
	\exp\left[\frac{i}{\hbar}\,\int_\gamma \la \phi, \t_{ij}^{-1}{\rm d}_\cO\t_{ij}\ra\right]
	=\exp\left[\frac{i}{\hbar}\, \la \phi, J_a\ra \int_\g {\rm d}_\cO \z_{ij}^a\right]
	=1\,.
	\label{triv b}
\end{equation}
Recall that the transition function becomes identity (i.e. $\zeta_{ij}^a$ vanish) at the end points of the path $\g$.
In this case, we find
\begin{equation}
	\int_{\g} {\rm d}_\cO \z_{ij}^a=\oint_{\G_{ij}}{\rm d}_G\z^a\,,
\end{equation}
where $\G_{ij}=\s_i(\g)\cdot \s_j(\g)^{-1}$ is the closed path lying in $G$\,.
The parameters $\z^a$ may or may not be periodic
depending on the nature of the generator $J_a$.
If a generator $J_a$ exponentiates to a $U(1)$
so $\z^a$ is periodic (with period $T_a$)\,,
then the integral,
\begin{equation}
	w_a=\frac1{T_a}\oint_{\G_{ij}} {\rm d}_G\z^a\in \mathbb Z\,,
	\label{quantisation condition}
\end{equation}
gives the number of times that the closed path $\G_{ij}$
winds the cycle associated with the $\zeta^a$ coordinate.
Therefore, the condition \eqref{triv b} requires
that each of $\frac{T_a}{2\pi\hbar} \la\phi,J_a\ra$
be an integer. Recall that the latter $\la\phi,J_a\ra$
are all zero in a nilpotent orbit.
Therefore, here only semisimple orbits are concerned.
We consider $\hbar$ as a fixed constant,
so $\la\phi,J_a\ra$ is quantised.
If a generator $J_b$  exponentiates
rather to an $\mathbb R$, then $\z^b\in\mathbb R$
and there cannot be any non-trivial winding of $\G_{ij}$.
Therefore, $\oint_{\G_{ij}} {\rm d}_G\z^b=0$
and no condition is imposed on $\la\phi,J_b\ra$.
From the above discussion, we see that
the quantisation selects
a certain discretum of coadjoint orbits
among an infinite continuum of 
semisimple coadjoint orbits.\footnote{This condition is also an example
of the mechanism of quantisation of coupling constants in field theory spelled out in e.g. \cite{Alvarez:1984es}.}.
This selection is in fact equivalent to the prequantisation condition of
the geometric quantisation:
if $\g$ is closed, we can take two disks $\S_i\subset U_i$ and $\S_j\subset U_j$ such that $\partial \S_i=\gamma=\partial \S_j$.
In such cases,
the difference of the action reduces to
\be
	\int_{\g}\la \phi,\t_{ij}{}^{-1}\,{\rm d}_\cO\t_{ij}\ra=\oint_{\S_{ji}}\o\,,
	\label{S dif}
\ee
where $\S_{ji}=\S_j\cup \bar\S_i$ 
(here, $\bar\S_i$ is the disk $\S_i$ with the opposite orientation) has the topology of a two sphere $S^2$.
Remark however that the quantisation of $\la \phi, J_a\ra$ takes place even when there is no $\S_i$ or $\S_j$ satisfying the condition. 

Clearly, the change of sections by a transition map $\t_{ij}$ can be interpreted as a gauge transformation. 
The role of this gauge symmetry will become more manifest when we reformulate the action as a constrained Hamiltonian system.
It is also worth noting that
the condition \eqref{quantisation condition} 
depends on the topology of $G$:
it changes if we change the Lie group $G$
by its one of covering groups.
Since the coadjoint orbits
of $G$ and its various covers
are all the same, the KKS symplectic form
$\omega$ is also the same.
However, the symplectic potential $\theta_i$ depends on the covering structure of the group.

%*******************%
\subsubsection*{Spin}
%*******************%

Typically, the components $\phi_a:=\la \phi,J_a\ra$
match the labels of particle species such as mass and spin.
Due to the mechanism described above, some of these labels
may be quantised: the (conventional) spin label
ought to be  quantised always, but sometimes
the mass label is quantised as well, e.g. in AdS spacetime.%
\footnote{Let us point out that the mass label 
of generic massive particles is quantised in AdS$_{d+1}$  
since the time translation forms a compact  subgroup $SO(2)$
of the AdS group $SO^+(2,d)$ (or its double cover). 
This is to be contrasted with the discrete mass level
of partially-massless particles, for which the value
of the mass is related to that of the spin and depth of the field.
For a continuous spectrum of mass for massive particles, 
one can replace AdS spacetime by its infinite cover CAdS.}
Then, what are the key differences between mass and spin
from the viewpoint of coadjoint orbits?
A key feature of the spin is that when quantised,
it leads to a finite-dimensional Hilbert space.
When a coadjoint orbit is compact,  
we will find that only a finite number of modes
survive upon imposing quantisation conditions,
and hence the associated Hilbert space is finite-dimensional.
Let us illustrate the issue with an example.
Consider a coadjoint orbit of $\mathfrak{so}(3)$ spanned by $J_i$.
Up to $SO(3)$ rotation,
there is only one type of coadjoint vector $\phi=s\,\cJ^3$
(here, $\cJ^a$ are the dual basis of $\mathfrak{so}(3)^*$
with $\langle \cJ^a, J_b \rangle = \delta^a_b$),
and the corresponding orbit is $S^2$ with radius $s$.
This orbit is two-dimensional, so the system
has one mechanical degree of freedom. When $S^2$ is quantised,
only integral  $s$ is allowed,
and the space of phase space functions
is reduced from the space of functions on $S^2$
to the space of spin-$s$ spherical harmonics on $S^2$.
Hence, the dimension of the Hilbert space
is reduced from $\infty$ to $2s+1$,
and the number of degrees of freedom
--- mathematically speaking twice
the Gelfand--Kirillov dimension --- is reduced from 2 to 0.
This reduction is a generic feature of compact coadjoint orbits
as they are associated with finite dimensional representations.
When a coadjoint orbit is non-compact,
a similar reduction of the number of modes 
may take place due to the presence of a compact subspace.

Let us comment here that the spinning particle action
with bosonic variables should not be confused
with the model of relativistic spherical top 
(see e.g. \cite{Hanson:1974qy} for the classical account
and also \cite{Steinhoff:2015ksa,Kim:2021rda}
and reference therein for recent developments).
For example, the spin degrees of freedom
of the four-dimensional massive spinning particle
(in the sense of the current paper)
are the coordinates of $S^2$, a $SO(3)$ coadjoint orbit,
whereas the spin degrees of freedom
of a spherical top are the coordinates
of the cotangent bundle $T^*SO(3)$.
The quantisation of the latter 
gives the infinite direct sum
of the tensor product of two spin $s$ 
representations, without any projection.
See e.g. \cite{Cho:1994gs}
for the description of a spinning particle
inspired by the spherical top model.
We postpone the relevant discussions
to the sequel paper
where we cover the issues of quantisation.

%*************************************%
\subsubsection*{Geometric quantisation}
%*************************************%

Let us conclude this section by pointing out
that the quantisation condition 
\eqref{S dif} also appears
in the context of geometric quantisation, where it is
known as the prequantisation condition (see e.g. 
\cite{Simms:1976dhp, Bates:1997kc, Kirillov2004, Moshayedi:2020spz, Wernli:2023pib}). In this approach
to quantisation, one aims at defining, from a symplectic
manifold $(\cM,\omega)$, a Hilbert space $\Hilbert$
and a quantisation map $\cQ: \Functions(\cM)\to\Op(\Hilbert)$
from functions on $\cM$ to linear operators on the Hilbert
space. This map should verify a few conditions,
the most constraining ones being that it defines
a morphism of Lie algebra between $\Functions(\cM)$
endowed with its Poisson bracket to $\Op(\Hilbert)$
endowed with the commutator, i.e.
\begin{equation}
    [\cQ(f), \cQ(g)] = -i\hbar\,\cQ(\{f,g\})\,,
    \qquad 
    \forall f,g \in \Functions(\cM)\,,
\end{equation}
which is usually referred to as the Dirac condition.
In order for such a map to be well-defined globally,
on top of obeying all conditions including Dirac's,
one is lead to introducing a linear connection $\nabla$
on a line bundle over $\cM$ (that is, a vector bundle
whose fibers are isomorphic to $\C$) whose curvature
is proportional to the symplectic form 
$\omega$.
The existence of a line bundle equipped with such
a connection requires that 
\begin{equation}
    \oint_\Sigma \omega \in 2\pi\hbar\,\mathbb Z\,,
\end{equation}
for any closed $2$-dimensional manifold of $\Sigma\subset\cM$.
In our case, the linear connection is simply the pullback
of the Maurer--Cartan form of $G$ on the coadjoint orbit
$\cO^G_\phi$, evaluated on $\phi$, which we have seen
is subject to the above condition, see \eqref{S dif}.
For more details, see e.g. \cite[Sec. 3]{Wernli:2023pib}.

%***************************************%
\subsection{Examples: $\mathfrak{so}(3)$,
$\mathfrak{so}(2,1)$, $\mathfrak{iso}(2)$
and $\mathfrak{iso}(1,1)$}
%***************************************%

For concrete examples,
let us consider  the coadjoint orbits of three-dimensional Lie 
groups 
$SO(3)$, $SO(2,1)$, $ISO(2)$, $ISO(1,1)$
and their simply connected counterparts
as well as their double covers:
note the isomorphisms
$\widetilde{SO}(3)\cong SU(2)$ and
$\widetilde{SO}^+(2,1)\cong SU(1,1)\cong SL(2,\mathbb R)\cong Sp(2,\mathbb R)$\,.
The example of $SL(2,\mathbb R)$ 
coadjoint orbits has been treated in
numerous papers, e.g. \cite{Witten:1987ty, Dzhordzhadze:1994np, Ashok:2022thd}.

Let us fix the convention first.
The Lie algebras  $\so(3)$ and $\so(2,1)$
are generated by $J_a$ ($a=1,2,3$) obeying
\begin{equation}
  [J_a, J_b] = \epsilon_{ab}{}^c\, J_c\,,
  \label{JJ com}
\end{equation}
where the Levi--Civita tensor $\epsilon_{abc}$ is defined with $\e_{123}=1$\,.
The Latin indices are raised and lowered with
the Euclidean metric for $\so(3)$, and with the Minkowski metric
$\eta={\rm diag}(-1,-1,1)$ for $\so(2,1)$\,.

The Lie algebras  $\mathfrak{iso}(2)$ and $\mathfrak{iso}(1,1)$
are generated by $P_a$ ($a=1,2$) and $J$ obeying
\begin{equation}
  [P_a, P_b] = 0\,,
  \qquad 
  [J,P_a]=\epsilon_{a}{}^b\,P_b\,,
\end{equation}
where the Levi--Civita tensor $\epsilon_{ab}$ is defined with $\e_{12}=1$\,.
The indices are raised and lowered with
the Euclidean metric for $\mathfrak{iso}(2)$, and with the Minkowski metric
$\eta={\rm diag}(1,-1)$ for $\mathfrak{iso}(1,1)$\,.
The Lie algebra $\mathfrak{iso}(2)$ can also be obtained from
$\so(3)$ or $\so(2,1)$ by contracting the $J_3$ generator,
whereas $\mathfrak{iso}(1,1)$ can be obtained
from $\so(2,1)$ by contracting the $J_1$ generator.

%*************************%
\subsubsection*{Geometries}
%*************************%

An arbitrary element in  $\so(3)^*$ or $\so(1,2)^*$
can be written as
\be
	\phi = j_a\,\cJ^a\,,
\ee
where $\cJ^a$ are the dual basis satisfying
$\la \cJ^a\,,J_b\ra=\delta^a_b$.

\begin{figure}[!ht]
  \centering
  \begin{tikzpicture}
  \shade[ball color = RoyalBlue!30] (0,0) circle (1.8cm);
     \shade[ball color = OliveGreen!30] (0,0) circle (1.3cm);
       \shade[ball color = BrickRed!30] (0,0) circle (0.8cm);
    \draw[thick,->] (0,0) -- (3,0) node[anchor=west] {$\cJ^1$};
    \draw[thick,->] (0,0) -- (0,3) node[anchor=west] {$\cJ^2$};
    \draw[thick,->] (0,0) -- (-2,-1.25) node[anchor=south east] {$\cJ^3$};
    \draw (0,0) circle (1.8cm);
    \draw (-1.8,0) arc (180:360:1.8 and 0.6);
    \draw[dashed] (1.8,0) arc (0:180:1.8 and 0.6);
    \fill[fill=black] (0,0) circle (1pt);
    \fill[fill=black] (0,0) circle (1pt);
    \fill[fill=black] (0,0) circle (1pt);
  \end{tikzpicture}
  \caption{Examples of coadjoint orbits of $SO(3)$,
  which are simply two-spheres of different radii.}
  \label{fig:orbits_SO(3)}
\end{figure}
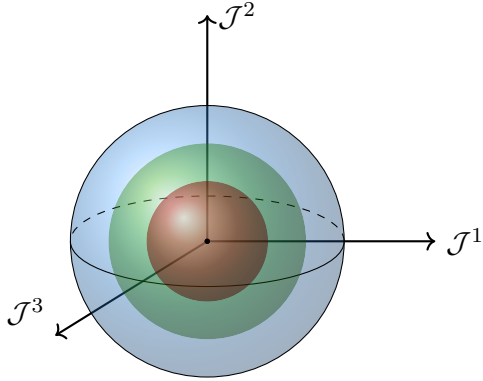

For $\so(3)^*$, any coadjoint vector $\phi$
can be rotated to the form, 
\be
	\phi=\sqrt{j^2}\,\cJ^3\,,
\ee
and it has a stabiliser $SO(2)$ generated by $J_3$\,.
The above is the representative of the coadjoint orbit
$\cO^{SO(3)}_\phi$ which is a two-sphere
$S^2\cong SO(3)/SO(2)$ with radius $\sqrt{j^2}$. 
The coadjoint space $\so(3)^*\cong \R^3$
is foliated by a continuum of spherical orbits
of different radii (see Figure \ref{fig:orbits_SO(3)}).
The stabilisers of each orbit are all
$\mathfrak{g}_\phi ={\rm span}\{J_3\} \simeq \uu(1)$.
Since the quotient algebra
$\mathfrak{g}^{\rm Ab}_\phi=\mathfrak{g}_\phi = \uu(1)$
is compact, the orbit is elliptic.

For $\so(2,1)^*$, depending on the value
of $j^2=j_a\,j^a$,\footnote{Note that here,
we are using the convention that the direction $3$
is the time-like one (usually denoted by $0$).}
a coadjoint vector $\phi$ can be rotated or boosted
to one of the three representatives:
\be
	\phi=\left\{ 
	\begin{array}{ccc}
	\pm\sqrt{j^2}\,\cJ^3\,, \qquad \qquad &[j^2>0,\ \pm j_3>0]\\
	\pm(\cJ^1+\cJ^3), \qquad \qquad &[j^2=0,\  \pm j_3>0]\\
	\sqrt{-j^2}\,\cJ^1\,, \qquad \qquad &[j^2<0] \\
	0\,, \qquad \qquad &[j_a=0]
	\end{array}
	\right..
\ee
The coadjoint vectors with $\pm$ signs
belong to two distinct coadjoint orbits
of $SO^+(2,1)$.
These two orbits form a single disconnected coadjoint orbit 
of $SO(2,1)$ as
they are mapped to each other by the ``time reversal'' transformation, forming the $\mathbb Z_2$ finite subgroup.
The coadjoint orbits with the above representative
vectors are all
given by two-dimensional 
quadratic surfaces,
\ba
	H^2(a)\eq \{(x,y,z)\in \R^3\,|\,-x^2-y^2+z^2=a\}\,.
    \label{2d surface}
\ea
The surface with $a<0$
is the one-sheeted hyperbolic
hyperboloid,
and the surface with $a>0$
is the two-sheeted elliptic hyperboloid.
The special case $a=0$
corresponds to the two-dimensional cone: $H^2(0)=C^2$.
When $a\ge 0$,
namely the two-sheeted hyperboloids
and the cone,
contain two disconnected parts:
the upper/lower hyperboloids
$H^2_\pm(a>0)=\{(x,y,z)\in H^2(a)\,|\,\pm z>0\}$
and the upper/lower cones
$C^2_\pm=\{(x,y,z)\in C^2\,|\,\pm z>0\}.$

The coadjoint orbit represented by the first $\phi$ has 
the stabiliser $SO(2)$ generated by $J_3$,
and it is an elliptic orbit since 
$\mathfrak{g}^{\rm Ab}_\phi={\rm span}\{J_3\}\simeq
\mathfrak{u}(1)$.
It has the geometry of the two-dimensional elliptic hyperboloid $H^2_\pm(j^2)\cong SO^+(2,1)/SO(2)$.
The second case has 
the stabiliser $\mathbb R$ generated by $J_1-J_3$,
and the orbit is nilpotent
since $\la\phi,J_1-J_3\ra=0$\,.
Its geometry is a two dimensional cone $C^2_\pm\cong SO^+(2,1)/\mathbb{R}$.
The third case has the stabiliser $SO^+(1,1)$ generated by $J_1$,
and the orbit is hyperbolic since 
$\mathfrak{g}^{\rm Ab}_\phi={\rm span}\{J_1\}\simeq
\mathbb R$.
The geometry is a two-dimensional hyperbolic hyperboloid $H^2(-j^2)\cong SO^+(2,1)/SO^+(1,1)$.
The last case has the entire $SO^+(2,1)$ as its stabiliser and 
the orbit is the single point at the origin.
The coadjoint space $\so(2,1)^*\cong \R^3$ is foliated by a continuum of  
hyperboloid-type orbits 
$H^2_\pm(j^2)$ and $H^2(-j^2)$
with different $j$'s,
 two conical orbits
$C_\pm^2$ and the origin (see Figure \ref{fig:orbits_SL(2,R)}).

Since the Lie algebras $\so(3)$ and $\so(2,1)$ are semisimple,
their coadjoint spaces can be identified with the adjoint spaces
through the Killing forms. This allows us to view
the coadjoint actions of Lie group elements
as mere rotations or boosts, that is,
the adjoint actions of $SO(3)$ or $SO^+(2,1)$.
In other words, we may as well study their adjoint orbits.
The adjoint representation of $J_3$ for both $SO(3)$
and $SO^+(2,1)$ is
\be
	{\rm ad}_{J_3}=\begin{pmatrix}
	0 & 1 & 0 \\
	-1 & 0 & 0 \\
	0 & 0 & 0
	\end{pmatrix}\,,
	\label{J3 orbit}
\ee
and it has the eigenvalues $+i, -i, 0$\,,
confirming that the orbit is elliptic.
On the other hand, the adjoint representations
of $J_1$ and $J_1+ J_3$ for $SO^+(2,1)$ are 
\be
	{\rm ad}_{J_1}=\begin{pmatrix}
	0 & 0 & 0 \\
	0 & 0 & 1 \\
	0 & 1 & 0
	\end{pmatrix}\,,
	\qquad 
	{\rm ad}_{J_1+ J_3}=\begin{pmatrix}
	0 &  1 & 0 \\
	- 1 & 0 & 1 \\
	0 & 1 & 0
	\end{pmatrix}\,,
\ee
and they have the eigenvalues $+1, -1, 0$
and $0,0,0$ respectively, confirming that
the corresponding orbits are hyperbolic and nilpotent,
respectively.
\begin{figure}[!ht]
  \centering
    \begin{tikzpicture}
    %% Ellipse %% 
     \draw[fill=OliveGreen!20] (0,4.45) ellipse (2.45 and 0.5);
    \draw[fill=BrickRed!20] (0,3.25) ellipse (3.2 and 0.5);
     \draw[fill=MidnightBlue!20] (0,2.15) ellipse (4 and 0.5);
    %%% Two-sheeted hyperboloid %%%
    \draw[fill=OliveGreen!30]
        plot[smooth] coordinates{(-2.45, 4.4) (0,2.9) (2.45,4.4)}
        arc (355:175:2.45 and 0.5);
     \draw[fill=OliveGreen!30]
        plot[smooth] coordinates{(-2.45, -4.4) (0,-2.9) (2.45,-4.4)}
        arc (360:180:2.45 and 0.5);
    %%% Cone %%%
    \draw[fill=BrickRed!30]
        (-3.2,3.2) -- (3.2,-3.2) arc (175:365:-3.2 and 0.5)
        -- (-3.2,-3.2) -- (3.2,3.2)
        arc (355:175:3.2 and 0.5);     
    %\draw[fill=OliveGreen!20] (0,-4.5) ellipse (2.5 and 0.5);
      %%% One-sheeted hyperboloid %%%
    \draw[fill=MidnightBlue!30]
        plot[smooth] coordinates {(-4,2.1) (-2.5,0) (-4,-2.1)}
        -- (-4,-2.1) arc (365:185:-4 and 0.5) --
        plot[smooth] coordinates {(4,-2.1) (2.5,0) (4,2.1)}
        arc (355:175:4 and 0.5);
    %%% Origin %%%
     \draw[color=red] (0,0) node {$\pmb\times$};
    %%% Axis %%%
    \draw[thick,->] (-4,0) -- (4,0) node[anchor=west] {$\mathcal J^1$};
    \draw[thick,->] (0,-5.5) -- (0,5.5) node[anchor=west] {$\mathcal J^3$};
    \draw[thick,->] (3.75,1.5) -- (-3.75,-1.5) node[anchor=south east] {$\mathcal J^2$};
    \end{tikzpicture}
    \caption{Example of coadjoint orbits of $SO(2,1)$:
    in \textcolor{MidnightBlue!80}{blue} the one-sheeted hyperboloid,
    in \textcolor{BrickRed!80}{red} the two disconnected upper and
    lower cones and in \textcolor{OliveGreen!80}{green} two one-sheeted
    hyperboloids.}
    \label{fig:orbits_SL(2,R)}
\end{figure}
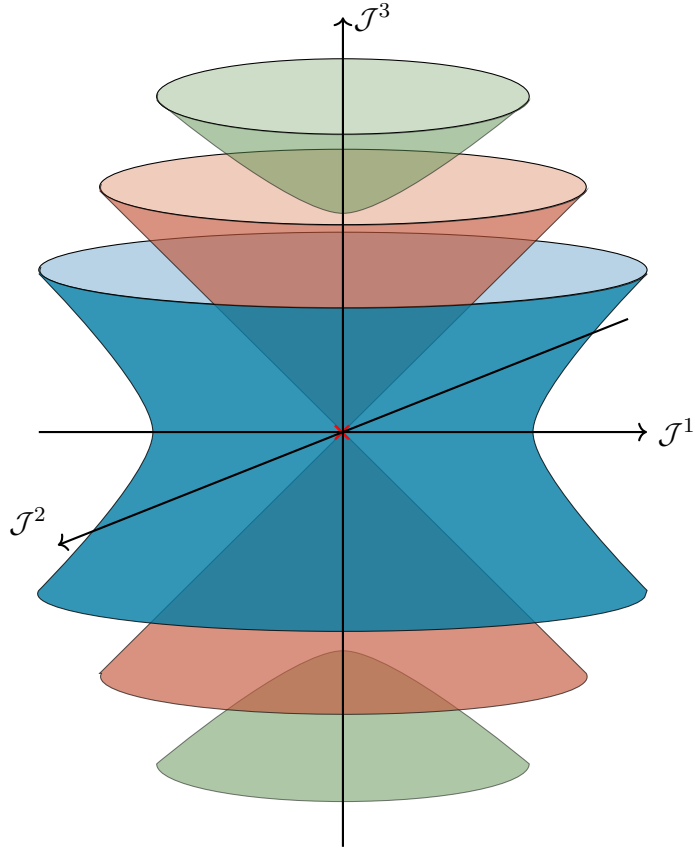

Let us move to the non-semisimple cases $ISO(2)$ and $ISO(1,1)$.
An arbitrary element of  $\mathfrak{iso}(2)^*$ or $\mathfrak{iso}(1,1)^*$ can be written as
\be
	\phi=p_a\,\cP^a+j\,\cJ\,,
\ee
where $\cP^a$ and $\cJ$ are the dual basis satisfying $\la \cP^a,P_b\ra=\delta^a_b$, $\la \cJ,J\ra=1$ and
$\la \cP^a,J\ra=0=\la \cJ,P_a\ra$\,.
The coadjoint action of $ISO(2)$ or $ISO^+(1,1)$ on $\phi$ is
\be
	{\rm Ad}^*_{e^{x^aP_a}\,\L}(p_a\,\cP^a+j\,\cJ)
	=p_a\,\L^a{}_b\,\cP^b+
	(j+\epsilon^{bc}\,p_a\,\L^a{}_b\,x_c)\,\cJ\,,
\ee
where $\L$ is the rotation or boost element in $SO(2)$ or $SO^+(1,1)$ generated by $J$.

\begin{figure}[!ht]
  \centering\small
  \begin{tikzpicture}

\draw[fill=RoyalBlue!20] (0,2.75) ellipse (2.5 and 0.5);
    %%% Inner cylinder %%%
    \draw[fill=OliveGreen!30]
        (-2,3) -- (-2,-2) arc (180:360:2 and 0.4)
        -- (2,-2) -- (2,3) arc (360:180:2 and 0.4);
    \draw[fill=OliveGreen!20] (0,3) ellipse (2 and 0.4);
    %%% Outer cylinder %%%
    \draw[fill=RoyalBlue!30]
        (-2.5,2.75) -- (-2.5,-1.5) arc (180:360:2.5 and 0.5)
        -- (2.5,-1.5) -- (2.5,2.75) arc (360:180:2.5 and 0.5);
    
    %%% Axis %%%
    \draw[thick,->] (0,0) -- (4,0) node[anchor=west] {$\cP_1$};
    \draw[thick,->] (0,-3) -- (0,4) node[anchor=west] {$\cJ$};
    \draw[thick,->] (0,0) -- (-3,-2) node[anchor=south east] {$\cP_2$};
    %%% Isolated points %%%
    \draw[color=BrickRed] (0,-0.2) node {$\pmb\times$};
    \draw[color=BrickRed] (0,1.5) node {$\pmb\times$};
    \draw[color=BrickRed] (0,1) node {$\pmb\times$};
    \draw[color=BrickRed] (0,-1) node {$\pmb\times$};

  \end{tikzpicture}
  \caption{Examples of coadjoint orbits of $ISO(2)$:
  two cylinders centered around the $\cJ$ axis of different
  radii in {\color{RoyalBlue} blue}
  and {\color{OliveGreen} green}, corresponding to orbits
  with $p^2\neq0$, and four isolated points on this same axis
  {\color{BrickRed} red} (randomly distributed),
  corresponding to orbits with $p^2=0$.}
  \label{fig:orbits_ISO(2)}
\end{figure}
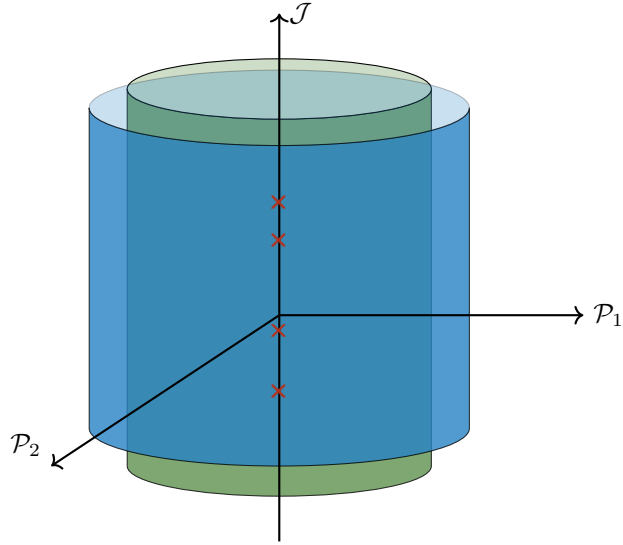

For the $\mathfrak{iso}(2)^*$ case, any coadjoint vector $\phi$ can be
transformed into
\be
	\phi=\left\{ 
	\begin{array}{ccc}
	\sqrt{p^2}\,\cP^1\,, \qquad \qquad &[p^2\neq 0]\\
	j\,\cJ, \qquad \qquad &[p^2=0]
	\end{array}
	\right.\,.
\ee
The first case has the stabiliser $\R$
generated by $P_1$, and the orbit is a two-dimensional 
cylinder $ISO(2)/\R \cong S^1 \times \R$
of radius $\sqrt{p^2}$.
The stabiliser of the second case is 
the entire Euclidean group $ISO(2)$, and
the orbit is a single point located on the $j$-axis.
Again the coadjoint space $\iso(2)^*\cong \R^3$
is foliated by a continuum of cylindrical orbits
and  a continuum of points on the $j$-axis,
see Figure \ref{fig:orbits_ISO(2)}.

For $ISO^+(1,1)$ case, any coadjoint vector $\phi$
can be transformed into
\be
	\phi=\left\{ 
	\begin{array}{ccc}
	\pm\sqrt{p^2}\,\cP^1\,, \qquad \qquad &[p^2>0,\ \pm p_1>0]\\
	\pm\sqrt{-p^2}\,\cP^2\,, \qquad \qquad &[p^2<0,\ \pm p_2>0]\\
	\pm\cP^1\pm'\cP^2, \qquad \qquad &[p^2=0,\  \pm p_1>0\,,\ \pm' p_2>0]\\
	j\,\cJ, \qquad \qquad & [p_a=0]
	\end{array}
	\right.,
\ee
where $\pm'$ means an independent sign possibilities with respect to $\pm$.
The stabilisers of the first three classes of the coadjoint orbits
are all $\mathbb R$, generated respectively by
$P_1, P_2$ and $\pm P_1\mp'P_2$.
The corresponding orbits are hyperbolic cylinders and conical cylinder.
The last case has the  entire $ISO^+(1,1)$ as its stabiliser,
and the orbit is a single point on the $j$-axis.
The coadjoint space $\mathfrak{iso}(1,1)^*\cong \mathbb R^3$ is 
foliated by a continuum of hyperbolic cylinder shaped orbits and one conical cylinder
(which is subdivided by four pieces of two half-planes) and a continuum of points on $j$-axis, see Figure \ref{fig:orbits_ISO(1,1)}.

\begin{figure}[!ht]
  \centering\small
  \begin{tikzpicture}

%%% Back cylinder %%%
    \draw[fill=RoyalBlue!20]
    plot[smooth] coordinates {(-3,3.5) (0,2.7) (3,3.5)} -- (3,-1)
    plot[smooth] coordinates {(3,-0.5) (0,-1.3) (-3,-0.5)} -- (-3,3.5);

    %%% Conical cylinder (behind) %%%
    \draw[fill=OliveGreen!20] (-3,3) -- (0,2) -- (0,-2) -- (-3,-1);
    \draw[fill=OliveGreen!20] (3,3) -- (0,2) -- (0,-2) -- (3,-1);
    
     %%% Left hyperbolic cylinder %%%
    \draw[fill=RoyalBlue!30] (-3,2) rectangle (-1.5,-2);
    \draw[fill=RoyalBlue!30] plot[smooth] coordinates {(-3,-1.4) (-1.5,-2) (-3,-2.6)};
    \draw[fill=RoyalBlue!30] plot [smooth] coordinates {(-3,2.6) (-1.5,2) (-3,1.4)};
    \draw[White, thick, opacity=1] (-3,3) -- (-3,-3);

    %%% Right hyperbolic cylinder %%%
    \draw[fill=RoyalBlue!30] (3,2) rectangle (1.5,-2);
    \draw[fill=RoyalBlue!30] plot[smooth] coordinates {(3,-1.4) (1.5,-2) (3,-2.6)};
    \draw[fill=RoyalBlue!30] plot [smooth] coordinates {(3,2.6) (1.5,2) (3,1.4)};
    
     %%% Conical cylinder (forefront) %%%
    \draw[fill=OliveGreen!30] (0,2) -- (-3,1) -- (-3,-3) -- (0,-2);
    \draw[fill=OliveGreen!30] (0,2) -- (3,1) -- (3,-3) -- (0,-2);

    %%% Front cylinder %%%
    \draw[fill=RoyalBlue!40]
    plot[smooth] coordinates {(-3,0.5) (0,1.3) (3,0.5)} -- (3,-3.5)
    plot[smooth] coordinates {(3,-3.5) (0,-2.7) (-3,-3.5)} -- (-3,0.5);

    \draw[White, thick, opacity=1] (-3,4) -- (-3,-4);
    \draw[White, thick, opacity=1] (3,4) -- (3,-4);
    
    %%% Axis %%%
    \draw[thick,->] (0,0) -- (4,0) node[anchor=west] {$\cP_1$};
    \draw[thick,->] (0,-4) -- (0,4) node[anchor=west] {$\cJ$};
    \draw[thick,->] (0,0) -- (-3.5,-1.5) node[anchor=south east] {$\cP_2$};
    
    %%% Isolated points %%%
    \draw[color=BrickRed] (0,-0.25) node {$\pmb\times$};
    \draw[color=BrickRed] (0,1.6) node {$\pmb\times$};
    \draw[color=BrickRed] (0,1) node {$\pmb\times$};
    \draw[color=BrickRed] (0,-1) node {$\pmb\times$};
    
  \end{tikzpicture}
  \caption{Examples of coadjoint orbits of $ISO(1,1)$:
  in {\color{RoyalBlue} blue} (a two-sheet) hyperbolic cylinder,
  in {\color{OliveGreen} green} a conical cylinder,
  and in {\color{BrickRed} red} four isolated points
  (randomly distributed).}
  \label{fig:orbits_ISO(1,1)}
\end{figure}
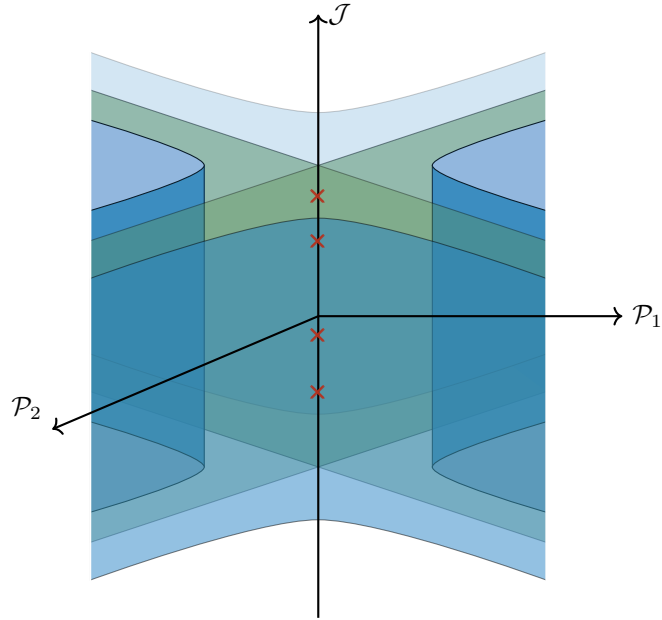

Recall that $ISO(2)$ and $ISO^+(1,1)$
can be obtained by a {I}n\"on\"u--Wigner contraction
of $SO(3)$ and $SO^+(2,1)$, respectively.
In fact, $ISO(2)$ can be obtained from either 
$SO(3)$ or $SO^+(2,1)$, and in the latter case, 
the generator $J_3$ needs to be contracted.
The spherical cylinder of $ISO(2)$ 
are the contractions of a sphere of $SO(3)$
as well as a one-sheeted hyperboloid of $ISO^+(1,1)$.
The hyperbolic cylinder and the conical cylinder
of $ISO^+(1,1)$ can be obtained
from a one- or two-sheeted hyperboloid 
and cone of $SO^+(2,1)$, respectively.
The isolated points on the $\cJ$ axis
can be also obtained by assigning a suitable scaling 
of $\phi$ under the contraction.
See \cite{Enayati:2022hed,Enayati:2023lld}
for more discussions about the contraction
of (A)dS orbits to Poincar\'e ones.

So far we have not considered the double cover
of $SO(3)$ or $SO^+(2,1)$ because they define
the same hypersurface in $\mathfrak{g}^*$\,:
they have the same  coadjoint representations.
As far as the geometries are concerned,
there is no difference. Below, we shall see that
the difference arises when considering
their symplectic potentials.

%************************************%
\subsubsection*{Symplectic structures}
%+***********************************%

Let us first have a closer look at the elliptic
coadjoint orbits of $SO(3)$ and $SO^+(2,1)$
and their double cover $SU(2)$ and $SU(1,1)$.
These coadjoint orbits are all represented by
the coadjoint vector $\phi=\sqrt{j^2}\,\cJ^3$
with the stabiliser $G_\phi=\{ \, \exp(\g\,J_3) \,\}$,
and can be described respectively by 
a sphere $S^2$ or an elliptic hyperboloid $H^2_+$.
To proceed the analysis, let us parameterise
a group element $g$ with the Euler angles $\x,\u,\z$ as
\be
	g(\x,\u,\z)=\exp(\x\,J_3)\,\exp(\u\,J_1)\,\exp(\z\,J_3)\,,
	\label{euler}
\ee
where the range of parameters $\x,\u,\z$
depends on the cases.
First, $\x$ always belongs to $[0,2\pi)$,
whereas $\u$ belongs to $[0,2\pi)$ for the compact case
and $[0,\infty)$ for the non-compact case.
Lastly, the range of $\z$ depends on 
whether the Lie groups associated with $\so(3)^*$
and $\so(2,1)^*$ are $SO(3)$ and $SO^+(2,1)$
or $SU(2)$ and $SU(1,1)$:
$\z\in [0,2\pi)$ in the former cases
while $\z\in [0,4\pi)$ for the latter cases.
With the periodic conditions on $\xi$ and $\zeta$,
this coordinate system is well-defined everywhere
except for the region near the north pole
$\mathsf N$ ($\u=0$) for both $\so(3)^*$ and $\so(2,1)^*$
and the south pole $\mathsf S$ ($\u=\pi$) for only $\so(3)^*$. 

A simple computation gives 
\be
	\la \phi,\Theta\ra
    = \la \sqrt{j^2}\,\cJ^3, g^{-1}{\rm d}_Gg\ra
    = \sqrt{j^2}\times \left\{
	  \begin{array}{cc}
	  \cos\u\,  {\rm d}_G \x+{\rm d}_G \z\,,
	  \qquad\qquad & [\,\so(3)\,]\\
	  \cosh\u\,  {\rm d}_G \x+{\rm d}_G \z\,,
	\qquad\qquad & [\,\so(2,1)\,]
	\end{array}
	\right..
\ee
On the coordinate chart \eqref{euler},
we choose different local sections $\sigma_i(\x,\u)$
which determine $\z_i$ as functions $\z_i(\x,\u)$
of the coadjoint orbit coordinates $\x,\u$\,:
\be
	\sigma_i(\x,\u)
    = \exp(\x\,J_3)\,\exp(\u\,J_1)\,\exp(\z_i(\x,\u)\,J_3)\,.
\ee
Then, by pulling back $\la \phi,\Theta\ra $ with $\sigma_i$,
we obtain the symplectic potential  $\theta_i$ as
\be
	\theta_i(\x,\u)= \sqrt{j^2}\times\left\{
	  \begin{array}{cc}
	  \cos\u\,  {\rm d}_\cO \x+{\rm d}_\cO\z_i(\x,\u)\,,
	  \qquad\qquad & [\,\so(3)\,]\\
	  \cosh\u\,  {\rm d}_\cO \x+{\rm d}_\cO \z_i(\x,\u)\,,
	   \qquad\qquad & [\,\so(2,1)\,]
	\end{array}\right..
\ee
Here, $\x$ is the azimuthal angle, and $\u$ is the inclination angle or rapidity 
of the coadjoint orbit $S^2$ or $H_{+}^2$, respectively.
The difference between two symplectic potentials $\theta_i$ is  
\be
	\theta_i=\theta_j+\sqrt{j^2}\,{\rm d}_\cO\z_{ij}\,,
\ee
where $\z_{ij}=\z_i-\z_j$ and $\t_{ij}=\exp(\z_{ij}\,J_3)$\,.
The transformation of the worldline action under 
the change of section by $\zeta_{ij}$ is
\be
	S_j=S_i+{\sqrt{j^2}}\int_\g {\rm d}_\cO\zeta_{ij}
\ee
and 
\be
	\int_\g {\rm d}_\cO\zeta_{ij}=
	\oint_{\G_{ij}}{\rm d}_G\zeta
	\in \left\{\begin{array}{cc}
	2\pi\,\mathbb Z & \qquad [SO(3)\ {\rm or}\ SO^+(2,1)]\\
	4\pi\,\mathbb Z & \qquad [SU(2)\ {\rm or}\ SU(1,1)]
	\end{array}
	\right.,
\ee
where $\mathbb Z$ corresponds to the set of possible numbers that the trajectory $\G_{ij}$ winds the cycle corresponding to the $\zeta$ coordinate.
The invariance of $e^{i\,S/\hbar}$ under this transformation leads to the quantisation of the orbit radius: 
\be
	\sqrt{j^2}=\hbar\,\ell\,,
	\qquad 
	\ell \in \left\{\begin{array}{cc}
	\mathbb N & \qquad [SO(3)\ {\rm or}\ SO^+(2,1)]\\
	\mathbb N/2 & \qquad [SU(2)\ {\rm or}\ SU(1,1)]
	\end{array}
	\right..
	\label{quant}
\ee
In the case of the $\so(3)^*$, we can cover the entire orbit $S^2$ with two charts $U_N=S^2-\{\mathsf S\}$ and $U_S=S^2-\{\mathsf N\}$.
By choosing the sections as
\be
	\zeta_N(\xi,\u)=-\xi\,, 
	\qquad \zeta_S(\xi,\u)=\xi\,,
	\qquad 
	\zeta_{NS}(\xi,\u)=-2\xi\,,
	\label{NS section}
\ee
the symplectic potential in Euler angles is well-defined in each chart, that is,
near the north pole $\mathsf N$ and the south pole $\mathsf S$.
For $SO(3)$ where $\zeta_{NS}\in[0,2\pi)$, 
the transition map $\xi\in S^1_{2\pi}\mapsto \zeta_{NS}\in S^1_{2\pi}$ 
winds twice.
For $SU(2)$ where $\zeta_{NS}\in[0,4\pi)$, 
the transition map $\xi\in S^1_{2\pi}\mapsto \zeta_{NS}\in S^1_{4\pi}$ 
winds once and this fiber bundle structure
corresponds to the Hopf fibration of $SU(2) \cong S^3$
over the two-sphere $S^2$.
In both cases, the difference of the worldline action under the change of the sections \eqref{NS section} is
\be
	\int_\g \sqrt{j^2}\,{\rm d}_\cO\zeta_{NS}
	=\int_0^{2\pi}\sqrt{j^2}\,(-2\,{\rm d}_\cO\xi)=-4\pi\,\sqrt{j^2}\,,
\ee
and this can be rewritten as the integral of the symplectic two-form over the orbit $S^2$\,:
\be
	\int_\g \sqrt{j^2}\,{\rm d}_\cO\zeta_{NS}
	=\int_\g (\theta_N-\theta_S)
	=\int_{\S_N\cup \bar\S_S=S^2}\o\,.
\ee
The above quantity should be $2\pi\,n$ for an integer $n$ in order for the path integral to be invariant under such a transformation,
and this is the prequantisation condition
in the context of geometric quantisation.
Note that the prequantisation condition is weaker than the condition of
the invariance of the action under a change of section: in the latter case we find \eqref{quant}
whereas the prequantisation condition does not give any restriction on the $H_{+}^2$ orbit of $\so(2,1)^*$
and it allows the half-integral radius for $SO(3)$ case.
See \cite{Witten:1987ty}
for related discussions.

Next, let us consider the nilpotent coadjoint orbit satisfying $j^2=0$ and $j_3>0$.
Any such vector can be rotated to $\cJ^1+\cJ^3$.
Again to proceed the analysis, we take the Iwasawa decomposition,
\be
	g(\x,\u,\z)=\exp(\x\,J_3)\,\exp(\u\,J_2)\,\exp(\z\,(J_1+ J_3))\,,
\ee
which is well adapted to the nilpotent orbit. 
Here, the ranges of the parameters  are $\xi\in [0,2\pi)$ and $\u,\z\in (-\infty,+\infty)$.
The one-form 
$\la \phi,\Theta\ra$ is
\be
	\la \phi,\Theta\ra= \la \cJ^1+\cJ^3, g^{-1}{\rm d}_Gg\ra =
	e^{\u}\,{\rm d}\xi+2\,{\rm d}_G \z\,.
\ee
Since $\zeta$ belongs to $\mathbb R$,
the difference of the action always vanishes, $\oint_{\G_{ij}}{\rm d}_G\zeta=0$\,.

Lastly, let us consider the hyperbolic coadjoint orbit $H^2(j^2)$ of $\so(2,1)^*$ given by $j^2<0$. 
Any vector in it  can be rotated to $\sqrt{-j^2}\,\cJ^1$,
and a convenient decomposition is
\be
	g(\x,\u,\z)=\exp(\x\,J_1)\,\exp(\u\,J_3)\,\exp(\z\,J_1)\,,
\ee
where both $\x$ and $\z$ belong to $(-\infty,\infty)$ and $\u$
belongs to $[0,2\pi)$ for $SO^+(2,1)$ and $[0,4\pi)$ for $SU(1,1)$\,. 
The one-form $\la\phi,\Theta\ra$ is
\be
	\la\phi,\Theta\ra= \la \sqrt{-j^2}\,\cJ^1, g^{-1}{\rm d}_Gg\ra =
	\sqrt{-j^2}\left(\cos\u\,{\rm d}_G\xi+{\rm d}_G\z\right).
\ee
Again $\z$ belongs to $\mathbb R$,
and the difference of the action always vanishes.  
Therefore, no condition is imposed on $\sqrt{-j^2}$.

About the cylindrical orbits of $\mathfrak{iso}(2)^*$
and the hyperbolic cylinder orbits of $\mathfrak{iso}(1,1)^*$, we use the decomposition,
\be
	g(\xi,\u,\zeta)=\exp(\xi\,P_2)\,\exp(\u\,J)\,\exp(\zeta\,P_1)\,,
\ee
for a Lie group element, where $\xi$ and $\zeta$ belongs to $\mathbb R =(-\infty,+\infty)$ and $\u$ belongs to $[0,2\pi)$
for $ISO(2)$ and $\mathbb R$ for $ISO(1,1)$.
The one-form $\la\phi,\Theta\ra$ is
\be
	\la\phi,\Theta\ra =\la \sqrt{p^2}\,\cP^1, g^{-1} {\rm d}_G g \ra
	=\sqrt{p^2}\times
	\left\{
	 \begin{array}{cc}
	  \sin\u\,  {\rm d}_G \x+{\rm d}_G\z\,,
	  \qquad\qquad & [\,\mathfrak{iso}(2)\,]\\
	  \sinh\u\,  {\rm d}_G\x+{\rm d}_G \z\,,
	   \qquad\qquad & [\,\mathfrak{iso}(1,1)\,]
	\end{array}\right..
\ee
The other $\mathfrak{iso}(1,1)^*$ orbits with $p^2<0$ are isomorphic to the ones with $p^2>0$.
For the $\mathfrak{iso}(1,1)^*$ orbit  which has the shape of a conical cylinder,
we use the decomposition,
\be
	g(\xi,\u,\zeta)=\exp\big(\xi\,(P_1+P_2)\big)\,\exp(\u\,J)\,\exp\big(\zeta\,(P_1-P_2)\big)\,.
\ee
The one-form  $\la \phi,\Theta\ra$ is
\be
	\la \phi,\Theta\ra =\la \cP^1+\cP^2, g^{-1} {\rm d}_G g \ra
	=e^\u\,{\rm d}_G\x+{\rm d}_G \zeta\,.
\ee
In all cases of $\mathfrak{iso}(2)^*$ and $\mathfrak{iso}(1,1)^*$,
 the coordinate $\z$ belongs to $\mathbb R$.
 Therefore, it has no contribution to the action under a change of section.

%***********************************%
\subsection{Phase space and dynamics}
\label{sec: phase space and dynamics}
%***********************************%

The coadjoint orbit is a symplectic space, so it can serve as a phase space of a mechanical system,
but it does not seem to provide a Hamiltonian at first glance.
Indeed, the examples that we have treated just above
did not show any Hamiltonian.
On the contrary, in the introduction, we showed briefly how a relativistic scalar particle action
can be obtained from a coadjoint orbit of Poincar\'e group.
The difference between the two cases is in different parameterizations of a Lie group element.
For concreteness, let us consider again the massive scalar orbit of Poincar\'e group
with $\phi=m\,\cP_0$\,.

First, let us consider the parameterization of a Lie group element given by the decomposition,
\be
	g=e^{y^a\,P_a}\,e^{v^a\,J_{0a}}\,e^{y^0\,P_0}\,R\,,
	\label{order 1}
\ee 
where $a=1,2, \ldots, d-1$ and $R \in SO(d-1)$\,.
When $d=2$, this choice reduces to the $\mathfrak{iso}(1,1)$ example we treated just above. 
Since all the stabiliser $G_\phi\cong \mathbb R\times SO(d-1)$ is present on the right side of $g$, it is well suited for the right quotient $ISO^+(1,d-1)/(\R\times SO(d-1))$. 
This choice leads to
\be
	\la \phi,\Theta\ra =p_a\,{\rm d}_Gy^a+m\,{\rm d}_G y^0\,,
\ee
where $p_a=m\,\frac{\sinh v}{v}\,v_a$.
Up to the boundary term ${\rm d}_G y^0$,
 we recover the canonical symplectic structure $p_a\,{\rm d}y^a$ but without any non-trivial Hamiltonian.
 The boundary term $m\,{\rm d}_G y^0$ might also be regarded as a constant Hamiltonian
 if we take $y^0$ as the proper time of the worldline. In any case, it is a static system.
 
 Instead, if we take the group element as 
 \be
 	g=e^{x^a\,P_a+x^0\,P_0}\, e^{v^a\,J_{0a}}\,R\,,
	\label{order 2}
\ee
we would find
\be
	\la \phi,\Theta\ra= p_a\,{\rm d}_Gx^a-\sqrt{m^2+p_a\,p^a}\,{\rm d}_Gx^0\,.
\ee
We can set $x^0$ as the proper time using a reparametrization of the worldline,
then we recover the familiar scalar particle Lagrangian with a non-trivial Hamiltonian.

Since different decompositions of $g$ correspond to different coordinate systems for $G$,
the two choices are related by a coordinate transformation,
\be
 	(x^0,x^a)=\left(\frac{\sqrt{m^2+p_a\,p^a}}{m}\,y^0, y^a+\frac{p^a\,y^0}m\right),
 \ee
 which can be easily obtained by reordering \eqref{order 2} into \eqref{order 1}.
This coordinate transformation --- which trivialize the particle dynamics ---
 is similar to the canonical transformation resulting in action-angle variables:
Hamiltonian in action-angle variables can be simply reabsorbed by
shifting the angle variable $\phi_i$ by the frequencies $w_i$: $\phi_i\to \phi_i-t\,w_i$.

As we could see from the above example,
the Hamiltonian action associated to a coadjoint orbit
always can  be written in the trivial form $p_i\,{\rm d}x^i$ without a Hamiltonian (up to a total derivative term),
at least locally (Darboux's theorem guarantees it).
Therefore, in order to interpret a coadjoint orbit action as
a relativistic spinning particle action, it is crucial to choose an appropriate set of coordinates.
And the appropriateness is the covariance of the system under the Lie group $G$.
This perspective resonates with the appropriate choice of a group decomposition in
nonlinear realisation where the distinction of broken symmetries and unbroken symmetries is important.
A good coordinate system may make a certain part of the symmetry manifest, but
it can never do so for the entire symmetry.
For the full manifest covariance, we need to involve additional variables
and make the system a constrained one.

%%%%%%%%%%%%%%%%%%%%%%%%%%%%%%%%%%%%%%%%
\section{Constrained Hamiltonian system}
\label{sec: constraint}
%%%%%%%%%%%%%%%%%%%%%%%%%%%%%%%%%%%%%%%%

Before moving to the reformulation of coadjoint orbit actions
as constrained systems,
let us review the standard formulation of constrained Hamiltonian systems
with an emphasis on its relation to coadjoint orbits.
We shall see in particular how a coadjoint orbit is related to second class constraints
whereas its stabiliser is related to first class constraints.

%********************************%
\subsection{Hamiltonian reduction}
%********************************%
When a Lie group $G$ acts on a symplectic space $\cM$
and the action is Hamiltonian, we can establish
a correspondence between coadjoint orbits in $\mathfrak{g}^*$
and a certain set of constrained surfaces 
(or the reduced phase spaces thereof) inside $\cM$.
For a better understanding of this perspective, 
let us review the relevant mathematical material. 
In the next subsection, we will recast the content of this subsection 
in terms of Hamiltonian mechanics.

A symplectic manifold $(\cM,\Omega)$ is equipped 
with the Poisson bracket,
\begin{equation}
	\{f,g\}=\Pi({\rm d}f,{\rm d}g)\,,
\end{equation}
where the Poisson bivector $\Pi$ is the inverse of the symplectic two-form $\Omega$\,:
in a coordinate system $\{y^\mu\}$,
the symplectic two-form
$\O=\O_{\m\n}(y)\,{\rm d}y^\m\wedge{\rm d}y^\n$ and 
Poisson bivector $\Pi=\Pi^{\m\n}(y)\,\frac{\partial}{\partial y^\m}\wedge\frac{\partial}{\partial y^\n}$
are related by $\O_{\m\n}(y)\,\Pi^{\n\r}(y)=\d^{\r}_{\m}$\,,
and we have $\{y^\m,y^\n\}=\Pi^{\m\n}(y)$\,.

Suppose that the vectors fields $V_a\in T\cM$ 
correspond to the generators $J_a$ of a Lie algebra $\mathfrak{g}$ 
satisfying $[J_a,J_b]=f_{ab}{}^c\,J_c$, in that they verify
the same relation
\begin{equation}
	[V_a,V_b]=f_{ab}{}^c\,V_c\,.
\end{equation}
If these vectors fields are Hamiltonian, that is to say, 
there exists a set of functions $\mu_a\in \Functions(\cM)$
obeying $i_{V_a}\Omega={\rm d}\m_a$,
then, these functions satisfy 
\begin{equation}
    \{\mu_a, \mu_b\} = f_{ab}{}^c\, \mu_c + \tau_{ab}\,,
    \label{mu mu}
\end{equation}
where $\tau_{ab} \in \Functions(\cM)$
is a central function, i.e. a function whose Poisson
bracket with any other function vanishes.%
\footnote{The condition that $V_a = \{\mu_a,-\}$ obey
$[V_a,V_b]=f_{ab}{}^c\,V_c$ is equivalent to
$\{\{\mu_a,\mu_b\} - f_{ab}{}^c\,\mu_c, -\}=0$, which in turn
imply that $\{\mu_a,\mu_b\} - f_{ab}{}^c\,\mu_c=\tau_{ab}$
where $\{\tau_{ab},f\}=0$ for any function 
$f\in\Functions(\cM)$. The Jacobi identity 
for the Poisson bracket implies that $\tau_{ab}$ is
a Chevalley--Eilenberg two-cocycle in the trivial module.
If the corresponding cohomology class is non-trivial,
then $\tau$ defines a central extension $\mathfrak g$.}
A consequence of the action of $\mathfrak g$ being Hamiltonian
is that the symplectic form $\Omega$ is preserved by infinitesimal 
diffeomorphism generated by the fundamental vector fields $V_a$,
\begin{equation}
    \cL_{V_a}\Omega = 0\,,
\end{equation}
as can be easily seen by using Cartan's homotopy formula.
Whenever $\tau$ vanishes, the co-moment map defined as
\begin{eqnarray}
	\mu^*\ :\ & \mathfrak{g}\ & \longrightarrow\ \Functions(\cM)\,,\nn
	& \xi=\xi^a\,J_a\ & \longmapsto\ \mu^*(\xi)
	= \langle \mu, \xi \rangle = \mu_a\,\xi^a\,,
\end{eqnarray}
is a Lie algebra morphism from $(\mathfrak{g}, [\cdot,\cdot])$
to $(\Functions(\cM),\{\cdot,\cdot\})$.
One can also assemble  $\mu_a$  into the moment map $\mu$\,,
\begin{eqnarray}
	\mu=\mu_a\,\cJ^a\ :\  & \cM\ & \longrightarrow\ \mathfrak{g}^*\,,\nn
	& y\ & \longmapsto\ \mu(y)=\mu_a(y)\,\cJ^a\,,
\end{eqnarray}
where $\cJ^a$ are the basis of $\mathfrak{g}^*$ dual to $J_a$\,: 
$\langle \cJ^a, J_b \rangle = \delta^a_b$\,. 
The moment and co-moment maps are related by
\begin{equation}
    \langle \mu(y), \xi \rangle = \mu^*(\xi)(y)\,,
\end{equation}
for any $\xi \in \mathfrak g$ and $y \in \cM$, 
so that one can think of the moment map $\mu$ 
as the dual of the co-moment map $\mu^*$ (and vice versa).

The pre-image of the element $\phi\in\mathfrak{g}^*$
under $\mu$\,,
\begin{equation}
	\mu^{-1}(\phi)=\{y\in \cM\,|\,\mu(y)=\phi\}\subset \cM\,,
\end{equation}
is not a symplectic submanifold of $\cM$:
using the inclusion map
$\iota_\phi : \mu^{-1}(\phi) \hookrightarrow \cM$,
one can pullback $\O$ onto  $\mu^{-1}(\phi)$ to get
the two-form $\iota_\phi^*\Omega$, which is degenerate
unless $G_\phi$ is trivial. If $G_\phi$  
acts freely and properly\footnote{Recall that
the action of a group $G$ on $\cM$ (denoted by
$\triangleright: G \times \cM \to \cM$) is called \emph{free}
if the stabiliser of any point $y \in \cM$
is trivial, meaning if $g \in G$ fixes a point $y$,
that is $g \triangleright y = y$, then it is the group identity, 
$g=\1$, necessarily. It is called \emph{proper}
if the inverse image of compact sets under the group action
are compact. These two conditions ensure that the quotient space
$\cM/G$ admits a structure of smooth manifold, and
$\cM \twoheadrightarrow \cM/G$ is a smooth principal $G$-bundle.}
on $\mu^{-1}(\phi)$, and $\phi \in \mathfrak g^*$
is a regular value of $\mu$,\footnote{A regular value
of $\mu: \cM \to \mathfrak g^*$ is an element $\phi\in\mathfrak g^*$
such that, for any point in its pre-image $y \in \mu^{-1}(\phi)$,
the pushforward
$(\mu_\ast)_y: T_y\cM \to T_\phi\mathfrak g^* \cong \mathfrak g^*$
is surjective. This implies that $\mu^{-1}(\phi)$
is a submanifold of $\cM$ (see e.g. \cite[Sec. 1.1.13.]{Ortega2013}).}
the quotient space,
\begin{equation}
	\cN_\phi:=\mu^{-1}(\phi)/G_\phi\,,
\end{equation}
is the base space of the principal $G_\phi$-bundle
$\mu^{-1}(\phi)$ with the projection
$\pi : \mu^{-1}(\phi) \rightarrow \cN_\phi$\,. 
Then, $\cN_\phi$ has a unique symplectic two-form $\omega$
satisfying $\pi^*\o=\i_\phi^*\Omega$:
we can use $\mu^{-1}(\phi)$ to compare the symplectic form
on the $G$-manifold $\cM$ and the reduced phase space $\cN_\phi$
by pulling them back with the inclusion and projection respectively,
as illustrated below.
\begin{equation}
    \begin{tikzcd}
        \mu^{-1}(\phi) \ar[r, hook, "\iota_\phi"]
        \ar[d, two heads, "\pi"] & \cM\\
        \cN_\phi &
    \end{tikzcd}
\end{equation}
This result is known as the Marsden--Weinstein--Meyers
theorem, see e.g. \cite{Ortega2013}.

Note that the moment map $\mu$ is equivariant
with respect to the $\mu^*(\mathfrak{g})$ action 
and the ${\rm ad}^*_{\mathfrak{g}}$ action\,: 
for any $\xi\in \mathfrak{g}$,
\begin{equation}
	\{ \mu^*(\xi),\mu\}={\rm ad}^*_{\xi}\,\mu\,.
\end{equation}
If the vector fields $V_a$ can be integrated,
the equivariance can be promoted to the Lie group $G$\,:
for any $g\in G$\,,
\begin{equation}
	\mu(g\triangleright y)= {\rm Ad}^*_g\mu(y)\,.
\end{equation}
As the moment map $\mu$ defines a $G$-equivariant homomorphism
from $\cM$ to $\mathfrak{g}^*$, the pre-image $\mu^{-1}(\phi)$ 
satisfies
\begin{equation}
	g\triangleright( \mu^{-1}(\phi) )
	=\mu^{-1}({\rm Ad}^*_{g}\phi)\,.
\end{equation}
This shows that $\mu^{-1}(\phi)$ is closed under
the action of the stabiliser $G_\phi$, and hence,
\begin{equation}
	\mu^{-1}(\cO^G_\phi) \cong \cO^G_\phi \times \mu^{-1}(\phi)\,.
\end{equation}
Note that the hypersurfaces $\mu^{-1}(\phi)$
and $\mu^{-1}(\cO^G_\phi)$ in $\cM$
have co-dimension $\dim\mathfrak{g}$
and $\dim\mathfrak{g}-\dim\cO^G_\phi=\dim\mathfrak{g}_\phi$,
respectively. The coadjoint space is foliated by the coadjoint orbits,
$\mathfrak{g}^*=\bigcup_{\phi\in \Phi}  \cO^G_\phi\,,$
with an infinite set $\Phi=\mathfrak g^*/G$ of representative vectors, 
and $\Phi$ can be further decomposed as 
$\Phi=\bigcup_{\varphi \in \Psi} \Phi_\varphi$ 
with a finite set $\Psi$ and infinite sets $\Phi_\varphi$
where $\varphi$ are stereotypical representative vectors.
% Here, ${\rm Aut}(G_\varphi)
% =N(G_\varphi)/Z(G_\varphi)$\,.}
The pre-image of the entire coadjoint space $\mathfrak{g}^*$, 
that is nothing but $\cM$, also admits the foliation,
\begin{equation}
	\cM=\mu^{-1}(\mathfrak{g}^*)\cong \bigcup_{\phi\in \Phi} \cO^G_\phi\times \mu^{-1}(\phi)
    \cong 
    \bigcup_{\varphi\in \Psi}
    G_\varphi\times 
    \bigcup_{\phi\in \Phi_\varphi}
    \cO^G_\phi\times \cN_\phi\,.
    \label{M decomp}
\end{equation}
Remark that both $\cO^G_\phi$ and $\cN_\phi$
are symplectic submanifold of $\cM$, 
whereas $G_\varphi$ is an isotropic one.
When $G$ is compact, the infinite set $\Phi$  
corresponds to $\mathfrak{h}^*/W$
where $\mathfrak h$ is the Cartan subalgebra
and $W$ the Weyl group. In plain words,
$\Phi$ is the set of orbits of $W$ in $\mathfrak{h}^*$.
Each $\Phi_\varphi$ corresponds to
either interior, boundary, or corner
regions of $\Phi$.

%********************************************%
\subsection{Constrained Hamiltonian mechanics}
\label{sec:hamiltonian}
%********************************************%

Let us rephrase the above discussion
in the framework of constrained Hamiltonian mechanics.  
The symplectic space $\cM$ is the embedding phase space endowed with the canonical structure $\O(y)$,
and the hypersurface $\mu^{-1}(\phi)$ is the constraint surface determined by
the Hamiltonian constraints,
\begin{equation}
	\chi_a(y) = \mu_a(y) - \phi_a \approx 0\,,
	\label{eq:constraints}
\end{equation}
and therefore has dimension $\dim\mu^{-1}(\phi)=\dim\cM-\dim\mathfrak g$.
The Poisson bracket of any two constraints then takes
the form
\begin{equation}
	\{\chi_a, \chi_b\} = f_{ab}{}^c\, \mu_c
	\approx f_{ab}{}^c\, \phi_c\,,
	\label{chichi}
\end{equation}
where $\approx$ denotes a weak equality, i.e. an equality
on the constraint surface $\mu^{-1}(\phi)$. 
Recall that in a constrained Hamiltonian system, one distinguishes
between first and second class constraints: the former
are constraints whose Poisson brackets with any other
constraint weakly vanish (i.e. they vanish on the constraint
surface) while the latter are constraints whose Poisson
brackets with at least one constraint does not vanish. 
To distinguish between first and second class constraints $\chi_a$, it
is convenient to introduce the notation
$\chi^*(\xi) := \xi^a\, \chi_a$ so that 
each constraint can be labeled by an element $\xi=\xi^a\,J_a$ of $\mathfrak{g}$.
It can also be understood as a shifted co-moment map $\chi^*=\mu^*-\phi$\,.
In this notation, the Poisson bracket \eqref{chichi} between any two constraints can be written as
\begin{equation}
	\{\chi^*(\xi), \chi^*(\zeta)\} \approx
	\langle \phi, [\xi, \zeta] \rangle= -\la {\rm ad}^*_\xi\phi, \zeta\ra\,,
\end{equation}
for any $\xi, \zeta \in \mathfrak{g}$.
For $\xi\in \mathfrak{g}_\phi$, 
the constraints $\chi^*(\xi)$ weakly commute with any other constraints,
and hence they are the first-class constraints.
The remaining constraints $\chi^*(\xi)$ with $\xi\notin\mathfrak{g}_\phi$
are the second-class constraints.
To recapitulate, the set of constraints, $\chi^*(\mathfrak{g})$,
is divided into the set of the first class constraints, $\chi^*(\mathfrak{g}_\phi)$,
and the set of  the second class constraints, $\chi^*(\mathfrak{g}/\mathfrak{g}_\phi)
\cong \chi^*(T_\phi\cO^G_\phi)$\,.

The quotient space $\cN_\phi=\mu^{-1}(\phi)/G_\phi$
is the physical phase space, i.e. the constraint surface reduced by the action of the gauge symmetry generated by
the first class constraints. The latter corresponds to the stabiliser
$\mathfrak{g}_\phi$, so that the reduced phase space $\cN_\phi$
has dimension,
\be
	\dim \cN_\phi = \dim \mu^{-1}(\phi)-\dim\mathfrak{g}_\phi= \dim \cM - \dim\cO^G_\phi -2\dim\mathfrak{g}_\phi\,,
	\label{dim count}
\ee
and one can confirm that each first and second class constraints remove 
respectively two and one dimension from the embedding phase space.

The  action corresponding to this phase space is
\begin{equation}
    S[y,A] = \int_I \vartheta(y)
    - \langle  \chi(y) , A\rangle\,,
    \label{eq:worldline_i}
\end{equation}
where 
$\vartheta$ is the symplectic potential of $\Omega$
satisfying $\Omega = -\rd\vartheta$. The Lagrange multiplier
$A \in \Omega^1(I,\mathfrak{g})$ is a worldline
one-form, valued in the Lie algebra $\mathfrak{g}$.
Note that   $y^\mu(t) \equiv y^\mu(\g(t))$ where $t \in I\subset \R$
is the worldline parameter and $\g\,:\,I\to \cM$ is 
the worldline, i.e. the (phase space) trajectory of a point particle in $\cM$.
Under the transformation generated by the gauge parameter $\l\in \O^0(I,\mathfrak{g})$,
\begin{equation}
    \delta_\l y^\mu = \{\chi^*(\l),y^\mu\}\,,
    \qquad
    \delta_\l A = \rd \l + [A,\l]\,,
\end{equation}
the action transforms as
\be
    \delta_\l S[y,A] = \int_I\rd\,\big(i_\l\,\vartheta(y) -\la \mu(y),\l\ra\big)+
   \la \phi,\rd \l+[A,\l]  \rangle\,.
\ee
Up to a total derivative,
the above reduces to the integral of $\la \phi, [A,\l]\ra$
and it vanishes only when $\l$
takes value in the isotropy subalgebra $\mathfrak{g}_\phi$.
This shows that only
the first class constraints associated with $\mathfrak{g}_\phi$
lead to gauge symmetries.
Under a finite gauge transformation $h\in G_\phi$, 
\be
    y\ \to\ y^h=h^{-1}\triangleright y\,,
	\qquad
    A\ \to\ A^h=h^{-1}(A+\rd)h\,,
\ee
the action changes as 
\ba
    S[y^h, A^h]-S[y,A]\eq  
	\int_I\vartheta(y^h)-\vartheta(y)-\la \chi,\rd h\,h^{-1}\ra \nn
	\eq \int_I \vartheta(y^h)
    - \vartheta(y) - \langle \mu(y),\rd h\,h^{-1} \rangle +\la \phi, h^{-1}\rd h\ra\,,
\ea
where the first three terms are the finite counterpart of the total derivative
$\rd(i_\l\vartheta-\la\mu(y),\l\ra)$ appearing for
infinitesimal gauge transformations.
The invariance of $\exp(\frac{i}{\hbar}\,S)$ under the above transformation
requires that the last term be proportional to $2\pi\,\hbar$ times an integer.
This leads precisely to the same quantisation condition on $\phi$ as in \eqref{quantisation condition}. One can also convert all the second class constraints into first class ones 
by introducing additional variables: see Appendix \ref{app:trick}.

%***************************************************%
\subsection{Example: Cotangent bundle of a Lie group}
%***************************************************%

An important class of examples of the above discussion
is the cotangent bundle $\cM=T^*G$ of a Lie group $G$,
which is a symplectic manifold as any cotangent bundle:
Locally, the symplectic form reads
\begin{equation}
	\Omega_{T^*G} = \rd \vartheta\,,
	\qquad
	\vartheta = p_\mu\, \rd x^\mu\,,
	\label{eq:symplectic_T*G}
\end{equation}
where $\{x^\mu\}$ are coordinates on $G$ and
$\{p_\mu\}$ are coordinates in the fiber
directions of $T^*G$. The Lie group $G$ acts
on its algebra of functions via left- or right-
invariant vector fields, which will be denoted by
\begin{equation}
	\rho_a = \rho_a{}^\mu(x)\,
	\tfrac{\partial}{\partial x^\mu}
	\in \Gamma(TG)\,,
\end{equation}
and where the Latin index $a$ refers to
a basis $\{J_a\}$ of the Lie algebra
$\mathfrak{g}$. These vector fields can be
lifted to functions on $T^*G$ via
\begin{equation}
	\mu_a(x,p) := \rho_a{}^\mu(x)\, p_\mu\,,
\end{equation}
which verifies
\begin{equation}
	\{\m_a, \m_b\} = f_{ab}{}^c\, \m_c\,,
\end{equation}
where $\{-,-\}$ denotes the Poisson bracket
associated with the symplectic two-form
\eqref{eq:symplectic_T*G}. With this data,
we can consider a constrained Hamiltonian
system of the type described previously,
whose corresponding worldline is given by
\begin{equation}
	S[x,p,A] = \int p_\mu\, \rd x^\mu
	- A^a\left(\rho_a{}^\mu(x)\,p_\mu
                        - \phi_a\right)\,.
\end{equation}
The constraints
\begin{equation}
	\chi_a(x,p) = \rho_a{}^\mu(x)\,p_\mu
    - \phi_a \approx 0\,,
\end{equation}
can be solved simply by
\begin{equation}
	p_\mu = e_\mu^a(x)\, \phi_a\,,
\end{equation}
where $e_\mu^a(x)$ are the components of
the left-invariant Maurer--Cartan form of $G$,
\begin{equation}
	g(x)^{-1}\rd g(x)
	= e^a_\mu(x)\,\rd x^\mu\, J_a
	\in \Omega^1(G,\mathfrak{g})\,,
\end{equation}
which are the inverse of the components
of the left-invariant vector fields.
Inserting the solution of the constraints
in the action, we recover the expression,
\begin{equation}
	S[x] = \int 
	\phi_a\, e^a_\mu(x)\, \rd x^\mu
	= \int \la \phi, g(x)^{-1} \rd g(x)\ra\,.
 \label{g inv g act}
\end{equation}
Remark that the constraint surface is 
\begin{equation}
    \mu^{-1}(\phi)
    = \{(g,{\rm Ad}_g^*\phi)\in T^*G\,|\,
    g\in G \}\cong G\,,
\end{equation}
where we used the fact that the cotangent bundle
of a Lie group is trivial,
$T^*G \cong G \times \mathfrak{g}^*$.
Further quotienting by the gauge symmetry generated by
the first class constraints, which is given by
the action of the isotropy group $G_\phi$, leads to
\begin{equation}
    \cN_\phi= \mu^{-1}(\phi)/G_\phi
   \cong \cO^G_\phi\,,
\end{equation}
i.e. the reduced phase space is nothing but
the coadjoint orbit of $\phi$.\footnote{In this simple case,
we can also verify the Marsden--Meyers--Weinstein
theorem explicitly. To do so, let us note that,
under the trivialization provided by the Maurer--Cartan
form, the tautological form on the cotangent bundle
$T^*G \cong G \times \mathfrak g^*$ reads
$\vartheta\rvert_{(g,\varphi)}
    = \langle \varphi, \Theta_g \rangle$
implying $\Omega_{T^*G}\rvert_{(g,\varphi)}
    = \rd\langle\varphi,\Theta_g\rangle$
at any point $(g,\varphi) \in G \times \mathfrak g^*$,
and where $\Theta \in \Omega^1(G,\mathfrak g)$
is the left-invariant Maurer--Cartan form of $G$.
We can now compare the pullback of the symplectic form
on the reduced phase space, which is the coadjoint orbit
$\cO^G_\phi$, by the projection
$\pi_\phi: G \twoheadrightarrow \cO^G_\phi:
    g \mapsto {\rm Ad}_g^*\phi$,
with the pullback of the symplectic form on the phase space
$T^*G \cong G \times \mathfrak g^*$ by the inclusion 
$\iota_\phi: G \hookrightarrow G \times \mathfrak g^*:
g \mapsto (g, \phi)$. We already computed the first
pullback in Section \ref{sec:KKS}, while the second one
simply amounts to the evaluation of $\Omega_{T^*G}$
at $\phi\in\mathfrak g^*$, so that we find
$\pi_\phi^*\Omega_{\cO^G_\phi}
    = \rd_G\langle\phi,\Theta\rangle
    = \iota_\phi^*\Omega_{T^*G}$ as expected.}
Applying the general story \eqref{M decomp} to this case,
we find  
\begin{equation}
    T^*G \cong  \bigcup_{\varphi\in \Psi}
    G_\varphi \times \bigcup_{\phi \in \Phi_\varphi}
    \cO^G_\phi \times \cO^G_\phi\,.
\end{equation}
Note that the quantisation of the above,
in the case when the Lie group $G$ is compact,
leads to the Peter--Weyl theorem,
\be 
    L^2(G) = \bigoplus_{\lambda}
    \pi^G_\lambda \otimes (\pi^G_\lambda)^*\,,
\ee
where $\pi_\lambda$ is the unitary irreducible
representations of $G$ labelled by $\lambda$
(its highest weight, since $G$ is assumed compact
here). The above decomposition of $T^*G$
can be recovered as its orbit space under the action
of $G \times G$ where the first factor acts
from the left and the second from the right
(see e.g. \cite{Baier2023} for a recent discussion
in that direction).
In the next section, we shall see a similar pattern
of decompositions, but involving coadjoint orbits
of two different Lie groups.

%%%%%%%%%%%%%%%%%%%%%%%%%%%%%%%%%%%%%%%%%%%%%%%%%%%%%%%%%%%%%%%%%%%%
\section{Manifestly covariant formulation of coadjoint orbit action}
\label{sec: cov action}
%%%%%%%%%%%%%%%%%%%%%%%%%%%%%%%%%%%%%%%%%%%%%%%%%%%%%%%%%%%%%%%%%%%%
In this section, we will explain how a mechanical system 
given by a coadjoint orbit $\cO^G_\phi$ of a Lie group $G$ 
can be realised as a constrained Hamiltonian system, 
where the constraints are associated with a different 
coadjoint orbit $\cO^{\tilde G}_{\tilde \phi}$ 
of a different Lie group $\tilde G$. Remark that
the analysis of the previous section applies to $\tilde G$,
associated with the \emph{gauge} symmetry of the system,
while $G$ is the \emph{global} symmetry.
We will show that the Lie groups $G$ and $\tilde G$ 
are dual in the sense of \emph{symplectic dual pairs} 
\`a la Weinstein \cite{Weinstein1983} 
(see also \cite[Chap. 4]{CannasDaSilva1999} 
for a textbook account). The quantised picture 
corresponds to Howe duality \cite{Howe1989i, Howe1989ii}, 
also known as dual pair correspondence 
(see also e.g. \cite{Prasad1993, Kudla1996, Adams2007, Rowe:2011zz, Rowe:2012ym}).

The construction  below can be understood 
as a method of obtaining a good coordinate system
for the coadjoint orbit $\cO^G_\phi$,
along the lines of the discussion 
in \ref{sec: phase space and dynamics}. More precisely,
we want to reformulate the system in such a way 
that its global symmetries are manifest. In other words, 
we want the global symmetry to be realised linearly, 
as opposed to a nonlinear realisation,
so that all the phase space variables
carry faithful representations of the global symmetry.
This can be achieved by using the definition 
of various matrix groups as Hamiltonian constraints.
In this set up, the phase space variables carry
the defining representations of the global symmetry,
as well as a representation of a gauge symmetry group.
We will find an exquisite relation between
the global and gauge symmetries of the system.

In the first two subsections \ref{classcial Lie group}
and \ref{semidirect product} we present the construction 
of worldline action for all classical Lie groups
and their semi-direct product with Abelian ideals.
The treatment here will be rather brief
as we consider the case of indefinite orthogonal group
and inhomogeneous orthogonal group in detail
in the subsections \ref{orthogonal groups} 
and \ref{inhomogeneous orthogonal}.
The other cases will be detailed
in the sequel paper \cite{partII}, 
along with twistor descriptions.
The readers who wish to focus on the case 
of (inhomogeneous) orthogonal groups may skip 
the first two subsections.

%*******************************%
\subsection{Classical Lie groups}
\label{classcial Lie group}
%*******************************%
Let us  consider the classical Lie groups
$GL(N,\mathbb F)$, $U(p,N-p)$, $O(p,N-p)$,
$O^*(2N)$, $O(N,\mathbb C)$, $Sp(2N,\mathbb R)$,
$Sp(p,N-p)$ and $Sp(2N,\mathbb C)$,
where, $\mathbb F=\mathbb R, \mathbb C$, 
or the quaternion $\mathbb H$. They are all reductive,
and include many physically relevant cases, 
such as Lorentz groups and (A)dS isometry groups:
the family $O(p,N-p)$ contains the AdS$_d$ symmetry 
$O(2,d-1)$ and the dS$_d$ symmetry $O(1,d)$.
The double covers of the orthochronous Lorentz groups 
$SO^+(2,1)$, $SO^+(3,1)$ and $SO^+(5,1)$
are isomorphic to $SL(2,\mathbb R)$, $SL(2,\mathbb C)$ 
and $SL(2,\mathbb H)$, the  simple parts of $GL(2,\mathbb F)$.
The double covers of the conformal groups $SO^+(2,3)$, 
$SO^+(2,4)$ and $SO^+(2,6)$ are isomorphic to $Sp(4,\mathbb R)$, 
$SU(2,2)$ and $O^*(8)$.

All classical Lie groups are subgroups of a general linear
group $GL(N, \mathbb F)$ with $\mathbb F=\mathbb R, \mathbb C$, 
defined by a quadratic equation --- expressing the fact 
that they preserve a certain bilinear form. We can introduce 
most of them in a unified fashion as in \cite{Howe:1992bv}:
let us define 
\begin{equation}
	B(b_{\sst (N)},\mathbb F)
	= \{\,A \in GL(N,\mathbb F) \mid 
    A^\dagger\,b_{\sst (N)}\,A = b_{\sst (N)}\,\}\,,
    \label{B group}
\end{equation}
where $b_{\sst (N)}$ is an element of $GL(N,\mathbb F)$\,, 
and $A^\dagger=(A^{t})^*$ with $t$ the matrix transpose 
and $*$ the conjugation of $\mathbb F$, which is the identity map
for $\mathbb R$, the complex conjugation for $\mathbb C$,
and the quaternionic conjugation for $\mathbb H$\,.
Up to a $GL(N,\mathbb F)$ transformation 
$b_{\sst (N)}\to T^\dagger\,b_{\sst (N)}\,T$,
we have essentially two possibilities:
either it is Hermitian or anti-Hermitian.
The Hermitian ones are equivalent
to $b_{\sst (N)}=\eta^{\sst (p,N-p)}$, 
the standard flat metric of $(p,N-p)$ signature.
The anti-Hermitian ones are equivalent 
to $b_{\sst (N)}=\Omega_{\sst (N)}$,
the canonical symplectic matrix.
See e.g. \cite[Prop. 9.3.2.]{Collingwood1993} 
for relevant discussions. The group $B(b_{\sst (N)},\mathbb F)$
either simply coincides with one of classical Lie groups
or is isomorphic to it:
\begin{eqnarray}
    & B(\eta^{\sst (p,N-p)},\mathbb R)=O(p,N-p)\,,
    \qquad
    & B(\Omega_{\sst (N)},\mathbb R)=Sp(N,\mathbb R)\,,\\
    & B(\eta^{\sst (p,N-p)},\mathbb C)=U(p,N-p)\,,
    \qquad 
    & B(\Omega_{\sst (N)},\mathbb C) \simeq U(\tfrac{N}2,\tfrac{N}2)\,,\\
    & B(\eta^{\sst (p,N-p)},\mathbb H) = Sp(p,N-p)\,,
    \qquad 
    & B(\O_{\sst (N)},\mathbb H) \simeq O^*(2N)\,.
\end{eqnarray}
In the right column, the cases with $\mathbb F=\mathbb R$
and $\mathbb C$ are defined only for even $N$.
For the case $\mathbb F=\mathbb H$, $N$ can be both even 
and odd because the second element of $\mathbb H$ 
can be seen as a two-dimensional symplectic matrix
so we can take $\O_{\sst (N)}=\mathbf{j}\,I_{\sst (N)}$.
Note that $B(\O_{\sst (N)},\mathbb C)$ are isomorphic 
to unitary groups because we can diagonalize $\O_{\sst (N)}$ 
as $i\,\eta^{\sst (N/2,N/2)}$. On the contrary,
$B(\O_{\sst(N)},\mathbb H)$ is not isomorphic to $Sp(N/2,N/2)$ 
but $O^*(2N)$, even though $\O_{\sst (N)}$ can still be diagonalised 
as $\mathbf{i}\,\eta^{\sst (N/2,N/2)}$. It is because $\mathbf{i}$,
the first element of the basis of $\mathbb H$, 
does not commute with a quaternionic matrix.

We can also define another class of classical Lie groups
by using the transpose $t$ at the place of the Hermitian conjugate
in the definition \eqref{B group}. Then, for $\mathbb F=\mathbb R$, 
this trivially coincides with $B(b_{\sst (N)},\mathbb R)$.
For $\mathbb F=\mathbb H$, this fails to form a group.
Only $\mathbb F=\mathbb C$, it defines a new classical Lie group,
\begin{equation}
    C(b_{\sst (N)}) = \{A\in GL(N,\mathbb C)\,|\,
    A^t\,b_{\sst (N)}\,A=b_{\sst (N)}\}\,.
\end{equation}
Again, up to a $GL(N,\mathbb C)$ transformation, 
we have two possibilities, $b_{\sst (N)}=I_{\sst (N)}$
and $b_{\sst (N)}=\Omega_{\sst (N)}$, corresponding to
\begin{equation}
    C(I_{\sst (N)}) = O(N,\mathbb C)\,,
    \qquad 
    C(\O_{\sst (N)}) = Sp(N,\mathbb C)\,.
\end{equation}
The latter case is defined only for even $N$.
Note that these Lie groups are not simple
as a real Lie group, but semisimple.

On the manifold $GL(N,\mathbb F)$, the components
of $X^a{}_{b}\in \mathbb F$ of an element $X \in  GL(N,\mathbb F)$
serve a natural and global coordinate system,
and a left-invariant vector field is given by
$V=V^a{}_b\,X^{c}{}_a\,\frac{\partial}{\partial X^c{}_b}$.
The action takes the simple form,
\begin{equation}
    S[X,P,A] = \frac12\int\tr_{N \times N}\left[P\,\rd X
    + A\,(P\,X - \phi) + (\rm conj)\right],
\end{equation}
where all three fields $X$, $P$ and $A$ as well as $\phi$
take value in $Mat_{N\times N}(\mathbb F)$,
and the symbol (conj) stands for the conjugate in $\mathbb F$.
Remark that adding the conjugate is trivial for $\mathbb F=\mathbb R$
as it duplicates the Lagrangian and only replaces
the factor $\tfrac12$ by $1$ in the end.
For $\mathbb F = \mathbb C$ and $\mathbb F = \mathbb H$
complementing the Lagrangian with the conjugate
is necessary to recover the $2(2N)^2$ and $2(4N)^2$ dimensional
symplectic potentials. Note that we can solve the constraint
algebraically to get
\begin{equation}
	S[X] = \frac12\int \tr_{N\times N} \left[\phi\,X^{-1} \rd X
    +({\rm conj})\right],
    \label{mat g inv}
\end{equation}
which is nothing but a matrix form of \eqref{g inv g act}.

If the coadjoint element $\phi$ is a rank $M$ matrix with $M \le N$,
the above action can be reduced to\footnote{To be more
precise, this reduction requires integrating out
non-dynamical variables corresponding
to the components of $X$ in the subspace
$Mat_{N \times (N-M)}(\mathbb F)$.}
 \be
	S[X,P,A] = \frac12\int\tr_{M\times M}\left[P\,\rd X
    + A\,(P\,X - \tilde\phi) + {(\rm conj)}\right],
\ee 
where the fields $X$, $P$ and $A$ takes value 
in $Mat_{N\times M}(\mathbb F)$,
$Mat_{M\times N}(\mathbb F)$
and $Mat_{M\times M}(\mathbb F)$, respectively.
Here $\tilde \phi$ also belongs to $Mat_{M\times M}(\mathbb F)$
and it is the $M\times M$ submatrix of a triangulation of $\phi$\,.
The resulting action describes a $G=GL(N,\mathbb F)$
coadjoint orbit $\cO^G_\phi$ as a reduced phase space
inside $\mathbb F^{2MN}$ where the constraints
are given by the moment maps
$\tilde\mu(X,P)=P\,X\in Mat_{M\times M}(\mathbb F)$
generating $\tilde G=GL(M,\mathbb F)$ under Poisson bracket.
Note also that the moment map 
$\mu(X,P)=X\,P\in Mat_{N\times N}(\mathbb F)$
associated with the original $GL(N,\mathbb F)$ symmetry
commutes with $\tilde\mu$\,. The constraints
$\tilde \chi(X,P)=\tilde\mu(X,P) - \tilde\phi \approx 0$
are associated with the $\tilde G=GL(M,\mathbb F)$
coadjoint orbit $\cO^{\tilde G}_{\tilde\phi}$.

Now, let us move to the classical Lie groups
$B(b_{\sst (N)},\mathbb F)$. Adding the definition \eqref{B group}
to the action \eqref{mat g inv} as a constraint,
we start with the action
\ba
	S[X,A] \eq \int \frac12 \tr_{N \times N}\left[\left(\phi\,
    X^{-1} \rd X\right)
    + A\,(X^\dagger\,b_{\sst (N)}\,X-b_{\sst(N)})+({\rm conj})\right]\nn
    & \cong & \int\tr_{N \times N}\left[\left(\phi\,
    b_{\sst (N)}^{\ \ -1}\,X^\dagger\,b_{\sst (N)}\,\rd X
    \right)
    + A\,(X^\dagger\,b_{\sst (N)}\,X - b_{\sst (N)})+({\rm conj})\right]\,,
\ea 
where $\cong$ means the equivalence up to a redefinition of $A$.
For any $\phi$, there exists a $T\in Mat_{N\times M}(\mathbb F)$
and $\tilde b_{\sst (M)}$ with $M\le N$ such that 
\be
    \phi = T\,\tilde b_{\sst (M)}\,T^\dagger\,b_{\sst (N)}\,.
\ee 
Here, $\tilde b_{\sst (M)}=\O_{\sst (M)}$ 
for $b_{\sst (N)}=\eta^{\sst (p,N-p)}$\,,
and $\tilde b_{\sst (M)}=\eta^{\sst (q,M-q)}$
for $b_{\sst (N)}=\O_{\sst (N)}$\,.
With a suitable redefinition of $X$ and $A$ in terms of $T$,
we can express the action as
\be
	S[X,A] = \int \frac12 \tr_{M \times M}\left[\left(\tilde b_{\sst (M)}\,
 X^\dagger\,b_{\sst (N)}\,\rd X\right)
 +A\left(X^\dagger\,b_{\sst (N)}\,X-\tilde b_{\sst (M)}^{\ \ -1}\,\tilde\phi\right)+{\rm (conj)}\right],
\ee
where $X$ takes value in $Mat_{N\times M}(\mathbb F)$ and 
$\tilde\phi\in Mat_{M\times M}$ is given by
\be 
   \tilde\phi = \tilde b_{\sst (M)}\,T^\dagger\,b_{\sst (N)}\,T\,.
\ee 
Note that the moment maps $\tilde \mu(X)
= X^\dagger\,b_{\sst (N)}\,X\in Mat_{M\times M}(\mathbb F)$ 
generates the dual symmetry
$\tilde G=B(\tilde b_{\sst (M)},\mathbb F)$ whereas
$\mu(X) = X\,\tilde b_{\sst (M)}\,X^\dagger \in Mat_{N \times N}(\mathbb F)$
generates the original symmetry $G=B(b_{\sst (N)},\mathbb F)$\,.

The classical Lie group $C(b_{\sst(N)})$ can be treated
in a very similar manner. We find 
\be
    S[X,A] = \int \frac12\,\tr_{M\times M}
    \left[\tilde b_{\sst (M)}\,X^t\,b_{\sst (N)}\,\rd X
    +A\left(X^t\,b_{\sst (N)}\,X
    -\tilde b_{\sst (M)}^{\ \ -1}\,\tilde\phi\right)
    + ({\rm conj})\right]\,,
\ee 
where the dual coadjoint vector $\tilde\phi$ is related to
the coadjoint vector $\phi$ through
\be 
    \phi = T\,\tilde b_{\sst (M)}\,T^t\,b_{\sst (N)},
    \qquad 
    \tilde\phi = \tilde b_{\sst (M)}\,T^t\,b_{\sst (N)}\,T\,,
\ee 
with a $T\in Mat_{N\times M}(\mathbb C)$\,.
Here, $\tilde b_{\sst (M)}=\O_{\sst (M)}$ 
for $b_{\sst (N)}=I_{\sst (N)}$\,,
and $\tilde b_{\sst (M)}=I_{\sst (M)}$
for $b_{\sst (N)}=\O_{\sst (N)}$\,.
The moment maps $\tilde \mu(X)=
X^t\,b_{\sst (N)}\,X\in Mat_{M\times M}(\mathbb C)$
generate the dual symmetry $\tilde G=C(\tilde b_{\sst (M)})$
whereas $\mu(X)=X\,\tilde b_{\sst (M)}\,X^t \in Mat_{N\times N}(\mathbb C)$
generates the original symmetry $G=C(b_{\sst (N)})$\,.

%************************************%
\subsection{Semi-direct product group}
\label{semidirect product}
%************************************% 
The above construction can be extended to a class
of non-reductive Lie groups $G$, which are given by
semi-direct product $G=I \rtimes H$ between a reductive group $H$
treated above, and an Abelian ideal $I$ carries a $H$-representation $\pi$.
Any element $g$ of $G$ can be denoted by $g=i\,h$ with $h\in H$
and $i\in I$\,. The semi-direct product rule can be deduced
from $h\,i=(\pi(h)\,i)\,h$\,. The adjoint and coadjoint actions
of an element $(a,h)\in G$ read 
\ba
	&& {\rm Ad}_{(e^a,h)}(\chi,\xi)
	= \big(\pi(h)(\chi+\pi(\xi)\,a),\, {\rm Ad}_h\xi\big)\,,\nn
	&& {\rm Ad}^*_{(e^a,h)}(\phi_I,\phi_H)
	= \big(\pi^*(h)\phi_I,\,
        \langle\pi^*(h)\phi_I, \pi(-)\,a\rangle_{\mathfrak i}
            + {\rm Ad}^*_h\phi_H\big)\,,
	\label{ad coad action}
\ea
where $\langle \phi_I, \pi(-)\,a \rangle_{\mathfrak i} \in \mathfrak{h}^*$
is defined such  that for any $\xi \in \mathfrak{h}$,
\be
	\big\la \la  \phi_I, \pi(-)\,a \ra_{\mathfrak i}, \xi\big\ra_{\mathfrak h}
	= \la  \phi_I, \pi(\xi)\,a \ra_{\mathfrak i}\,.
	\label{coad h}
\ee
From the coadjoint action
${\rm Ad}^*_{(i,{\rm id})}(\phi_I,\phi_H)
=(\phi_I, \la\phi_I,\pi(-)\,a\ra_{\mathfrak i}+\phi_H)$
and the property $\big\la \la\phi_I,\pi(-)\,a\ra_{\mathfrak i}+\phi_H,\xi\big\ra_{\mathfrak h}
=-\la \pi^*(\xi)\,\phi_I,a\ra_{\mathfrak i}+\la \phi_H,\xi\ra_{\mathfrak h}$,
we can see that, if $\xi$ does not belong to the little group algebra $\mathfrak{h}_{\phi_I}$ of $\phi_I$,
we can set  $\la \phi_H,\xi\ra_{\mathfrak h}$ to zero with a suitable choice of $a$.  	
This means that a coadjoint vector $(\phi_I, \phi_H)$ can be
always rotated in a way that $\phi_H$ belongs to 
$\mathfrak{h}_{\phi_I}^*$\,. 
It is also known that a coadjoint orbit
of a semi-direct product group has the structure
of a fiber bundle with the `momentum orbit',
i.e. the orbit of $H$ on $\phi_I$, as the base manifold
and the direct product of the cotangent space 
of the momentum orbit  times the coadjoint orbit of $\phi_H$
under the `little group' $H_{\phi_I}$ as the fiber,
see e.g. \cite{Rawnsley1975, Baguis1998}
or \cite{Oblak:2016eij, Arathoon2019} and references therein.

The Maurer--Cartan element reads
\be
	g^{-1}\,{\rm d}_G\,g = (e^a\,h)^{-1}\,{\rm d}_G\,(e^a\,h)
	= \pi(h^{-1})\,{\rm d}_I\,a + h^{-1}{\rm d}_H\,h\,,
\ee
and the coadjoint orbit action is
\be
	S[g] = \int \langle \phi, g^{-1}\,{\rm d}_G\,g \rangle_{\mathfrak g}
	=\int \left[\langle \phi_I,
    \pi(h^{-1})\,{\rm d}_I\,a\rangle_{\mathfrak i}
    + \langle \phi_H,h^{-1}{\rm d}_H\,h\rangle_{\mathfrak h}\right]\,,
\ee
where the coadjoint vector $\phi$ is split
into $\phi=\phi_H + \phi_I$ with $\phi_H \in \mathfrak{h}^*$
and $\phi_I \in \mathfrak{i}^*$\,.
We can treat the second part of the action
as in the subsection \ref{classcial Lie group}.

%**************************************%
\subsection{Orthogonal groups}
\label{orthogonal groups}
%**************************************%

In this section, we reconsider the coadjoint orbit actions
of the indefinite orthogonal groups $O(p,N-p)$ with more details.
From the definition, the action is given by
\ba
	S[X,A] \eq \int \tr_{N\times N}\left[\phi\,X^{-1}\,{\rm d}X
    + A\,(X^t\,\eta\,X-\eta)\right]\nn
	& \cong & \int \tr_{N\times N}\left[\phi\,
    \eta^{-1}\,X^t\,\eta\,{\rm d}X + A\,(X^t\,\eta\,X-\eta)\right],
\ea
where $\eta$ is the flat metric of $(p,N-p)$ signature.
The matrix $\phi\,\eta^{-1}$ is antisymmetric, 
and suppose that its rank is $2M \le N$.
Then we can always find a rectangular matrix
$T^a{}_\beta \in Mat_{N \times M}(\R)$ such that
\be
	\phi^{ab} = T^a{}_\alpha\,T^b{}_\beta\,\Omega^{\alpha\beta}\,,
	\label{phi}
\ee
where $\Omega^{\a\b}$ is the symplectic matrix of rank $2M \le N$
and $\phi^{ab}=(\phi\,\eta^{-1})^{ab} = \phi^a{}_c\,\eta^{cb}$

Here, the indices $a,b=1,\ldots, N$ while $\a,\b=1,\ldots,2M$. 
We can append to $T$ a $N\times (N-2M)$ matrix $R$
so that they jointly form a matrix $(T\,R) \in GL(N,\mathbb R)$\,.
Introducing the indices $\bar\alpha,\bar\beta=2M+1,\ldots, N$,
we can consider the redefinition,
$A=(T\,R)\,A'\,(T\,R)^t$ or in components
\be	
    A^{ab} = A'^{\alpha\beta}\,T^a{}_\alpha\,T^b{}_\beta
    + A'^{\bar\alpha\beta}\,R^a{}_{\bar\alpha}\,T^b{}_\beta 
    + A'^{\alpha\bar\beta}\,T^a{}_{\alpha}\,R^b_{\bar\beta} 
    +A'^{\bar\alpha\bar\beta}\,R^a{}_{\bar\alpha}\,R^b{}_{\bar\beta}\,, 
\ee
then also 
\be
    X'^a{}_\alpha = X^{a}{}_b\,T^b{}_\alpha\,,
    \qquad 
    \bar X'^a{}_{\bar\alpha} = X^a{}_b\,R^b{}_{\bar\alpha}\,.
\ee
By removing the prime from the variables,
the action can be written as 
\ba
	S[X,A] \eq \int \Omega^{\alpha\beta}\,X^a{}_{\beta}\,{\rm d}X_{a\alpha}
	+A^{\a\b}\left(X_{c\b}\,X^c{}_\a-\tilde\phi_{\a\b}\right)\nn 
    && \quad +\,2\,A^{\alpha\bar\beta} 
    \left(\bar X_{c\bar\b}\,X^c{}_\a-\varphi_{\a\bar\b}\right)
    +A^{\bar\a\bar\b} 
    \left(\bar X_{c\bar\b}\,\bar X^c{}_{\bar\a}-\varphi_{\bar\a\bar\b}\right),
\ea
where the Latin indices are lowered by $\eta_{ab}$
and $\tilde\phi_{\a\b}$, $\varphi_{\a\bar\b}$,
and $\varphi_{\bar\a\bar\b}$ are given by
\be
	 \tilde\phi_{\a\b} = T^a{}_\a\,T_{a\b}\,,
  \quad 
  \varphi_{\a\bar\b} = T^a{}_\a\,R_{a\bar\b}\,,
  \quad 
  \varphi_{\bar\a\bar\b} = R^a{}_{\bar\a}\,R_{a\bar\b}\,.
	 \label{varphi}
\ee
The constraints given by $A^{\a\bar\b}$ and $A^{\bar\a\bar\b}$ 
can be algebraically solved for the variables $\bar X^a{}_{\bar\a}$,
and this results in a constant factor. Discarding this factor, 
the final form of the action is simply
\be 
    S[X,A] = \int \Omega^{\alpha\beta}\,X^a{}_{\beta}\,{\rm d}X_{a\alpha}
	+ A^{\alpha\beta}\left(X_{c\beta}\,X^c{}_\alpha
    -\tilde\phi_{\alpha\beta}\right)\,.
 	\label{O action}
\ee 
The constraints are given by
the momentum maps $\tilde \mu_{\b\a}=X_{c\b}\,X^c{}_\a$
closed under the Poisson bracket as 
\be
    \{\tilde \mu_{\a\b},\tilde\mu_{\g\d}\}
    = \Omega_{\beta\gamma}\,\tilde\mu_{\alpha\delta}
    + \Omega_{\alpha\gamma}\,\tilde\mu_{\beta\delta}
    + \Omega_{\alpha\delta}\,\tilde\mu_{\beta\gamma}
    + \Omega_{\beta\delta}\,\tilde\mu_{\alpha\gamma}\,,
    \label{eq:moment_sp}
\ee
and hence defines the dual Lie algebra
$\tilde{\mathfrak{g}}=\mathfrak{sp}(2M,\mathbb R)$\,.
Since the constraints $\tilde\chi_{\a\b}
=\tilde\mu_{\a\b}-\tilde\phi_{\a\b}\approx 0$
are given with a constant shift
$\tilde\phi_{\a\b}$\,,
they are a mixture of the first
and the second class constraints.
According to the general results
presented in Section \ref{sec:hamiltonian},
the first class constraints
are the linear combinations
$\tilde\chi^*(\xi)=\xi^{\a\b}\,\tilde\chi_{\a\b}$
satisfying
\be 
    \{\tilde\chi^*(\xi),\tilde\chi_{\g\d}\}\,
    \approx\, 2\,\xi^{\a\b}
    (\Omega_{\beta\gamma}\,\tilde\phi_{\alpha\delta}
    + \Omega_{\beta\delta}\,\tilde\phi_{\alpha\gamma})=0
    \qquad \forall \g,\,\d\,.
\ee 
This forms
a subalgebra 
$\tilde{\mathfrak{g}}_{\tilde\phi}
\subset \tilde{\mathfrak{g}}=\mathfrak{sp}(2M,\mathbb R)$\,,
whose structure
is determined by $\tilde\phi_{\a\b}$
hence by $\phi^{ab}$.
The matrix 
$\tilde\phi_{\a\b}
=\O_{\a\g}\,\tilde\phi^{\g}{}_\b$
then $\tilde\phi^{\a}{}_\b$
corresponds to the coadjoint vector of $\tilde{\mathfrak{g}}^*=
\mathfrak{sp}(2M,\mathbb R)^*$\,.
The remaining constraints
are the second class ones
corresponding to
the dual coadjoint orbit
$\cO^{\tilde G}_{\tilde\phi}=\tilde G/\tilde G_{\tilde\phi}$.

This construction clearly exhibits
the intimate relation between
$\phi^{a}{}_b$ and $\tilde\phi^{\a}{}_{\b}$:
they are both given by
two different  contractions of $T^a{}_\b$:
\be 
    \phi^a{}_b=T^a{}_\a\,\O^{\a\b}\,T^c{}_\b\,\eta_{cb}\,,
    \qquad 
    \tilde\phi^\a{}_\b
    =\O^{\a\g}\,T^a{}_\g\,
    \eta_{ab}\,T^b{}_\b\,.
    \label{phi varphi rel}
\ee 
Here, it is worth emphasising that the components $\tilde\phi^\a{}_\b$
are determined by $\phi^a{}_b$,
up to a $Sp(2M,\R)$ transformation.
The choice of a coadjoint element $\phi^a{}_b$ itself is also fixed up to a $O(p,N-p)$ transformation.
In other words, the choice of the matrix $T^a{}_\b$ is determined up to a $O(p,N-p)\times Sp(2M,\R)$ transformation.
In matrix form, 
the relations \eqref{phi varphi rel} read
\be
	\phi=T\,\Omega\,T^t\,\eta,
 \qquad 
 \tilde\phi=\O\,T^t\,\eta\,T\,,
\ee
and they satisfy the same invariant equations,
\be
	\label{In}
 \tr(\phi^n)=I_n=\tr(\tilde\phi^n)\,,
\ee
for any natural number $n$.
Allowing $\phi$ and $\tilde\phi$ to vary,
the above are polynomial functions in $\mathfrak g^*$
and $\tilde{\mathfrak{g}}^*$ respectively,
which commute (with respect to the Poisson bracket)
with any other functions, i.e. they are Casimir functions
of $\mathfrak{g}^*$ and $\tilde{\mathfrak{g}}^*$
respectively. The previous identity therefore
tells us that evaluating these Casimir functions
on two dual coadjoint orbits of $G$ and $\tilde G$
yields the same result. This is another `classical'
counterpart of a feature present in the dual pair correspondence:
the values of the Casimir operators of two dual groups,
on a pair of representations which are dual to one another,
are related \cite{Itoh2003}. In this last case, however,
the relation between the values of the two Casimir 
operators involves a rank-dependant shift
stemming from 
a quantum mechanical
ordering issue.

%*****************************************%
\subsection{Inhomogeneous orthogonal group}
\label{inhomogeneous orthogonal}
%*****************************************%

The coadjoint orbits of inhomogeneous orthogonal  group $IO(p,N-p)$ can be classified as follows.
Let $P_a$ and $J_{ab}$ the generators of the Lie algebra and $\cP^a$ and $\cJ^{ab}$ their duals. A coadjoint  vector is given by $\phi=\phi_{I\,a}\,\cP^a+
\phi_{H\,ab}\,\cJ^{ab}$.
If a coadjoint orbit has $\phi_{I\,a}=0$, then it reduces to that of the subgroup $O(p,N-p)$, 
which, for $p=1$, might be interpreted as the dS group of one lower dimensions.
Therefore, we focus on the coadjoint orbits with non-vanishing $\phi_I$.
As we have seen below \eqref{coad h}, $\phi_H$ can be chosen in
the dual space of the little group algebra associated with $\phi_I$:
$O(p-1,q)$, $IO(p-1,q-1)$ and $O(p,q-1)$ for $\phi_I^2>0$, $\phi_I^2=0$ and $\phi_I^2<0$.
The classification of $\phi_H$ simply follows that of the
coadjoint orbits of the corresponding little group algebra.

Let us apply the general method outlined previously to $IO(p,N-p)$ for which $H$ as the indefinite orthogonal group $O(p,N-p)$ and $I$ as the translation group, 
$\mathbb R^{N}$ carrying a vector representation
of $H$. 
The resulting action reads
\ba
	S[x, \Sigma, A] \eq \int \left[\phi_I{}_a\,
    (\Sigma^{-1})^a{}_b\,{\rm d}x^b +\phi_H{}^a{}_b\,
    (\Sigma^{-1})^b{}_c\,{\rm d}\Sigma^c{}_a\right] \nn
	&\cong & \int \left[\phi_I{}_a\,\Sigma_b{}^a\,{\rm d}x^b
	+ \phi_H{}^{ab}\,\Sigma_{cb}\,{\rm d}\Sigma^c{}_a
    + A^{ab}(\Sigma_{cb}\,\Sigma^{c}{}_a-\eta_{ba})\right]\,,
\ea
where we denote elements of the homogeneous group $H$ by $\Sigma$,
and elements of the Abelian ideal $I$ by $x$.
We first decompose the Lagrange multiplier as $A^{ab}=\phi_I{}^a\,\phi_I{}^b\,B+\phi_I{}^{(a}\,B^{b)}+B^{ab}$,
then skew-diagonalise and normalize $\phi_H$ as $\phi_H{}^{ab}=T^{a}{}_\a\,T^{b}{}_\b\,\Omega^{\a\b}$.
Here again, $\a, \b =1,\ldots, 2M$ where $2M$ is the rank of $\phi_H$. 
 Finally, 
by substituting
$(B,B^a,B^{ab})$ with
$(A,A^a,A^{ab})$ again, 
and discarding
non-dynamical variables, we 
reduce the action as 
\ba
	S[x, p, \Sigma, A] & = & \int \Big[\,p_a\,{\rm d}x^a
	+ \Omega^{\a\b}\,\Sigma_{c\b}\,{\rm d}\Sigma^c{}_\a
	+ A\,(p_a\,p^a-\tilde\phi_C) \nn 
	&& \hspace{40pt} +\,A^\a\,(p_a\,\Sigma^a{}_\a-\tilde\phi_{I\,\a})
	+ A^{\a\b}\,(\Sigma_{c\b}\,\Sigma^c{}_\a
    -\tilde\phi_{H\,\b\a})\Big]\,,
	\label{inhomo Lorentz}
\ea
where $p_a=\Sigma_a{}^b\,\phi_I{}_b$
and 
\be     \tilde\phi_{H\,\a\b}=T^a{}_\a\,T^b{}_\b\,
\eta_{ab}\,,
\quad 
\tilde\phi_{I\,\a}=\phi_{I\,a}\,T^a{}_\a,
\quad 
\tilde\phi_{C}=\phi_{I\,a}\,\phi_I{}^a\,.
\ee
Here, the dual algebra is associated with the moment maps
$\mu_{\a\b} = \Sigma_{c\a}\,\Sigma^c{}_\b$,
$\mu_\a = p_a\,\Sigma^a{}_\a$ and $\mu = p_a\,p^a$
satisfying \eqref{eq:moment_sp} and 
\be
	\{ \mu_{\a\b},\mu_\g\}= 2\,\mu_{(\alpha}\,\Omega_{\beta)\gamma}\,,
	\qquad 
	\{\mu_\a,\mu_\b\}= \Omega_{\alpha\beta}\,\mu\,,
\ee
where $\mu$ is the center. This Lie algebra
is isomorphic to $\heis_{2M} \niplus \sp(2M,\R)$
with dimension $(M+1)(2M+1)$,
the semi-direct sum of the Heisenberg
and the symplectic algebra. To recapitulate,
\begin{equation}
    \phi_H=T\,\Omega\,T^t\,\eta\,,
    \qquad 
    \tilde \phi_H=\Omega\,T^t\,\eta\,T\,,
    \qquad 
    \phi_I\,T=\tilde \phi_I\,,
\end{equation}
and we find that the following three quantities
\begin{equation}
    \tr(\phi_H{}^n)=I_n=\tr(\tilde \phi_H{}^n)\,,
    \qquad
    (\phi_I|\phi_I)
    =J_2=\tilde \phi_C\,,
\end{equation}
and
\begin{equation}
    (\phi_I|\phi_H{}^{n}|\phi_I{})
    =J_{n+2}
    =\la \tilde \phi_I|\tilde \phi_H{}^{n-1}|{\tilde \phi_I}\ra
    \qquad [n\ge 1]\,,
\end{equation}
where
\begin{equation}
    (v|A|w)=v_a\,A^a{}_b\,\eta^{bc}\,w_c\,,
   \qquad 
    \la v|A|w\ra=v_\a\,A^\a{}_\b\,\O^{\b\g}\,w_\g\,,
\end{equation}
relating the same traces of powers
of $\phi=(\phi_H,\phi_I)$ and its dual
$\tilde\phi=(\tilde\phi_H,\tilde\phi_I)$.
However, $I_n$ and $J_n$ are invariant only
under the homogeneous part of the group,
$G_H$ or $\tilde G_H$, except for $J_2$.
The higher order Casimir invariant functions 
$C_{2(n+1)}$ for the full Poincar\'e algebra
can be constructed  using the Pauli--Lubanski
tensors $W_{(n+1)}$ given by,
\begin{equation}
    W_{(n+1)}{}^{a_1\cdots a_{d-2n-1}}=\dfrac1{2^n\,n!}\,\ve^{a_1\ldots a_d}\,\f_H\,_{a_{d-2n}a_{d-2n+1}}\,\cdots\,\f_H\,_{a_{d-2}a_{d-1}}\, \f_I\,_{a_d}\,,
\end{equation}
as (see e.g. \cite{Kuzenko:2020ayk})
\begin{equation}
    C_{2(n+1)} =\frac1{(d-2n-1)!}\,
    W_{(n+1)}{}^{a_1\cdots a_{d-2n-1}}\,
    W_{(n+1)\,a_1\cdots a_{d-2n-1}}\,.
\end{equation}
These invariant functions can be expressed
in terms of $I_n$ and $J_n$\,,
and they are also invariant under the dual group.
For our purpose, it is sufficient to consider
the first two,
\begin{equation}
    C_2=-J_2\,,
    \qquad 
    C_4=-J_4+\frac12\,I_2\,J_2\,.
\end{equation}

%%%%%%%%%%%%%%%%%%%%%%%%%%%%%%%%%%%%%%%%%%%
\section{Vectorial description of particles
in Minkowski space}
\label{sec: Minkowski}
%%%%%%%%%%%%%%%%%%%%%%%%%%%%%%%%%%%%%%%%%%%

In Section \ref{inhomogeneous orthogonal},
we have already presented the derivation of covariant actions 
from coadjoint orbits of inhomogeneous Lorentz groups. 
Let us resume our analysis with the action \eqref{inhomo Lorentz},
\ba
	S[x,p,\Sigma,A] &=& \int \Big[\,p_a\,{\rm d}\,x^a
	+\Omega^{\a\b}\,\S_{c\b}\,{\rm d}\,\S^c{}_\a
	+A\,(p_a\,p^a-\tilde\phi_C)\nn 
	&&\hspace{45pt}	+\,A^\a\,(p_a\,\S^a{}_\a-\tilde\phi_{I\,\a})
	+A^{\a\b}\,(\S_{c\b}\,\S^c{}_\a-\tilde\phi_{H\,\b\a})
	\Big]\,,
	\label{Ham act Mink}
\ea
where the vector indices $a$, $b$ run from $0$ to $d-1$,
and the spin-variable indices $\a,\b$ run from $1$ to $2M$.
Recall that $2M$ is the rank of $\phi_H$. 
In case of the usual spinning particle, $M$ corresponds essentially
to the number of rows of the Young diagram
of the (mixed-symmetry) tensor field
associated with the spinning particle under consideration.
The Hamiltonian constraints correspond to
a coadjoint orbit of the dual algebra
\begin{equation}
    \tilde{\mathfrak g} =\heis_{2M}\niplus\sp(2M,\R)\,,
\end{equation}
generated by
\begin{equation}
    T=p^2\,, \qquad M_\a=p_a\,\S^a{}_\alpha\,,
    \qquad S_{\alpha\beta}=\S_{a\alpha}\,\S^a{}_\beta\,.
\end{equation}
The dual coadjoint vector $\tilde\phi \in \tilde{\mathfrak{g}}^*$
is given by
\be
	\tilde\phi = \tilde\phi_C\,\cT + \tilde\phi_{I\,\alpha}\,\cM^\alpha
    + {\tilde\phi}_{H\,\alpha\b}\,\cS^{\alpha\beta}\,.
\ee
We can integrate out $p_a$ from the Hamiltonian type
action \eqref{Ham act Mink} to get a Polyakov-type
Lagrangian action,
\be
	L=-\frac1{2\,e}(\dot x^a+\l^\a\,\S^a{}_\a)^2
	-\l^\a\,\phi_{I\,\a}-\frac e2\,\phi_I{}^2
	+\O^{\a\b}\,\Sigma_{a\a}\,\dot \S^{a}{}_\b
    +\l^{\a\b}\,(\S_{a\b}\,\S^{a}{}_{\a}-{\tilde\phi}_{H\b\a})\,,
\ee
with $A=\frac{e}2\,{\rm d}t$, $A^\a=\l^\a\,{\rm d}t$ and $A^{\a\b}=\l^{\a\b}\,{\rm d}t$.
If the matrix $\S_{a\a}\,\S^{a}{}_\b\simeq {\tilde\phi}_{H\a\b}$ 
is invertible with the inverse $\D^{\a\b}$,
we can also remove $\l^\a$ in terms of its equation of motion to get
\ba
	L\eq -\frac{1}{2\,e}\left(\dot x^a{}^2
	-\dot x^a\,\S_{a\a}\,\D^{\a\b}\,\S_{b\b}\,\dot x^b\right)
	-\frac{e}2 \left(\tilde\phi_C-\tilde\phi_{I\a}\,\D^{\a\b}\,
 \tilde\phi_{I\b}\right)
	\nn
	&&+\,
	\S_{a\b}\left(\O^{\a\b}\,\dot\S^a{}_\a+\dot x^a\,\D^{\a\b}\,
 \tilde\phi_{I\a}\right)
	+\l^{\a\b}\,(\S_{a\b}\,\S^{a}{}_{\a}-{\tilde\phi}_{H\b\a})\,.
\ea
In the case where $\lambda^\alpha$ cannot be solved,
it is associated with a first class constraint.
Finally, when $\tilde\phi_C-\tilde\phi_{I\a}\,\D^{\a\b}\,\tilde\phi_{I\b}\neq0$,
we can solve $e$ out to find the Nambu-type action,
\ba
	L\eq -\sqrt{\left(\tilde\phi_C-\tilde\phi_{I\a}\,\D^{\a\b}\,
 \tilde\phi_{I\b}\right)     \left(\dot x^a{}^2
	-\dot x^a\,\S_{a\a}\,\D^{\a\b}\,\S_{b\b}\,\dot x^b\right)}
	\nn
	&&+\,
	\S_{a\b}\left(\O^{\a\b}\,\dot\S^a{}_\a+\dot x^a\,\D^{\a\b}\,\tilde\phi_{I\a}\right)
	+\l^{\a\b}\,(\S_{a\b}\,\S^{a}{}_{\a}-{\tilde\phi}_{H\b\a})\,.	
\ea
Even though the above action contains all the parameters $\tilde\phi_C, \tilde\phi_{I\a}$ and ${\tilde\phi}_{H\a\b}$,
the dependence on $\tilde\phi_{I\a}$ is in fact irrelevant because $\tilde\phi_{I\a}$ are non-trivial
only when $\S_{a\a}\,\S^{a}{}_\b$ is not invertible. The above type of the action
has been derived in \cite{Lyakhovich:1998ij} for
massive spinning particles.
One may even convert the spin variables into
Lagrangian \cite{Kuzenko:1994vh, Kuzenko:1995aq}
ending up with a double square root type action.

Below, we present more details of the scalar particles
($M=0$) and the spinning particles ($M=1$),
along with the classification 
of coadjoint orbits of the Poincar\'e group.
See \cite{Kosinski:2020jmd, Lahlali:2021nrf}
for explicit characterisations of the
Poincar\'e group orbits.
See also \cite{Figueroa-OFarrill:2023qty}
for a discussion of the coadjoint orbits of the Carroll group.

%***************************%
\subsection{Scalar particles}
%***************************%

In the scalar particle case, we have $\phi_{H\,ab}=0$,
that is ${\tilde\phi}_{H\,\a\b}=0$,
and the action gets simplified to the familiar form,
\be
	S[x,p,A] = \int p_a\,{\rm d}x^a
    +A\,(p_a\,p^a-\tilde\phi_{C})\,.
\ee
The dual algebra in this case is merely 
% $\tilde{\mathfrak{g}}={\rm Span}\{T=p^2\}\simeq \mathbb R$.
$\tilde{\mathfrak{g}} = \mathbb R$, generated by $T=p^2$.
The dual coadjoint orbit is given by ${\tilde\phi}=\tilde\phi_C\,\cT$
and the stabiliser is the dual algebra itself
$\tilde{\mathfrak{g}}_{\tilde\phi}=\tilde{\mathfrak{g}} = \R$.
Depending on the signature of the vector $\tilde\phi_{I\,a}$,
we have massive, massless and tachyonic particles
with mass squared given by $\phi_I{}^2$.
The corresponding stabilisers are
\begin{subequations}
\begin{equation} 
    \mathfrak{iso}(1,d-1)_{m\,\cP^0}= \R_{P_0} \oplus \so(d-1)_{[1,\dots,d-1]}\,,
\end{equation}
\begin{equation}
     \mathfrak{iso}(1,d-1)_{E\,\cP^+}
     = \R_{P_-} \oplus \iso(d-2)_{[-;1,\dots,d-2]}\,,
\end{equation}
\begin{equation}
    \mathfrak{iso}(1,d-1)_{\mu\,\cP^{d-1}}=  \R_{P_{d-1}} \oplus \so(1,d-2)_{[0,1,\dots,d-2]}\,,
\end{equation}
\end{subequations}
where the subscripts refer to a basis for the stabilisers
of the representative considered here,
see Appendix \ref{app:notations} for details on this notation.
The dimensions of these coadjoint orbits are all $2(d-1)$
implying that they describe $d$-dimensional particles.
Indeed, the Hilbert space corresponding to these particles
will consist of wave functions on a $(d-1)$-dimensional Cauchy surface, which is a Lagrangian submanifold of the phase space.
The dual coadjoint orbits are all zero-dimensional as each of them is a single point.
Remark that the massless coadjoint orbit is nilpotent
and it is dual to the trivial orbit.
In the dimension counting \eqref{dim count},
we have $\dim\,\cM = 2d$ (here $\cM=T^*\R^d$), $\dim\,{\cO}^{\tilde G}_{\tilde\phi}=0$ and $\dim\,\tilde{\mathfrak{g}}_{\tilde\phi}=1$,
and hence $\dim\,\cN_{\tilde\phi} = \dim\,\cO^G_\phi=2(d-1)$.

%*****************************%
\subsection{Spinning particles}
\label{Poincare spinning}
%*****************************%
In the spinning particle case with $M=1$,
we relabel once again $(\S^a{}_{\chi},\S^a{}_\pi)$,
the two non-trivial vectors in $\S^a{}_\a$, as $(\chi^a,\pi^a)$ 
(note here we use $\chi, \pi$ to denote both the indices
and the vectors). The resulting action reads
\ba
	&&S[x,p,\chi,\pi,A]=\int \Big[\,p \cdot {\rm d}x
	+ \pi \cdot {\rm d}\chi + A\,(p^2-\tilde\phi_{C})
	+ A^\chi\,(p\cdot \chi-\tilde\phi_{I\chi})
	+ A^\pi\,(p\cdot \pi-\tilde\phi_{I\pi}) \nn 
	&&\hspace{45pt}	
	+\,A^{\chi\chi}\,(\chi^2-{\tilde\phi}_{H\chi\chi})
	+A^{\chi\pi}\,(\chi\cdot\pi-{\tilde\phi}_{H\chi\pi})
	+A^{\pi\pi}\,(\pi^2-{\tilde\phi}_{H\pi\pi})\Big]\,,
\ea
where we used the shorthand notation
$v \cdot w = v^a\,w_a$ for contraction of Lorentz indices.
When this particle action is quantized, 
a part\footnote{In case of second class constraints, only half of the constraints can be consistently imposed at the quantum level.}
of the constraints will be associated with the field equations:
with $p_\m\,\sim\,\partial/\partial x^\m\ \text{and}\ \p_\m\,\sim\, \partial/\partial\c^\m$,
the terms $p^2, p\cdot\c, p\cdot\p, \c\cdot\p\ \text{and}\ \p^2$ will be mapped to the
d'Alembertian, gradient, divergence, spin number and trace operators respectively.
\footnote{In fact, it seems reasonable to expect that such an interpretation holds,
in some sense, for the three types of constraints appearing in the general action \eqref{Ham act Mink}:
the first one is generically a mass-shell constraint, the second one correspond to a type of transversality condition,
and the third one to symmetry and trace condition,
to be imposed on the tensor field describing the particle.}
More details of this association will be treated in our forthcoming paper.

Below, we shall provide the classification
of the coadjoint orbits $\cO^G_\phi$ of Poincar\'e 
algebra with $M=1$ and the corresponding coadjoint orbit
$\cO^{\tilde G}_{\tilde\phi}$ of the dual algebra
$\tilde{\mathfrak{g}} = \heis_2 \niplus \sp(2,\R)$.
Note that we will always assume that the labels
of the representative vectors are generic:
they are non-vanishing and different unless stated otherwise.
From the vector, 
\be
	{\tilde\phi}=\tilde\f_C\ \cT+
	\tilde\phi_{I\,\chi}\,\cM^\chi +\tilde\phi_{I\,\pi}\,\cM^\pi+ 
	{\tilde\phi}_{H\,\chi\chi}\,\cS^{\chi\chi}
	+2\,{\tilde\phi}_{H\,\chi\pi}\,\cS^{\chi\pi}
    +{\tilde\phi}_{H\,\pi\pi}\,\cS^{\pi\pi}\,,
\ee
one can read off the parameters which appear in the particle action.
The dimensions of $\cO^G_\phi$ is 
$\dim\iso(1,d-1)-\dim\,\mathfrak{g}_\phi
=\frac{d(d+1)}2-\dim\,\mathfrak{g}_\phi\,,$
whereas in the dimension counting \eqref{dim count} we find
\ba
	{\rm dim}\,\cO^G_\phi= {\rm dim}\,\cN_{\tilde\phi}
	\eq {\rm dim}\,\cM -{\rm dim}\,{\cO}^{\tilde G}_{\tilde\phi}-2\,{\rm dim}\,\tilde{\mathfrak{g}}_{\tilde\phi} \nn
	\eq {\rm dim}\,\cM -{\rm dim}\,\tilde{\mathfrak{g}}
	-{\rm dim}\,\tilde{\mathfrak{g}}_{\tilde\phi} =2(2d-3)-{\rm dim}\,\tilde{\mathfrak{g}}_{\tilde\phi} \,.
	\label{poin dim count}
\ea
As we shall see below, $\dim\,\tilde{\mathfrak g}_{\tilde\phi}$
is either $2$ or $4$ in the $M=1$ case. Therefore,
the corresponding phase spaces have dimensions
either $2(2d-4)$ or $2(2d-5)$.
Comparing these with the phase space dimensions
of scalar particles, $2(d-1)$,
they are greater by $2(d-3)$ or $2(d-4)$.
As a mechanical system, one may interpret these dimensions
directly as the number of degrees of freedom
but here lies a subtlety. As we have discussed
with the example of the compact coadjoint orbit $S^2$ of $SU(2)$,
a compact phase space does not contribute
to the continuous degrees of freedom but only discrete labels.
In the usual spinning particle case, the additional dimensions
can be  understood as `spin-orbits',
which often correspond to the compact part
of the fiber of the coadjoint orbit viewed as
a fiber bundle over the momentum orbit \cite{Rawnsley1975, Baguis1998},
contributing again only some discrete labels.

\medskip

In the following, we present the classification
of coadjoint orbits with $M=1$.

\begin{itemize}
\item 
A massive spinning particle corresponds
to the coadjoint orbit $\cO^G_\phi$
with representative and stabiliser given by
\begin{equation}\label{eq:massive_spinning}
	\phi = m\,\cP^0 + s\,\cJ^{12}\,,
    \qquad 
    \mathfrak{g}_\phi = \R_{P_0} \oplus \uu(1)_{J_{12}}
    \oplus \so(d-3)_{[3, \dots, d-1]}\,.
\end{equation}
The coadjoint orbit $\cO^G_\phi
\simeq \frac{IO(1,d-1)}{\R \times O(2)\times O(d-3)}$
can be viewed as a fiber bundle
with the massive scalar coadjoint orbit
$\frac{IO(1,d-1)}{{\mathbb R}\times O(d-1)}$
(which is the cotangent bundle of the momentum orbit)
as the base space and the spin-orbit
$\frac{O(d-1)}{O(2)\times O(d-3)}$ as the fiber.
The latter, the real Grassmannian ${\rm Gr}_{\R}(2,d-1)$,
is a compact manifold, and hence contributes only
to discrete degrees of freedom, when quantised.
With the choice of  the indices $\chi=1, \pi=2$,  
the dual coadjoint orbit $\cO^{\tilde G}_{\tilde\phi}$
is characterised by
\begin{equation}
	\tilde\phi = -m^2\,\cT + \cS^{\chi\chi}
    + s^2\,\cS^{\pi\pi}\,,
    \qquad 
    \tilde{\mathfrak{g}}_{\tilde\phi}
    = \R_{T} \oplus \uu(1)_{S_{\pi\pi}
    +s^2\, S_{\chi\chi}}\,.
\end{equation}
The quadratic and quartic Casimir functions 
of this orbit are given by
\begin{equation}
    C_2=m^2\,,
    \qquad
    C_4=m^2\,s^2\,,
\end{equation}
and, up to a shift (that should originate
from an ordering issue when quantising) reproduces
the value of the Casimir operators of the Poincar\'e
group on the irrep corresponding to a massive spinning
particle. See \cite{Lyakhovich:1998ij}
for the derivation of a related worldline action
for a massive spinning particle in flat spacetime.

\item
A massless spinning particle corresponds to the coadjoint
orbit $\cO^G_\phi$ with representative and stabiliser
\begin{equation}
	\phi=E\,\cP^++s\,\cJ^{12}\,,
    \qquad
    \mathfrak g_\phi = \big(\heis_2
    \niplus \uu(1)_{J_{12}}\big)
    \oplus \iso(d-4)_{[-;3, \dots, d-2]}\,,
\end{equation}
and where the Heisenberg algebra is generated by
\begin{equation}
    -E\,J_{-2}+\,s\,P_1, 
    \qquad 
    E\,J_{-1}+\,s\,P_2
    \qquad \text{and} \qquad
    P_-\,.
\end{equation}
With the indices $\chi=1, \pi=2$, we find the dual
coadjoint orbit $\cO^{\tilde G}_{\tilde\phi}$, 
characterised by
\begin{equation}
	\tilde\phi = \cS^{\chi\chi} + s^2\,\cS^{\pi\pi}\,,
    \qquad 
	\tilde{\mathfrak{g}}_{\tilde\phi}
	=\heis_2 \niplus \uu(1)\,.
\end{equation}
with stabiliser generated by $T$, $M_\pi$, $M_\chi$,
and $S_{\pi\pi}+s^2\,S_{\chi\chi}$.
Note that $E$ is not a proper label for the orbit $\cO^G_\phi$
because rescaling of $E$ does not change the orbit.
We can verify this in the dual coadjoint orbit
$\cO^{\tilde G}_{\tilde\phi}$: the coadjoint vector $\tilde\phi$
does not depend on $E$. The Casimir functions
of this orbit vanish: $C_2=0=C_4$\,.

\item A continuous spinning particle corresponds
to the coadjoint orbit $\cO^G_\phi$ with representative
and stabiliser
\begin{equation}\label{eq:continuous_flat}
	\phi = E\,\cP^+ + \epsilon\,\cJ^{-1}\,,
    \qquad 
	\mathfrak{g}_\phi = \R_{P_-}
    \oplus \R_{2\,\epsilon\,P_+ + E J_{-1}}
    \oplus \so(d-3)_{[2, \dots, d-2]}\,.
\end{equation}
With the indices $\chi=1, \pi=-$, we find the dual coadjoint orbit $\cO^{\tilde G}_{\tilde\phi}$ is characterised by
\begin{equation}
	\tilde\phi = -E\,\epsilon\,\cM^{\pi} + \,\cS^{\chi\chi}\,,
    \qquad 
	\tilde{\mathfrak{g}}_{\tilde\phi}
    = \R_{T} \oplus \R_{2\,M_{\p}-E\,\e\,S_{\c\c}}\,.
\end{equation}
Here again $E$ and $\e$ are not proper labels for $\cO_\phi^G$,
but only the combination $\varepsilon^2=E\,\epsilon$ is
(note that the sign of $\cM^\pi$ term
can be changed by a coadjoint action, and that
the square symbol should not be understood literally,
i.e. $\varepsilon^2$ may be either positive or negative).
Hence, the particle action involves only one parameter, 
$\varepsilon$.  The Casimir functions of this orbit
take the values
\begin{equation}
    C_2=0\,,
    \qquad
    C_4=4\,\varepsilon^4 .
\end{equation}
Remark that the massive and massless spinning particles
share the same spin part $s\,\cJ^{12}$
and $\cS^{\chi\chi}+s^2\cS^{\pi\pi}$
in the coadjoint vectors $\phi$ and $\tilde\phi$.
We may refer to this as space-like spin.
In the continuous spin case, the spin part of the coadjoint orbits
are $\e\,\cJ^{-1}$ and $\cS^{\chi\chi}$ are null-vectors,
so we may refer this as light-like spin.
Note that a particle action for continuous spin
fields was discussed in \cite{Mourad:2004fg, Edgren:2005gq, Buchbinder:2018soq}, 
which involves
4 first class constraints
corresponding to
Wigner's equations \cite{Wigner:1963wwt},
whereas
our system involves
2 first and 4 second class constraints,
which can be viewed
as a partially gauge fixed version of the former.
See e.g. \cite{Mourad:2005rt,Mourad:2006xk,Bekaert:2005in,Alkalaev:2017hvj, Alkalaev:2018bqe}
for related works.

\item 
There are three sub-categories for a tachyonic spinning particle.
The first case is the space-like spin coadjoint orbit $\cO^G_\phi$
with representative and stabiliser given by
\begin{equation}
	\phi = \mu\,\cP^{d-1} + s\,\cJ^{12}\,,
    \qquad
    \mathfrak{g}_\phi = \R_{P_{d-1}} \oplus \uu(1)_{J_{12}}
    \oplus \so(1,d-4)_{[0,3,\dots, d-2]}\,.
\end{equation}
With the indices $\chi=1, \pi=2$, we find the dual coadjoint orbit $\cO^{\tilde G}_{\tilde\phi}$ is characterised by
\begin{equation}
	\tilde\phi = \mu^2\,\cT + \cS^{\chi\chi}
    +s^2\,\cS^{\pi\pi}\,,
    \qquad
    \tilde{\mathfrak{g}}_{\tilde\phi}
    = \R_{T} \oplus \uu(1)_{S_{\pi\pi}+s^2\,S_{\chi\chi}}\,.
\end{equation}
Here, one can also note that the spin part shares
the same structure as the massive and massless spinning case.
The Casimir functions of this orbits are
\begin{equation}\label{eq:Casimir_spinning_tachyon}
    C_2=-\mu^2\,,
    \qquad
    C_4=-\mu^2 s^2 .
\end{equation}
and are related to those of a massive spinning
orbit \eqref{eq:massive_spinning} by setting $m=i\mu$,
in accordance with our interpretation as a tachyonic
spinning orbit.

\item
The second case is the time-like spin coadjoint orbit $\cO^G_\phi$,
with representative and stabiliser
\begin{equation}
	\phi = \mu\,\cP^{d-1} + \nu\,\cJ^{01}\,,
	\label{Mink st}
    \qquad 
    \mathfrak{g}_\phi = \R_{P_{d-1}} \oplus \R_{J_{02}}
    \oplus \so(d-3)_{[2, \dots, d-2]}\,.
\end{equation}
With the indices $\chi=0, \pi=1$, we find the dual coadjoint orbit $\cO^{\tilde G}_{\tilde\phi}$ is characterised by  
\begin{equation}
	\tilde\phi = \mu^2\,\cT + \cS^{\chi\chi} - \nu^2\cS^{\pi\pi}\,,
    \qquad
    \tilde{\mathfrak{g}}_{\tilde\phi} 
    = \R_{T} \oplus \R_{S_{\pi\pi}-\nu^2\,S_{\chi\chi}}\,.
\end{equation}
The Casimir functions of this orbit are given by
\begin{equation}
    C_2=-\mu^2\,,
    \qquad
    C_4=\mu^2 \nu^2\,,
\end{equation}
and one can notice that they are related to
those of the tachyonic spinning orbit
with space-like \eqref{eq:Casimir_spinning_tachyon}
by setting $s=i\,\nu$.

\item
The last case is the light-like spin
coadjoint orbit $\cO^G_\phi$ with representative
and stabiliser
\begin{equation}
	\phi=\mu\,\cP^{d-1}+\epsilon\,\cJ^{-2}\,,
	\label{Mink sl}
    \qquad
	\mathfrak{g}_\phi = \R_{P_{d-1}}
    \oplus \R_{J_{+2}} \oplus \iso(d-4)_{[+;3, \dots, d-2]}\,.
\end{equation}
With the indices $\chi=2, \pi=-$, we find that the dual
coadjoint orbit $\tilde\cO^{\tilde G}_{\tilde\phi}$
is characterised by
\begin{equation}
	\tilde\phi = \mu^2\,\cT + \,\cS^{\chi\chi}\,,
    \qquad
	\tilde{\mathfrak{g}}_{\tilde\phi}
    = \R_{T} \oplus \R_{S_{\pi\pi}}\,.
\end{equation}
The Casimir functions of this orbit read 
\begin{equation}
    C_2=-\m^2\,,
    \qquad
    C_4=0\,.
\end{equation}
\end{itemize}
Remark that except for the continuous spin particles,
all other particles are described by the action
with two parameters $\CasimirMass$ and $\CasimirSpin$
in the end:
\begin{eqnarray}
	S[x,p,\chi,\pi,A] & = & \int \Big[\,p_a\,{\rm d}\,x^a
	+\pi_a\,{\rm d}\,\chi^a +A\,(p^2+\CasimirMass)
	+A^\chi\,p\cdot \chi +A^\pi\,p\cdot \pi \nn
	&& \hspace{30pt} +\,A^{\chi\chi}\,(\chi^2-1)
	+A^{\chi\pi}\,\chi\cdot\pi
	+A^{\pi\pi}\,(\pi^2-\,\CasimirSpin)\Big]\,.
\end{eqnarray}
Massive, massless and tachyonic particles of spin $s$
are described by $\CasimirSpin=s^2$ and positive,
zero and negative values of $\CasimirMass$, respectively.
The tachyonic particles of time-like and light-like spins
are described by a negative and zero $\CasimirSpin$, 
respectively, and a negative $\CasimirMass$.

Let us also recapitulate the system of constraints
of the above particle action.
Each of constraints are associated with the generators of the dual algebra as
\be\begin{gathered}
  \c^*(T) = p^2+\CasimirMass\,,\qquad \c^*(M_\c)=p\cdot\c\,,\qquad \c^*(M_\p)=p\cdot\p\,, \\
  \c^*(S_{\c\c})=\c^2-1\,,\qquad \c^*(S_{\c\p})=\c\cdot\p\,,\qquad \c^*(S_{\p\p})=\p^2-\CasimirSpin\,,
\end{gathered}\ee
and they are either first or second class constraints depending on the value of $\CasimirMass$ and $\CasimirSpin$.
For the massive and tachyonic cases with $\CasimirMass\neq0$, 
we have two first class constraints,
\be
  \c^*(T)\,,\qquad \c^*(S_{\p\p}+\CasimirSpin\,S_{\c\c})\,,
\ee
forming the Lie algebra $\R\oplus\mathfrak{u}(1)$,
while the other $4$ constraints are second class.
For the massless case with $\CasimirMass=0$, we have four first class constraints,
\be
  \c^*(T)\,,\qquad \c^*(S_{\p\p}+\CasimirSpin\,S_{\c\c})\,,\qquad \c^*(M_\c)\,,\qquad \c^*(M_\p)\,,
\ee
forming the Lie algebra $\mathfrak{heis}_2 \niplus \mathfrak{u}(1)$,
while the other $2$ constraints are second class.

%*************************************************%
\subsection{Spinning particles with mixed symmetry}
%*************************************************%
The coadjoint orbits with higher $M$ correspond typically
to spinning particles with mixed symmetry, characterised
by a $M$-row Young diagram. In the following, we shall provide 
the representative vectors $\phi$ of the coadjoint orbits
with higher $M$ and their stabilisers $\mathfrak{g}_\phi$.
The stabilisers of the dual algebra
$\tilde{\mathfrak{g}}_{\tilde\phi}$ are always isomorphic
to the $d$-independent part of $\mathfrak{g}_\phi$.

The coadjoint orbit of a mixed symmetry spinning particle is described by a representative vector
where the space-like spin $s\,\cJ^{12}$ is replaced by % $s_1\,\cJ^{12}+\cdots+s_M\,\cJ^{2M-1\,2M}$\,.
\begin{equation}
    s_1\,\cJ^{12}+s_2\,\cJ^{34}+\cdots+s_M\,\cJ^{2M-1\,2M}\,,
    \qquad 
    M \leq [\tfrac d2]\,,
\end{equation}
where $[x]$ denotes the integer part of $x$,
and where we can also assume that 
\begin{equation}
    s_1 \geq s_2 \geq \dots \geq s_M\,,
\end{equation}
without loss of generality. If $s_k \in \mathbb N$,
then this defines a Young diagram. In order to take
into account the possibility that several consecutive
rows of the diagram have the same length, i.e.
\begin{equation}
    \ell_1 := s_{1}=\cdots =s_{h_1}\,,
    \qquad
    \ell_2 := s_{h_1+1}=\cdots=s_{h_1+h_2}\,,
\end{equation}
and so on, it is convenient to describe
the diagram in terms of blocks of width $\ell_k$
and height $h_k$, as illustrated in Figure \ref{fig:block} below.
\begin{figure}
    \centering
    \begin{tikzpicture}
        \draw[thick] (0,4) rectangle (5,2)
            node[pos=0.5] {$\leftarrow\hspace{40pt}\ell_1\hspace{40pt}\rightarrow$};
        \draw[dashed] (5,4) -- (5.5,4);
        \node at (5.5,3.85) {$\uparrow$};
        \node at (5.5,3) {$h_1$};
        \node at (5.5,2.15) {$\downarrow$};
        \draw[dashed] (5,2) -- (5.5,2);
        \draw[thick] (0,2) rectangle (3.5,0)
            node[pos=0.5] {$\leftarrow\qquad\ell_2\qquad\rightarrow$};
        \draw[dashed] (3.5,0) -- (5.5,0);
        \node at (5.5,1.85) {$\uparrow$};
        \node at (5.5,1) {$h_2$};
        \node at (5.5,0.15) {$\downarrow$};
        \draw[thick, dashed] (0,0) -- (0,-1);
        \draw[thick, dashed] (3.5,0) -- (2,-1);
        \draw[thick] (0,-1) rectangle (2,-2.5)
            node[pos=0.5] {$\leftarrow\ \ell_p\ \rightarrow$};
        \draw[dashed] (2,-1) -- (5.5,-1);
        \node at (5.5,-1.15) {$\uparrow$};
        \node at (5.5,-1.75) {$h_p$};
        \node at (5.5,-2.35) {$\downarrow$};
        \draw[dashed] (2,-2.5) -- (5.5,-2.5);
    \end{tikzpicture}
    \caption{Young diagram presented in block form.}
    \label{fig:block}
\end{figure}
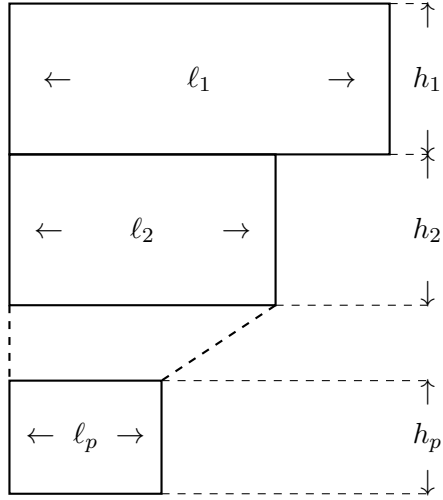
The stabilisers of such spinning particles are given by
\begin{equation}
    \mathfrak g_\phi = \R \oplus \mathfrak{u}(h_1)
    \oplus \dots \oplus \mathfrak{u}(h_p)
    \oplus \left\{
        \begin{aligned}
            \so(d-1-2M) \qquad
            & \quad \text{[massive],}\\
            \so(1,d-2-2M) \quad
            & \quad \text{[tachyonic],}
        \end{aligned}
    \right.
\end{equation}
and in the massless case, 
\begin{equation}
    \mathfrak g_\phi = \heis_{2M} \niplus \Big(\mathfrak{u}(h_1)
    \oplus \dots \oplus \mathfrak{u}(h_p)\Big) \oplus \iso(d-2-2M)\,,
\end{equation}
where $h_1+\dots+h_p=M$. 

In the case of the coadjoint orbits
with light-like or time-like spin, 
simply the $M-1$ space-like spins
are added on top of the former.
We therefore find the stabilisers
\begin{equation}
    \mathfrak{g}_\phi = \R \oplus \R
    \oplus \uu(h_1) \oplus \dots
    \oplus \uu(h_p) \oplus \left\{
    \begin{aligned}
        \so(d-1-2M)
        & \quad \text{[continuous spin} \\
        & \qquad \text{and time-like tachyonic],}\\
        \iso(d-2-2M)
        & \quad \text{[light-like tachyonic],}
    \end{aligned}
   \right.
\end{equation}
where $h_1+\dots+h_p=M-1$.

%*************************%
\subsection{Null particles}
%*************************%
The last class of particles with Poincar\'e symmetry 
are what we refer to as `null' particles,
corresponding to the coadjoint orbit with $\phi_I=0$\,.
Clearly, the condition $\phi_I=0$
trivializes the ideal --- translational --- part of 
the Poincar\'e algebra, and hence such orbits
are simply identical to Lorentz coadjoint orbits.
Upon quantisation, these coadjoint orbits
would correspond to the unfaithful representations
of Poincar\'e with $p_\m=0$ and the little group $SO(1,d-1)$,
hence again reduces to unitary irreducible representations
of Lorentz group. Since the Lorentz group can be viewed
as dS group of one lower dimensions, the classification
of null coadjoint orbits are the same as 
the classification of dS coadjoint orbits.
The only differences are in the interpretation.
Even though the null coadjoint orbits
seem somewhat dull in their defining nature, 
they may capture some important peculiarities
of Poincar\'e symmetry because analogous dull orbits
are not present for (A)dS symmetry. In fact,
the null particles can be interpreted as
the `soft limit' of massless particles.
We shall come back to this point in 
Section \ref{sec: inclusion ss}.
Let us conclude this section by remarking that,
in a sense, these null particles can be viewed
as a kind of flat space analogue of the AdS singleton.
See  \cite{Fronsdal:1986ui, Ponomarev:2022ryp, Bekaert:2022oeh}
for more serious proposals concerning this issue.

%%%%%%%%%%%%%%%%%%%%%%%%%%%%%%%%%%%%%%%%%%%%%%%%%%%%%%%%%%%
\section{Vectorial description of particles in (A)dS space}
\label{sec: AdS}
%%%%%%%%%%%%%%%%%%%%%%%%%%%%%%%%%%%%%%%%%%%%%%%%%%%%%%%%%%%

For the  covariant description of various particles
in dS and AdS spaces,
we begin with the covariant action \eqref{O action}
of the orthogonal groups $O(p,N-p)$:
$O(1,d)$ for dS and $O(2,d-1)$ for AdS.
The indices $\amb A, \amb B$ take values $0,1,\ldots, d-1$
and $\bullet$ ($\bullet=d$ for dS
and $\bullet=0'$ for AdS).
The metric $\eta_{\amb{AB}}$is ${\rm diag}(-1,+1,\ldots,+1,\s)$
where $\s=+1$ for dS and $-1$ for AdS.
Relabeling the variables as
\be
    X^{\amb A}{}_\bullet = X^{\amb A}\,,
    \qquad X^{\amb A}{}_\beta\,\phi_{P}{}^\beta
    = \tfrac12\,P^{\amb A}\,,
	\qquad X^{\amb A}{}_\alpha
    = \Sigma^{\amb A}{}_\alpha\,,
    \qquad [\alpha \ge 3]\,,
\ee
the action can be expressed in a more familiar form,
\ba
    S[X,P,\Sigma,A]
	\eq \int \Big[P \cdot {\rm d}X
	+ \Omega^{\alpha\beta}\,\S_\beta \cdot {\rm d}\S_{\alpha}\nn
	&&\quad +\,A^{XX}(X^2-{\tilde\phi}_{XX})
    + A^{PP}(P^2-{\tilde\phi}_{PP})
    + A^{XP}(X \cdot P-{\tilde\phi}_{XP})\\
	&&\quad +\,A^{X\a}(X \cdot \S_\a-{\tilde\phi}_{X\a})
    +A^{P\a}(P \cdot \S_\a-{\tilde\phi}_{P\a})
	+A^{\a\b}\big(\S_\b \cdot \S_\a-{\tilde\phi}_{\a\b}\big)\Big]\,,\nonumber
	\label{Ham act AdS}
\ea
where $X^{\amb A}$ and $P_{\amb A}$ will play the role
of the ambient space position and momentum.
Here again we used the notation
$v \cdot w = v^{\amb A}\,w_{\amb A}$ for contraction
of ambient indices. The Hamiltonian constraints
are associated with the dual algebra,\footnote{Note that
in this section, we use the convention
that the rank of the dual group is $M+1$,
different from the convention used
in Section \ref{orthogonal groups}
where the rank was $M$.}
\begin{equation}
	\tilde{\mathfrak{g}} = \mathfrak{sp}(2(M+1),\R)\,,
\end{equation}
generated by
\begin{subequations}
\begin{equation}
    T := P^2\,, \quad U := X^2\,, \quad V := X \cdot P\,,
\end{equation}
\begin{equation}
    M_\a := P \cdot \S_\a\,,
    \quad N_\a := X \cdot \S_\a\,,
    \quad S_{\a\b} := \S_{\a} \cdot \S_\b\,.
\end{equation}
\end{subequations}
Note that the dimensions of $\tilde{\mathfrak{g}}$ is $(M+1)(2M+3)$,
and differs from the dimension of the dual algebra of Poincar\'e
$\heis_{2M} \niplus \mathfrak{sp}(2M,\mathbb R)$,
which is $(M+1)(2M+1)$, by $2(M+1)$.\footnote{Let us remark
that the dual of the Poincar\'e algebra appears as a subalgebra
of the In\"on\"u--Wigner contraction
of $\mathfrak{sp}\big(2(M+1),\mathbb R\big)$ that preserves
an $\mathfrak{sp}(2M,\mathbb R)$ subalgebra. More precisely,
this contraction yields a semi-direct sum
$\mathfrak{sp}(2M,\mathbb R) \inplus \mathfrak{n}_{2M}$
where $\mathfrak{n}_{2M}$ is a nilpotent Lie algebra,
made out of two copies of $\heis_{2M}$ and
a central term, and these two Heisenberg algebras
only commute with one another up to this central term.}
This corresponds to the number of additional constraints necessary
to bring the ambient space to the intrinsic (A)dS.
In the following, the parameters in the action can be read off from 
the dual coadjoint vector ${\tilde\phi} \in \tilde{\mathfrak{g}}^*$,
\be
	{\tilde\phi}={\tilde\phi}_{PP}\, \cT+
	{\tilde\phi}_{XX}\,\cU+{\tilde\phi}_{XP}\,\cV
	+{\tilde\phi}_{P\a}\,\cM^\a+{\tilde\phi}_{X\a}\,\cN^\a+
	 {\tilde\phi}_{\a\b}\,\cS^{\a\b}\,.
\ee

Similarly to the Minkowski case, we can integrate out $P_{\amb A}$
from the Hamiltonian type action \eqref{Ham act AdS} 
to get a Polyakov-type Lagrangian action,
\ba
	L \eq -\frac1{2\,e}\,(\cD_t\,X^{\amb A}+\l^\a\,\S^{\amb A}{}_\a)^2
	-\l^\a\,{\tilde\phi}_{P\a}-\frac e2\,{\tilde\phi}_{PP}
	+\O^{\a\b}\,\Sigma_{\a} \cdot \dot \S_\b\nn
    &&+\,\l^{\a\b}\,(\S_{\a} \cdot \S_\b-{\tilde\phi}_{H\b\a})
    +\rho\,(X^2-{\tilde\phi}_{XX})-\tau\,{\tilde\phi}_{XP}
    +\t^\a\,(X \cdot \S_{\a}-{\tilde\phi}_{X\a})\,,
\ea
where $\cD_t\,X^{\amb A}=\dot X^{\amb A}+\t\,X^{\amb A}$
and the components of the gauge fields are
$A^{XX}=\rho\,{\rm d}t$,
$A^{PP}=\frac e2\,{\rm d}t$, 
$A^{XP}=\tau\,{\rm d}t$,
$A^{X\a}=\t^\a\,{\rm d}t$,
$A^{P\a}=\l^\a\,{\rm d}t$
and 
$A^{\a\b}=\l^{\a\b}\,{\rm d}t$.
As we shall see, we can always choose
a representative vector with ${\tilde\phi}_{XP}=0$.
In such a case, the equation for $\tau$ simply
reduces to $\t\,X^2+X\cdot\dot X + \l^\a\,\S_\a\cdot X=0$.
If $\S_{\a} \cdot X \simeq {\tilde\phi}_{X\a}=0$ and $X^2\simeq {\tilde\phi}_{XX}\neq 0$, 
we can remove $\tau$ to get
\be   
    \cD_t X^{\amb A} = \bar\cD_t X^{\amb A}
    = \dot X^{\amb A}
    -\frac{X \cdot \dot X}{X^2}\,X^{\amb A}\,.
\ee 
Note that this expression is the pullback to the worldline
of the ambient lift of an (A)dS covariant derivative.
If the matrix 
$\S_{\a} \cdot \S_\b \simeq {\tilde\phi}_{\a\b}$ 
is invertible with the inverse $\D^{\a\b}$,
we can remove $\l^\a$ to get
\ba
	L \eq -\frac{1}{2\,e}\left(\bar\cD_t X^2
	-\bar\cD_t X \cdot \S_{\a}\,\D^{\a\b}\,
    \S_{\b} \cdot \bar\cD_t X\right)
	-\frac{e}2 \left({\tilde\phi}_{PP}-{\tilde\phi}_{P\a}\,
    \D^{\a\b}\,{\tilde\phi}_{P\b}\right)\nn
	&& +\,\S_{\b} \cdot \left(\O^{\a\b}\,\dot\S_\a
    + \bar\cD_t X\,\D^{\a\b}\,{\tilde\phi}_{P\b}\right)
	+\l^{\a\b}\,(\S_{\a} \cdot \S_\b-{\tilde\phi}_{\b\a})\nn
    && +\,\rho\,(X^2-{\tilde\phi}_{XX}) + \t^\a\,X \cdot \S_{\a}\,,
\ea
and for ${\tilde\phi}_{PP}
-{\tilde\phi}_{P\a}\,\D^{\a\b}\,{\tilde\phi}_{P\b}\neq0$,
we can further remove $e$ to get the Nambu-type action,
\ba
	L \eq -\sqrt{\left({\tilde\phi}_{PP}
    -{\tilde\phi}_{P\a}\,\D^{\a\b}\,{\tilde\phi}_{P\b}\right)
    \left(\bar\cD_t X^2
	-\bar\cD_t X \cdot \S_{\a}\,\D^{\a\b}\,
    \S_{\b} \cdot \bar\cD_t X\right)}\nn
	&& +\,\S_{\b} \cdot \left(\O^{\a\b}\,\dot\S_\a
    +\bar\cD_t X\,\D^{\a\b}\,\phi_{I\b}\right)
	+\l^{\a\b}\,(\S_{\a} \cdot \S_\b-{\tilde\phi}_{\b\a})\nn
    &&+\, \rho\,(X^2-{\tilde\phi}_{XX})+\t^\a\,X \cdot \S_{\a}\,.
\ea
For the usual spinning particle type,
a similar type of action
has been derived in \cite{Kuzenko:1994ju,Kuzenko:1995aq}.

In the following, we present the coadjoint vectors
of (A)dS algebra using a basis which singles out the Lorentz subalgebra
and with remaining, \emph{transvection}, generators
defined as
\begin{equation}
	P_a=\tfrac1\ell\,J_{\bullet a}\,,
	\qquad 
	[P_a,P_b]=\tfrac\sigma{\ell^2}\,J_{ab}\,,
    \qquad
    [a,b=0,1,\dots,d-1]\,,
\end{equation}
where $\ell$ is the (A)dS radius and its dual
$\cP^a=\ell\,\cJ^{\bullet a}$\,, in order to make
the analogy with the Minkowski case manifest.
From now on, we set $\ell=1$ for simplicity.
As we shall see below, many cases can be viewed
as the (A)dS counterparts of the Poincar\'e orbits,
but there are also several cases which
\emph{do not} have a Poincar\'e analogue.
For the purpose of comparison between them, 
it will be convenient to parameterise the Casimir
functions in terms of $I_2$ and $I_4$
(defined in eq. \eqref{In} previously) as
\begin{equation}
    C_2=\sigma\,\tfrac12\,I_2\,,
    \qquad 
    C_4=\tfrac18\,(I_2)^2-\tfrac14\,I_4\,.
\end{equation}
This will also be useful to compare with
the results in literature. 
For explicit characterisation of some (A)dS
coadjoint orbits, see e.g. 
\cite{Kosinski:2022qrf,Enayati:2022hed,Enayati:2023lld}.

%***************************%
\subsection{Scalar particles}
%***************************%
In the scalar particle case,
we can always set ${\tilde\phi}_{XX}=\s$
and ${\tilde\phi}_{XP}=0$\,, so that 
the action depends only on ${\tilde\phi}_{PP}$\,:
\begin{equation}
	S[X,P,A] = \int \Big[P \cdot {\rm d}X
	+A^{XX}(X^2-\s)+A^{PP}(P^2-{\tilde\phi}_{PP})
    +A^{XP}\,X \cdot P\Big]\,,\
\end{equation}
where the Hamiltonian constraints are associated with the dual algebra $\sp(2,\R)$.
From the constraint $X^2=\s$, we can naturally interpret $X^{\amb A}$ as the ambient space coordinate for (A)dS spacetime. The condition $X \cdot P=0$ can be understood as a fixed homogeneity,
and finally $P^2={\tilde\phi}_{PP}$ is the mass-shell constraint.
See e.g. \cite{Engquist:2005yt} for related analysis and discussions.
The coadjoint orbits of the dual $\sp(2,\R) \simeq \so(2,1)$
are given by the two-dimensional surfaces
$H^2(\sigma\,{\tilde\phi}_{PP})$.
For more details, let us introduce a $\sigma$-dependent
notation for the one-dimensional Lie group $I(\sigma)$,
\be
	I(+1) = U(1)\,, \qquad I(-1) = \R\,,
\ee
and the associated Lie algebra $\mathfrak{i}(\sigma)$:
$\mathfrak{i}(+1) = \uu(1)$ and $\mathfrak{i}(-1) = \R$\,.
In the following, we match each one of the three types
of (A)dS scalar orbits --- massive, massless and tachyonic ---
with one of the three types of $\sp(2,\R) \simeq \so(2,1)$
orbits $H^2(a)$ defined in \eqref{2d surface}.
\begin{itemize}
\item The (A)dS orbit of the massive particle is
given by
\begin{equation}
    \phi = m\,\cP^0\,,
    \qquad
    \mathfrak{g}_\phi = \mathfrak{i}(-\sigma)_{P_0}
    \oplus \so(d-1)_{[1,\dots, d-1]}\,,
\end{equation}
while the  dual $\mathfrak{sp}(2,\mathbb R)$ orbit is given by 
\begin{equation}
    \tilde\phi = -m^2\,\cT + \sigma\,\cU\,,
    \qquad 
    \tilde G_{\tilde\phi} = I(-\sigma)\,,
\end{equation}
has the geometry of hyperboloid $H^2(-\s\,m^2)$.
\item The massless particle orbit has representative and stabiliser,
\begin{equation}
    \phi = E\,\cP^+\,,
    \qquad 
    \mathfrak{g}_\phi = \R_{P_-} \oplus \iso(d-2)_{[-;1,\dots,d-2]}\,.
\end{equation}
The dual orbit is characterised by
\begin{equation}
    \tilde\phi = \sigma\,\cU\,,
    \qquad 
    \tilde G_{\tilde\phi} = \R
\end{equation}
and corresponds to the cone $C^2$.
\item Lastly, the tachyonic particle orbit is given by
\begin{equation}
    \phi=\mu\,\cP^1\,,
    \qquad 
    \mathfrak{g}_\phi = \mathfrak{i}(\sigma)_{P_1}
    \oplus \so(1,d-2)_{[0,2,\dots,d-1]}\,.
\end{equation}
The dual orbit has
\begin{equation}
    \tilde\phi = \mu^2\,\cT + \sigma\,\cU\,,
    \qquad 
    \tilde G_{\tilde\phi} = I(\sigma)\,,
\end{equation}
and corresponds to the hyperboloid $H^2(\sigma\,\mu^2)$.
\end{itemize}
Remark that the map of massive and tachyonic
coadjoint orbits of spacetime symmetry to the one-sheet
and two-sheet hyperbolic coadjoint orbits of $\sp(2,\R)$
works oppositely for AdS and dS.
Note that $C_2={\tilde\phi}_{XP}^2-{\tilde\phi}_{XX}\,{\tilde\phi}_{PP}$ 
is a constant on these orbits. Massive scalar orbits in dS 
and tachyonic scalar orbits in AdS  
have positive $C_2$, and they can be given by
different representative vectors with ${\tilde\phi}_{PP}=0$\,:
$({\tilde\phi}_{PP},{\tilde\phi}_{XX},{\tilde\phi}_{XP})=(0,1,m)$ or $(0,-1,\mu)$. 
 Massive scalar orbits in AdS and tachyonic scalar orbits in dS
 have negative $C_2$, and they do not contain
 a vector with ${\tilde\phi}_{PP}=0$\,.
 However, if we insist on it naively, they could be given by the complex vectors 
$({\tilde\phi}_{PP},{\tilde\phi}_{XX},{\tilde\phi}_{XP})=(0,-1,i\,m)$ or $(0,1,i\,\mu)$\,.
This seemingly ill-defined choice of coadjoint vector
makes  sense after quantisation: since $P_{\amb A}=-i\,\hbar\,\partial/\partial X^{\amb A}$,
it defines a homogeneity condition with a real degree of homogeneity and
hence corresponds to a more standard way to describe a AdS field using ambient space. The two choices are related by a 
complexified global transformation --- a $Sp(2,\mathbb C)$ rotation.

%*****************************%
\subsection{Spinning particles}
%*****************************%
Let us move to the spinning case with $M=1$.
Relabelling the non-trivial elements $(\S^{\amb A}{}_{\chi},\S^{\amb A}{}_\pi)=(\chi^{\amb A},\pi^{\amb A})$,
we find 
\ba
	S[X,P,\chi,\pi,A] \eq \int \Big[P \cdot \rd X
    + \pi \cdot \rd\chi + A^{XX}(X^2-\s)
    + A^{PP}(P^2-{\tilde\phi}_{PP}) + A^{XP}\,X \cdot P\nn
	&&\quad +\,A^{X\pi}(X \cdot \pi-{\tilde\phi}_{X\pi})
    +A^{X\chi}(X \cdot \chi-{\tilde\phi}_{X\chi}) \nn
	&& \hspace{75pt} +A^{P\pi}(P \cdot \pi-{\tilde\phi}_{P\pi})
    + A^{P\chi}(P \cdot \chi-{\tilde\phi}_{P\chi}) \nn
	&& \quad +\,A^{\pi\pi}\big(\pi^2-{\tilde\phi}_{\pi\pi}\big)
	+A^{\chi\pi}\big(\chi\cdot\pi-{\tilde\phi}_{\chi\pi}\big)
	+A^{\chi\chi}\big(\chi^2-{\tilde\phi}_{\chi\chi}\big)\Big]\,,
\ea
where the Hamiltonian constraints are associated
with the dual algebra $\sp(4,\R)$\,. Note that here
$\chi^{\amb A}$ and $\pi_{\amb A}$ are $(d+1)$-dimensional vectors.
Comparing the dimension counting \eqref{poin dim count}, 
we find the same result as in the Minkowski case:
\be
	\dim\,\cO_\phi = \dim\,\cM - \dim\,\tilde{\mathfrak g}
    - \dim\,\tilde{\mathfrak g}_\phi
	= 2(2d-3) - \dim\,\tilde{\mathfrak g}_{\tilde\phi}\,,
\ee
where the increase of dimension in $\cM$ is compensated
by that of $\tilde{\mathfrak{g}}$. Compared to the Minkowski
particles and also to the (A)dS scalar particles,
the association of (A)dS coadjoint orbits with the spinning
particles in (A)dS is more subtle.
Therefore, we first provide the classification
using the simple terminologies that distinguish
the causal properties of the momenta and spins. 
\begin{itemize}
\item The coadjoint orbit with time-like momenta
and space-like spins has representative and stabiliser
\begin{equation} \label{eq:ads time time}
	\phi = m\,\cP^0 + s\,\cJ^{12}\,,
    \qquad 
	\mathfrak{g}_\phi =\mathfrak{i}(-\s)_{P_0}
    \oplus \uu(1)_{J_{12}} \oplus \so(d-3)_{[3,\dots,d-1]}\,.
\end{equation}
Here, we take the non-trivial indices as $\chi=1, \pi=2$,
and find
\begin{equation}
	{\tilde\phi} = -\,m^2\,\cT + \sigma\,\cU
    + \cS^{\chi\chi} + s^2\,\cS^{\pi\pi}\,,
    \qquad 
	\tilde{\mathfrak{g}}_{\tilde\phi}
    = \mathfrak{i}(-\sigma) \oplus \uu(1)\,,
\end{equation}
where the stabiliser is generated by
$T-\sigma\,m^2\,U$ and $s^2\,S_{\chi\chi}+S_{\pi\pi}$.
The Casimir functions are given by
\begin{equation}\label{eq:AdS time space casimir}
    C_2 = m^2 - \sigma\,s^2\,,
    \qquad
    C_4 = m^2 s^2\,,
\end{equation}
which, again up to a dimension-dependent shift,
agree with those of the $\so(2,d-1)$-irreps
corresponding to massive spinning particles.
See \cite{Kuzenko:1994vh, Kuzenko:1995aq}
for the derivation of a related worldline action
of massive spinning particle in AdS. Note also 
that the above set of constraints is the same
as the ones identified and used in the treatment
of massive and (partially-)massless mixed-symmetry
fields in AdS$_d$ using BRST techniques
and the ambient space approach
\cite{Alkalaev:2009vm, Alkalaev:2011zv}.

For dS, the mass parameter is a positive real $m>0$,
and all these orbits correspond 
to massive spinning particles.

For AdS, the quantisation condition requires $m\in\N$,
and the value $m=s$ is singular because
in that case we have different stabilisers,
\begin{equation}
	\mathfrak{g}_\phi = \uu(1,1)
    \oplus \so(d-3)_{[3,\dots,d-1]}\,,
    \qquad 
    \tilde{\mathfrak{g}}_{\tilde\phi} = \uu(1,1)\,,
\end{equation}
where the $\uu(1,1)$ subalgebras are generated by
\begin{equation}
    P_1-J_{02}\,, \qquad P_2+J_{01}\,,
    \qquad P_0+J_{12}\,, \qquad P_0-J_{12}\,,
\end{equation}
and
\begin{equation}
    T + s^2\,U\,, \qquad S_{\chi\chi} + s^2\,S_{\pi\pi}\,,
    \qquad M_\chi +\,N_\pi\,, \qquad M_\pi - s^2\,N_\chi\,,
    \label{eq:massless ads}
\end{equation}
respectively. Therefore, in AdS, we can interpret the case with $m>s$
as massive spinning particles, and $m=s$
as massless spinning particle.

One may expect that the coadjoint orbits with $m=s-1, s-2,\ldots, 1$ 
correspond to the partially massless representations in AdS,
with conformal weights $\D=s+d-4, s+d-5, \ldots, d-2$.
Up to the quantum shift, $\D=m+d-3$, 
the labels of these representations seem to match those of the coadjoint orbits  with $m=s-1, s-2,\ldots, 1$.
However, these representations are not unitary
--- they are unitary only in dS ---
and  one may conclude that this class of coadjoint orbits
do lead to non-unitary representations upon quantisations.
We believe that this is not the case for the following reasons.

The coadjoint orbits with $m<s$ rather lead
to an unfamiliar class of \emph{unitary} representations
which are not of the lowest energy type:
see our forthcoming paper \cite{partII} for explicit
construction of such representations.
Luckily, the $d=3$ case can give us good lessons about this issue:
using the isomorphism $\so(2,2) \simeq \so(2,1) \oplus \so(2,1)$,
we can decompose the $\so(2,2)$ orbit 
as a product of two  $\so(2,1)$ orbits
(see Appendix \ref{app: so(2,2)} for a dictionary).
The massive spinning particle orbits with $m>s$ correspond  
to the products of two elliptic hyperboloids $H^2_\pm$
of radius $j_{\rm\sst L}=\frac{m+s}2>0$ and  $j_{\rm\sst R}=\frac{m-s}2>0$
(see \eqref{2d surface} for 
the definition of $H^2_\pm$).
Since the $H^2_\pm(j^2)$ orbit corresponds  
the lowest/highest weight representation $\cD^\pm_{j}$
with the lowest/highest weight $\pm j$\,, the massive
spinning orbits correspond to the representations
$(\cD^\pm_{j_{\rm\sst L}} \otimes \cD^\pm_{j_{\rm\sst R}})
\oplus (\cD^\pm_{j_{\rm\sst R}}\otimes \cD^\pm_{j_{\rm\sst L}})$.
At the level of representations, we find the decomposition,
\begin{equation}
	\cD^\pm_{-\frac{t-1}2} = \cD_{\frac{t-1}2}
    \oplus \cD^\pm_{\frac{t+1}2}\,,
\end{equation}
when $2j$ becomes a non-positive integer $-(t-1)$
with $t \geq 1$.
Here, $\cD_{\frac{t-1}2}$ is the $t$-dimensional representation.
And, for $m=s-t$, the quotient representations 
\begin{equation}
	(\cD^{\pm}_{s-\frac{t-1}2} \otimes \cD_{\frac{t-1}2})
    \oplus
    (\cD_{\frac{t-1}2} \otimes \cD^{\pm}_{s-\frac{t-1}2})\,,
\end{equation}
correspond to partially massless representations of depth $t$,
which are non-unitary for $t>1$. 
Here, the non-unitarity is due to the finite-dimensional 
representation $\cD_{\frac{t-1}2}$ of $\so(1,2)$ algebra.
Therefore, one might confirm once again that the orbits
with $m=0,\ldots,s-2$ lead to non-unitary representations.
However, this is not correct because the non-unitary
representation $\cD_{\frac{t-1}2}$ should arise from $S^2$,
while the orbits with $m=s-t+1$ is given by
the product space,
\begin{equation}
    \left[H^2_\pm\big((s-\tfrac{t-1}2)^2\big)
    \times H^2_\mp\big((\tfrac{t-1}2)^2\big)\right]
    \cup \left[H^2_\mp\big((\tfrac{t-1}2)^2\big)
    \times H^2_\pm\big((s-\tfrac{t-1}2)^2\big)\right]\,.
\end{equation}
This coadjoint orbit would correspond
to the \emph{unitary} representation,
\begin{equation}
	(\cD^{\pm}_{s-\frac{t-1}2}
    \otimes \cD^{\mp}_{\frac{t-1}2})
    \oplus (\cD^{\mp}_{\frac{t-1}2}
	\otimes \cD^{\pm}_{s-\frac{t-1}2})\,,
\end{equation}
whose particle interpretation is unclear
for the moment.
In Section \ref{sec: BdS}, we propose an interpretation
for this type of orbits.

We may understand this issue from a different angle:
the $O(2,2)$ group has two discrete symmetries, the
time reversal sending $m\to -m$ and the parity sending $s \to -s$.
In terms of $O(1,2)\times O(1,2)$ it corresponds to $(j_{\rm\sst L},j_{\rm\sst R})\to(-j_{\rm\sst R},-j_{\rm\sst L})$
and $(j_{\rm\sst L},j_{\rm\sst R})\to (j_{\rm\sst R},j_{\rm\sst L})$.
We can consider yet another automorphism sending $(j_{\rm\sst L},j_{\rm\sst R})\to (j_{\rm\sst L},-j_{\rm\sst R})$
or equivalently $(m, s)\to (s,m)$,
which is not an element of $O(2,2)$.
Note that this ``inversion'' --- up to a dimension related shift which would arise upon quantisation --- has been used within the context of conformal field theory \cite{Caron-Huot:2017vep}.
Therefore,  the coadjoint orbits with $m<s$ in any dimensions
would also correspond to unitary representations, which are somehow mixed
with the usual massive spinning particle through the inversion.

\item The coadjoint orbits with light-like momenta
and space-like spins are given by
\begin{equation}
    \phi = E\,\cP^+ + s\,\cJ^{12}\,,
    \qquad 
	\mathfrak{g}_\phi = \R_{P_-} \oplus \uu(1)_{J_{12}}
    \oplus \iso(d-4)_{[-;3,\dots,d-2]}\,,
\end{equation}
while, with the indices $\chi=1, \pi=2$, we find
that the dual orbit is characterised by
\begin{equation}
	{\tilde\phi} = \sigma\,\cU + \cS^{\chi\chi}
    + s^2\,\cS^{\pi\pi}\,,
    \qquad 
	\tilde{\mathfrak{g}}_{\tilde\phi} =\R_{T}
    \oplus \uu(1)_{S_{\pi\pi}+s^2S_{\chi\chi}}\,.
\end{equation}
The Casimir functions of this orbit read
\begin{equation}
   C_2=-\sigma\,s^2 \,, \qquad C_4=0\,.
\end{equation}
In comparison with the Minkowski case, these (A)dS orbits 
seem to be related to the massless spinning particles,
but we have already seen that for AdS, the massless spinning particle
is associated with $\phi=s\,\cP^0+s\,\cJ^{12}$.
In fact, we see that the stabiliser
$\tilde{\mathfrak{g}}_{\tilde\phi} = \R \oplus \uu(1)$
is smaller than that of Minkowski,
$\heis_2 \niplus \uu(1)$\,: the former
has dimension $2$ and the latter has $4$. Therefore,
these orbits are too big for a massless spinning particle,
and they just correspond to the end point of the spectrum
of the massive and tachyonic spinning particles in dS and AdS,
respectively.

\item The coadjoint orbits with light-like momenta
and spins are given by
\begin{equation}
	\phi = E\,\cP^+ + \epsilon\,\cJ^{-1}\,.
\end{equation}
Here again, $E$ and $\e$ are not separately good parameters 
but the combination $\varepsilon^2 = E\,\epsilon$
is (we can always set $E\,\e>0$ by a suitable rotation).
By analogy with the Minkowski case,
the corresponding action can be interpreted as
the action for continuous spin particles in (A)dS. 

In dS, the coadjoint vector \eqref{conti spin AdS} 
actually belongs to a massive spinning orbits with
$\varepsilon\in \sqrt{2}\,\N$\,. Rescaling $\phi$
with $J_{+-}$ we can set
\begin{equation}
    \phi = \varepsilon\,\left(\cJ^{\bullet +}
    +\cJ^{-1}\right)
    = \tfrac\varepsilon2\,\big(\cJ^{\bullet 0}
    -\cJ^{10} +\cJ^{\bullet d-1}
    +\cJ^{1\,d-1}\big)\,.
\end{equation}
Note here that only when the components $\bullet$ 
and 1 have the same signature, that is only in dS,
we can perform a $\pi/2$-rotation
in the $\bullet$--$1$ plane to get
\begin{eqnarray}
    \phi \eq \frac\varepsilon{\sqrt{2}}\,
    \left(\cJ^{\bullet 0} + \cJ^{1\,d-1}\right) \nn
    & \simeq & \frac\varepsilon{\sqrt{2}}\,
    \left(\cP^0 + \cJ^{12}\right),
\end{eqnarray}
where we interchanged the coordinate $d-1$
with the coordinate 2 by a rotation to get
a canonical form. In AdS, this cannot be done,
so the coadjoint orbit given by \eqref{conti spin AdS}
is a genuinely new one.

Lagrangians for continuous particles have been constructed
by Metsaev in \cite{Metsaev:2016lhs}
where only the case of AdS is shown to be unitary.
Our orbit classification is consistent with this result.
Below, we shall see that the continuous spin particle
in AdS belongs to a larger class of particle species
with two labels, which are also consistent with the result
of Metsaev. We shall come back to this point shortly below.

Now, focusing on the AdS case with $\sigma=-1$,
we find that the stabiliser of $\phi$ is
\begin{equation}
	\mathfrak{g}_\phi = \uu(1)_{P_0+J_{1\,d-1}}
    \oplus \R_{P_{d-1} + J_{01}}
    \oplus \so(d-3)_{[2,\dots,d-2]}\,,
\end{equation}
with the indices $\chi=1, \pi=-$, we find
that the dual coadjoint orbit is characterised by
\begin{equation}
	\tilde\phi = -\,\cU +\,\cS^{\chi\chi}
    -\,\varepsilon^2\, \cM^\pi\,, 
    \qquad
	\tilde{\mathfrak{g}}_{\tilde\phi}
    = \uu(1) \oplus \R\,,
    \label{conti spin AdS}
\end{equation} 
with stabiliser generated by
$T-\,S_{\pi\pi}+2\,\varepsilon^2\, N_\chi$
and $2\,M_{\chi}+\varepsilon^2\,(U-S_{\chi\chi})$.
Note that the sign of $\cM^\pi$ term 
is not important as it can be changed by
a conjugation, and the stabiliser
contains a $\uu(1)$ subalgebra leading
to the quantisation of $\varepsilon \in \N$.
The Casimir functions on this orbit take
the values
\begin{equation}\label{eq:ads cont casimir}
    C_2=0\,, \qquad C_4=4\,\varepsilon^4\,,
\end{equation}
which is identical (up to a multiplicative factor)
to that of the continuous spin orbit identified
in the Poincar\'e case --- in accordance
with our interpretation as the orbit corresponding
to continuous spin particle as defined by Metsaev.
The dual coadjoint vector $\tilde\phi$
provides the worldline action
for the continuous spin particle in AdS,
which is a simple ambient space
generalisation of the Minkowski one.

\item 
Coadjoint orbits with space-like momenta
have three subcases.
First, the coadjoint orbit with space-like spin is given by 
\begin{equation}
	\phi = \mu\,\cP^{d-1} + s\,\cJ^{12}\,,
    \qquad 
	\mathfrak{g}_\phi = \mathfrak{i}(\sigma)_{P_{d-1}}
    \oplus \uu(1)_{J_{12}} \oplus \so(1,d-4)_{[0,3,\dots,d-2]}\,.
\end{equation}
Here we take the non-trivial indices
as $\chi=1$ and $\pi=2$, to find for the dual orbit
\begin{equation}
	\tilde\phi = \mu^2\,\cT + \sigma\,\cU + \cS^{\chi\chi}
    + s^2\,\cS^{\pi\pi}\,,
    \qquad 
	\tilde{\mathfrak{g}}_{\tilde\phi}
    = \mathfrak{i}(\sigma)_{T + \sigma\,\mu^2\,U}
    \oplus \uu(1)_{s^2\,S_{\chi\chi}-S_{\pi\pi}}\,.
\end{equation}
The Casimir functions of this orbit are given by
\begin{equation}\label{eq:ads space space casimir}
    C_2=-\mu^2-\sigma\,s^2\,, \qquad C_4=\mu^2s^2\,,
\end{equation}
and one can notice that they agree with those
of a massive spinning orbit upon setting
$m=i\mu$.
Since $\mathfrak{i}(+1) = \uu(1)$ for dS,
the value of $\mu$ should be quantised: $\mu \in \N$\,.
When $\mu=s$, the situation becomes singular and 
the stabilisers of the pair of dual orbits
are respectively enhanced to
\begin{equation}
	\mathfrak{g}_\phi =\uu(2)
    \oplus \so(1,d-4)_{[0,3,\dots,d-2]}\,,
    \qquad 
    \tilde{\mathfrak{g}}_{\tilde\phi} = \uu(2)\,,
\end{equation}
where the $\uu(2)$ subalgebras are generated by
\begin{equation}
    P_2 + J_{1d-1}\,, \qquad P_1 - J_{2 d-1}\,,
    \qquad P_{d-1} + J_{12}\,, \qquad P_{d-1} - J_{12}\,,
\end{equation}
and 
\begin{equation}
	T + s^2\,U\,, \qquad s^2\,S_{\chi\chi} + S_{\pi\pi}\,,
    \qquad M_\chi - N_\pi\,, \qquad M_\pi + s^2\,N_\chi\,,
\end{equation}
respectively. This special case actually corresponds
to the massless spinning particles in dS.
The other lower values of $\mu=1,2,\ldots,s-1$
correspond to the partially massless spinning particles.
The remaining values $\mu=s+1,s+2,\ldots$ might correspond
to the spinning tachyons, but there is a subtlety here:
Since $\cP^{d-1}=\cJ^{d\,d-1}$ and $\cJ^{12}$ can be interchanged
by a finite rotation, there is  no genuine difference
between the parameters $\mu$ and $s$. For this reason,
we can simply assume that the smaller one among two is $\mu$
and  the greater one is $s$\,: the equal case $\mu=s$
corresponds to the massless case. In this interpretation,
there is no coadjoint action for tachyonic spinning particle in dS.
This would mean in turn that there is no unitary irrep
of spinning tachyons in dS.
In AdS, $\mu$ is a real parameter and all of them
correspond to spinning tachyons.

\item Second, the coadjoint orbit with space-like momenta
and time-like spins is given by 
\begin{equation}
\label{orbit mu nu}
	\phi = \mu\,\cP^{d-1} + \nu\,\cJ^{01}\,,
    \qquad 
	\mathfrak{g}_\phi = \mathfrak{i}(\sigma)_{P_{d-1}}
    \oplus \R_{J_{01}} \oplus \so(d-3)_{[2,\dots,d-2]}\,.
\end{equation}
Taking the non-trivial indices as $\chi=0$ and $\pi=1$,
we find for the dual orbit
\begin{equation}
	\tilde\phi = \mu^2\,\cT - \cU + \cS^{\chi\chi}
    - \nu^2\,\cS^{\pi\pi}\,.
\end{equation}
Again, the above case corresponds to a new case
only in AdS, because in dS it is the same
as the massive spinning case with $m=\nu$
and $s=\mu$. The stabiliser is given by
\begin{equation}
    \tilde{\mathfrak{g}}_{\tilde\phi}
    = \R_{T - \mu^2\,U}
    \oplus \R_{\nu^2\,S_{\chi\chi} - S_{\pi\pi}}\,.
\end{equation}
This is the AdS analogue of the tachyonic particle
with time-like spin \eqref{Mink st} in Minkowski,
in accordance with the fact that the Casimir
functions of this orbit are given by
\begin{equation}\label{eq:ads space time casimir}
    C_2=-\mu^2-\nu^2\,, \qquad C_4=\mu^2\,\nu^2\,,
\end{equation}
and hence obtained from the massive spinning orbit
\eqref{eq:ads time time} by setting $m=i\mu$ and $s=i\nu$.
Since $\cP^{d-1}$ and $\cJ^{01}$
are in the same conjugacy class,
we can assume $\mu\ge \nu$.
When $\mu=\nu$, we find yet another enhancement
of the stabilisers,
\begin{equation} \label{eq:short casimir}
	\mathfrak{g}_\phi = \mathfrak{gl}(2,\R)
    \oplus \so(d-3)_{[2,\dots,d-2]}\,,
    \qquad 
    \tilde{\mathfrak{g}}_{\tilde\phi} = \mathfrak{gl}(2,\R)\,,
\end{equation}
where the $\mathfrak{gl}(2,\R)$ subalgebras are generated by
\begin{equation}
    P_0+J_{1d-1}\,, \qquad P_1+J_{0d-1}\,,
    \qquad P_{d-1}-J_{01}\,, \qquad P_{d-1}+J_{01}\,,
\end{equation}
and
\begin{equation}
    T - \nu^2\,U\,, \qquad \nu^2\,S_{\chi\chi} - S_{\pi\pi}\,,
    \qquad M_\chi + N_\pi\,, \qquad M_\chi + \nu^2\,N_\pi\,,
\end{equation}
respectively. 
We may refer to this case 
as \emph{short tachyon}.\footnote{We will refer
to the representations having a relatively
smaller/larger size as short/long representations.
On the other hand, when we refer to the orbits,
we will use more often the geometric adjectives, small/large.}
This exotic case can be better understood
in terms of $\so(2,2)$, again.
A generic coadjoint orbit with space-like momenta
and time-like spins is mapped
to the $\so(2,1) \oplus \so(2,1)$ coadjoint orbit,
$H^2(-(\mu+\nu)^2) \times H^2(-(\m-\n)^2)$\,. For $\mu=\nu$,
remark that the last factor becomes a point and not the cone,
as the latter correspond to another orbit to be discussed below.

\item Finally, the coadjoint orbit with space-like momenta
and light-like spins is given by 
\be
	\phi = \mu\,\cP^{d-1} + \epsilon\,\cJ^{-2}\,,
\ee
which again gives a new orbit only in AdS: in dS,
it is equivalent to the light-like momenta
and space-like spins. 
The stabiliser is
\begin{equation}
	\mathfrak{g}_\phi = \R_{P_{d-1}} \oplus \R_{J_{+2}}
    \oplus \iso(d-4)_{[+;3,\dots,d-2]}\,,
\end{equation}
and taking the non-trivial indices as $\chi=2$ and $\pi=-$,
we find for the dual orbit
\begin{equation}
	\tilde\phi = \mu^2\,\cT - \cU + \cS^{\chi\chi}\,,
    \qquad 
	\tilde{\mathfrak{g}}_{\tilde\phi} = \R_{T - \mu^2\,U}
    \oplus \R_{S_{\pi\pi}}\,.
\end{equation}
The Casimir functions of this orbit,
\be
    C_2=-\m^2\,,\qquad
    C_4=0\,.
\ee
take the same values as those of the tachyonic particle
with light-like spin \eqref{Mink sl} in Minkowski, and hence can be considered as its anti-de Sitter analogue.
\end{itemize}

Remark once again that apart from the continuous spin particles,
all other particles are described by the action
with two parameters $\CasimirMass$ and $\CasimirSpin$ as
\begin{eqnarray}
	S[X,P,\pi,\chi,A] \eq \int \Big[P \cdot {\rm d}X
    + \pi \cdot {\rm d}\chi +A^{XX}\,(X^2-\s)
    +A^{PP}\,(P^2+\CasimirMass)\nn
    && \qquad +\,A^{XP}\,X \cdot P + A^{X\pi}\,X\cdot\pi
    +A^{X\chi}\,X\cdot\chi +A^{P\pi}\,P\cdot\pi
    + A^{P\chi}\,P\cdot\chi \nn
	&& \qquad +\,A^{\pi\pi}\left(\pi^2-\CasimirSpin\right)
	+A^{\chi\chi}\left(\chi^2-1\right)
	+A^{\chi\pi}\,\chi\cdot\pi\Big]\,.
\end{eqnarray}
Massive, massless and tachyonic particles of spin $s$
are  described by $\CasimirSpin=s^2$ and positive, zero
and negative values of $\CasimirMass+\sigma\,\CasimirSpin$,
respectively. Note that we have a spin-dependent shift
and this quantity is different from
the Casimir invariant $C_2=\cC_M-\s\,\cC_S$.
For $\CasimirMass=-\s\,\cC_S$, the gauge symmetry
is enhanced for $\CasimirSpin>0$ in both AdS and dS
but for $\CasimirSpin<0$ only the AdS case
shows this gauge symmetry enhancement.
The tachyonic particles of time-like and light-like spins
are described by a negative and zero $\CasimirSpin$,
respectively, and a negative $\CasimirMass+\s\,\cC_S$.
However, in dS, seemingly tachyonic particles
are all equivalent to the massive cases,
except for the scalar case.

In AdS spacetime, besides the coadjoint orbits
associated with spinning particles, we have
three additional classes of coadjoint orbits.

%*************************************************%
\subsection{Particles with entangled mass and spin}
%*************************************************%

In the previous section, we have seen three special points
where coadjoint orbits become small:
the AdS massless particle given by
$\phi=s\,(\cP^0+\cJ^{12})$,
the AdS \emph{short} tachyon with time-like spin
given by $\phi=\nu\,(\cP^{d-1}+\cJ^{02})$,
and dS massless particle 
(or \emph{short} tachyon with space-like spin)
given by $\phi=s\,(\cP^{d-1}+\cJ^{12})$.
Their stabilisers are $\uu(1,1) \oplus \so(d-3)$,
$\mathfrak{gl}(2,\R) \oplus \so(d-3)$
and $\uu(2) \oplus \so(1,d-4)$ respectively.
We can add to 
this representative vector $\phi$
a new `spin' vector taken from the dual
of the stabiliser algebra. We may limit ourselves
to take this vector from the first part
of the stabiliser (meaning the $d$-independent part),
because taking other  spin components from the latter part
will be interpreted as mixed symmetry ones.
It turns out the dS short tachyon
(or equivalently the massless spinning particle) becomes
either a non-short tachyon or it changes the spin 
depending on whether the `spin' vector is taken  
from the dual of $\su(2)$ or the central $\uu(1)$
in $\uu(2) \subset \mathfrak{g}_\phi$\,.
Also in the AdS cases, if we add an elliptic vector
of $\su(1,1)^*$ or a hyperbolic vector
of $\mathfrak{sl}(2,\R)^*$,
we do not find new coadjoint orbits
but the ones equivalent to non-small coadjoint orbits
which we already considered. Similarly, taking 
the `spin' vector from the center will end up
changing the label of the small orbits.

A new coadjoint orbit with AdS symmetry can be obtained
either from a massless one, which is elliptic, by adding 
a hyperbolic vector $\nu\,(\cP^1-\cJ^{02})\in\uu(1,1)^*$ 
or from a short tachyon with time-like spin,
which is hyperbolic, by adding an elliptic vector
$s\,(\cP^0+\cJ^{12}) \in\mathfrak{gl}(2,\R)^* $.
In either ways, the resulting orbit is given by
\be \label{eq:one parm}
    \phi = s\,(\cP^0+\cJ^{12})+\nu\,(\cP^1-\cJ^{02})\,,
\ee
and has the stabiliser,
\be
    \mathfrak{g}_\phi =
    \uu(1)_{P_0+J_{12}}
    \oplus \R_{P_1-J_{02}} \oplus \so(d-3)_{[3,\dots,d-1]}\,.
\ee
The dual coadjoint orbit is given by
\be
    \tilde\phi = -\cU + \cS^{\chi\chi}
    +2\,s\,\nu\,\cM^\pi
    +(s^2-\nu^2)\,\big(\cT
    +\cS^{\pi\pi}\big)\,,
\ee
with stabiliser,
\begin{equation}
    \tilde{\mathfrak{g}}_{\tilde\phi} = \mathfrak{u}(1) \oplus \R\,,
\end{equation}
generated by 
\begin{equation}
    \begin{aligned}
    &T- S_{\pi\pi}
    -2\,s\,\n\,N_\chi
    - (s^2-\n^2)\,
    (U+S_{\chi\chi})\,
    ,\\
    & \qquad\text{and}\qquad
    M_\pi
    -\tfrac{s\,\nu}{2}\,U
    - (s^2-\nu^2)\,N_\chi
    -\tfrac{s^2-\nu^2}{2\,s\,\nu}\,S_{\pi\pi}
    -\tfrac{(2\,s^2-\nu^2)(s^2-2\,\nu^2)}{2\,s\,\nu}\,S_{\chi\chi}\,.
    \end{aligned}
\end{equation}
The Casimir functions 
of this orbit are given by
\begin{equation}\label{eq:one parm casimir}
    C_2=2(s^2-\n^2)\,,
    \qquad
    C_4=(s^2+\n^2)^2\,.
\end{equation}
Depending on the sign of $s-\nu$, 
the corresponding particle
could be interpreted either massive ($s>\nu$),
massless ($s=\nu$) or tachyonic ($s<\nu$),
but with a rather strange spin.
In fact, it reduces to
the continuous spin particle
with $\phi=2s\,(\cP^++\cJ^{-2})$
in the massless case.
We may interpret these particles
as massive, massless and tachyonic particles
of continuous spin.

Two other orbits can be obtained in a similar fashion
by adding a nilpotent vector
proportional to $\e$, taken from  $\mathfrak{u}(1,1)^*$
and $\mathfrak{gl}(2,\mathbb R)^*$, respectively.
Firstly, the coadjoint orbit given by
\begin{equation}\label{eq:entangled massless}
    \phi = s\,(\cP^0+\cJ^{12})
    + \epsilon\,(\cP^0+\cP^1-\cJ^{12}-\cJ^{02})\,,
\end{equation}
with stabiliser
\begin{equation}
    \mathfrak{g}_\phi = \R_{P_1-J_{02}+2J_{12}}
    \oplus \uu(1)_{P_0+J_{12}} \oplus \so(d-3)_{[3,\dots,d-1]}\,,
\end{equation}
has the dual orbit given by the representative,
\begin{equation}
    \tilde\phi = -\cU-\cN^\chi
    -4\,s^2\,(\cS^{\pi\pi} + \cM^\pi)\,,
    \qquad
    \tilde{\mathfrak{g}}_{\tilde\phi}
    = \R_{M_\pi + 4\,s^2\, N_\chi}
    \oplus \uu(1)_{T+4\,s^2\, S_{\chi\chi}}\,.
\end{equation}
The Casimir functions of this orbit are 
\begin{equation}
    \label{C values ent}
    C_2=2\,s^2 \,, \qquad C_4=s^4\,,
\end{equation}
and they coincide with those
of massless spin $s$ orbit.
This orbit can be understood as follows.
When the mass value $m$
of the massive orbit \eqref{eq:ads time time}
tends to the shortening point $m=s$,
the $2(2d-4)$ dimensional massive orbit
splits into two: the massless orbit
of dimension $2(2d-5)$ and a $2(2d-4)$-dimensional
remnant orbit, corresponding to the one given by
\eqref{eq:entangled massless}. Let us contemplate
this issue in terms of representations.
The massive spinning orbit
would correspond to the irrep $\cD(m+d-3,s)$,
which in the massless limit splits into 
the massless irrep $\cD(s+d-3,s)$
and a massive one of one lower spin
$\cD(s+d-2,s-1)$ (see e.g. \cite{Heidenreich:1980xi}
and also \cite{Girardello:2002pp} for a proposal
wherein this splitting could lead massless higher spin
fields to become massive).
The Casimir operator eigenvalues
of these two irreps are identical.
In this reasoning, quantisation of the orbit
\eqref{eq:entangled massless} may give rise to
$\cD(s+d-2,s-1)$. At the same time
the latter irrep can certainly arise from
the massive spinning orbit of mass $s+1$
and spin $s-1$, with $C_2=2(s^2+1)$
and $C_4=(s^2-1)^2$\,, which are slightly different
from \eqref{C values ent}. This reflects the fact
that the quantisation of the orbit
\eqref{eq:entangled massless} is rather peculiar.
We expect that this is a common feature
of the orbits which contains a nilpotent part in it.

The above phenomenon can be better understood
from the $d=3$ case, where the orbit  
\eqref{eq:entangled massless}
corresponds to 
$(H_+^2(s^2) \times C^2_+) \cup 
(C^2_+ \times H_+^2(s^2))$.
On the other hand, the orbit of mass $s+1$
and spin $s-1$ corresponds to
$(H_+^2(s^2)\times H^2_+(1)) \cup 
(H^2_+(1) \times H_+^2(s^2))$.
The $O(2,1)$ orbit $H_+^2(1)$
can be quantised to result in the irrep $\cD^+_1$
with vanishing Casimir. The nilpotent orbit $C^2_+$ 
admits a one-parameter family of quantisation
\cite{Vasiliev:1989re}, and gives $\cD^+_{\lambda}$
with $\lambda>0$.\footnote{Here, we consider
the Fock model of deformed oscillator,
i.e. the representation space is the space
of excited oscillator states of the Fock vacuum.}
Therefore, the orbit \eqref{eq:entangled massless}
can be quantised to $\cD(s+\lambda, s-\lambda)$
with a continuous spin label $s-\lambda$.
In $3d$, all spin eigenstates are one-dimensional,
and the spin number is quantised only for
the global consistency of $O(2,2)$, i.e. as a result
of requiring to have a UIR of the \emph{group}.
In higher dimensions, for a non-(half-)integral spin,
the number of spin states cannot be finite
and the corresponding fields will have infinitely
many components. In other words, no `spin projection'
takes place. Note that for $0<\lambda<1$,
the discrete series representation $\cD^+_\l$ mixes
with the complementary series representation
which could arise by quantising
$C^2_+ \cup \{0\} \cup C^2_-$.\footnote{The complementary
series representation might be obtained from the deformed
oscillators \cite{Vasiliev:1989re} by considering
a Segal--Bargmann model instead of the Fock model.}
Here, the inclusion of the origin $\{0\}$ indicates
that the massless spin $s$ orbit is also contained in it.
The Metsaev's infinite-component field 
\cite{Metsaev:2019opn} seems to provide
the first quantised description of the above case.
See Appendix \ref{sec: Metsaev} for related discussions.
Among the one-parameter possibility of quantisation
of the remnant orbit, the discrete series representation
with $\lambda=1$, i.e. $\cD(s+1,s-1)$ in 3d, is consistent
with the splitting phenomenon of the long massive
spin $s$ representation into a massless spin $s$
and a massive spin $s-1$ representations.

Secondly, the coadjoint orbit given by
 \be \label{eq:entangled short}
    \f=\n\,(\cP^{2}+\cJ^{01})
    +\e\,(\cP^0+\cP^{2}
    -\cJ^{21}-\cJ^{01}),
\ee 
with stabiliser
\be 
    \mathfrak{g}_\phi = \R_{P_0+2J_{01}+J_{12}}
    \oplus \R_{P_2+J_{01}} \oplus \so(d-3)_{[3,\dots,d-1]}\,,
\ee 
has the dual orbit given by
\begin{equation}
    \tilde\phi = -\cU-\cN^\chi
    +4\,\nu^2\,(\cS^{\pi\pi} + \cM^\pi)\,,
    \qquad 
    \tilde{\mathfrak{g}}_{\tilde\phi}
    = \R_{M_\pi -4\,\nu^2\, N_\chi}
    \oplus \R_{T-4\,\nu^2 \,S_{\chi\chi}}\,.
\end{equation}
The Casimir functions of this orbit are 
\begin{equation}\label{eq:entangled short casimir}
    C_2 = -2\,\nu^2 \,, \qquad C_4 = \nu^4\,,
\end{equation}
which coincides with
those of the short tachyon orbit.
This orbit corresponds again to
the $2(2d-4)$ dimensional remnant
of the shortening phenomenon.

As briefly mentioned above,
Metsaev constructed 
a Lagrangian for
infinite component fields 
having independent 
quadratic and quartic Casimir values
\cite{Metsaev:2019opn} (see also \cite{Metsaev:2016lhs, Metsaev:2017cuz, Metsaev:2017ytk, Metsaev:2018moa, Metsaev:2021zdg} for further developments).
The model contains two constants
parameterising the Casimir values
and is divided into several subcases
depending on the regions
of these constants.
For all these subcases, 
the Lagrangian was generally referred 
to as continuous spin in AdS.
Comparing this 
work of Metsaev  with our classification,
the various subcases of AdS continuous spin
in \cite{Metsaev:2019opn}
corresponds to various coadjoint orbits
identified in this paper.
See Appendix \ref{sec: Metsaev} for more details.

%********************************************%
\subsection{Particles in bitemporal AdS space}
%********************************************%
\label{sec: BdS}

The coadjoint orbits with vanishing momenta, but space-like spins are given by
\begin{equation}
	\phi = m\,\cJ^{12}\,,
    \qquad
	\mathfrak{g}_\phi = \uu(1)_{J_{12}}
    \oplus \so(2,d-3)_{[0,0',3,\dots,d-1]}\,,
\end{equation}
and the dual coadjoint orbit is characterised by
\be
	{\tilde\phi}=m^2\,\cT+\cU\,,
    \qquad 
	\tilde{\mathfrak{g}}_{\tilde\phi}
   =\mathfrak{u}(1)_{T+m^2\,U}\,.
\ee
The corresponding action has the form,
\be
	S[X,P,A] = \int P \cdot {\rm d}X
	+ A^{XX}(X^2-1) + A^{PP}(P^2-m^2) + A^{XP}\,X \cdot P\,,
	\label{2t scalar}
\ee
where the ambient space condition is 
given with the opposite sign $X^2=+1$\,.
This means that the corresponding particle 
lives in a spacetime with two temporal directions.
Let us refer to this
spacetime as Bitemporal Anti de Sitter in short BdS.
Note that this case exists only for AdS because the analogue in dS is essentially the same as the  tachyonic scalar.
Moreover, the analogue orbit with time-like spins is equivalent to the  tachyonic scalar in AdS and the massive scalar in dS.

In the regard of BdS physics, let us consider the coadjoint orbit given by
\be
    \f=E\,\cJ^{1+}\,,
    \qquad
% \ee
% with the stabiliser,
% \be
	\mathfrak{g}_\phi = \R_{J_{-1}}
    \oplus \mathfrak{iso}(1,d-3)_{[-;0',2,\dots,d-2]}\,,
\ee
which is dual to the orbit characterised by
\be
	\tilde\phi = \cU\,,
    \qquad
	\tilde{\mathfrak{g}}_{\tilde\phi} = \R_{T}\,.
\ee
This orbit can be interpreted as 
``massless'' scalar in BdS,
while the previous one as massive scalar
(with $C_2=m^2>0$) in BdS.
Remark that the scalar tachyon in AdS with $\phi=\mu\,\cP^1$ can be equally interpreted
as a scalar tachyon in BdS,
and hence 
it will be more
useful to group
particles
with $\mathfrak{so}(2,d-1)$ symmetry into AdS particles,
 BdS particles
 and tachyons.
We shall comment more on
the intriguing relations between these three species
in the next section.
We admit that
there is no concrete physical context for the BdS particles (even tachyons).
However, we find useful
to use these concepts with physics flavor in analysing 
the mathematical objects that are coadjoint orbits.

We may add space-like spins to the massive
or massless scalars in BdS.
The orbit of
the massive space-like spin particle in BdS is determined by
the coadjoint vector,
\be     
    \phi = m\,\cJ^{12} + s\,\cJ^{34}\,.
\ee 
For $m > s$, the stabiliser is
\begin{equation} 
    \mathfrak{g}_\phi =
    \uu(1)_{J_{12}} \oplus \uu(1)_{J_{34}}
    \oplus \so(2,d-5)_{[0',0,5,\dots,d-1]} \,,
\end{equation}
and the dual coadjoint orbit is characterised by
\begin{equation}
    \tilde \phi = m^2\,\cT + \cU
    + s^2\,\cS^{\pi\pi} + \cS^{\chi\chi}\,,
    \qquad 
    \tilde{\mathfrak{g}}_{\tilde\phi}
    = \uu(1)_{T+m^2\,U}
    \oplus \uu(1)_{S_{\pi\pi} + s^2\,S_{\chi\chi}}\,.
\end{equation}
The Casimir functions of this orbit coincide
with those of the massive spinning orbit
\eqref{eq:AdS time space casimir}.

For $m=s$, the stabiliser and the dual stabiliser
are enhanced to 
\begin{equation}
	\mathfrak{g}_\phi = \uu(2)
    \oplus \so(2,d-5)_{[0',0,5,\dots,d-1]}\,,
    \qquad \text{and} \qquad
    \tilde{\mathfrak{g}}_{\tilde\phi} = \uu(2)\,,
    \label{eq:massless bds}
\end{equation}
where the $\uu(2)$ subalgebras are generated by
\begin{equation}
    J_{13} + J_{24}\,, \qquad J_{14}-J_{23}\,,
    \qquad J_{12} + J_{34}\,, \qquad J_{12} - J_{34}\,,
\end{equation}
and 
\begin{equation}
    T + s^2\,U\,, \qquad S_{\pi\pi} + s^2\,S_{\chi\chi}\,, 
    \qquad M_\pi + s^2\,N_\chi\,, \qquad M_\chi-N_\pi\,,
\end{equation}
respectively, and hence we can interpret
this as a massless spinning BdS particle.
We can also consider a BdS particle with
light-like momentum and space-like spin
determined by the coadjoint vector,
\begin{equation}
    \phi = E\,\cJ^{1+} + s\,\cJ^{23}\,,
    \qquad
    \mathfrak{g}_\phi =\R_{J_{-1}}
    \oplus \uu(1)_{J_{23}}
    \oplus \iso(1,d-5)_{[-;0',4,\dots,d-2]}
\end{equation}
with its  dual coadjoint orbit characterised by
\begin{equation}
    \tilde \phi= \cU + s^2\,\cS^{\pi\pi}
    + \cS^{\chi\chi}\,,
    \qquad
    \tilde{\mathfrak{g}}_{\tilde\phi}
    = \uu(1)_{S_{\pi\pi} + s^2\,S_{\chi\chi}}
    \oplus \R_{M_\pi + s^2\,N_\chi}\,.
\end{equation}
The Casimir functions of this orbit
are given by
\begin{equation}
    C_2 = s^2\,, \qquad C_4 = 0\,.
\end{equation}

In fact, the coadjoint orbits of time-like momenta
and space-like spins with $m < s$,
the ones having the risk of confusion
with the partially massless particles,
can also be regarded as spinning BdS particles.
Changing the role of $(X,P)$ and $(\chi,\pi)$
and $m$ and $s$, the action becomes
\begin{eqnarray}
	S[X,P,\chi,\pi,A] \eq \int \Big[P_a\,{\rm d}X^a
    + \pi_a\,{\rm d}\chi^a
	+A^{XX}(X^2-1)+A^{PP}(P^2-\,m^2)\nn
	&& \quad + A^{XP}\,X \cdot P +\,A^{X\pi}\,X\cdot\pi
    +A^{X\chi}\,X\cdot\chi + A^{P\pi}\,P\cdot\pi
    \label{2t spin}\\
	&& \qquad + A^{P\chi}\,P\cdot\chi
    +\,A^{\pi\pi}\left(\pi^2-s^2\right)
	+A^{\chi\chi}\left(\chi^2+1\right)
	+A^{\chi\pi}\,\chi\cdot\pi\Big]\,. \nonumber
\end{eqnarray}
The above action can be interpreted
as the action for a particle of mass $m$
and space-like spin $s$ in BdS.
Note here that the space-like spin is given by
$s\,\cJ^{0'0}$ which is inequivalent
to the space-like spin considered
in the previous two cases. In fact,
we have more types of spins in BdS.
We can exclude time-like spins in BdS
because they can be interpreted as tachyonic
particles. Otherwise, light-like
or doubly-light-like spins in BdS give us
a new class of orbits.
 
Firstly, let us consider the light-like spin
given by $\e\,\cJ^{1+}$. 
In the case of massive BdS particles,
the light-like spin is simply equivalent
to the massless BdS particles with space-like spin.
On the other hand, the massless BdS particles
with light-like spin are new ones. The orbit
is given by
\begin{equation}
    \phi = E\,\cJ^{1+} + \epsilon\,\cJ^{2+'}\,,
\end{equation}
with the stabiliser
(for the choice $E=\e=1$)
\begin{equation}
	\mathfrak{g}_\phi = \R_{J_{1-}-J_{2-'}}
    \oplus \so(2)_{J_{12}-2J_{-+'}-2J_{+-'}}
    \inplus \heis_{2(d-4)}
    \niplus \so(d-5)_{[3,\ldots,d-3]}\,,
\end{equation}
where $\heis_{2(d-4)}$ is generated by
\begin{equation}
    J_{1-}+J_{2-'},\quad 
    J_{1-'}+J_{2-},\quad 
    J_{i-},\quad  J_{i-'},
    \quad J_{--'}\,,
    \qquad [i=3,\dots,d-3]\,.
\end{equation}
Here, we take the non-trivial indices
as $\chi = 2$, $\pi = +$ to find 
that the dual coadjoint orbit is given by
\begin{equation}
	\tilde\phi =\cU+\cS^{\c\c}\,,
    \qquad 
    \tilde{\mathfrak{g}}_{\tilde\phi}
    = \R_{T+S_{\pi\pi}}
    \oplus \iso(2)_{T-S_{\p\p}, M_{\p}, N_{\p}-M_{\c}}\,.
\end{equation}
Note that in this case  $E$ and $\epsilon$ can be 
rescaled independently leaving no label for this orbit,
and hence it is a nilpotent orbit,
with vanishing Casimir functions: $C_2=0=C_4$\,.

Secondly, we can consider doubly-light-like spin.
In the massive case, the orbit is characterised by
the following representative and its stabiliser,
\begin{equation}
	 \phi = m\,\cJ^{12} + \epsilon\,\cJ^{++'}\,,
  \qquad 
  \mathfrak{g}_\phi = \uu(1)_{J_{12}}
  \oplus [\sp(2,\R) \oplus \so(d-5)_{[3,\dots,d-3]}]
  \inplus \heis_{2(d-5)}\,,
\end{equation}
where the $\sp(2,\R)$ subalgebra is generated by
\begin{equation}
    J_{+-}-J_{+'-'}\,, \qquad J_{+-'}\,, \qquad J_{-+'}\,,
\end{equation}
and the Heisenberg subalgebra by
\begin{equation}
	J_{-i}\,, \qquad J_{-'i}\,, \qquad J_{--'}\,,
    \qquad\quad
    [i=3,\dots,d-3]\,.
\end{equation}
The dual orbit is given by
\begin{equation}
	\tilde\phi = m^2\,\cT + \cU\,,
    \qquad
	\tilde{\mathfrak{g}}_{\tilde\phi}
	= \uu(1)_{U+m^2\,T}
    \oplus \sp(2,\R)_{S_{\pi\pi}, S_{\pi\chi}, S_{\chi\chi}}\,.
\end{equation}
The  Casimir functions of this orbit read
\begin{equation}
    C_2 = m^2\,, \qquad C_4 = 0\,.
\end{equation}

Finally, the massless doubly-light-like spinning
BdS particle is given by
\be 
    \phi=E\,\cJ^{1+}+
    \e \,\cJ^{-+'}\,,
\ee
with the stabiliser (for the choice $E=\e=1$),
\be 
    \mathfrak{g}_\phi = \R_{J_{-'-}}
    \oplus \R_{J_{1-}-2\,J_{+-'}}
    \oplus \iso(d-4)_{[-';2,\dots,d-3]}\,. 
\ee
The dual orbit and dual stabiliser are given by
\begin{equation}
	\tilde\phi = \cU + \cM^{\pi}\,,
    \qquad
	\tilde{\mathfrak{g}}_{\tilde\phi}
    = \R_{T-N_\chi} \oplus \R_{S_{\chi\chi}}\,.
\end{equation}
Again, the above orbit is nilpotent
as can be seen from the fact
that $E$ and $\epsilon$ can be rescaled independently,
and the Casimir functions vanish: $C_2=0=C_4$.

%**********************************************%
\subsection{Conformal particles on the boundary}
%**********************************************%

For the AdS algebra $\mathfrak{so}(2,d-1)$,
we have yet another class of coadjoint orbits,
which are very \emph{small} compared to others.
Consider the coadjoint orbit given by 
\be
	\phi = \epsilon\,\cJ^{++'}\,,
\ee
where $\pm'$ is the lightcone coordinate
from $0'$ and $d-2$. The stabiliser is
\begin{equation}
	\mathfrak{g}_\phi = \big(\sp(2,\R)
    \oplus \so(d-3)_{[1,\dots,d-3]}\big)
    \inplus \mathfrak {heis}_{2(d-3)}\,,
\end{equation}
where the $\sp(2,\R)$ subalgebra is generated by
\begin{equation}
    J_{+-}-J_{+'-'}\,, \qquad J_{+-'}\,, \qquad J_{-+'}\,,
\end{equation}
and the Heisenberg subalgebra by
\begin{equation}
    J_{-i}\,, \qquad J_{-'i}\,, \qquad J_{--'}\,,
    \qquad\quad 
    [i=1,\dots,d-3]\,.
\end{equation}
Interestingly, the dual coadjoint orbit is trivial:
\begin{equation}
	\tilde\phi = 0\,, 
    \qquad
	\tilde{\mathfrak{g}}_{\tilde\phi}
    = \sp(2,\R)_{T,U,V}\,.
\end{equation}
Here, we take $X^{\amb A}{}_+=X^{\amb A}$
and $X^{\amb A}{}_{+'}=P^{\amb A}$
to find the action
\be
	S[X,P,A] = \int \left[P \cdot {\rm d}X
    + A^{XX}\,X^2 + A^{PP}\,P^2
    + A^{XP}\,X \cdot P\right]\,.
\ee
The constraint $X^2=0$ and the homogeneity condition
$X\cdot P=0$ tells that the particle leaves
on a $(d-1)$-dimensional section of the cone $X^2=0$,
and it corresponds to the conformal particle
in $(d-1)$-dimensions. Indeed, we can see
that the dimension of the coadjoint orbit is $2(d-2)$.
See e.g. \cite{Engquist:2005yt}
for related analysis and discussions.
When quantised, this leads to the scalar singleton 
representation. See \cite{Brylinski1998, Fronsdal2009, Bekaert:2011js, Joung:2014qya} for the references. 
As we shall see below, there are also 
conformal spinning particles on the boundary.
In order to understand them, we need to discuss
about mixed symmetry cases, first.

%*************************************************%
\subsection{Spinning particles with mixed symmetry}
\label{subsec:mixed sym}
%*************************************************%

Similarly to the Poincar\'e case, the coadjoint orbits
of (A)dS algebra with higher $M$ correspond typically 
to spinning particles with mixed symmetry of $M$-row
Young diagram. Interestingly, we find the classical
analogues of various subtleties of mixed symmetry
representations in (A)dS algebra. In the following,
we present the representative vectors $\phi$
of the coadjoint orbits with higher $M$
and their stabilisers $\mathfrak{g}_\phi$.
The dual stabilisers $\tilde{\mathfrak{g}}_{\tilde\phi}$
are isomorphic to the $d$-independent part
of $\mathfrak{g}_\phi$.

In AdS$_d$, the massive and massless spinning
particles are given by
\begin{equation}
    \phi = m\,\cP^0 +s_1\,\cJ^{12} + \cdots
    + s_M\,\cJ^{2M-1\,2M}\,,   
    \label{mixed (A)dS}
\end{equation}
where $m \in \N$,
 $s_1 \ge s_2 \ge \cdots \ge s_M$, 
and
 $M \leq [\tfrac{d-1}2]$\,.
The massless point corresponds to $m=s_1$
and hence it is massive if $m > s_1$.
For  $m < s_1$,
we interpret
the coadjoint orbits as 
mixed-symmetry spinning particles in BdS, rather than
partially-massless ones
for a similar reason we explained
in the symmetric spinning case.
We will use the block notation
introduced in the previous section, where $h_k$
denotes the height of the $k$th block, of width
$\ell_k=s_{h_1+\dots+h_{k-1}+1}$, and
$h_1 + \cdots + h_p = M$. When $m=\ell_n$,
the stabiliser becomes
\begin{equation}
    \mathfrak{g}_\phi = \uu(h_1) \oplus \cdots
    \oplus \uu(1,h_n) \oplus \cdots \oplus \uu(h_p)
    \oplus \so(d-1-2M)\,,
\end{equation}
for $1 \leq n \leq p$. Therefore, we do have
a rich variety of exotic class of particles
living in BdS. The tachyonic particles with space-like
spins are generalised to 
\be 
    \phi = \mu\,\cP^{d-1}
    + s_1\,\cJ^{12} + \cdots + s_M\,\cJ^{2M-1\,2M}\,,
\label{mixed T}
\ee
with 
\be 
    \mathfrak{g}_\phi = \R \oplus \uu(h_1)
    \oplus \cdots \oplus \uu(h_p) \oplus \so(d-1-2M)\,,
\ee
and the coadjoint orbit given by 
\be 
    \phi = E\,\cP^{+} + s_1\,\cJ^{12}
+ \cdots + s_M\,\cJ^{2M-1\,2M}\,,
\label{mixed E}
\ee
with 
\be 
    \mathfrak{g}_\phi = \R \oplus \uu(h_1)
    \oplus \cdots \oplus \uu(h_p) \oplus \iso(d-2-2M)\,,
\ee
corresponds to the end point of the tachyonic spectrum.

In dS, the massive spinning particle is given again by
\eqref{mixed (A)dS} but
with $m \in \R$ and the stabiliser is
\be 
    \mathfrak{g}_\phi = \R \oplus \uu(h_1)
    \oplus \cdots \oplus \uu(h_p) \oplus \so(d-1-2M)\,.
\ee 
The coadjoint orbit given by \eqref{mixed E}
with stabiliser
\be 
    \mathfrak{g}_\phi = \R \oplus \uu(h_1)
    \oplus \cdots \oplus \uu(h_p) \oplus \iso(d-2-2M)\,,
\ee 
corresponds to the end point of the massive spectrum.
The coadjoint orbits given by the representative vector  
\eqref{mixed T}
with $\mu \in \N$ contain (partially-)massless spinning particles
 of mixed symmetry.
As discussed in the $M=1$ case,
the generator $\cP^{d-1}=\cJ^{d\,d-1}$ is
not different from any of $\cJ^{2k\,2k+1}$
and we assume that $\mu\le s_M$.
The equality $\mu=s_M$ corresponds to the massless case
whereas other cases with $\mu<s_M$ correspond
to the partially-massless cases.
Note that there are no spinning tachyons in dS.

Let us compare
our results with
the pattern of massless mixed-symmetry representations in (A)dS.
In AdS, such representations are known
to be unitary only when the gauge parameter
has the symmetry of the gauge field Young diagram
with one box removed at the bottom of the \emph{first} block
\cite{Metsaev:1995re, Metsaev:1998xg} (see also \cite{Boulanger:2008kw, Boulanger:2008up, Skvortsov:2009nv, Skvortsov:2009zu} for more details
on mixed-symmetry fields). In dS, unitarity
requires the gauge parameter to have the symmetry
of the gauge field Young diagram where one removes
$t$ boxes from the \emph{last} block:
here $t$ is the depth
of the partially-massless field.
The mass parameter
of these fields will depend on the length
of the block affected by the gauge symmetry, but not the other blocks.
This distinction seem to be reflected in the classes
of coadjoint orbits corresponding to (partially-)massless fields
in AdS or dS that we have identified here.
In this comparison,
it is important to take into account
the coadjoint orbits of BdS particles,
which would lead to unfamiliar 
classes of unitary representations.

In the case of the coadjoint orbits
with light-like or time-like spin, 
simply the $M-1$ space-like spins are added on top of the light-like or time-like spin.
The continuous spin particle exists 
only in AdS and has
the stabiliser 
$\mathfrak{g}_\phi= \mathbb R\oplus \mathbb R
\oplus \mathfrak{u}(h_1)\oplus \cdots \oplus \mathfrak{u}(h_p)\oplus
\mathfrak{so}(d-1-2M)\,.$
The tachyon with time-like spin has
the stabiliser 
\be 
    \mathfrak{g}_\phi= \mathfrak{i}(\s) \oplus 
\mathbb R
\oplus \mathfrak{u}(h_1)\oplus \cdots \oplus \mathfrak{u}(h_p)\oplus
\mathfrak{so}(d-1-2M)\,.
\ee 
The tachyon with light-like spin has the stabiliser
\be 
    \mathfrak{g}_\phi= \mathbb R\oplus \mathbb R
\oplus \mathfrak{u}(h_1)\oplus \cdots \oplus \mathfrak{u}(h_p)\oplus
\mathfrak{iso}(d-2-2M)\,.
\ee
In AdS, there are yet another class of coadjoint orbits
given by \be 
\phi=\e\,\cJ^{++'}+s_1\,\cJ^{12}+\cdots+s_M\,\cJ^{2M-1\,2M}\,,
\ee
and the stabilisers
\be 
    \mathfrak{g}_\phi= 
\mathfrak{u}(h_1)\oplus \cdots \oplus \mathfrak{u}(h_p)\oplus [\mathfrak{sp}(2,\mathbb R)\oplus
\mathfrak{so}(d-3-2M)]\inplus \heis_{2(d-3-2M)}\,.
\ee 

In AdS, besides the above orbits,
we have also mixed-symmetry extension of the coadjoint orbits with
entangled labels.
To the `mass' vector $\phi=s(\cP^0+\cJ^{12}+\cdots +\cJ^{2h-1\,2h})$ with the stabiliser $\mathfrak{u}(1,h)$,
 we can append a `spin' vector taken from $\uu(1,h)^*$.
Like in the $h=1$ case, any elliptic vector would not lead to a new orbit, but hyperbolic or nilpotent ones will
result in new coadjoint orbits.

Finally, let us comment about the particular 
case of maximal rank massless spinning particle
in an odd $D$-dimensional AdS spacetime
with 
\be         \phi=s(\cP^0+\cJ^{12}+\cdots+\cJ^{D-2\,D-1})\,.
\ee 
It stabiliser in $\mathfrak{so}(2,D-1)$ is $\mathfrak{g}_\phi
= \mathfrak{u}(1,\frac{D-1}2)$\,,
and 
the dimension of the coadjoint orbit
is $\frac{(D-1)(D+1)}4$\,.
This is to be compared with
the maximal rank
massless spinning particle in an even $d$ dimensional Minkowski 
space
with 
$\phi=E\,\cP^0+s(\cJ^{12}+\cdots+\cJ^{d-3\,d-2})$\,.
The stabiliser in $\mathfrak{iso}(1,d-1)$ is
 $\mathfrak{g}_\phi
=
\heis_{d-2}
\oplus \mathfrak{u}(\frac{d-2}2)$\,, and 
the dimension of the coadjoint orbit
is $\frac{d(d+2)}4$\,.
By matching $D=d+1$, we find that 
the two orbits have the same dimensions. 
The phenomena can be 
understood as the classical counterpart of the
peculiar branching rule of spinning singletons
\cite{Angelopoulos:1980wg, Angelopoulos:1997ij, Angelopoulos:1999bz}.

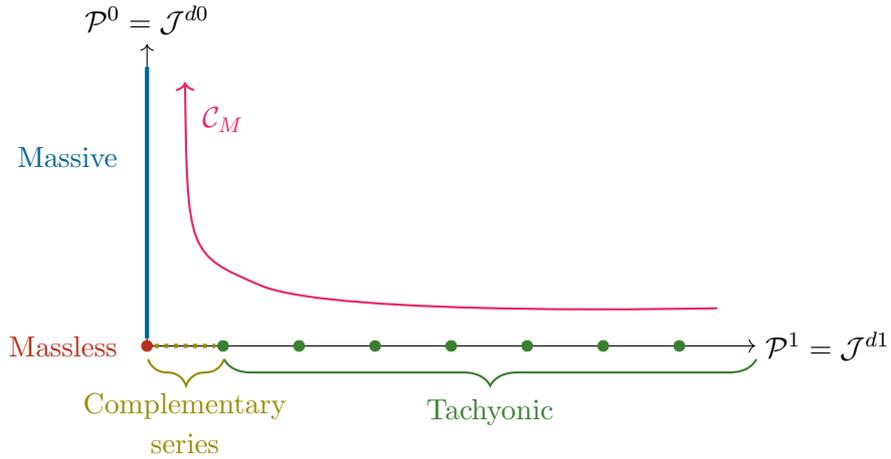
\begin{figure}[ht!]
\centering
\begin{tikzpicture}
\draw[<-, thick,WildStrawberry]
        plot[smooth] coordinates {(0.5,3.5) (1.5,0.8) (7.5,0.5)};
\node at (1,3) {\color{WildStrawberry} $\cC_M$};
\draw [->] (0,0) -- (8,0);
\node [right] at (8,0) {$\cP^1=\cJ^{d1}$};
\draw [->] (0,0) -- (0,4);
\node [above] at (0,4) {$\cP^0=\cJ^{d0}$};
\draw [MidnightBlue,ultra thick] (0,0.1) -- (0,3.7);
\draw [olive, dotted, ultra thick] (0.1,0) -- (0.9,0);
\draw [BrickRed,fill=BrickRed] (0,0) circle(.07);
\draw [OliveGreen,fill=OliveGreen] (1,0) circle(.07);
\draw [OliveGreen,fill=OliveGreen] (2,0) circle(.07);
\draw [OliveGreen,fill=OliveGreen] (3,0) circle(.07);
\draw [OliveGreen,fill=OliveGreen] (4,0) circle(.07);
\draw [OliveGreen,fill=OliveGreen] (5,0) circle(.07);
\draw [OliveGreen,fill=OliveGreen] (6,0) circle(.07);
\draw [OliveGreen,fill=OliveGreen] (7,0) circle(.07);

\draw [thick, OliveGreen,decorate,decoration={brace,amplitude=10pt,mirror},xshift=0.4pt,yshift=-5pt](1,0) -- (8,0) node[black,midway,yshift=-20pt] {\color{OliveGreen} 
Tachyonic};
\node [left] at (-0.25,2.5) {\color{MidnightBlue} Massive};
\node [left, BrickRed] at (-0.25,0) {Massless};
\draw [thick, olive,decorate,decoration={brace,amplitude=10pt,mirror},xshift=0.4pt,yshift=-5pt](0,0) -- (1,0);
\node at (0.5,-0.8) {\color{olive} Complementary};
\node at (0.5,-1.3) {\color{olive} series};
\end{tikzpicture}
\caption{Scalar particles in dS}
\label{fig:scalar_dS}
\end{figure}

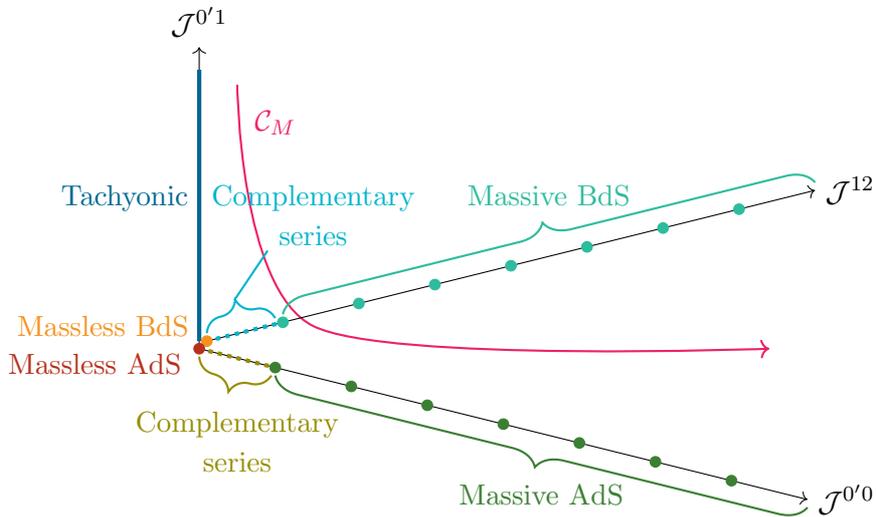
\begin{figure}[ht!]
\centering
\begin{tikzpicture}

\draw[->, thick,WildStrawberry]
        plot[smooth] coordinates {(0.5,3.5) (1.5,0.3) (7.5,0)};
\node at (1,3) {\color{WildStrawberry} $\cC_M$};

\draw [->] (0.1,0.1) -- (8.1,2.1);
\node [right] at (8.1,2.1) {$\cJ^{12}$};
\draw [->] (0,0) -- (8,-2);
\node [right] at (8,-2) {$\cJ^{0'0}$};
\draw [->] (0,0) -- (0,4);
\node [above] at (0,4) {$\cJ^{0'1}$};
\draw [MidnightBlue,ultra thick] (0,0.1) -- (0,3.7);

\draw [olive, dotted, ultra thick] (0,0) -- (1,-0.25);
\draw [Turquoise, dotted, ultra thick] (0.1,0.1) -- (1.1,0.35);

\draw [BurntOrange,fill=BurntOrange] (0.1,0.1) circle(.07);
\draw [BrickRed,fill=BrickRed] (0,0) circle(.07);
\draw [OliveGreen,fill=OliveGreen] (1,-0.25) circle(.07);
\draw [OliveGreen,fill=OliveGreen] (2,-0.5) circle(.07);
\draw [OliveGreen,fill=OliveGreen] (3,-0.75) circle(.07);
\draw [OliveGreen,fill=OliveGreen] (4,-1) circle(.07);
\draw [OliveGreen,fill=OliveGreen] (5,-1.25) circle(.07);
\draw [OliveGreen,fill=OliveGreen] (6,-1.5) circle(.07);
\draw [OliveGreen,fill=OliveGreen] (7,-1.75) circle(.07);

\draw [SeaGreen,fill=SeaGreen] (1.1,0.35) circle(.07);
\draw [SeaGreen,fill=SeaGreen] (2.1,0.6) circle(.07);
\draw [SeaGreen,fill=SeaGreen] (3.1,0.85) circle(.07);
\draw [SeaGreen,fill=SeaGreen] (4.1,1.1) circle(.07);
\draw [SeaGreen,fill=SeaGreen] (5.1,1.35) circle(.07);
\draw [SeaGreen,fill=SeaGreen] (6.1,1.6) circle(.07);
\draw [SeaGreen,fill=SeaGreen] (7.1,1.85) circle(.07);

\draw [thick, olive,decorate,decoration={brace,amplitude=10pt,mirror},yshift=-3pt](0,0) -- (0.95,-0.25); 
\node at (0.5,-1) {\color{olive} Complementary};
\node at (0.5,-1.5) {\color{olive} series};

\draw [thick, OliveGreen,decorate,decoration={brace,amplitude=10pt,mirror},yshift=-3pt](1,-0.25) -- (8,-2) node[black,midway,yshift=-20pt] {\color{OliveGreen} 
Massive AdS};

\draw [thick, Turquoise,decorate,decoration={brace,amplitude=10pt},yshift=3pt]
(0.1,0.1) -- (1,0.3);
\node at (1.5,2) {\color{Turquoise} Complementary};
\node at (1.5,1.5) {\color{Turquoise} series};
\draw [thick, Turquoise] (0.46,0.64) -- (0.9,1.3);

\draw [thick, SeaGreen,decorate,decoration={brace,amplitude=10pt},yshift=3pt](1.1,0.35) -- (8.1,2.1) node[black,midway,yshift=20pt] {\color{SeaGreen}  Massive BdS};

\node [left] at (0,2) {\color{MidnightBlue} Tachyonic};
\node [left, BrickRed] at (-0.1,-0.2) {Massless AdS};
\node [left, BurntOrange] at (0,0.3) {Massless BdS};
\end{tikzpicture}
\caption{Scalar particles in AdS}
\label{fig:scalar_AdS}
\end{figure}

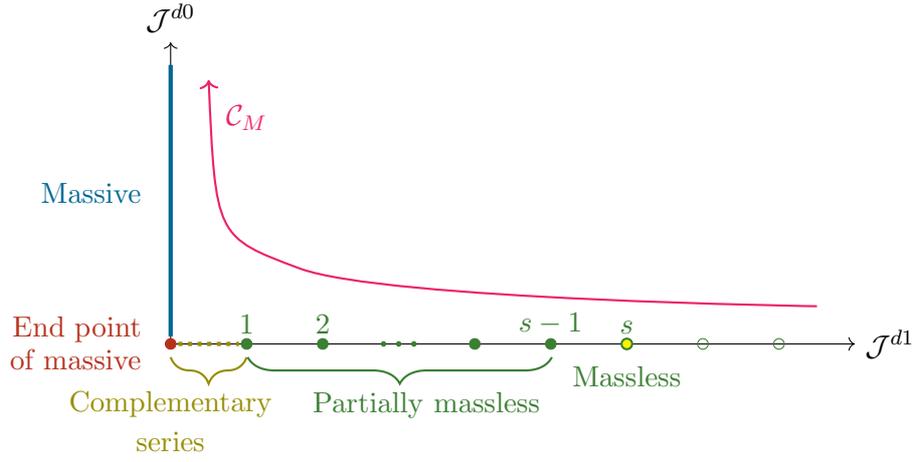
\begin{figure}[ht!]
\centering
\begin{tikzpicture}
\draw[<-, thick,WildStrawberry]
        plot[smooth] coordinates {(0.5,3.5) (1.7,1) (8.5,0.5)};
\node at (1,3) {\color{WildStrawberry} $\cC_M$};
\draw [->] (0,0) -- (9,0);
\node [right] at (9,0) {$\cJ^{d1}$};
\draw [->] (0,0) -- (0,4);
\node [above] at (0,4) {$\cJ^{d0}$};
\draw [MidnightBlue,ultra thick] (0,0.1) -- (0,3.7);
\draw [olive, dotted, ultra thick] (0.1,0) -- (0.9,0);
\draw [BrickRed,fill=BrickRed] (0,0) circle(.07);
\draw [thick, olive,decorate,decoration={brace,amplitude=10pt,mirror},xshift=0,yshift=-5pt](0,0) -- (1,0);
\node at (0,-0.8) {\color{olive} Complementary};
\node at (0,-1.3) {\color{olive} series};
\draw [OliveGreen,fill=OliveGreen] (1,0) circle(.07)
node[above] {$1$};
\draw [OliveGreen,fill=OliveGreen] (2,0) circle(.07)
node[above] {$2$};
\draw [OliveGreen,fill=OliveGreen] (2.8,0) circle(.03);
\draw [OliveGreen,fill=OliveGreen] (3,0) circle(.03);
\draw [OliveGreen,fill=OliveGreen] (3.2,0) circle(.03);
\draw [OliveGreen,fill=OliveGreen] (4,0) circle(.07);
\draw [OliveGreen,fill=OliveGreen] (5,0) circle(.07)
node[above] {$s-1$};
\draw [OliveGreen,thick, fill=yellow] (6,0) circle(.07)
node[above] {$\phantom{1}s\phantom{1}$};
\draw [OliveGreen,thin] (7,0) circle(.07);
\draw [OliveGreen,thin] (8,0) circle(.07);

\draw [thick, OliveGreen,decorate,decoration={brace,amplitude=10pt,mirror},xshift=0.4pt,yshift=-5pt](1,0) -- (5,0) node[black,midway,xshift=10pt,yshift=-18pt] {\color{OliveGreen} 
Partially massless};
\node[above] at (6,-0.7) {\color{OliveGreen} Massless};
\node [left] at (-0.25,2) {\color{MidnightBlue} Massive};
\node [left, BrickRed] at (-0.25,0.2) {End point};
\node [left, BrickRed] at (-0.25,-0.2) {of massive};
\end{tikzpicture}
\caption{Spin $s$ particles in dS}
\label{fig:spinning_dS}
\end{figure}

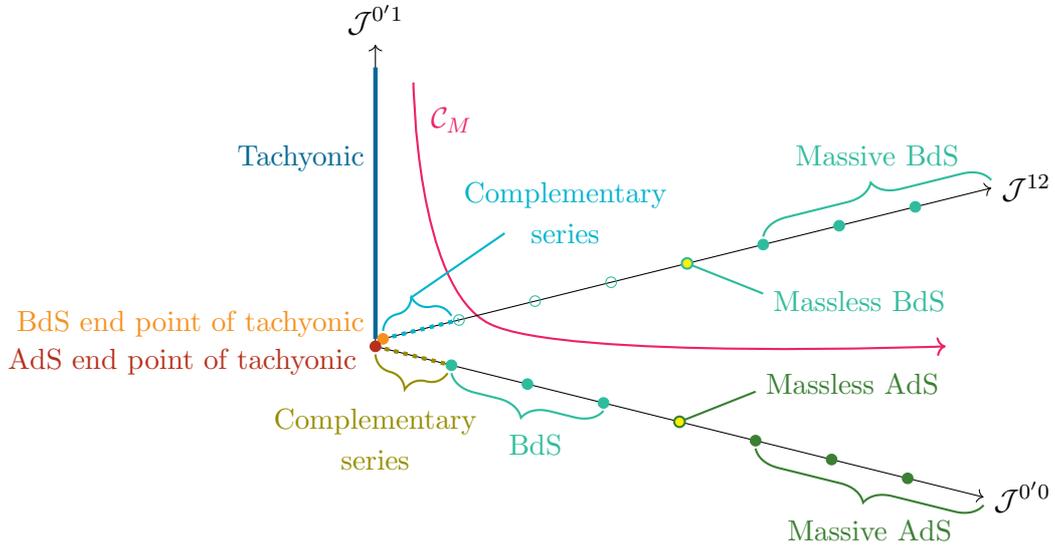
\begin{figure}[ht!]
\centering
\begin{tikzpicture}
\draw[->, thick,WildStrawberry]
        plot[smooth] coordinates {(0.5,3.5) (1.5,0.3) (7.5,0)};
\node at (1,3) {\color{WildStrawberry} $\cC_M$};
\draw [->] (0.1,0.1) -- (8.1,2.1);
\node [right] at (8.1,2.1) {$\cJ^{12}$};
\draw [->] (0,0) -- (8,-2);
\node [right] at (8,-2) {$\cJ^{0'0}$};
\draw [->] (0,0) -- (0,4);
\node [above] at (0,4) {$\cJ^{0'1}$};
\draw [MidnightBlue,ultra thick] (0,0.1) -- (0,3.7);
\draw[SeaGreen,thick] (4.1,1.1)--(5.1,0.7);
\node[right] at (5.1,0.6) {\color{SeaGreen} Massless BdS};
\draw[OliveGreen,thick] (4,-1)--(5,-0.6);
\node[right] at (5,-0.5) {\color{OliveGreen} Massless AdS};

\draw [olive, dotted, ultra thick] (0,0) -- (1,-0.25);
\draw [Turquoise, dotted, ultra thick] (0.1,0.1) -- (1.1,0.35);

\draw [BurntOrange,fill=BurntOrange] (0.1,0.1) circle(.07);
\draw [BrickRed,fill=BrickRed] (0,0) circle(.07);
\draw [SeaGreen,fill=SeaGreen] (1,-0.25) circle(.07);
\draw [SeaGreen,fill=SeaGreen] (2,-0.5) circle(.07);
\draw [SeaGreen,fill=SeaGreen] (3,-0.75) circle(.07);
\draw [OliveGreen,thick, fill=yellow] (4,-1) circle(.07);
\draw [OliveGreen,fill=OliveGreen] (5,-1.25) circle(.07);
\draw [OliveGreen,fill=OliveGreen] (6,-1.5) circle(.07);
\draw [OliveGreen,fill=OliveGreen] (7,-1.75) circle(.07);

\draw [SeaGreen,thin] (1.1,0.35) circle(.07);
\draw [SeaGreen,thin] (2.1,0.6) circle(.07);
\draw [SeaGreen,thin] (3.1,0.85) circle(.07);
\draw [SeaGreen,thick, fill=yellow] (4.1,1.1) circle(.07);
\draw [SeaGreen,fill=SeaGreen] (5.1,1.35) circle(.07);
\draw [SeaGreen,fill=SeaGreen] (6.1,1.6) circle(.07);
\draw [SeaGreen,fill=SeaGreen] (7.1,1.85) circle(.07);

\draw [thick, olive,decorate,decoration={brace,amplitude=10pt,mirror},yshift=-3pt](0,0) -- (0.95,-0.25); 
\node at (0,-1) {\color{olive} Complementary};
\node at (0,-1.5) {\color{olive} series};

\draw [thick, OliveGreen,decorate,decoration={brace,amplitude=10pt,mirror},yshift=-3pt](5,-1.25) -- (8,-2) node[black,midway,yshift=-20pt] {\color{OliveGreen} 
Massive AdS};

\draw [thick, SeaGreen,decorate,decoration={brace,amplitude=10pt,mirror},yshift=-3pt](1,-0.25) -- (3,-0.75) node[black,midway,xshift=3pt,yshift=-20pt] {\color{SeaGreen} 
BdS};

\draw [thick, Turquoise,decorate,decoration={brace,amplitude=10pt},yshift=3pt]
(0.1,0.1) -- (1,0.3);
\node at (2.5,2) {\color{Turquoise} Complementary};
\node at (2.5,1.5) {\color{Turquoise} series};
\draw [thick, Turquoise] (0.46,0.64) -- (1.7,1.5);

\draw [thick, SeaGreen,decorate,decoration={brace,amplitude=10pt},yshift=3pt](5.1,1.35) -- (8.1,2.1) node[black,midway,yshift=20pt] {\color{SeaGreen}  Massive BdS};

\node [left] at (0,2.5) {\color{MidnightBlue} Tachyonic};
\node [left, BrickRed] at (-0.1,-0.2) {AdS end point of tachyonic};
\node [left, BurntOrange] at (0,0.3) {BdS end point of tachyonic};
\end{tikzpicture}
\caption{Spin $s$ particles in AdS}
\label{fig:spinning_AdS}

\end{figure}

Let us conclude this section
with the figures which
summarise the spectra of
scalar 
(Figures \ref{fig:scalar_dS}
and \ref{fig:scalar_AdS})
and spinning (symmetric
space-like) particles in (A)dS
(Figures \ref{fig:spinning_dS}
and \ref{fig:spinning_AdS}), 
both in terms of the representatives
and the `mass squared' $\CasimirMass$.
We also indicated the regions
excluded by the quantisation condition
which nevertheless
should be associated to a class of 
unitary and irreducible representations
usually referred to as the complementary series.

%%%%%%%%%%%%%%%%%%%%%%%%%%%%%%%%%%%%%%%%%%%%
\section{Inclusion structure and soft limit}
\label{sec: inclusion}
%%%%%%%%%%%%%%%%%%%%%%%%%%%%%%%%%%%%%%%%%%%%

A coadjoint orbit may be contained in 
the closure of a larger coadjoint orbit. 
The simplest example is that
the inclusion of the origin, the trivial orbit,
in the closure of the conical nilpotent orbit of $O(2,1)$
(see Figures \ref{dS Hasse} and \ref{AdS Hasse} below).
The structure of inclusion
is well understood
for nilpotent orbits.
In the following, we briefly review 
some of the well-known results about the inclusion 
structure of nilpotent orbits,
and associate them with
the classifications carried out in this paper.
Physically the included
smaller orbit can be
understood as the soft or boundary limit
of the larger orbit:
the former
can be obtained from the latter by
taking a limit sending
a point in the phase space
to its boundary.
We also  discuss 
the analogous phenomena in semisimple orbits.

%**************************************************%
\subsection{Inclusion structure of nilpotent orbits}
%**************************************************%

\begin{figure}[ht]
    \centering
    \begin{tikzpicture}
        \draw[LimeGreen] (0,0)--(1.75,0)--(3.5,1)--(5.25,0)--(7,0)--(8.75,1)--(10.75,0)--(13,0);
        \draw[LimeGreen] (1.75,0)--(3.5,-1)--(5.25,0);
        \draw[LimeGreen] (7,0)--(8.75,-1)--(10.75,0);
        \node at (0,0) 
        {${\scriptsize\gyoung(;,;,;,;,;,;,;,;)}$};
        \node at (1.75,0) 
        {${\scriptsize\gyoung(;;,;;,;,;,;,;)}$};
        \node at (3.5,-1) 
        {${\scriptsize\gyoung(;;;,;,;,;,;,;)}$};
        \node at (3.5,1) 
        {${\scriptsize\gyoung(;;,;;,;;,;;)}$};
        \node at (5.25,0) 
        {${\scriptsize\gyoung(;;;,;;,;;,;)}$};
        \node at (7,0) 
        {${\scriptsize\gyoung(;;;,;;;,;,;)}$};
        \node at (8.75,1) 
        {${\scriptsize\gyoung(;;;;,;;;;)}$};
        \node at (8.75,-1) 
        {${\scriptsize\gyoung(;;;;;,;,;,;)}$};
        \node at (10.75,0) 
        {${\scriptsize\gyoung(;;;;;,;;;)}$};
        \node at (13,0) 
        {${\scriptsize\gyoung(;;;;;;;,;)}$};
    \end{tikzpicture}
    \caption{Nilpotent orbits of $\mathfrak{so}_{\mathbb C}(8)$
    and its inclusion structure: orbits on the left are contained in the closure of orbits on the right,
    to which they are related by a {\color{LimeGreen} green} line}.
    \label{so8 C}
\end{figure}
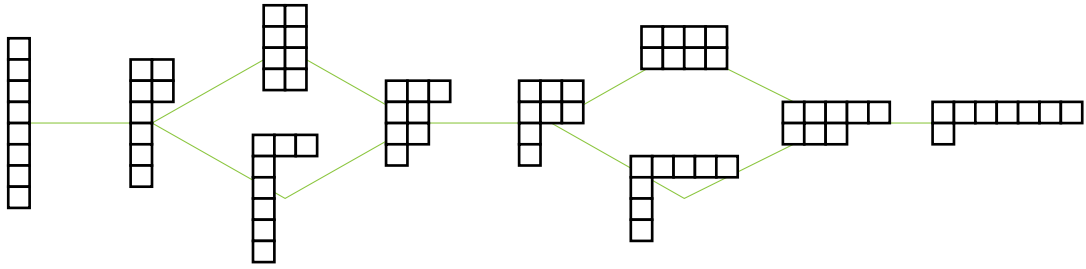

Nilpotent orbits of a complex Lie algebra
have a rich inclusion structure, which can be
described by a Hasse diagram.
See e.g. \cite{Hanany:2016gbz, Cabrera:2017ucb, Hanany:2017ooe, Hanany:2018uzt, Hanany:2018xth} 
for recent progress 
in the study of supersymmetric moduli spaces,
using nilpotent orbits.
Let us limit the scope of our discussion to
the $\mathfrak{so}_{\mathbb C}(n)$ case. 
Its nilpotent orbits are
in one-to-one correspondence with Young diagrams of $n$ boxes,
where rows of even lengths appear with even multiplicities (see e.g. \cite[Chap. 5.1]{Collingwood1993}),
and an orbit corresponding to 
the Young diagram $\mathbb Y_1$
contains an orbit corresponding to $\mathbb Y_2$ in its closure if and 
only if
$\mathbb Y_2$ can be obtained from $\mathbb Y_1$ by repeatedly moving a box from the right edge of one row to a lower row.
See the example of $n=8$ cases 
depicted in Figure \ref{so8 C}.

For the real form $\mathfrak{so}(p,n-p)$, the possible signed Young diagrams 
are composed of $n$ boxes and the distribution
of $+$ and $-$ signs corresponds to the signature $(p,n-p)$,
such that  the first box of 
even-length rows are labelled by a plus sign.
The inclusion structure of real coadjoint orbits is also given by the same rule as the complex case
but in terms of the signed Young diagrams.
These (signed) Young diagrams can also be used to compute
the dimension of the associated orbit.  The dimension
formula is given by
\begin{equation}
    \dim\cO_{[h_1,\dots,h_k]}
    = \dim\mathfrak{so}(p,n-p)
    -\tfrac12\,\sum_{i=1}^k h_i\big(h_i+(-1)^i\big)\,,
\end{equation}
where $h_i$ denote the height of the $i$-th column
of the signed Young diagram.

\begin{figure}[ht]
\centering
\begin{tikzpicture}
    \draw[LimeGreen] (0,0)--(2.8,0);
    \node at (0,0) {\scriptsize\gyoung(;-,|7+)};
    \node at (0,-1.5) {$\phi=0$};
     \node at (3,0) {\scriptsize\gyoung(;+;-;+,|5+)};
    \node at (3,-1.3) {$\phi=\cP^+$};
\end{tikzpicture}
\caption{Inclusion structure of dS nilpotent orbits}
\label{dS Hasse}
\end{figure}
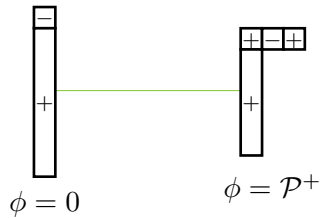

Let us enumerate the possible
signed Young diagrams for $\mathfrak{so}(1,d)$
and $\mathfrak{so}(2,d-1)$,
and show their inclusion structures.
For the dS algebra,
only two signed Young diagram are possible:
see Figure \ref{dS Hasse}.
Here, the first one is the trivial orbit
and the second one is
the massless scalar orbit
with the representative vector $\phi=\cP^+$\,.

The nilpotent coadjoint orbits of the Poincar\'e algebra and their
inclusion structure is
the same as the dS case: 
there are only two nilpotent orbits,
the trivial one given by $\phi=0$
and the massless scalar given by $\phi=\cP^+$.
The former is included in 
the closure of the latter.

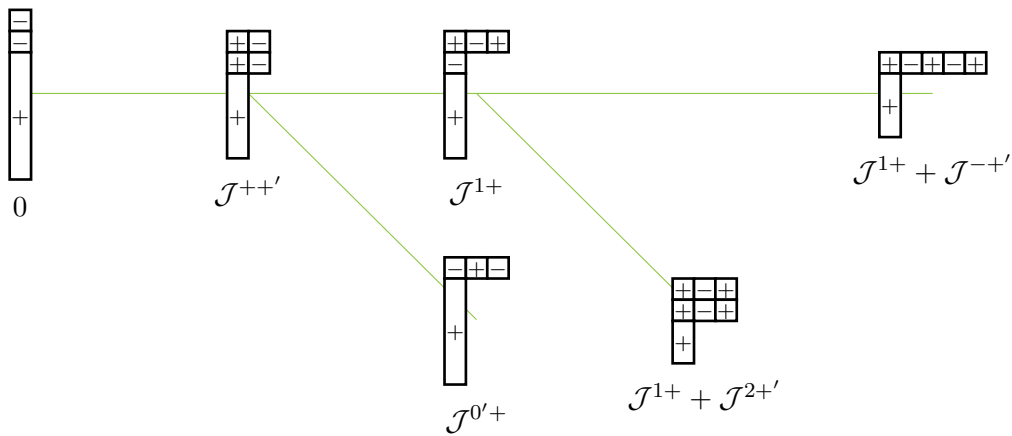
\begin{figure}[ht]
    \centering
\begin{tikzpicture}
    \draw[LimeGreen] (0,0)--(12,0);
    \draw[LimeGreen] (3,0)--(6,-3);
    \draw[LimeGreen] (6,0)--(9,-3);
    \node at (0,0)
    {\scriptsize\gyoung(;-,;-,|6+)};
    \node at (0,-1.5) {$0$};
     \node at (3,0)
    {\scriptsize\gyoung(;+;-,;+;-,|4+)};
    \node at (3,-1.3) {$\cJ^{++'}$};
     \node at (6,-3)  
    {\scriptsize\gyoung(;-;+;-,|5+)};
     \node at (6,-4.3) {$\cJ^{0'+}$};
   \node at (6,0)
    {\scriptsize\gyoung(;+;-;+,;-,|4+)};
     \node at (6,-1.3) {$\cJ^{1+}$};
   \node at (9,-3)
    {\scriptsize\gyoung(;+;-;+,;+;-;+,|2+)};
    \node at (9,-4) {$\cJ^{1+}+\cJ^{2+'}$};
   \node at (12,0)
    {\scriptsize\gyoung(;+;-;+;-;+,|3+)};
     \node at (12,-1) {$\cJ^{1+}+\cJ^{-+'}$};
\end{tikzpicture}
\caption{Inclusion structure of AdS nilpotent orbits}
\label{AdS Hasse}
\end{figure}

For the AdS algebra, we find six possible
signed Young diagrams:
see Figure \ref{AdS Hasse}.
We also provided the representative
vectors of the corresponding orbits.
Let us explain this inclusion structure in words. 
The trivial orbit is contained in the
closure of conformal scalar orbit given by $\cJ^{++'}$,
which is the minimal nilpotent orbit.
The conformal scalar is 
included both in the closure of
massless scalar in AdS with $\cJ^{0'+}$
and massless scalar in BdS with $\cJ^{1+}$.
The former is not contained anywhere, whereas
the latter is included both 
in the closure of massless light-like spin particle in BdS with $\cJ^{1+}+\cJ^{2+'}$
and in the massless doubly-light-like spin particle in BdS with
$\cJ^{1+}+\cJ^{-+'}$.

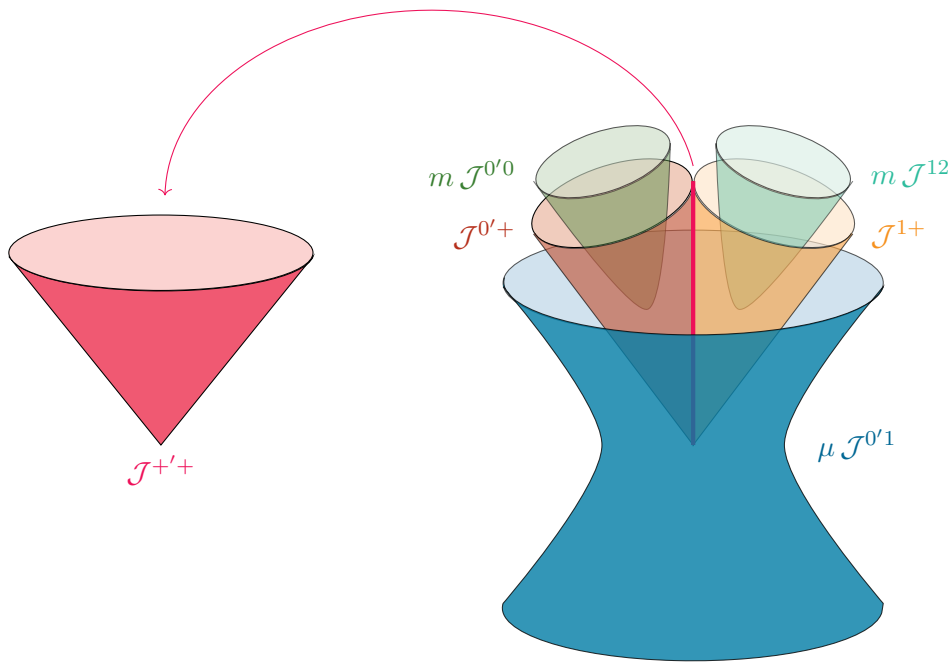
\begin{figure}[ht!]
  \centering
   
    \begin{tikzpicture}
   %%% Cone %%%
    \draw[fill=OrangeRed!80]
        (-9,2.5) -- (-7,0) -- (-5,2.5)
        arc(355:185:2 and 0.5); 
    \draw[fill=OrangeRed!20] (-7,2.55) ellipse (2 and 0.5);
 
  %% Ellipse %%
  \draw[fill=MidnightBlue!20,opacity=0.65] (0,2.15) ellipse (2.5 and 0.7);
    \draw[fill=BrickRed!20] [rotate=18.4] (0,3.38) ellipse (1.1 and 0.5);
     \draw[fill=BurntOrange!20,opacity=0.65] [rotate=-18.4] (0,3.38) ellipse (1.1 and 0.5);
     \draw[fill=OliveGreen!20,opacity=0.65] 
     [rotate=18.4] (0.06,3.94) ellipse (0.92 and 0.4);
     \draw[fill=SeaGreen!20,opacity=0.65]  [rotate=-18.4] (-0.06,3.94) ellipse (0.92 and 0.4);
    %%% Two-sheeted hyperboloid %%%
    \draw[fill=OliveGreen!80, opacity=0.5]
        plot[smooth] coordinates{(-2.1, 3.5) (-0.6,1.8) (-0.3,4)}
        {[rotate=18.4] arc (355:185:0.92 and 0.4)};
    \draw[fill=SeaGreen!80, opacity=0.5]
        plot[smooth] coordinates{(2.1, 3.5) (0.6,1.8) (0.3,4)}
        {[rotate=-18.4] arc (185:355:0.92 and 0.4)};
    %%% Cone %%%
    \draw[fill=BrickRed!80, opacity=0.65]
        (-2.1,2.8) -- (0,0) -- (0,3.5)
        {[rotate=18.4] arc(355:185:1.1 and 0.5)};     
    \draw[fill=BurntOrange!80, opacity=0.65]
        (2.1,2.8) -- (0,0) -- (0,3.5)
        {[rotate=-18.4] arc(185:355:1.1 and 0.5)};
    %% intersection %%
    \draw[ultra thick, OrangeRed] (0,0) -- (0,3.5);    
    %% One sheeted hyperboloid %%
    \draw[fill=MidnightBlue!80, opacity=0.8]
        plot[smooth] coordinates {(-2.5,2.1) (-1.2,0) (-2.5,-2.1)}
        -- (-2.5,-2.1) arc (365:185:-2.5 and 0.7) --
        plot[smooth] coordinates {(2.5,-2.1) (1.2,0) (2.5,2.1)}
        arc (355:175:2.5 and 0.7);

    \node[below] at (-7,0)
    {\color{OrangeRed} $\cJ^{+'+}$};
    \node[right] at (1.5,0)
    {\color{MidnightBlue}
    $\mu\,\cJ^{0'1}$};
    \node[right] at (2.2,2.8)
    {\color{BurntOrange}
    $\cJ^{1+}$};
    \node[left] at (-2.2,2.8)
    {\color{BrickRed}
    $\cJ^{0'+}$};
    \node[right] at (2.2,3.6)
    {\color{SeaGreen}
    $m\,\cJ^{12}$};
    \node[left] at (-2.2,3.6)
    {\color{OliveGreen}
    $m\,\cJ^{0'0}$};

    \draw[OrangeRed, ->]
    (0,3.7)
        arc(10:179:3.5 and 2.5);
    
    \end{tikzpicture}
    \caption{Scalar nilpotent orbits of AdS group
    and their adjacent semisimple orbits}
    \label{fig:inclusion}
\end{figure}

The inclusion structure can be intuitively understood by
the action of Lorentz boosts on the representative vector.
Two different Lorentz boosts act on the $\pm$ and $\pm'$ components of the vectors, and they can scale the vector down (or up).
In this way, one can easily understand 
$\cP^+$ can be scaled down to $0$ under the infinite boost along the $\pm$ directions in the dS and Poincar\'e cases.
In AdS, $\cJ^{++'}$ can be scaled down
to 0 by either boosts $\pm$ or $\pm'$.
Also, $\cJ^{1+}+\cJ^{2+'}$
and $\cJ^{1+}+\cJ^{-+'}$ can
be scaled down to $\cJ^{1+}$
under the $\pm'$ boost.
In order to get $\cJ^{++'}$
from $\cJ^{1+}$, 
we need to boost in the $0'-1$ plane to get $+'$ while renormalising  the vector
with the $\pm$ boost.
We can do the same for $\cJ^{0'+}$ to get $\cJ^{++'}$.
See Figure \ref{fig:inclusion},
where we used the same colors
as in Figure \ref{fig:scalar_AdS},
for a cartoon picture of
the inclusion of the conformal scalar orbit
$\cJ^{+'+}$
in the intersection of 
two massless scalar orbits $\cJ^{0'+}$
and $\cJ^{1+}$.
The closure of the latter two nilpotent orbits 
correspond to the massless limit 
$m\to 0$ 
of the semisimple orbits
$m\,\cJ^{0'0}$ and $m\,\cJ^{12}$.
The union of these two nilpotent orbit closures is
the $\mu\to0$ limit 
of the semisimple orbit $\mu\,\cJ^{0'1}$\,.

Reasoning in terms of boosts in lightcone coordinates
is also useful in understanding 
the nilpotent nature of the above orbits.
Nilpotent orbits should not have any labels. Therefore, any coefficients in a representative vector of nilpotent orbit
should be adjustable by a suitable boost.
One can convince oneself that
the above are all possible such
vectors up to rotations.

Remark that 
we had initially introduced
the representative vector of massive scalar as $\phi=E\,\cP^+$
while we removed the $E$ dependence
in this section as it can be rescaled to any number. And the rescaling $E$ to 0 is how we understand that the trivial orbit is included in the massless scalar orbit.
In this context,
the rescaling $E\to 0$ 
can be interpreted as the soft limit,
which unfortunately leaves nothing (trivial orbit)  in the scalar particle case (see more discussion in the following section).
In the case of the inclusion of
the conformal  scalar with $\phi=\cJ^{++'}$
in the massless scalar in AdS
with $\phi=\cJ^{0'+}$,
the mutual boosts in the $0'-1$, and $\pm$ planes can be understood
as the AdS boundary limit. 
It is interesting to note that
the boundary limit of massless scalar in BdS also leads to the conformal scalar.
Thinking in terms of the ambient space,
these boundary limits can be understood from the fact that the infinite regions
of both AdS ($X^2=-1$) and BdS ($X^2=1$) approach to the cone ($X^2=0$),
whose section can be viewed as the
conformal boundary.

%****************************************************%
\subsection{Inclusion structure of semisimple orbits}
%****************************************************%
\label{sec: inclusion ss}

There also exist a class of semisimple orbits
which contain smaller orbits in their closure.
These semisimple orbits are closely related
to nilpotent ones:
 their representative vectors can be obtained 
from that of a nilpotent orbit $\cO^G_{\phi_{\rm N}}$
by adding a representative vector of a 
semisimple orbit  lying in the stabiliser algebra $\mathfrak{g}_{\phi_{\rm N}}$.\footnote{This corresponds to adding a vector in the normal directions of the orbit.}
In this way, the inclusion nature is completely
controlled by the nilpotent part while
the semisimple part is simply a spectator.
To be more concrete, let us consider a nilpotent
orbit given by $\phi_{\rm N}$,
which includes $m$ sub-nilpotent orbits
\be
    \phi_{\rm N}^{\sst (0)}=0,
    \quad \phi^{\sst (1)}_{\rm N},
    \quad \ldots \quad, \quad  
    \phi^{\sst (m-1)}_{\rm N},
    \quad \phi^{\sst (m)}_{\rm N}=\phi_{\rm N}\,,
\ee 
with an inclusion structure, say,\footnote{Note that
here, it should be understood that an orbit
$\cO_{\phi^{(k)}_{\rm N}}$ is included
in the \emph{Zariski closure} of the next orbit
$\cO_{\phi^{(k+1)}_{\rm N}}$, see e.g. \cite{Collingwood1993}
for more details.}
\be 
  \cO_{\phi^{(0)}_{\rm N}}
    \subset \cO_{\phi^{ (1)}_{\rm N}}
    \subset \cdots \subset
    \cO_{\phi^{(m)}_{\rm N}}\,.
    \label{simple is}
\ee
Then for any semisimple orbit of
$\phi_{\rm S}\in \mathfrak{g}_{\phi_{\rm N}}^*$,
we have the inclusion structures,
\be 
   \cO_{\phi^{(0)}_{\rm S}}
    \subset \cO_{\phi^{ (1)}_{\rm S}}
    \subset \cdots \subset
    \cO_{\phi^{(m)}_{\rm S}}\,,
\ee 
where $\phi^{\sst (i)}_{\rm S}$ are given by
\be
    \phi^{\sst (i)}_{\rm S}
    =\phi^{\sst (i)}_{\rm N}
    +\phi_{\rm S}\,,
    \qquad [i=0,1,\ldots,m]\,.
\ee 
In \eqref{simple is}, we have considered
the simplest inclusion structure
which can be depicted by a diagram of a simple line,
but the same should hold for any more non-trivial
inclusion structures.

Let us consider the example of $\phi_{\rm N}=\cP^+$\,.
In dS and Poincar\'e, the closure of $\cO_{\cP^+}$
contains only the trivial orbit, but in AdS, it also
contains the orbit of the conformal scalar $\cO_{\cJ^{++'}}$.
We can add a spin
$\phi_{\rm S}=s\,\cJ^{12}\in \mathfrak{g}_{\cP^+}^*$
to these orbits to find semisimple orbits with representative,
\be 
    \phi^{\sst (0)}_{\rm S} = s\,\cJ^{12}\,,
    \qquad 
    \phi^{\sst (1)}_{\rm S} = \cJ^{++'}+s\,\cJ^{12}\,,
    \qquad 
    \phi^{\sst (2)}_{\rm S} = \cP^{+}+s\,\cJ^{12}\,,
\ee 
where $\phi^{\sst (1)}_{\rm S}$ is present only for AdS.
Let us first focus on the Poincar\'e case,
where $\phi^{\sst (2)}_{\rm S}=\cP^++s\,\cJ^{12}$ 
corresponds to the massless spin $s$ particle.
We find that its soft limit gives
the null particle with $\phi^{\sst (0)}_{\rm S}=s\,\cJ^{12}$.
The orbit of the latter has dimensions
$2(d-2)$ which could be understood as a particle in one lower dimension, similar
to the conformal scalar on the boundary.
Hopefully, this simple observation might give a new
insight on the issues of soft particles, BMS,
and celestial CFT etc. Note that the representations of the BMS group in three dimensions have been constructed using the orbit method \cite{Barnich:2014kra, Barnich:2015uva}.

In the case of dS and AdS, the soft limit
of the orbit $\phi^{\sst (1)}_{\rm S}=\cP^++s\,\cJ^{12}$
still leads to the orbit $\phi^{\sst (0)}_{\rm S}=s\,\cJ^{12}$\,,
but we need to interpret differently.
First of all, the latter orbit has dimensions $2(d-1)$,
greater than that of Poincar\'e. 
And the starting orbit $\phi^{\sst (1)}_{\rm S}=\cP^++s\,\cJ^{12}$
cannot be interpreted as a massless one,
but the end point of massive/tachyonic spin $s$ in dS/AdS, 
respectively. As we shall comment in the discussion section,
this end point will be even shielded by complementary series
representation, so would become an interior point of
 massive/tachyonic spectrum (see Figure \ref{fig:spinning_AdS}).
Moreover, $\phi^{\sst (0)}_{\rm S}=s\,\cJ^{12}$
can be interpreted as a tachyonic scalar in dS for $\so(1,d)$
and a massive scalar in BdS for $\so(2,d-1)$.
It is intriguing that in dS, a massive spin $s$
particle with a specially tuned mass 
would contain a tachyonic scalar
in the boundary of its phase space.
It is also intriguing that in AdS,
a tachyonic spin $s$ contains a BdS scalar,
though we have already seen that in AdS tachyonic 
particles can live both AdS and BdS.
A possibility is that due to the quantum shift,
the quantisation of this end point corresponds
to the opposite bound of the spectrum window
associated with the complementary series representations.
Then, the inclusion of the scalar orbit
may be interpreted as the development
of  the scalar gauge symmetry
in the zero mass limit of massive/tachyonic fields
in dS/AdS. In dS case, this corresponds to the maximal depth partially massless spin $s$ field
which appears at the lightest mass end point 
of massive spectrum.

Similarly, the soft limit of the orbit
$\phi_{\rm S}^{\sst (1)}=s\,(\cP^0+\cJ^{12})
+ \epsilon\,(\cP^0+\cP^1-\cJ^{12}-\cJ^{02})$
leads to the orbit of massless spin $s$ given by
$\phi_{\rm S}^{\sst (0)}=s\,(\cP^0+\cJ^{12})$.
As we discussed previously
below \eqref{eq:entangled massless}, 
the former orbit can be quantised
with one free parameter, and gives rise
to a field with infinitely many components
(that is, no spin projection). In the limit
where this parameter goes to a special value,
massive spin $s-1$ and massless spin $s$
field appear besides the remaining infinite-component field
(see \cite{Metsaev:2019opn} for an explicit description).
Therefore, the inclusion of the small massless spin $s$ orbit
in the large orbit \eqref{eq:entangled massless}
can be interpreted again as the splitting
of a long representation into short ones.
The analogous discussions can be made also for
the counterpart orbits of dS as well as for the BdS orbits.

We can also consider the mixed-symmetry analogues,
the massless AdS orbits given by
$\phi^{\sst (0)}_{\rm S}=\ell_1(\cP^0+\cJ^{12}+\cdots+
    \cJ^{2h_1-1\,2h_1})+\cdots$
with the stabiliser
$\mathfrak{g}_\phi = \uu(1,h_1) \oplus \cdots$.
Here, $\cdots$ denotes the part
which depends on additional spin components.
The algebra $\uu(1,h_1)$ has only one non-trivial
nilpotent orbit $\phi_{\rm N}^{\sst (1)}$
associated with the Young diagram
\begin{equation}
   \scriptsize\gyoung(;-;+,|4+)
\end{equation}
and by adding it to the 
original coadjoint vector $\phi_{\rm S}^{\sst (0)}$,
we obtain an orbit $\phi_{\rm S}^{\sst (1)}= \phi_{\rm N}^{\sst (1)}+\phi_{\rm S}^{\sst (0)}$
having the same dimensions as 
the massive orbit given by
$m\,\cP^0+\ell_1\,(\cJ^{12}+\cdots+\cJ^{2h_1-1\,2h_1})+\cdots$
with the stabiliser $\uu(1)\oplus\uu(h_1)\oplus \cdots$.
Classically, the large orbit $\phi_{\rm S}^{\sst (1)}$
includes the small one $\phi_{\rm S}^{\sst (0)}$.
The quantisation of the orbit 
$\phi_{\rm S}^{\sst (1)}$ would again
involve a free parameter and
a phenomenon analogous
to the symmetric case would take a place.
Remark that
when $d=2h_1+1$, the short massless representation is described by a field living on the $d-1$ dimensional boundary of AdS$_d$.

%%%%%%%%%%%%%%%%%%%%
\section{Conclusion}
\label{sec: conclusion}
%%%%%%%%%%%%%%%%%%%%
%**********************************%
\subsection{Summary and discussions}
%**********************************%

In this paper, we studied  the construction of worldline
particle actions starting from a coadjoint orbit
of the isometry group. The construction is based
on the KKS symplectic structure and the action is
given by the associated symplectic potential.
In order for the path integral quantisation of this action
to be well-defined, a part of the labels of particles,
such as the spin labels, are quantised.
Focusing on the classical Lie groups, we reformulate
the coadjoint orbit actions into constrained Hamiltonian ones, 
where the definition of the group gives rise to a mixture
of first and second class constraints. The constraints
appear as moment maps --- for the dual Lie algebra ---
shifted by some constants. As such, coadjoint orbits
of the dual group are defined  with the labels given by
the aforementioned constants.
In this way,
we find 
pairs of coadjoint orbits
$(\cO^G_\phi, \cO^{\tilde G}_{\tilde\phi})$,
where for a given $G$,
a choice of  
coadjoint orbit 
$\cO^G_\phi$
defines 
the dual group $\tilde G$ together with 
 its coadjoint orbit
$\cO^{\tilde G}_{\tilde\phi}$.

The mathematical 
structures underlying 
the above pairs of coadjoint orbits 
fall within the set-up of
the \emph{symplectic dual pair} \cite{Weinstein1983}.%
\footnote{A symplectic dual pair consists in a pair
of Poisson manifolds, say $\cP_1$ and $\cP_2$,
each equipped with a Poisson map $\pi_i: \cM\to\cP_i$
from the same symplectic manifold $(\cM,\Omega)$,
such that $\{\pi^*_1(f_1),\pi_2^*(f_2)\}_\cM = 0$,
for any $f_i \in \Functions(\cP_i)$,
where $\{-,-\}_\cM$ denotes the Poisson bracket on $\cM$
induced by the symplectic form $\Omega$. In other words,
the pullback of the algebra of functions on the Poisson
manifolds $\cP_1$ and $\cP_2$ commute with one another
in $\cM$. For more details, the interested reader
may consult \cite[Chap. 9]{CannasDaSilva1999}
or \cite[Chap. 11]{Ortega2013}.}
Relevant to us is the case of a symplectic manifold
$(\cM,\Omega)$ equipped with the Hamiltonian actions
of two Lie groups, say $G$ and $\tilde G$,
\begin{equation}
    \begin{tikzcd}
        & (\cM,\Omega) \ar[dl, swap, "\mu"]
        \ar[dr, "\tilde\mu"] & \\
        \mathfrak{g}^* && \tilde{\mathfrak{g}}^*
    \end{tikzcd}
    \label{symp dual pair}
\end{equation}
such that the actions commute with one another,
\begin{equation}
    \{\mu^*(f),\tilde\mu^*(g)\}_\cM = 0\,,
    \qquad 
    \forall\,f\in \mathscr C^\infty(\mathfrak{g}^*)\,,
    \quad 
    \forall\, g \in \Functions(\tilde{\mathfrak{g}}^*)\,,
\end{equation}
and such that the pre-image $\mu^{-1}(\phi)$
for a fixed $\phi\in \mathfrak{g}^*$
is a single $\tilde G$-orbit, and vice-versa.
Such actions are called `mutually transitive'
\cite{Skerritt2019}, and establish
a one-to-one correspondence between $G$-orbits 
in $\mu(\cM)$ and $\tilde G$-orbits in $\tilde\mu(\cM)$.
More precisely, the reduced phase space
at $\phi\in\mathfrak{g}^*$ is symplectomorphic
to a coadjoint orbit of $\tilde G$,
\begin{equation}
    \mu^{-1}(\phi)/G_\phi
    \cong \cO^{\tilde G}_{\tilde\phi}\,,
\end{equation}
where the representative
$\tilde\phi \in \tilde{\mathfrak{g}}^*$
is simply the image of a point $x \in \mu^{-1}(\phi)$
under the moment map for $\tilde G$,
i.e. $\tilde\phi=\tilde\mu(x)$.

The correspondence spelled out in this work admits
a description in terms of symplectic dual pairs,
wherein one considers the cotangent bundle 
$T^*Mat_{N \times M}(\mathbb F)
\cong \mathbb F^{2(N \times M)}$
of $N \times M$ matrices with coefficients
in $\mathbb F=\R, \C$ or $\mathbb H$ as the embedding
symplectic manifolds, equipped with two commuting actions
of reductive groups $G$ and $\tilde G$ (as suggested
by the examples worked out in \cite{Skerritt2019},
which seem to fall in the class of models considered
in this paper). The foliation of
$T^*Mat_{N\times M}(\mathbb F)$ under the action
of $G \times \tilde G$ gives
\begin{equation}
    T^*Mat_{N\times M}(\mathbb F)
    \cong \bigcup_{\phi} \cO^G_\phi
    \times \cO^{\tilde G}_{\tilde\phi(\phi)}\,,
    \label{geom cor}
\end{equation}
where the summation for $\phi$ is over
all coadjoint orbits present in this decomposition
and the dual coadjoint orbit 
with the representative element $\tilde\phi(\phi)$
is uniquely specified by $\phi$. The above 
is reminiscent of the foliation of the cotangent bundle 
$T^*G$ under the left and right action $G \times G$,
which leads to the Peter--Weyl theorem upon quantisation:
\begin{equation}
    T^*G \cong \bigcup_{\phi} \cO^G_\phi \times \cO^G_\phi
    \quad \xrightarrow{\rm\sst quantisation} \quad 
    L^2(G)\cong \bigoplus_{\l} \pi_\l^G
    \otimes (\pi_\l^G)^*\,.
    \label{TG LG}
\end{equation}
Note that the summation of $\phi$ over
all distinct coadjoint orbits is transmuted
into the summation of $\l$ over all distinct unitary 
irreducible representations.
By analogy with \eqref{TG LG}, one can understand
that the geometrical correspondence \eqref{geom cor}
is the classical analogue of the reductive dual
pair correspondence \cite{Howe1989i, Howe1989ii},
\begin{equation}
    W_{NM} \cong \bigoplus_{\lambda} \pi_\lambda^G
    \otimes \pi_{\tilde\lambda(\lambda)}^{\tilde G}\,,
\end{equation}
which consists in a bijection between irreducible
representations $\lambda$ and $\tilde\lambda$ 
of $G$ and $\tilde G$, appearing in the decomposition
of the oscillator representation $W_{NM}$
(i.e. the metaplectic representation) of $Sp(2NM,\R)$.
The representation $W_n$ is known to arise
as the quantisation of the minimal nilpotent orbit
of $Sp(2n,\R)$, which is simply the flat symplectic
manifold $(\R^{2n}\backslash\{0\})/\Z_2$
\cite{Brylinski1998}. Therefore, the irreps
appearing in the dual pair correspondence should arise
from the quantisation of pairs of coadjoint orbits
of $G \times \tilde G \subset Sp(2n,\R)$ embedded
in the minimal orbit of $Sp(2n,\R)$. See \cite{Kazhdan1978,
Adams1987, Przebinda1993, Daszkiewicz1997, Pan2010}
and references therein for works in this direction.

Let us come back to the content of this paper:
by focusing  our attention to the Poincar\'e and (A)dS groups,
we derived manifestly covariant actions for various particle
species in Minkowski and (A)dS spacetime. In (A)dS case,
the manifest covariance is realised using ambient space coordinates.
In the Poincar\'e case, the classification of coadjoint orbits is essentially the same as the Wigner classification.
In the (A)dS cases, it turned out that
there are far less
types of coadjoint orbits
of $O(1,d)$ compared to $O(2,d-1)$.
Limiting our attention to the cases of no more than one spin label, we could survey all possibilities.
Here, by `spin', we mean any additional label besides the mass,
the principal label.
Compared to the classification of unitary irreducible representations, the same task for coadjoint orbits is so much easier, and it allowed us to see 
the landscape of all the available 
particle species with Poincar\'e and (A)dS symmetry, assuming
a one-to-one correspondence between quantisable coadjoint orbits
and unitary irreducible representations.
We found that in dS there is no tachyonic particle except for the scalar.
On the other hand, in AdS, not only tachyons but also more exotic entities appear. We interpret them as the particles living in bitemproal AdS (BdS) as its worldline action involves the constraint $X^2=+1$. It is interesting to note that tachyons 
can be interpreted as either AdS or BdS particles,  hence
bridging the spectrum of the ordinary AdS particles and the BdS ones.
Even though BdS particles are exotic, they may play certain roles in CFT, such as in the inversion formula. 
The existence of the scalar and spinning conformal particles is another peculiarity of AdS,
whose closest counterpart might be the presence of null particles in Poincar\'e with vanishing momenta.
In Poincar\'e and AdS, we also identified 
coadjoint orbits that should correspond to continuous
spin particles, which is part of a two-parameter 
family of orbits in the AdS case. On top of that,
we found many more interesting classes of coadjoint orbits
whose particle/field interpretation either consistent
with the existing literature or yet to be described.

%******************%
\subsection{Outlook}
%******************%

In this paper, we focused on the (inhomogeneous)
orthogonal group, but our construction can be equally
applied to other classical Lie groups.
In the sequel paper, we shall cover
these other cases with their applications
to the twistor formulation of worldline particles.

Various issues related to quantisation 
need to be better understood.
First, we need to define a proper measure for
the path integral, but once this is done we expect that the constrained Hamiltonian system
can be easily quantised because all the constraints are quadratic. As mentioned earlier, 
the quantisation of these dual pairs of coadjoint orbits
should lead to the dual pair of unitary and irreducible 
representations appearing in Howe duality.
The natural setting for the latter is the Fock space
generated by several families of bosonic oscillators,
with which one can realise the action of a dual pair
of reductive groups \cite{Rowe:2012ym, Basile:2020gqi}.
This is consistent with the fact
that the constraints considered in our Hamiltonian
systems are obtained from moment maps for the group
actions of dual pairs (in the above sense), and that 
these moment maps are all quadratic. 
Moreover, when thought of in terms of the dual pair correspondence,
these constraints have a clear representation theoretical interpretation,
from which one can hope to derive some intuition.
This should be particularly useful when tackling the quantisation of some of the models
identified here, whose spacetime interpretation may not be fully transparent.
We also expect that the comparison between particle species appearing for different values
of the cosmological constant will be made easier due to the uniform treatment presented at the classical level
in this paper.

However, there  still remain a few important issues
to be clarified. One important such issue
is how the degrees of freedom associated to the spin
are projected to a finite dimensional space.
As we discussed with the example of $SU(2)$,
this can happen for a compact coadjoint orbit,
but understanding the same mechanism for a non-compact orbit
will be an important task.
In certain cases, such as massive spinning orbits
where the coadjoint orbits have a bundle structure
where the fiber is a compact coadjoint orbit,
it is easy to understand the mechanism. However,
in other cases, such as massless coadjoint orbits,
it is not easy to grasp the precise mechanism.
Especially in the dS partially massless cases,
the ``spin orbits'' are non-compact while we still expect
the projection to takes place.

Another important issue is how the classical
equations of motion for the first-quantised fields
arise from the Hamiltonian constraints
that we derived. We already see that the constraints
ought to be associated with a set of equations 
similar to that of Fierz equations.
By properly quantising the system with BRST symmetry,
we may get a gauge invariant equation from the BRST charge
(see  e.g. \cite{Alkalaev:2009vm, Alkalaev:2011zv}
and references therein).
When the spin degrees of freedom are to be projected,
we should be able to get the equations which define
tensor fields of a finite rank.

Finally, we would like to remark about the unitary
irreducible representations which would arise by quantising
the coadjoint orbits.
We have seen that some labels need to be quantised
for a well-defined path integral.
These labels will get some $d$-dependent constant shift
due to the ordering of the quantised variables:
e.g. the quadratic Casimir of $SU(2)$ gets shifted
from $s^2$ to $s(s+1)$.
In the end of the quantisation procedure, we expect
to recover most of unitary irreducible representations,
with notable exception of the
complementary series ones. 
The latter seems to arise generically 
from a coadjoint orbit containing
a non-trivial nilpotent part.
Quantisation of such an orbit
involves deformation parameters
which are associated with the labels
of the complementary series representations.
This type of orbits generically appear
at the end point of  a continuous family of orbits
(such as massive dS and tachyonic AdS),
or more generally whenever certain class of orbits
becomes small in a limit that one of its label
goes to a shortening point as we have seen
in the massless spin $s$ example.

\acknowledgments{
T.B. is grateful to Isma\"el Ahlouche Lahlali,
Pierre Bieliavsky and Nicolas Boulanger for discussions
on coadjoint orbits and their geometric quantization.
E.J. is grateful to 
Kyung-Sun Lee, Karapet Mkrtchyan and Junggi Yoon for discussions
related to this study.
T.O. is grateful to
Sang-Eon Bak and Kyung-Sun Lee
for a productive discussion.
The work of T.B. was supported by the European Union’s
Horizon 2020 research and innovation program under the Marie 
Sklodowska Curie grant agreement No 101034383.
The work of E.J. and T.O. was supported by the National Research Foundation of Korea (NRF) grant funded by the Korea government (MSIT) (No. 2022R1F1A1074977).
This work and its early versions have been presented in several 
scientific events supported by Asia Pacific Center for Theoretical Physics (APCTP). 
We appreciate APCTP for the support.

\appendix
%%%%%%%%%%%%%%%%%%%%%%%%%%%%%%%%%%%
\section{Conventions and notations}
\label{app:notations}
%%%%%%%%%%%%%%%%%%%%%%%%%%%%%%%%%%%

\begin{itemize}

    \item Our convention for the Lie algebras
    $\so(1,d)$ and $\so(2,d-1)$ is that
    they are generated by antisymmetric generators
    $J_{\amb{AB}}=-J_{\amb{BA}}$ with $\amb{A,B}=0,1,\dots,d-1,\bullet$,
    subject to
    \begin{equation}
        [J_{\amb{AB}}, J_{\amb{CD}}] = \eta_{\amb{AC}}\,J_{\amb{BD}}
        - \eta_{\amb{BC}}\,J_{\amb{AD}} - \eta_{\amb{AD}}\,J_{\amb{BC}}
        + \eta_{\amb{BD}}\,J_{\amb{AC}}\,,
    \end{equation}
    where $\eta_{\amb{AB}}={\rm diag}(-1,1,\dots,1,\sigma)$
    and $\sigma=+1$ for $\so(1,d)$ or $\sigma=-1$
    for $\so(2,d-1)$.

    \item 
    Defining $P_a:=\tfrac1\ell\,J_{\bullet a}$,
    with $a=0,1,\dots,d-1$, the above Lie bracket
    reads
    \begin{subequations}
        \begin{equation}
            [J_{ab}, J_{cd}] = \eta_{ac}\,J_{bd}
            - \eta_{bc}\,J_{ad} - \eta_{ad}\,J_{bc}
            + \eta_{bd}\,J_{ac}\,,
        \end{equation}
        \begin{equation}
            [J_{ab}, P_c] = \eta_{cb}\,P_{a}
            -\eta_{ca}\,P_{b}\,,
            \qquad 
            [P_a, P_b] = \tfrac\sigma{\ell^2}\,J_{ab}\,,
        \end{equation}
    \end{subequations}
    where $\eta_{ab}={\rm diag}(-1,1,\dots,1)$
    is the Minkowski metric and $\ell$ the (A)dS radius.
    In plain words, the generators $J_{ab}$
    span the Lorentz subalgebra $\so(1,d-1)$
    common to $\so(1,d)$ and $\so(2,d-1)$,
    while $P_a$ are the (A)dS `transvection' generators,
    the non-commutative counterpart of translations
    in flat space. 
    \item     
    Sending the (A)dS radius to infinity,
    $\ell\to\infty$, implements the In\"on\"u--Wigner
    contraction of $\so(1,d)$ or $\so(2,d-1)$
    to the Poincar\'e algebra, $\iso(1,d-1)$,
    whose Lie bracket is almost the same as above
    except for the fact that $P_a$ are now genuine
    generators of translation and hence Abelian.
    
    \item 
    The convention for 
    the lightcone indices $\pm$ is as follows. 
For any generator $V_a$ with a vector index
and its dual $\cV^a$, we set $V_{\pm}=V_0\pm V_{d-1}$
and $\cV^\pm=\frac12(\cV^0\pm\cV^{d-1})$.

\item 
In order to compactly encode the information
about the stabilisers of the various representatives,
we use the following notation. 
First, given an indefinite orthogonal algebra
$\so(p,q)$, we denote a subalgebra $\so(m,n)$
with $m \leq p$ and $n \leq q$, by
\begin{equation}
    \so(m,n)_{[I]} := {\rm span}\{J_{ij}\,,\
    i,j \in I\} \subset \so(p,q)\,,
\end{equation}
where $I$ denotes a subset of values for the indices
carried by the generators of $\so(p,q)$. Similarly,
we denote a subalgebra $\iso(m,n)$ by
\begin{equation}
    \iso(m,n)_{[a \pm b;I]}
    := {\rm span}\{J_{ai} \pm J_{bi}, J_{ij}\,,\
    i,j \in I\} \subset \so(p,q)\,,
\end{equation}
i.e. the indices $a \pm b$ specifies which combination
of generators form the Abelian ideal of translation
for the subalgebra $\so(m,n)_{[I]}$.
Finally, for low-dimensional subalgebras
$\mathfrak h \subset \mathfrak g$ (typically,
one-dimensional), we write
\begin{equation}
    \mathfrak h_{\mathsf t_1, \dots, \mathsf t_k}
    := {\rm span}\{\mathsf t_1, \dots, \mathsf t_k\}
    \subset \mathfrak g\,,
\end{equation}
where $\mathsf t_i$ denotes a basis of $\mathfrak g$.

\item We often omit the word ``coadjoint'' and simply say ``orbit'' to refer to a coadjoint orbit.

\end{itemize}

%%%%%%%%%%%%%%%%%%%%%%%%%%%%%%%%%%%%%%%%%%%%%%%%
\section{Conversion of second class constraints}
\label{app:trick}
%%%%%%%%%%%%%%%%%%%%%%%%%%%%%%%%%%%%%%%%%%%%%%%%
In this appendix, we spell out a simple way of converting
the second class constraints appearing in the Hamiltonian
action considered in Section \ref{sec: constraint} into
first class ones. First, let us point out that the dimension
of the reduced phase space, given in \eqref{dim count},
can be also expressed as
\begin{equation}
	\dim \cN_\phi = \dim\cM + \dim\cO^G_\phi
    -2\,\dim\mathfrak{g}\,,
\end{equation}
which suggests a way of converting
the second class constraints appearing for $\phi\neq0$
into first class ones. Doing so would amount to
extending the phase space of our worldline model from $\cM$
to $\cM \times \cO^G_\phi$ with the symplectic structure
$\O(y) - \o(\varphi)$, and potential
\begin{equation}
	\vartheta(y)+\la \phi, g_\varphi^{-1}\,\rd g_\varphi\ra\,,
\end{equation}
where $\varphi={\rm Ad}_{g_\varphi}^*\phi$ is a point on $\cO^G_\phi$ and
$g_\varphi$ is  a section $\cO^G_\phi \hookrightarrow G$.
In this extended phase space $\cM \times \cO^G_\phi$,
we impose the constraints,
\begin{equation}
    \chi_a(y,\varphi) = \mu_a(y)- \varphi_a \approx 0\,,
    \label{eq:constraints_extended}
\end{equation}
where $\varphi_a=\la \varphi, J_a\ra$ is a function on $\cO^G_\phi$.
The Poisson bracket between two such constraints
is
\be
	\{ \chi_a(y,\varphi),\chi_b(y,\varphi)\}_{\cM\times \cO^G_\phi}
	=\{ \chi_a(y,\varphi),\chi_b(y,\varphi)\}_{\cM}
	- \{ \chi_a(y,\varphi),\chi_b(y,\varphi)\}_{\cO^G_\phi}\,.
\ee
The first  Poisson brackets gives
\be
	\{ \chi_a(y,\varphi),\chi_b(y,\varphi)\}_{\cM}
	=\{ \mu_a(y),\mu_b(y)\}_{\cM}
	=f_{ab}{}^c\,\mu_a(y)\,, 
\ee
whereas the second Poisson bracket reduces to
\be
	\{ \chi_a(y,\varphi),\chi_b(y,\varphi)\}_{\cO^G_\phi}
	=\{ \varphi_a,\varphi_b\}_{\cO^G_\phi}\,.
\ee
The Poisson bracket on $\cO^G_\phi$ can be obtained
from the symplectic structure $\o$,
but also deduced directly from that of $\mathfrak{g}^*$.
The entire coadjoint space $\mathfrak{g}^*$ 
is not a symplectic manifold but is endowed
with a Poisson structure: for any two functions $f, g$
on $\mathfrak{g}^*$,
\begin{equation}
  \{f, g\}_{\mathfrak{g}^*} = f_{ab}{}^c\, x_c\,
  \frac{\partial f}{\partial x_a}\,
  \frac{\partial g}{\partial x_b}\,,
  \label{eq:KKS_Poisson}
\end{equation}
Here, $x=x_a\,\cJ^a$ is a vector in $\mathfrak{g}^*$\,.
The pullback of the above by the inclusion
$\i_{\phi}\,:\,\cO^G_\phi\hookrightarrow \mathfrak{g}^*$, \footnote{Note that the inclusion of $\cO^G_\phi$ in $\mathfrak g^*$ is a moment map for the coadjoint action of $G$.} 
gives the Poisson structure on $\cO^G_\phi$ as  
\be
	\i_{\phi}^*(\{ f, g\}_{\mathfrak{g}^*})=\{ \i_{\phi}^* f, \i_{\phi}^*g \}_{\cO^G_\phi}\,.
\ee
Since the pullback of the coordinate function $x_a$ is $\varphi_a$\,: $\i_{\phi}^*x_a=\varphi_a$, we find 
\be
	\{\varphi_a,\varphi_b\}_{\cO^G_\phi}=f_{ab}{}^c\,\varphi_c\,.
\ee
Combining the two Poisson brackets,
we find the constraints $\chi_a(y,\varphi)$ are all of first class type as $\{\chi_a,\chi_b\}_{\cM\times \cO^G_\phi}\approx 0$.
The particle action
corresponding to the extended constrained phase space is
\begin{equation}
    S[y,\varphi,A] = \int \vartheta(y)
    + \langle \phi, g_\varphi^{-1}{\rm d}g_\varphi \rangle
    - \langle A, \chi(y,\varphi)\rangle\,,
\end{equation}
which can be rewritten as
\begin{equation}
    S[y,\varphi,A] = \int \vartheta_A(y)+ \langle \phi, g_\varphi^{-1} D_A g_\varphi \rangle\,,
\end{equation}
in terms of 
\ba
	\vartheta_A=\vartheta-\la\mu(y),A\ra\,, \qquad
	g_\varphi^{-1}\,D_A g_\varphi=g_\varphi^{-1}(\rd+A)g_\varphi\,,
\ea
which are  separately invariant under 
\begin{equation}
    \delta_\l y^\mu = \{\mu^*(\l), y^\mu\}\,,
    \qquad
    \delta_\l A = \rd\l + [A,\l]\,,
    \qquad
    \delta_\l g_\varphi=-\l\,g_\varphi\,,
\end{equation}
up to a total derivative term. Here, the gauge parameter $\l\in\Omega^0(I,\mathfrak{g})$ 
takes values in the full Lie algebra $\mathfrak{g}$.
Under a change of section $g_\varphi \to g_\varphi\,h_\varphi$, the action 
changes exactly as in the coadjoint orbit action.
Therefore, we find a consistent result of quantisation condition for $\phi$.

Note that the above procedure to convert second class constraints
to first class ones (for the type of constrained Hamiltonian
systems discussed here), is an application of what is known as
the `shifting trick' in the context of symplectic reduction,
see e.g. \cite[Chap. 6.5]{Ortega2013}
or \cite[Chap. 6.3]{Dwivedi2019}.

%************************************************%
\section{Classification of the $O(n)$ coadjoint orbits}
\label{app:O(n)}
%************************************************%

As explained in Section \ref{sec:coadjoint},
the classification of
coadjoint orbits of $O(n)$
can be obtained 
using the bijection between adjoint and coadjoint orbits:
the conjugacy classes of $\so(n)$
correspond to
the elements of the
Cartan subalgebra
up to Weyl reflections.
The classification of orbits 
of the indefinite
orthogonal groups
is much more involved, but
has been
carried out in \cite{Burgoyne1977, Djokovic1983}.
To give an intuitive picture of this classification problem, 
let us review some details about the $O(n)$ case.

Let $J_{ab}$ the generators
of $\so(n)$ and $\cJ^{ab}$ their duals. 
Any representative coadjoint vector can be written as 
$\phi =\phi_{ab}\,\cJ^{ab}$\,.
Since $\phi_{ab}$ is an antisymmetric matrix, we can skew-block-diagonalize it 
(that is, bring it into the form
$\phi=\sum_{k=1}^{[\frac n2]}\phi_{2k-1\,2k}\,\cJ^{2k-1\,2k}
=\phi_{12}\,\cJ^{12}+\phi_{34}\,\cJ^{34}+\cdots$) by an orthogonal transformation, which 
is the same as the adjoint action.
This is one of the key differences
of the $O(n)$ case from the $O(p,n-p)$ ones because
the latter cannot be skew-block-diagonalized by an adjoint action.

In the $O(n)$ case,
we can furthermore set $|\phi_{12}|\ge |\phi_{34}|\ge\ldots$
by $\pi/2$-rotations. Finally, we can perform a $\pi$-rotation
in the $(2k)$--$(2k+1)$ plane to flip the sign
of $\phi_{2k-1\,2k}$. Continuing this procedure,
we can set $\phi_{2k-1\,2k}\ge0$ for $k=1,\ldots, [(n+1)/2]-1$:
only the sign of $\phi_{n-1\,n}$ for even $n$ cannot be adjusted in this way. In summary, for $n=2r$ or $n=2r+1$, 
we can always set a representative coadjoint vector as
\be
 	\phi=\sum_{k=1}^{r}\ell_k\,\cJ^{2k-1\,2k}\,,
	\qquad \ell_1 \ge \cdots \ge \ell_{r-1} \ge |\ell_r|\,,
\ee
where $\ell_r\ge0$ for $n=2r+1$.
Note that this standard representative is an element
of the Cartan subalgebra of $\so(n)$, in accordance
with the previous remark that (co)adjoint orbits 
of compact Lie groups are in correspondence
with (equivalence classes of) Cartan subalgebra elements.
For $\ell_k$'s satisfying
\be
    \ell_1 = \cdots = \ell_{h_1}
    > \ell_{h_1+1} = \cdots = \ell_{h_1+h_2}
    > \cdots > \ell_{h_1+\cdots+h_{p-1}+1}
    = \cdots = \ell_{h_1+\cdots+h_{p}}\,,
    \label{ell cond}
\ee
the stabiliser $G_\phi$ is isomorphic to
\be
    U(h_1) \times U(h_2) \times \cdots \times U(h_{p-1})
    \times O\big(n-2M\big)\,,
    \qquad 
    M := h_1 + \dots + h_p\,.
\ee
The derived algebra of $\mathfrak{g}_\phi$ in this case is 
\begin{equation}
    [\mathfrak{g}_\phi, \mathfrak{g}_\phi]
    = \su(h_1) \oplus \dots \su(h_p) \oplus \so(n-2M)\,,
\end{equation}
and therefore
\begin{equation}
   \mathfrak{g}^{\rm Ab}_\phi = \underbrace{\uu(1)
    \oplus \dots \oplus \uu(1)}_{M\ \text{times}}\,,
\end{equation}
is compact. According to the discussion
in Section \ref{sec:worldline_action}, quantisable
coadjoint orbits will correspond to those having $\ell_k\in\N$.
One can easily imagine that
after quantising these orbits, the label $(\ell_1,\ldots ,\ell_{r})$ for coadjoint orbits becomes the typical Young diagram label for the finite dimensional representations: $\ell_i$ is the number of boxes in the $i$-th row.

%%%%%%%%%%%%%%%%%%%%%%%%%%%%%%%%%%%%
\section{Summary of coadjoint orbits
of the Poincar\'e and (A)dS groups}
\label{app:tables}
%%%%%%%%%%%%%%%%%%%%%%%%%%%%%%%%%%%%

In this section, we present the summary of the coadjoint orbit data we
have obtained in Section \ref{sec: Minkowski} and \ref{sec: AdS}.
In the case of Poincar\'e, we left out the null particles as they coincide with those of dS in one lower dimensions: in terms of particle actions, we need to interpret $X$ and $P$ as a part of spin variables.
In the case of AdS, we omitted 
for simplicity the coadjoint orbits
where the spin and mass are entangled.
The symbols M$_+$, M$_-$ and  M$_0$
indicate the orbits of time-like,
space-like and light-like momenta.
Similarly,
S$_+$, S$_-$, S$_0$ and S$_{00}$
indicate the orbits of time-like,
space-like, light-like 
and doubly-light-like spin.

\begin{table}[ht!]
    \centering\small
    \begin{tblr}{%
        hlines = {1pt},
        vlines = {1pt},
        hline{1,2,11} = {2pt},
        vline{1,2,6} = {2pt},
        rows = {ht=0.75cm},
        columns = {halign=c}
        }
        & $\phi$    
        & $\mathfrak g_\phi$
        & $\tilde\phi$
         & $\tilde{\mathfrak{g}}_{\tilde\phi}$
        \\
        M${}_+$
        & $m\,\cP^0$
        & $\mathbb R \oplus \mathfrak{so}(d-1)$
         & $-m^2\,\cT$
        & $\mathbb R$ \\
        M${}_0$
        & $\cP^+$
        & $\mathbb R \oplus \mathfrak{iso}(d-2)$
        & $0$
        & $\mathbb R$ \\
        M${}_-$
        & $\mu\,\cP^{d-1}$
        & $\mathbb R \oplus \mathfrak{so}(1,d-2)$
        & $\mu^2\,\cT$
        & $\mathbb R$ \\
        M${}_+$S${}_{+}$
        & $m\,\cP^0 + s\,\cJ^{12}$
        & $\mathbb R \oplus \mathfrak u(1)
        \oplus \mathfrak{so}(d-3)$
        & $-m^2\,\cT + s^2\,\cS^{\pi\pi}
         + \cS^{\chi\chi}$
        & $\mathbb R \oplus \mathfrak u(1)$ \\
        m$_0$s$_+$
        & $\cP^+ + s\,\cJ^{12}$
        & $\big(\heis_2 \niplus \mathfrak u(1)\big)
        \oplus \mathfrak{iso}(d-4)$
        & $s^2\,\cS^{\pi\pi}
        +\cS^{\chi\chi} $
        & $\heis_2 \niplus \mathfrak u(1)$ \\
        M$_0$S$_0$
        & $\varepsilon\left(\cP^+ + \cJ^{-1}\right)$
        & $\mathbb R \oplus \mathbb R \oplus \mathfrak{so}(d-3)$
        & $-\varepsilon^2\,\cM^\pi+\cS^{\chi\chi}$
        & $\mathbb R \oplus \mathbb R$ \\
        M$_-$S$_+$
        & $\mu\,\cP^{d-1} + s\,\cJ^{12}$
        & $\mathbb R \oplus \mathfrak u(1)
        \oplus \mathfrak{so}(1,d-4)$
        & $\mu^2\,\cT  + s^2\,\cS^{\pi\pi}+ \cS^{\chi\chi}$
        & $\mathbb R \oplus \mathfrak u(1)$ \\
        M$_-$S$_0$
        & $\mu\,\cP^{d-1} +\cJ^{-2}$
        & $\mathbb R \oplus \mathbb R \oplus \mathfrak{iso}(d-4)$
        & $\mu^2\,\cT + \cS^{\chi\chi}$
        & $\mathbb R \oplus \mathbb R$ \\
        M$_-$S$_-$
        & $\mu\,\cP^{d-1} + \nu\,\cJ^{01}$
        & $\mathbb R \oplus \mathbb R \oplus \mathfrak{so}(d-3)$
        & $\mu^2\,\cT  - \nu^2\,\cS^{\pi\pi}+ \cS^{\chi\chi}$
        & $\mathbb R \oplus \mathbb R$ \\
    \end{tblr}
    \caption{Summary of the data defining a coadjoint
    orbit of the Poincar\'e group 
     and its dual 
    except for the null particles.}
    \label{tab:poincare}
\end{table}

\begin{table}[ht!]
    \centering\small
    \begin{tblr}{%
        hlines = {1pt},
        vlines = {1pt},
        hline{1,2,9} = {2pt},
        vline{1,2,6} = {2pt},
        rows = {ht=0.75cm},
        columns = {halign=c}
        }
        & $\phi$ 
        & $\mathfrak g_\phi$
        & $\tilde\phi$
        & $\tilde{\mathfrak{g}}_{\tilde\phi}$ \\
        M$_+$
        & $m\,\cP^0$
        & $\R \oplus \so(d-1)$
        & $\cU-m^2\,\cT$
        & $\R$ \\
        M$_0$
        & $\cP^+$
        & $\R \oplus \iso(d-2)$
        & $\cU$
        & $\R$ \\
        M$_-$
        & $\mu\,\cP^1$
        & $\uu(1) \oplus \so(1,d-2)$
        & $\cU+\mu^2\,\cT$
        & $\uu(1)$ \\
        \SetCell[r=1]{m}
         M$_+$S$_+$
        & $m\,\cP^0 + s\,\cJ^{12}$
        & $\R \oplus \uu(1) \oplus \so(d-3)$
        & \SetCell[r=1]{m} {$\cU + \cS^{\chi\chi}$\\
        $-m^2\,\cT + s^2\,\cS^{\pi\pi}$}
        & $\R \oplus \uu(1)$ \\
         M$_0$S$_+$
        & $\cP^+ + s\,\cJ^{12}$
        & $\R \oplus \uu(1) \oplus \iso(d-4)$
        & $\cU + \cS^{\chi\chi}
        + s^2\,\cS^{\pi\pi}$
        & $\R \oplus \uu(1)$ \\
        M$_-$S$_+$
        & $\mu\,\cP^{d-1} + s\,\cJ^{12}$
        & $\uu(1) \oplus \uu(1) \oplus \so(1,d-4)$
        & \SetCell[r=1]{m} {$\cU + \cS^{\chi\chi}$\\
        $+\mu^2\,\cT + s^2\,\cS^{\pi\pi}$}
        & $\uu(1) \oplus \uu(1)$ \\
        m$_-$s$_+$
        & $s\,(\cP^{d-1} + \cJ^{12})$
        & $\uu(2) \oplus \so(1,d-4)$
        & \SetCell[r=1]{m} {$\cU+\cS^{\chi\chi}$\\
        $+s^2\,(\cT + \cS^{\pi\pi})$}
        & $\uu(2)$ \\
    \end{tblr}
    \caption{Summary of the data defining a coadjoint
    orbit of the dS group and its dual.}
    \label{tab:dS}
\end{table}

\begin{table}[ht!]
    \centering\small
    \begin{tblr}{%
        hlines = {1pt},
        vlines = {1pt},
        hline{1,2,22} = {2pt},
        vline{1,2,6} = {2pt},
        rows = {ht=0.75cm},
        columns = {halign=c}
        }
        & $\phi$ 
        & $\mathfrak g_\phi$
        & $\tilde\phi$
        & $\tilde{\mathfrak{g}}_{\tilde\phi}$ \\
        M$_+$
        & $m\,\cP^0$
        & $\uu(1) \oplus \so(d-1)$
        & $-\,\cU-m^2\,\cT$
        & $\uu(1)$ \\
       M$_0$
        & $\cP^+$
        & $\R \oplus \iso(d-2)$
        & $-\,\cU$
        & $\R$ \\
        M$_-$
        & $\mu\,\cP^1$
        & $\R \oplus \so(1,d-2)$
        & $-\,\cU+\mu^2\,\cT $
        & $\R$ \\
        S$_+$
        &
        $m\,\cJ^{12}$
        &
        $\uu(1)\oplus\so(2,d-3)$
        &
        $\cU+m^2\,\cT$
        &
        $\uu(1)$
        \\
        S$_0$
        &
        $\cJ^{1+}$
        &
        $\R\oplus\iso(1,d-3)$
        &
        $\cU$
        &
        $\R$
        \\
        S$_{00}$
        &
        $\cJ^{++'}$
        &
        \SetCell[r=1]{m} {
        $[\sp(2,\mathbb R)\oplus \so(d-3)]$\\
        $\inplus \heis_{2(d-3)}$}
        & $0$
        &
        $\sp(2,\mathbb R)$
        \\
        M$_+$S$_+$
        & $m\, \,\cP^0 + s\,\cJ^{12}$
        & $\uu(1) \oplus \uu(1) \oplus \so(d-3)$
        & \SetCell[r=1]{m} {$-\,\cU+ \cS^{\chi\chi}$\\
        $-m^2\,\cT + s^2\,\cS^{\pi\pi}$}
        & $\uu(1) \oplus \uu(1)$ \\
        m$_+$s$_+$
        & $s\,(\cP^0 + \cJ^{12})$
        & $\mathfrak{u}(1,1) \oplus \mathfrak{so}(d-3)$
        & \SetCell[r=1]{m} {$-\,\cU+\cS^{\chi\chi}$\\
        $+s^2\,(-\cT+ \cS^{\pi\pi})$}
        & $\mathfrak{u}(1,1)$ \\
        M$_0$S$_+$
        & $\cP^+ + s\,\cJ^{12}$
        & $\R \oplus \uu(1) \oplus \iso(d-4)$
        & $-\,\cU+s^2\,\cS^{\pi\pi}+\cS^{\chi\chi}$
        & $\R \oplus \uu(1)$ \\
        M$_0$S$_0$
        & $\ve\,(\cP^+ + \cJ^{-1})$
        & $\R \oplus \uu(1) \oplus \so(d-3)$
        & $-\,\cU- \ve^2\,\cM^\pi+\cS^{\chi\chi}
        $
        & $\R \oplus \uu(1)$ \\
        M$_-$S$_+$
        & $\mu\,\cP^{d-1} + s\,\cJ^{12}$
        & $\R \oplus \uu(1) \oplus \so(1,d-4)$
        & \SetCell[r=1]{m} {$-\,\cU+\cS^{\chi\chi}$\\
        $+\mu^2\,\cT+s^2\,\cS^{\pi\pi}$}
        & $\R \oplus \uu(1)$ \\
        M$_-$S$_0$
        & $\mu\,\cP^{d-1} + \cJ^{-2}$
        & $\mathbb R \oplus \mathbb R \oplus \mathfrak{iso}(d-4)$
        & $-\,\cU+\mu^2\,\cT  + \cS^{\chi\chi}$
        & $\mathbb R \oplus \mathbb R$ \\
        M$_-$S$_-$
        & $\mu\,\cP^{d-1} + \nu\,\cJ^{01}$
        & $\mathbb R \oplus \mathbb R \oplus \mathfrak{so}(d-3)$
        & \SetCell[r=1]{m} {$-\,\cU+ \cS^{\chi\chi}$\\
        $+\mu^2\,\cT - \nu^2\,\cS^{\pi\pi}$}
        & $\mathbb R \oplus \mathbb R$ \\
        m$_-$s$_-$
        & $\mu\,(\cP^{d-1} + \cJ^{01})$
        & $\mathfrak{gl}(2, \mathbb R) \oplus \mathfrak{so}(d-3)$
        & \SetCell[r=1]{m} {$-\,\cU+ \cS^{\chi\chi}$\\
        $+\mu^2\,(\cT- \cS^{\pi\pi})$}
        & $\mathfrak{gl}(2, \mathbb R)$
       \\
        S$_+$S$_+$
        &
        $m\,\cJ^{12}+s\,\cJ^{34}$
        &
        $\uu(1)\oplus\uu(1)\oplus\so(2,d-5)$
        & \SetCell[r=1]{m} {$\cU+\cS^{\chi\chi}$\\
        $+m^2\,\cT+s^2\,\cS^{\pi\pi}$}
        &
        $\uu(1)\oplus \uu(1)$
        \\
        s$_+$s$_+$
        &
        $s(\cJ^{12}+\cJ^{34})$
        &
        $\uu(2)\oplus\so(2,d-5)$
        & $\cU+\cS^{\chi\chi}
        +s^2\,(\cT+\cS^{\pi\pi})$
        &
        $\uu(2)$\\
        S$_0$S$_+$
        &
        $\cJ^{1+}+s\cJ^{34}$
        &
        $\R\oplus\iso(1,d-5)$
        &     $\cU+
        s^2\,\cS^{\pi\pi}
        +\cS^{\chi\chi}$
        &
        $\R\oplus \uu(1)$
        \\
        S$_0$S$_0$
        &
        $\cJ^{1+}+\cJ^{2+'}$
        & \SetCell[r=1]{m}
        {$\R\oplus \so(2)\inplus \heis_{2(d-4)}$
        \\
        $\niplus \so(d-5)$}
        &    
        $\cU+\cS^{\chi\chi}$
        &
        $\R\oplus \iso(2)$
        \\
        S$_+$S$_{00}$
        &
        $m\,\cJ^{12}+\cJ^{++'}$
        &
        \SetCell[r=1]{m} {
        $[\sp(2,\mathbb R)\oplus \so(d-5)]$
        \\$\inplus \heis_{2(d-5)}
        \oplus\uu(1)$}
        & 
        $\cU+m^2\,\cT$
        &
        $\uu(1)\oplus \uu(1)$\\
        S$_0$S$_{00}$
        &
        $\cJ^{1+}+\cJ^{-+'}$
        &
        $\R\oplus\R\oplus \iso(d-4)$
        & 
        $\cU+\cM^\pi$
        &
        $\R\oplus\R$
        \\       
    \end{tblr}
    \caption{Summary of the data defining a coadjoint
    orbit of the AdS group and its dual
    except for the cases with entangled mass and spin.}
    \label{tab:AdS}
\end{table}

\clearpage
\newpage

%%%%%%%%%%%%%%%%%%%%%%%%%%%%%%%%%%%%%%%%%%%%%%%%%%%%%%%%
\section{Particles with $\mathfrak{so}(2,2)$ symmetry}
\label{app: so(2,2)}
%%%%%%%%%%%%%%%%%%%%%%%%%%%%%%%%%%%%%%%%%%%%%%%%%%%%%%%%

In this appendix, we take advantage of the semisimple
nature of the AdS$_3$ isometry algebra
$\so(2,2) \simeq \so(2,1)_{\sst\rm L} \oplus \so(2,1)_{\sst\rm R}$ to describe
the (co)adjoint orbits of $SO(2,2)$ as a product
of $SO(2,1)$ orbits, for which we have a clear geometrical picture.
The generators of  $\so(2,1)_{\sst\rm L}$
and $\so(2,1)_{\sst\rm R}$ are given by
\be 
    J^{\sst\rm L}_a=\frac{\frac12\,\e_{a}{}^{bc}\,J_{bc}+P_a}2,\qquad 
    J^{\sst\rm R}_a
    =\frac{\frac12\,\e_{a}{}^{bc}\,J_{bc}-
   P_a}2\,,
\ee 
where $J_a^{\rm\sst L/R}$ satisfies
\eqref{JJ com} 
with the metric ${\rm diag}(-,+,+)$
and the Levi--Civita symbol
with $\e_{012}=1$.
The corresponding dual generators are 
\be 
    \cJ_{\sst\rm L}^a=\frac12\,\e^{a}{}_{bc}\,\cJ^{bc}+\cP^a,\qquad 
    \cJ_{\sst\rm R}^a
    =\frac12\,\e^{a}{}_{bc}\,\cJ^{bc}-
   \cP^a\,,
   \label{so22 so21 dic}
\ee 
and we take the representatives
for each coadjoint orbits as
\be 
    \phi_{\sst \rm L}=
    \left\{ 
	\begin{array}{c}
    \pm j_{\sst \rm L}\,
    \cJ^0_{\sst \rm L}\\
    \pm \cJ^+_{\sst \rm L} \\
    k_{\sst \rm L}\, \cJ^2_{\sst \rm L}
    \end{array}\right.\,,
    \qquad 
    \phi_{\sst \rm R}=
    \left\{ 
	\begin{array}{c}
    \mp j_{\sst \rm R}\,
    \cJ^0_{\sst \rm R}\\
    \mp \cJ^+_{\sst \rm R} \\
    -k_{\sst \rm R}\, \cJ^2_{\sst \rm R}
    \end{array}\right.\,.
\ee 
Note that for a more intuitive picture,
we take an inverted picture for $\so(2,1)_{\sst\rm R}$:
the upper/lower elliptic two-sheeted hyperboloids are associated with $\pm j_{\sst \rm L}\,
    \cJ^0_{\sst \rm L}$
and   $\mp j_{\sst \rm R}\,
    \cJ^0_{\sst \rm R}$,
and the upper/lower cones
are associated with 
$\pm \cJ^+_{\sst \rm L}$
 and $\mp \cJ^+_{\sst \rm R}$\,.
The representative in $\so(2,2)^*$ basis
is simply given by $\phi=\phi_{\sst\rm L}
+\phi_{\sst\rm R}$ with \eqref{so22 so21 dic}.
In the usual massive spinning case,
the mass and spin labels are 
related to 
the labels
$j_{\sst\rm L}$ and $j_{\sst\rm R}$
of the elliptic two-sheeted hyperboloids
as
\be
    m=j_{\sst \rm L}+j_{\sst \rm R}\,,
    \qquad 
    s=j_{\sst \rm L}-j_{\sst \rm R}\,,
\ee
and we have similar relations
in the tachyonic case,
\be
    \mu=k_{\sst \rm L}+k_{\sst \rm R}\,,
    \qquad 
    -\nu=k_{\sst \rm L}-k_{\sst \rm R}\,.
\ee
In Table \ref{tab:so(2,2)},  we have collected
representatives of the coadjoint orbits of $SO^+(2,2)$
and arranged them in a `multiplication table'
 to highlight
the product structure of the corresponding orbits,
in terms of coadjoint orbits of $SO^+(2,1)$.

For the $\so(2,1)$ coadjoint orbits,
we have a good understanding
on their quantisations.
The elliptic and hyperbolic orbits
of $\so(2,1)$ (as classical phase spaces)
give rise to the discrete and principal series representations of $\so(2,1)$ (as Hilbert spaces),
respectively (see e.g. \cite[Sec. 2(b)]{Witten:1987ty}). 
The nilpotent orbit gives the minimal representation which lies at 
the end point of the principal representation.
From these data,
we can also understand the quantisation
 the $\so(2,2)$ coadjoint orbits:
tensor products of two $\so(2,1)$ representations give 
the representations of $\so(2,2)$
(see e.g. \cite{Boulanger:2014vya,Joung:2015jza}
for relevant discussions).
For instance, the tensor products of two discrete series of representations
describe the familiar massive
and massless
spinning particles in AdS$_3$,
together with the unfamiliar
spinning particles in BdS.
Tensor
products of a 
principal series representation, whose spectrum is unbounded (it is not
of lowest weight type), with any other representation
give other exotic types of particles with $\so(2,2)$ symmetry.

\begin{landscape}
\begin{table}[ht!]
    \centering\footnotesize
    \begin{tblr}{%
        hlines = {1pt},
        vlines = {1pt},
        hline{1,2,8} = {2pt},
        vline{1,2,8} = {2pt},
        rows = {ht=1.25cm},
        columns = {halign=c}
        }
       L$\backslash$R
& 
        \hyperboloid{0.5}
        & \bowlUp{0.6}
        & \bowlDown{0.6}
        & \coneUp{0.6}
        & \coneDown{0.6}
        & $\pmb\times$ \\
        \hyperboloid{0.5}
        & $\mu\,\cP^2 + \nu\,\cJ^{01}$
        & \SetCell[r=1]{c} {$k_{\sst\rm L}\,(\cP^2 - \cJ^{01})$\\
            $+\,j_{\sst\rm R}\,(\cP^0 - \cJ^{12})$}
        & \SetCell[r=1]{c} {$k_{\sst\rm L}\,(\cP^2 - \cJ^{01})$\\
            $-\,j_{\sst\rm R}\,(\cP^0 - \cJ^{12})$}
        & \SetCell[r=1]{c} {$k_{\sst\rm L}\,(\cP^2 - \cJ^{01})$\\
            $+(\cP^+ -\cJ^{+2})$}
        & \SetCell[r=1]{c} {$k_{\sst\rm L}\,(\cP^2 - \cJ^{01})$\\
            $-(\cP^+ -\cJ^{+2})$}
        & {$k_{\sst\rm L}\,(\cP^2 - \cJ^{01})$} \\
        \bowlUp{0.6}
        & \SetCell[r=1]{c} {$j_{\sst\rm L}\,(\cP^0 + \cJ^{12})$\\
             $+\,k_{\sst\rm R}\,(\cP^2 + \cJ^{01})$}
        & $m\,\cP^0 + s\,\cJ^{12}$
        & $s\,\cP^0 + m\,\cJ^{12}$
        & \SetCell[r=1]{c} {$j_{\sst\rm L}\,(\cP^0 + \cJ^{12})$ \\
           $+(\cP^+ -\cJ^{+2})$}
        & \SetCell[r=1]{c} {$j_{\sst\rm L}\,(\cP^0 + \cJ^{12})$ \\
           $-(\cP^+ -\cJ^{+2})$}
        & {$j_{\sst\rm L}\,(\cP^0 + \cJ^{12})$} \\
        \bowlDown{0.6}
      & \SetCell[r=1]{c} {$-j_{\sst\rm L}\,(\cP^0 + \cJ^{12})$\\
             $+\,k_{\sst\rm R}\,(\cP^2 + \cJ^{01})$}
        & $-\,s\,\cP^0 -m\,\cJ^{12}$
        & $-m\,\cP^0 -s\,\cJ^{12}$
        & \SetCell[r=1]{c} {$-\,j_{\sst\rm L}\,(\cP^0 + \cJ^{12})$ \\
           $+(\cP^+ -\cJ^{+2})$}
        & \SetCell[r=1]{c} {$-\,j_{\sst\rm L}\,(\cP^0 + \cJ^{12})$ \\
           $-(\cP^+ -\cJ^{+2})$}
        & {$-\,j_{\sst\rm L}\,(\cP^0 + \cJ^{12})$} \\
        \coneUp{0.6}
        & \SetCell[r=1]{c} {$
        (\cP^++\cJ^{+2})$ \\
        $+\,k_{\sst\rm R}(\cP^2 + \cJ^{01})$}
        & \SetCell[r=1]{c} {$
        (\cP^++\cJ^{+2})$ \\
        $+\,j_{\sst\rm R}(\cP^0 - \cJ^{12})$}
        & \SetCell[r=1]{c} {$
        (\cP^++\cJ^{+2})$ \\
        $-\,j_{\sst\rm R}(\cP^0 - \cJ^{12})$}
        & $2\,\cP^+$
        & $2\,\cJ^{+2}$
        & $\cP^+ + \cJ^{+2}$ \\
        \coneDown{0.6}
        & \SetCell[r=1]{c} {$
        -(\cP^++\cJ^{+2})$ \\
        $+\,k_{\sst\rm R}(\cP^2 + \cJ^{01})$}
        & \SetCell[r=1]{c} {$
        -(\cP^++\cJ^{+2})$ \\
        $+\,j_{\sst\rm R}(\cP^0 - \cJ^{12})$}
        & \SetCell[r=1]{c} {$
        -(\cP^++\cJ^{+2})$ \\
        $-\,j_{\sst\rm R}(\cP^0 - \cJ^{12})$}
        & $-2\,\cJ^{+2}$
        & $-2\,\cP^{+}$
        & $-(\cP^+ + \cJ^{+2})$ \\
        $\pmb\times$ 
        & $k_{\sst\rm R}\,(\cP^2 + \cJ^{01})$
        & $j_{\sst\rm R}\,(\cP^0 - \cJ^{12})$
        & $-j_{\sst\rm R}\,(\cP^0 - \cJ^{12})$
        & $\cP^+ - \cJ^{+2}$
        & $-(\cP^+ - \cJ^{+2})$
        & $0$ \\
    \end{tblr}
    \caption{Multiplication table for orbits
    of $SO^+(2,1)$ and $SO^+(2,2)$}
    \label{tab:so(2,2)}
\end{table}

\end{landscape}

\section{Comparison with Metsaev's work}
\label{sec: Metsaev}

Through a series of work
\cite{Metsaev:2016lhs, Metsaev:2017cuz, Metsaev:2017ytk, Metsaev:2018moa, Metsaev:2019opn, Metsaev:2021zdg},
Metsaev constructed and studied an infinite-component 
field theory Lagrangian,
in terms of which he could identify
many novel elementary fields.
In this section, we attempt to make a correspondence
between such fields and the coadjoint orbits 
classified in this paper. 
A proper quantisation of each of these orbits
can make this correspondence precise eventually,
but for the moment we only make a preliminary assessment.

In \cite{Metsaev:2019opn}, Metsaev classified different fields
in AdS according to their quadratic and quartic Casimir
values. These Casimir values should coincide with
those of coadjoint orbits up to a quantum shift,
which would arise from an ordering issue
and depends on dimensions.
Discarding the shift, the Metsaev's parameterisation
of $C_2$ and $C_4$ are
\be
    \label{eq:Metsaev casimir}
    C_2  =  p^2+q^2\,,\qquad 
    C_4  =  p^2\,q^2\,,
\ee
where $p$ and $q$ are complex numbers.
Basically, $p$ and $q$ are related to the mass $m$
and spin $s$, or their analogues (see below).
Imposing a unitarity of field theory Lagrangian,
possible values of $p$ and $q$ are further restricted
and there are six classes.
Using the same enumeration symbol 
as in \cite{Metsaev:2019opn} for different classes, 
we have
\begin{enumerate}[label=\roman*.]

\item \underline{$\Re p=0$\,, $\Re q=0$ ($p=i\,\mu$\,, $q=i\,\nu$):}\\
This case corresponds to
the orbit $\phi=\mu\,\cP^{d-1}+\nu\,\cJ^{01}$ with Casimirs \eqref{eq:ads space time casimir}. 
In our classification, we found a shortening condition $\mu=\nu$
where a small orbit $\phi=\mu\,(\cP^{d-1}+\cJ^{01})$ \eqref{eq:short casimir},
together with a large remnant orbit 
$\phi=\mu\,(\cP^{d-1}+\cJ^{01})+
\e\,(\cP^{0}+\cP^{d-1}-\cJ^{d-1\,1}-\cJ^{01})$ appear, but there is no
analogue of this  in Metsaev's result.

\item  \underline{$p^*=q$ ($p=s+i\,\nu$\,, $q=s-i\,\nu$):}\\
This case corresponds to
the orbit  
$\phi=s\,(\cP^0+\cJ^{12})+\nu\,(\cP^{d-1}-\cJ^{02})$
with Casimir \eqref{eq:one parm casimir}.

\item \underline{$p^*=-q$ ($p=s+i\,\nu$\,, $q=-s+i\,\nu$):}\\
This case corresponds to the orbit
$\phi=s\,(\cP^0-\cJ^{12})+\nu\,(\cP^{d-1}+\cJ^{02})$
which can be obtained from the previous case by
a $\pi$-rotation in (2--3) plane.
Hence, according to our classification, this
case is equivalent to the previous one for $d\ge 4$.
For $d=3$, they are different but related by the parity map.

\item \underline{$\Re p=0$\,, $\Im q=0$ ($p=i\,\mu$\,, $q=s$):}\\
This case corresponds 
to the orbit $\phi=\mu\,\cP^{d-1}+s\,\cJ^{12}$
with Casimir
\eqref{eq:ads space space casimir}.  
In our classification, any integer values are allowed for $s$, but in Metsaev's result
only a small interval near $0$ is allowed for $q$.
When $q$ is on the boundary of the interval,
the field becomes reducible.
This reducible point seems related to the 
$\nu\to 0$ limit of $\phi=\mu\,\cP^{d-1}+\nu\,\cJ^{01}$ where
a short scalar tachyon orbit $\phi=\mu\,\cP^{d-1}$
appear together with
a large remnant orbit $\phi=\mu\,\cP^{d-1}
+\e\,\cJ^{-1}$.
The small interval may correspond 
to the complementary series representation
arising from
a quantisation of the orbit $\phi=\mu\,\cP^{d-1}
+\e\,\cJ^{-1}$ containing a singularity with a deformation parameter.

\item \underline{$\Im p=0$\,, $\Re q=0$
($p=m$\,, $q=i\,\nu$):}\\
This case is also further restricted such that
only a small interval near $0$ is allowed for $p$
and the field becomes reducible when $p$ takes
the boundary value of the interval.
This seems again related to the $\mu\to0$ limit
of $\phi=\mu\,\cP^{d-1}+\nu\,\cJ^{01}$ where
a short scalar tachyon orbit $\phi=\nu\,\cJ^{01}
\simeq \nu\,\cP^{d-1}$
appear together with
a large remnant orbit $\phi=E\,\cP^{+}
+\nu\,\cJ^{01}$
(which is of a different class from the orbit $\phi=\mu\,\cP^{d-1}
+\e\,\cJ^{-1}$ appeared in the previous case).
Again, the small interval may correspond 
to the complementary series representation
arising from
a quantisation of singular orbit $\phi=E\,\cP^{+}
+\nu\,\cJ^{01}$ with a deformation parameter.

\item \underline{$\Im p=0$\,, $\Im q=0$ ($p=m$\,, $q=s$):}\\
This case is divided into several sub cases, and
all such cases seem related to the 
shortening condition $m=s$ where
a small massless orbit $\phi=s\,(\cP^{0}+\cJ^{12})$
appears together with a large remnant one
$\phi=s\,(\cP^{0}+\cJ^{12})+\e\,(\cP^0+\cP^1-\cJ^{02}-\cJ^{12})$.
The latter orbit again contains a singularity
and its quantisation may involve a deformation parameter.
In such a case, the spin projection does not take place,
and $s$ does not need to be quantised either.
Therefore, this will lead to
a small interval either only one among $p$ and $q$
or for both of $p$ and $q$ near the shortening point
given by an integer $m=s$. The Metsaev's results treat
the cases with $p$ and $q$ exchanged as different.
This may correspond again to the orbits related by a parity map.
\end{enumerate}

In this section, we discussed a possible link between
the results of Metsaev and our coadjoint orbit classification.
Metsaev's results cover a part of coadjoint orbits
and they are often related to a deformation quantisation
of the orbit with a non-trivial nilpotent part.
Let us conclude this section with a disclaimer
that the above discussion is rather a speculation
for the moment. We will revisit this issue in the sequel paper,
and hopefully provide more evidences for the statements
we made in this section.

\bibliographystyle{JHEP}
\bibliography{biblio}

\providecommand{\href}[2]{#2}\begingroup\raggedright\begin{thebibliography}{100}

\bibitem{Kirillov2004}
A.~A. Kirillov, \emph{Lectures on the orbit method}, vol.~64 of \emph{Graduate
  Studies in Mathematics}.
\newblock American Mathematical Soc., 2004.

\bibitem{Basile:2020gqi}
T.~Basile, E.~Joung, K.~Mkrtchyan and M.~Mojaza, \emph{{Dual Pair
  Correspondence in Physics: Oscillator Realizations and Representations}},
  \href{http://dx.doi.org/10.1007/JHEP09(2020)020}{\emph{JHEP} {\bf 09} (2020)
  020}, [\href{http://arxiv.org/abs/2006.07102}{{\tt 2006.07102}}].

\bibitem{Weinstein1983}
A.~Weinstein, \emph{{The local structure of Poisson manifolds}},
  \href{http://dx.doi.org/10.4310/jdg/1214437787}{\emph{Journal of differential
  geometry} {\bf 18} (1983) 523--557}.

\bibitem{Howe1989i}
R.~Howe, \emph{{Remarks on classical invariant theory}},
  \href{http://dx.doi.org/10.1090/S0002-9947-1989-0986027-X}{\emph{Transactions
  of the American Mathematical Society} {\bf 313} (1989) 539--570}.

\bibitem{Howe1989ii}
R.~Howe, \emph{{Transcending classical invariant theory}},
  \href{http://dx.doi.org/10.2307/1990942}{\emph{Journal of the American
  Mathematical Society} {\bf 2} (1989) 535--552}.

\bibitem{Martin1959}
J.~Martin, \emph{{Generalized classical dynamics, and the ‘classical
  analogue’of a Fermioscillator}},
  \href{http://dx.doi.org/10.1098/rspa.1959.0126}{\emph{Proceedings of the
  Royal Society of London. Series A. Mathematical and Physical Sciences} {\bf
  251} (1959) 536--542}.

\bibitem{Gershun:1979fb}
V.~D. Gershun and V.~I. Tkach, \emph{{Classical and Quantum Dynamics of
  Particles with Arbitrary Spin}}, {\emph{JETP Lett.} {\bf 29} (1979)
  288--291}.

\bibitem{Howe:1988ft}
P.~S. Howe, S.~Penati, M.~Pernici and P.~K. Townsend, \emph{{Wave Equations for
  Arbitrary Spin From Quantization of the Extended Supersymmetric Spinning
  Particle}}, \href{http://dx.doi.org/10.1016/0370-2693(88)91358-5}{\emph{Phys.
  Lett. B} {\bf 215} (1988) 555--558}.

\bibitem{Casalbuoni:1976tz}
R.~Casalbuoni, \emph{{The Classical Mechanics for Bose-Fermi Systems}},
  \href{http://dx.doi.org/10.1007/BF02729860}{\emph{Nuovo Cim. A} {\bf 33}
  (1976) 389}.

\bibitem{Berezin:1976eg}
F.~A. Berezin and M.~S. Marinov, \emph{{Particle Spin Dynamics as the Grassmann
  Variant of Classical Mechanics}},
  \href{http://dx.doi.org/10.1016/0003-4916(77)90335-9}{\emph{Annals Phys.}
  {\bf 104} (1977) 336}.

\bibitem{Brink:1976sz}
L.~Brink, S.~Deser, B.~Zumino, P.~Di~Vecchia and P.~S. Howe, \emph{{Local
  Supersymmetry for Spinning Particles}},
  \href{http://dx.doi.org/10.1016/0370-2693(76)90115-5}{\emph{Phys. Lett. B}
  {\bf 64} (1976) 435}.

\bibitem{Barducci:1976qu}
A.~Barducci, R.~Casalbuoni and L.~Lusanna, \emph{{Supersymmetries and the
  Pseudoclassical Relativistic electron}},
  \href{http://dx.doi.org/10.1007/BF02730291}{\emph{Nuovo Cim. A} {\bf 35}
  (1976) 377}.

\bibitem{Galvao:1980cu}
C.~A.~P. Galvao and C.~Teitelboim, \emph{{Classical Supersymmetric Particles}},
  \href{http://dx.doi.org/10.1063/1.524603}{\emph{J. Math. Phys.} {\bf 21}
  (1980) 1863}.

\bibitem{Siegel:1988ru}
W.~Siegel, \emph{{Conformal Invariance of Extended Spinning Particle
  Mechanics}}, \href{http://dx.doi.org/10.1142/S0217751X88001132}{\emph{Int. J.
  Mod. Phys. A} {\bf 3} (1988) 2713--2718}.

\bibitem{Marcus:1994mm}
N.~Marcus, \emph{{Kahler spinning particles}},
  \href{http://dx.doi.org/10.1016/0550-3213(95)00056-X}{\emph{Nucl. Phys. B}
  {\bf 439} (1995) 583--596}, [\href{http://arxiv.org/abs/hep-th/9409175}{{\tt
  hep-th/9409175}}].

\bibitem{Gorbunov:1998id}
I.~V. Gorbunov and S.~L. Lyakhovich, \emph{{Hidden supersymmetry and Berezin
  quantization of N=2, D = 3 spinning superparticles}},
  \href{http://dx.doi.org/10.1063/1.532861}{\emph{J. Math. Phys.} {\bf 40}
  (1999) 2230--2253}, [\href{http://arxiv.org/abs/hep-th/9809104}{{\tt
  hep-th/9809104}}].

\bibitem{Kuzenko:1995mg}
S.~M. Kuzenko and Z.~V. Yarevskaya, \emph{{Conformal invariance, N extended
  supersymmetry and massless spinning particles in anti-de Sitter space}},
  \href{http://dx.doi.org/10.1142/S0217732396001648}{\emph{Mod. Phys. Lett. A}
  {\bf 11} (1996) 1653--1664}, [\href{http://arxiv.org/abs/hep-th/9512115}{{\tt
  hep-th/9512115}}].

\bibitem{Fedoruk:2006jm}
S.~Fedoruk, E.~Ivanov and J.~Lukierski, \emph{{Massless higher spin D=4
  superparticle with both N=1 supersymmetry and its bosonic counterpart}},
  \href{http://dx.doi.org/10.1016/j.physletb.2006.08.032}{\emph{Phys. Lett. B}
  {\bf 641} (2006) 226--236}, [\href{http://arxiv.org/abs/hep-th/0606053}{{\tt
  hep-th/0606053}}].

\bibitem{Fedoruk:2006it}
S.~Fedoruk and E.~Ivanov, \emph{{Master Higher-spin particle}},
  \href{http://dx.doi.org/10.1088/0264-9381/23/17/006}{\emph{Class. Quant.
  Grav.} {\bf 23} (2006) 5195--5214},
  [\href{http://arxiv.org/abs/hep-th/0604111}{{\tt hep-th/0604111}}].

\bibitem{Bastianelli:2007pv}
F.~Bastianelli, O.~Corradini and E.~Latini, \emph{{Higher spin fields from a
  worldline perspective}},
  \href{http://dx.doi.org/10.1088/1126-6708/2007/02/072}{\emph{JHEP} {\bf 02}
  (2007) 072}, [\href{http://arxiv.org/abs/hep-th/0701055}{{\tt
  hep-th/0701055}}].

\bibitem{Bastianelli:2008nm}
F.~Bastianelli, O.~Corradini and E.~Latini, \emph{{Spinning particles and
  higher spin fields on (A)dS backgrounds}},
  \href{http://dx.doi.org/10.1088/1126-6708/2008/11/054}{\emph{JHEP} {\bf 11}
  (2008) 054}, [\href{http://arxiv.org/abs/0810.0188}{{\tt 0810.0188}}].

\bibitem{Bonezzi:2012cf}
R.~Bonezzi, \emph{{$U(N)$ spinning particles and higher spin fields on Kaehler
  backgrounds}}, {\emph{TSPU Bulletin} {\bf 2012} (2012) 32--36},
  [\href{http://arxiv.org/abs/1210.2585}{{\tt 1210.2585}}].

\bibitem{Bastianelli:2014lia}
F.~Bastianelli, R.~Bonezzi, O.~Corradini and E.~Latini, \emph{{Massive and
  massless higher spinning particles in odd dimensions}},
  \href{http://dx.doi.org/10.1007/JHEP09(2014)158}{\emph{JHEP} {\bf 09} (2014)
  158}, [\href{http://arxiv.org/abs/1407.4950}{{\tt 1407.4950}}].

\bibitem{Bette:1984qt}
A.~Bette, \emph{{On a point - like relativistic and spinning particle}},
  \href{http://dx.doi.org/10.1063/1.526463}{\emph{J. Math. Phys.} {\bf 25}
  (1984) 2456}.

\bibitem{Lyakhovich:1996we}
S.~L. Lyakhovich, A.~Y. Segal and A.~A. Sharapov, \emph{{A Universal model of D
  = 4 spinning particle}},
  \href{http://dx.doi.org/10.1103/PhysRevD.54.5223}{\emph{Phys. Rev. D} {\bf
  54} (1996) 5223--5238}, [\href{http://arxiv.org/abs/hep-th/9603174}{{\tt
  hep-th/9603174}}].

\bibitem{Kuzenko:1994ju}
S.~M. Kuzenko, S.~L. Lyakhovich and A.~Y. Segal, \emph{{A Geometric model of
  arbitrary spin massive particle}},
  \href{http://dx.doi.org/10.1142/S0217751X95000735}{\emph{Int. J. Mod. Phys.
  A} {\bf 10} (1995) 1529--1552},
  [\href{http://arxiv.org/abs/hep-th/9403196}{{\tt hep-th/9403196}}].

\bibitem{Fedoruk:2014vqa}
S.~Fedoruk and J.~Lukierski, \emph{{Massive twistor particle with spin
  generated by Souriau\textendash{}Wess\textendash{}Zumino term and its
  quantization}},
  \href{http://dx.doi.org/10.1016/j.physletb.2014.04.059}{\emph{Phys. Lett. B}
  {\bf 733} (2014) 309--315}, [\href{http://arxiv.org/abs/1403.4127}{{\tt
  1403.4127}}].

\bibitem{Mezincescu:2013nta}
L.~Mezincescu, A.~J. Routh and P.~K. Townsend, \emph{{Supertwistors and massive
  particles}}, \href{http://dx.doi.org/10.1016/j.aop.2014.04.007}{\emph{Annals
  Phys.} {\bf 346} (2014) 66--90}, [\href{http://arxiv.org/abs/1312.2768}{{\tt
  1312.2768}}].

\bibitem{Routh:2015ifa}
A.~J. Routh and P.~K. Townsend, \emph{{Twistor form of massive 6D
  superparticle}},
  \href{http://dx.doi.org/10.1088/1751-8113/49/2/025402}{\emph{J. Phys. A} {\bf
  49} (2016) 025402}, [\href{http://arxiv.org/abs/1507.05218}{{\tt
  1507.05218}}].

\bibitem{Mezincescu:2015apa}
L.~Mezincescu, A.~J. Routh and P.~K. Townsend, \emph{{Twistors and the massive
  spinning particle}},
  \href{http://dx.doi.org/10.1088/1751-8113/49/2/025401}{\emph{J. Phys. A} {\bf
  49} (2016) 025401}, [\href{http://arxiv.org/abs/1508.05350}{{\tt
  1508.05350}}].

\bibitem{Shirafuji:1983zd}
T.~Shirafuji, \emph{{Lagrangian Mechanics of Massless Particles With Spin}},
  \href{http://dx.doi.org/10.1143/PTP.70.18}{\emph{Prog. Theor. Phys.} {\bf 70}
  (1983) 18--35}.

\bibitem{Bengtsson:1987ap}
A.~K.~H. Bengtsson, I.~Bengtsson, M.~Cederwall and N.~Linden, \emph{{Particles,
  Superparticles and Twistors}},
  \href{http://dx.doi.org/10.1103/PhysRevD.36.1766}{\emph{Phys. Rev. D} {\bf
  36} (1987) 1766}.

\bibitem{Bengtsson:1987si}
I.~Bengtsson and M.~Cederwall, \emph{{Particles, Twistors and the Division
  Algebras}}, \href{http://dx.doi.org/10.1016/0550-3213(88)90667-0}{\emph{Nucl.
  Phys. B} {\bf 302} (1988) 81--103}.

\bibitem{Townsend:1991sj}
P.~K. Townsend, \emph{{Supertwistor formulation of the spinning particle}},
  \href{http://dx.doi.org/10.1016/0370-2693(91)91326-Q}{\emph{Phys. Lett. B}
  {\bf 261} (1991) 65--70}.

\bibitem{Howe:1992bv}
P.~S. Howe and P.~C. West, \emph{{The Conformal group, point particles and
  twistors}}, \href{http://dx.doi.org/10.1142/S0217751X92003057}{\emph{Int. J.
  Mod. Phys. A} {\bf 7} (1992) 6639--6664}.

\bibitem{Cederwall:2000km}
M.~Cederwall, \emph{{Geometric construction of AdS twistors}},
  \href{http://dx.doi.org/10.1016/S0370-2693(00)00552-9}{\emph{Phys. Lett. B}
  {\bf 483} (2000) 257--263}, [\href{http://arxiv.org/abs/hep-th/0002216}{{\tt
  hep-th/0002216}}].

\bibitem{Bars:2005ze}
I.~Bars and M.~Picon, \emph{{Single twistor description of massless, massive,
  AdS, and other interacting particles}},
  \href{http://dx.doi.org/10.1103/PhysRevD.73.064002}{\emph{Phys. Rev. D} {\bf
  73} (2006) 064002}, [\href{http://arxiv.org/abs/hep-th/0512091}{{\tt
  hep-th/0512091}}].

\bibitem{Arvanitakis:2016wdn}
A.~S. Arvanitakis, L.~Mezincescu and P.~K. Townsend, \emph{{Pauli-Lubanski,
  Supertwistors, and the Superspinning Particle}},
  \href{http://dx.doi.org/10.1007/JHEP06(2017)151}{\emph{JHEP} {\bf 06} (2017)
  151}, [\href{http://arxiv.org/abs/1601.05294}{{\tt 1601.05294}}].

\bibitem{Arvanitakis:2016vnp}
A.~S. Arvanitakis, A.~E. Barns-Graham and P.~K. Townsend,
  \emph{{Anti\textendash{}de Sitter Particles and Manifest (Super)Isometries}},
  \href{http://dx.doi.org/10.1103/PhysRevLett.118.141601}{\emph{Phys. Rev.
  Lett.} {\bf 118} (2017) 141601}, [\href{http://arxiv.org/abs/1608.04380}{{\tt
  1608.04380}}].

\bibitem{Arvanitakis:2017cpk}
A.~S. Arvanitakis, A.~E. Barns-Graham and P.~K. Townsend, \emph{{Twistor
  description of spinning particles in AdS}},
  \href{http://dx.doi.org/10.1007/JHEP01(2018)059}{\emph{JHEP} {\bf 01} (2018)
  059}, [\href{http://arxiv.org/abs/1710.09557}{{\tt 1710.09557}}].

\bibitem{Buchbinder:2018soq}
I.~L. Buchbinder, S.~Fedoruk, A.~P. Isaev and A.~Rusnak, \emph{{Model of
  massless relativistic particle with continuous spin and its twistorial
  description}}, \href{http://dx.doi.org/10.1007/JHEP07(2018)031}{\emph{JHEP}
  {\bf 07} (2018) 031}, [\href{http://arxiv.org/abs/1805.09706}{{\tt
  1805.09706}}].

\bibitem{Buchbinder:2019gbq}
I.~L. Buchbinder, S.~Fedoruk and A.~P. Isaev, \emph{{Infinite Spin Particles
  and Superparticles}},
  \href{http://dx.doi.org/10.1007/978-981-15-7775-8_6}{\emph{Springer Proc.
  Math. Stat.} {\bf 335} (2019) 83--96}.

\bibitem{Buchbinder:2019iwi}
I.~L. Buchbinder, S.~Fedoruk and A.~P. Isaev, \emph{{Twistorial and space-time
  descriptions of massless infinite spin (super)particles and fields}},
  \href{http://dx.doi.org/10.1016/j.nuclphysb.2019.114660}{\emph{Nucl. Phys. B}
  {\bf 945} (2019) 114660}, [\href{http://arxiv.org/abs/1903.07947}{{\tt
  1903.07947}}].

\bibitem{Buchbinder:2019sie}
I.~L. Buchbinder, S.~Fedoruk and A.~P. Isaev, \emph{{Massless infinite spin
  (super)particles and fields}},
  \href{http://dx.doi.org/10.1134/S0081543820030049}{\emph{Proc. Steklov Inst.
  Math.} {\bf 309} (2020) 46--56}, [\href{http://arxiv.org/abs/1911.00362}{{\tt
  1911.00362}}].

\bibitem{Buchbinder:2021bgv}
I.~L. Buchbinder, S.~A. Fedoruk and A.~P. Isaev, \emph{{Twistor formulation of
  massless 6D infinite spin fields}},
  \href{http://dx.doi.org/10.1016/j.nuclphysb.2021.115576}{\emph{Nucl. Phys. B}
  {\bf 973} (2021) 115576}, [\href{http://arxiv.org/abs/2108.04716}{{\tt
  2108.04716}}].

\bibitem{Souriau2012}
J.-M. Souriau, \emph{Structure of dynamical systems: a symplectic view of
  physics}, vol.~149.
\newblock Springer Science \& Business Media, 2012.

\bibitem{Tod:1977vf}
K.~P. Tod, \emph{{Some Symplectic Forms Arising in Twistor Theory}},
  \href{http://dx.doi.org/10.1016/0034-4877(77)90074-X}{\emph{Rept. Math.
  Phys.} {\bf 11} (1977) 339--346}.

\bibitem{Balachandran:1983oit}
A.~P. Balachandran, G.~Marmo, B.~S. Skagerstam and A.~Stern, \emph{{Gauge
  Theories and Fibre Bundles - Applications to Particle Dynamics}}, vol.~188.
\newblock 1983,
  \href{http://dx.doi.org/10.1007/3-540-12724-0\_1}{10.1007/3-540-12724-0\_1}.

\bibitem{Kuzenko:1994vh}
S.~M. Kuzenko, S.~L. Lyakhovich, A.~Y. Segal and A.~A. Sharapov, \emph{{Anti-de
  Sitter spinning particle and two sphere}},
  \href{http://arxiv.org/abs/hep-th/9411162}{{\tt hep-th/9411162}}.

\bibitem{Kuzenko:1995aq}
S.~M. Kuzenko, S.~L. Lyakhovich, A.~Y. Segal and A.~A. Sharapov, \emph{{Massive
  spinning particle on anti-de Sitter space}},
  \href{http://dx.doi.org/10.1142/S0217751X96001589}{\emph{Int. J. Mod. Phys.
  A} {\bf 11} (1996) 3307--3330},
  [\href{http://arxiv.org/abs/hep-th/9509062}{{\tt hep-th/9509062}}].

\bibitem{Lyakhovich:1998ij}
S.~L. Lyakhovich, A.~A. Sharapov and K.~M. Shekhter, \emph{{Massive spinning
  particle in any dimension. 1. Integer spins}},
  \href{http://dx.doi.org/10.1016/S0550-3213(98)00617-8}{\emph{Nucl. Phys. B}
  {\bf 537} (1999) 640--652}, [\href{http://arxiv.org/abs/hep-th/9805020}{{\tt
  hep-th/9805020}}].

\bibitem{Lyakhovich:1998ve}
S.~L. Lyakhovich, A.~A. Sharapov and K.~M. Shekhter, \emph{{Massive spinning
  particle in any dimension. 2. (Half)integer spins}},
  \href{http://arxiv.org/abs/hep-th/9811003}{{\tt hep-th/9811003}}.

\bibitem{Andrzejewski:2020qxt}
K.~Andrzejewski, C.~Gonera, J.~Gonera, P.~Kosinski and P.~Maslanka,
  \emph{{Spinning particles, coadjoint orbits and Hamiltonian formalism}},
  \href{http://dx.doi.org/10.1016/j.nuclphysb.2022.115664}{\emph{Nucl. Phys. B}
  {\bf 975} (2022) 115664}, [\href{http://arxiv.org/abs/2008.09478}{{\tt
  2008.09478}}].

\bibitem{Gorbunov:1999jg}
I.~V. Gorbunov, V.~A. Dolgushev and S.~L. Lyakhovich, \emph{{Galileo particle
  of nonzero spin}}, \href{http://dx.doi.org/10.1007/BF02509967}{\emph{Russ.
  Phys. J.} {\bf 42} (1999) 168--178}.

\bibitem{Duval:2014ppa}
C.~Duval and P.~A. Horvathy, \emph{{Chiral fermions as classical massless
  spinning particles}},
  \href{http://dx.doi.org/10.1103/PhysRevD.91.045013}{\emph{Phys. Rev. D} {\bf
  91} (2015) 045013}, [\href{http://arxiv.org/abs/1406.0718}{{\tt 1406.0718}}].

\bibitem{Batlle:2017cfa}
C.~Batlle, J.~Gomis, L.~Mezincescu and P.~K. Townsend, \emph{{Tachyons in the
  Galilean limit}},
  \href{http://dx.doi.org/10.1007/JHEP04(2017)120}{\emph{JHEP} {\bf 04} (2017)
  120}, [\href{http://arxiv.org/abs/1702.04792}{{\tt 1702.04792}}].

\bibitem{Ivanov:1999fwa}
E.~Ivanov and S.~Krivonos, \emph{{$N=1$ $D=2$ supermembrane in the coset
  approach}},
  \href{http://dx.doi.org/10.1016/S0370-2693(99)00314-7}{\emph{Phys. Lett. B}
  {\bf 453} (1999) 237--244}, [\href{http://arxiv.org/abs/hep-th/9901003}{{\tt
  hep-th/9901003}}].

\bibitem{Ivanov:1999gy}
E.~Ivanov, \emph{{Diverse PBGS patterns and superbranes}},  in \emph{{14th Max
  Born Symposium: New Symmetries and Integrable Systems}}, pp.~206--217, 9,
  1999.
\newblock \href{http://arxiv.org/abs/hep-th/0002204}{{\tt hep-th/0002204}}.
\newblock \href{http://dx.doi.org/10.1142/9789812793263_0018}{DOI}.

\bibitem{Bellucci:1998mk}
S.~Bellucci, E.~Ivanov and S.~Krivonos, \emph{{Partial breaking of N=1 D = 10
  supersymmetry}},
  \href{http://dx.doi.org/10.1016/S0370-2693(99)00753-4}{\emph{Phys. Lett. B}
  {\bf 460} (1999) 348--358}, [\href{http://arxiv.org/abs/hep-th/9811244}{{\tt
  hep-th/9811244}}].

\bibitem{Bellucci:2000bd}
S.~Bellucci, E.~Ivanov and S.~Krivonos, \emph{{Superworldvolume dynamics of
  superbranes from nonlinear realizations}},
  \href{http://dx.doi.org/10.1016/S0370-2693(00)00529-3}{\emph{Phys. Lett. B}
  {\bf 482} (2000) 233}, [\href{http://arxiv.org/abs/hep-th/0003273}{{\tt
  hep-th/0003273}}].

\bibitem{Bellucci:2002ji}
S.~Bellucci, E.~Ivanov and S.~Krivonos, \emph{{AdS / CFT equivalence
  transformation}},
  \href{http://dx.doi.org/10.1103/PhysRevD.66.086001}{\emph{Phys. Rev. D} {\bf
  66} (2002) 086001}, [\href{http://arxiv.org/abs/hep-th/0206126}{{\tt
  hep-th/0206126}}].

\bibitem{Gomis:2006xw}
J.~Gomis, K.~Kamimura and P.~C. West, \emph{{The Construction of brane and
  superbrane actions using non-linear realisations}},
  \href{http://dx.doi.org/10.1088/0264-9381/23/24/010}{\emph{Class. Quant.
  Grav.} {\bf 23} (2006) 7369--7382},
  [\href{http://arxiv.org/abs/hep-th/0607057}{{\tt hep-th/0607057}}].

\bibitem{Gomis:2006wu}
J.~Gomis, K.~Kamimura and P.~C. West, \emph{{Diffeomorphism, kappa
  transformations and the theory of non-linear realisations}},
  \href{http://dx.doi.org/10.1088/1126-6708/2006/10/015}{\emph{JHEP} {\bf 10}
  (2006) 015}, [\href{http://arxiv.org/abs/hep-th/0607104}{{\tt
  hep-th/0607104}}].

\bibitem{Gomis:2007zz}
J.~Gomis, K.~Kamimura and P.~C. West, \emph{{Non-linear realizations, super
  branes and kappa symmetry}},
  \href{http://dx.doi.org/10.1002/prop.200610350}{\emph{Fortsch. Phys.} {\bf
  55} (2007) 731--735}.

\bibitem{Gomis:2021irw}
J.~Gomis, E.~Joung, A.~Kleinschmidt and K.~Mkrtchyan, \emph{{Colourful
  Poincar\'e symmetry, gravity and particle actions}},
  \href{http://dx.doi.org/10.1007/JHEP08(2021)047}{\emph{JHEP} {\bf 08} (2021)
  047}, [\href{http://arxiv.org/abs/2105.01686}{{\tt 2105.01686}}].

\bibitem{Batlle:2023zhp}
C.~Batlle, V.~Campello and J.~Gomis, \emph{{Particle realization of
  Bondi-Metzner-Sachs symmetry in 2+1 space-time}},
  \href{http://arxiv.org/abs/2307.13984}{{\tt 2307.13984}}.

\bibitem{Alekseev:1988vx}
A.~Alekseev, L.~D. Faddeev and S.~L. Shatashvili, \emph{{Quantization of
  symplectic orbits of compact Lie groups by means of the functional
  integral}}, \href{http://dx.doi.org/10.1016/0393-0440(88)90031-9}{\emph{J.
  Geom. Phys.} {\bf 5} (1988) 391--406}.

\bibitem{Alekseev:1988tj}
A.~Y. Alekseev and S.~L. Shatashvili, \emph{{Propagator for the Relativistic
  Spinning Particle via Functional Integral Over Trajectories}},
  \href{http://dx.doi.org/10.1142/S0217732388001859}{\emph{Mod. Phys. Lett. A}
  {\bf 3} (1988) 1551--1559}.

\bibitem{Alekseev:1988ce}
A.~Alekseev and S.~L. Shatashvili, \emph{{Path Integral Quantization of the
  Coadjoint Orbits of the Virasoro Group and 2D Gravity}},
  \href{http://dx.doi.org/10.1016/0550-3213(89)90130-2}{\emph{Nucl. Phys. B}
  {\bf 323} (1989) 719--733}.

\bibitem{Aratyn:1990dj}
H.~Aratyn, E.~Nissimov, S.~Pacheva and A.~H. Zimerman, \emph{{Symplectic
  actions on coadjoint orbits}},
  \href{http://dx.doi.org/10.1016/0370-2693(90)90420-B}{\emph{Phys. Lett. B}
  {\bf 240} (1990) 127--132}.

\bibitem{Alekseev:2015hda}
A.~Alekseev, O.~Chekeres and P.~Mnev, \emph{{Wilson surface observables from
  equivariant cohomology}},
  \href{http://dx.doi.org/10.1007/JHEP11(2015)093}{\emph{JHEP} {\bf 11} (2015)
  093}, [\href{http://arxiv.org/abs/1507.06343}{{\tt 1507.06343}}].

\bibitem{Alekseev:2018pbv}
A.~Alekseev and S.~L. Shatashvili, \emph{{Coadjoint Orbits, Cocycles and
  Gravitational Wess\textendash{}Zumino}},
  \href{http://arxiv.org/abs/1801.07963}{{\tt 1801.07963}}.

\bibitem{Schubert:1996jj}
C.~Schubert, \emph{{An Introduction to the worldline technique for quantum
  field theory calculations}}, {\emph{Acta Phys. Polon. B} {\bf 27} (1996)
  3965--4001}, [\href{http://arxiv.org/abs/hep-th/9610108}{{\tt
  hep-th/9610108}}].

\bibitem{Bastianelli:2009mw}
F.~Bastianelli, O.~Corradini, P.~A.~G. Pisani and C.~Schubert, \emph{{Worldline
  Approach to QFT on Manifolds with Boundary}},  in \emph{{9th Conference on
  Quantum Field Theory under the Influence of External Conditions (QFEXT 09):
  Devoted to the Centenary of H. B. G. Casimir}}, pp.~415--420, 2010.
\newblock \href{http://arxiv.org/abs/0912.4120}{{\tt 0912.4120}}.
\newblock \href{http://dx.doi.org/10.1142/9789814289931_0051}{DOI}.

\bibitem{Bastianelli:2008vh}
F.~Bastianelli, O.~Corradini, P.~A.~G. Pisani and C.~Schubert, \emph{{Scalar
  heat kernel with boundary in the worldline formalism}},
  \href{http://dx.doi.org/10.1088/1126-6708/2008/10/095}{\emph{JHEP} {\bf 10}
  (2008) 095}, [\href{http://arxiv.org/abs/0809.0652}{{\tt 0809.0652}}].

\bibitem{Bonezzi:2017mwr}
R.~Bonezzi, \emph{{Induced Action for Conformal Higher Spins from Worldline
  Path Integrals}},
  \href{http://dx.doi.org/10.3390/universe3030064}{\emph{Universe} {\bf 3}
  (2017) 64}, [\href{http://arxiv.org/abs/1709.00850}{{\tt 1709.00850}}].

\bibitem{Bastianelli:2022pqq}
F.~Bastianelli, R.~Bonezzi and M.~Melis, \emph{{Gauge-invariant coefficients in
  perturbative quantum gravity}},
  \href{http://dx.doi.org/10.1140/epjc/s10052-022-11119-w}{\emph{Eur. Phys. J.
  C} {\bf 82} (2022) 1139}, [\href{http://arxiv.org/abs/2206.13287}{{\tt
  2206.13287}}].

\bibitem{Bastianelli:2023oca}
F.~Bastianelli, F.~Comberiati, F.~Fecit and F.~Ori, \emph{{Six-dimensional
  one-loop divergences in quantum gravity from the $\mathcal{N}=4$ spinning
  particle}},  \href{http://arxiv.org/abs/2307.09353}{{\tt 2307.09353}}.

\bibitem{Bastianelli:2023oyz}
F.~Bastianelli and M.~D. Paciarini, \emph{{Worldline path integrals for the
  graviton}},  \href{http://arxiv.org/abs/2305.06650}{{\tt 2305.06650}}.

\bibitem{Albonico:2022pmd}
G.~Albonico, Y.~Geyer and L.~Mason, \emph{{From Twistor-Particle Models to
  Massive Amplitudes}},
  \href{http://dx.doi.org/10.3842/SIGMA.2022.045}{\emph{SIGMA} {\bf 18} (2022)
  045}, [\href{http://arxiv.org/abs/2203.08087}{{\tt 2203.08087}}].

\bibitem{Howe:1989vn}
P.~S. Howe, S.~Penati, M.~Pernici and P.~K. Townsend, \emph{{A Particle
  Mechanics Description of Antisymmetric Tensor Fields}},
  \href{http://dx.doi.org/10.1088/0264-9381/6/8/012}{\emph{Class. Quant. Grav.}
  {\bf 6} (1989) 1125}.

\bibitem{Casalbuoni:2014ofa}
R.~Casalbuoni and J.~Gomis, \emph{{Conformal symmetry for relativistic point
  particles}}, \href{http://dx.doi.org/10.1103/PhysRevD.90.026001}{\emph{Phys.
  Rev. D} {\bf 90} (2014) 026001}, [\href{http://arxiv.org/abs/1404.5766}{{\tt
  1404.5766}}].

\bibitem{Casalbuoni:2014qia}
R.~Casalbuoni and J.~Gomis, \emph{{Conformal symmetry for relativistic point
  particles: an addendum}},
  \href{http://dx.doi.org/10.1103/PhysRevD.91.047901}{\emph{Phys. Rev. D} {\bf
  91} (2015) 047901}, [\href{http://arxiv.org/abs/1412.6903}{{\tt 1412.6903}}].

\bibitem{Rempel:2016elp}
T.~Rempel and L.~Freidel, \emph{{Bilocal model for the relativistic spinning
  particle}}, \href{http://dx.doi.org/10.1103/PhysRevD.95.104014}{\emph{Phys.
  Rev. D} {\bf 95} (2017) 104014}, [\href{http://arxiv.org/abs/1609.09110}{{\tt
  1609.09110}}].

\bibitem{Grigoriev:2021bes}
M.~Grigoriev, A.~Meyer and I.~Sachs, \emph{{A toy model for background
  independent string field theory}},
  \href{http://dx.doi.org/10.1007/JHEP05(2022)020}{\emph{JHEP} {\bf 05} (2022)
  020}, [\href{http://arxiv.org/abs/2106.07966}{{\tt 2106.07966}}].

\bibitem{Boffo:2023fsz}
E.~Boffo, \emph{{Spinning particles and background fields}},  in
  \emph{{CORFU2022: 22th Hellenic School and Workshops on Elementary Particle
  Physics and Gravity}}, 4, 2023.
\newblock \href{http://arxiv.org/abs/2304.12909}{{\tt 2304.12909}}.

\bibitem{Eastwood:2002su}
M.~G. Eastwood, \emph{{Higher symmetries of the Laplacian}},
  \href{http://dx.doi.org/10.4007/annals.2005.161.1645}{\emph{Annals Math.}
  {\bf 161} (2005) 1645--1665},
  [\href{http://arxiv.org/abs/hep-th/0206233}{{\tt hep-th/0206233}}].

\bibitem{Joung:2014qya}
E.~Joung and K.~Mkrtchyan, \emph{{Notes on higher-spin algebras: minimal
  representations and structure constants}},
  \href{http://dx.doi.org/10.1007/JHEP05(2014)103}{\emph{JHEP} {\bf 05} (2014)
  103}, [\href{http://arxiv.org/abs/1401.7977}{{\tt 1401.7977}}].

\bibitem{Michel2011}
J.-P. Michel, \emph{{Higher symmetries of the Laplacian via quantization}},
  \href{http://dx.doi.org/10.5802/aif.2891}{\emph{Annales de l'Institut
  Fourier} {\bf 64} (2014) 1581--1609},
  [\href{http://arxiv.org/abs/1107.5840}{{\tt 1107.5840}}].

\bibitem{Joung:2015jza}
E.~Joung and K.~Mkrtchyan, \emph{{Partially-massless higher-spin algebras and
  their finite-dimensional truncations}},
  \href{http://dx.doi.org/10.1007/JHEP01(2016)003}{\emph{JHEP} {\bf 01} (2016)
  003}, [\href{http://arxiv.org/abs/1508.07332}{{\tt 1508.07332}}].

\bibitem{Rowe:2011zz}
D.~J. Rowe, J.~Repka and M.~J. Carvalho, \emph{{Simple unified proofs of four
  duality theorems}}, \href{http://dx.doi.org/10.1063/1.3525978}{\emph{J. Math.
  Phys.} {\bf 52} (2011) 013507}.

\bibitem{Rowe:2012ym}
D.~J. Rowe, M.~J. Carvalho and J.~Repka, \emph{{Dual pairing of symmetry groups
  and dynamical groups in physics}},
  \href{http://dx.doi.org/10.1103/RevModPhys.84.711}{\emph{Rev. Mod. Phys.}
  {\bf 84} (2012) 711--757}, [\href{http://arxiv.org/abs/1207.0148}{{\tt
  1207.0148}}].

\bibitem{partII}
T.~Basile, E.~Joung and T.~Oh, \emph{{Manifestly Covariant Worldline Actions
  from Coadjoint Orbits, Part II: Twistorial description and Quantization}}, .

\bibitem{Metsaev:2019opn}
R.~R. Metsaev, \emph{{Light-cone continuous-spin field in AdS space}},
  \href{http://dx.doi.org/10.1016/j.physletb.2019.04.041}{\emph{Phys. Lett. B}
  {\bf 793} (2019) 134--140}, [\href{http://arxiv.org/abs/1903.10495}{{\tt
  1903.10495}}].

\bibitem{Brugues:2004an}
J.~Brugues, T.~Curtright, J.~Gomis and L.~Mezincescu, \emph{{Non-relativistic
  strings and branes as non-linear realizations of Galilei groups}},
  \href{http://dx.doi.org/10.1016/j.physletb.2004.05.024}{\emph{Phys. Lett. B}
  {\bf 594} (2004) 227--233}, [\href{http://arxiv.org/abs/hep-th/0404175}{{\tt
  hep-th/0404175}}].

\bibitem{Gomis:2004pw}
J.~Gomis, K.~Kamimura and P.~K. Townsend, \emph{{Non-relativistic
  superbranes}},
  \href{http://dx.doi.org/10.1088/1126-6708/2004/11/051}{\emph{JHEP} {\bf 11}
  (2004) 051}, [\href{http://arxiv.org/abs/hep-th/0409219}{{\tt
  hep-th/0409219}}].

\bibitem{Brugues:2006yd}
J.~Brugues, J.~Gomis and K.~Kamimura, \emph{{Newton-Hooke algebras,
  non-relativistic branes and generalized pp-wave metrics}},
  \href{http://dx.doi.org/10.1103/PhysRevD.73.085011}{\emph{Phys. Rev. D} {\bf
  73} (2006) 085011}, [\href{http://arxiv.org/abs/hep-th/0603023}{{\tt
  hep-th/0603023}}].

\bibitem{Casalbuoni:2008iy}
R.~Casalbuoni, J.~Gomis, K.~Kamimura and G.~Longhi, \emph{{Space-time Vector
  Supersymmetry and Massive Spinning Particle}},
  \href{http://dx.doi.org/10.1088/1126-6708/2008/02/094}{\emph{JHEP} {\bf 02}
  (2008) 094}, [\href{http://arxiv.org/abs/0801.2702}{{\tt 0801.2702}}].

\bibitem{Gomis:2011dw}
J.~Gomis and K.~Kamimura, \emph{{Schrodinger Equations for Higher Order
  Non-relativistic Particles and N-Galilean Conformal Symmetry}},
  \href{http://dx.doi.org/10.1103/PhysRevD.85.045023}{\emph{Phys. Rev. D} {\bf
  85} (2012) 045023}, [\href{http://arxiv.org/abs/1109.3773}{{\tt 1109.3773}}].

\bibitem{Bergshoeff:2014jla}
E.~Bergshoeff, J.~Gomis and G.~Longhi, \emph{{Dynamics of Carroll Particles}},
  \href{http://dx.doi.org/10.1088/0264-9381/31/20/205009}{\emph{Class. Quant.
  Grav.} {\bf 31} (2014) 205009}, [\href{http://arxiv.org/abs/1405.2264}{{\tt
  1405.2264}}].

\bibitem{Bergshoeff:2015wma}
E.~Bergshoeff, J.~Gomis and L.~Parra, \emph{{The Symmetries of the Carroll
  Superparticle}},
  \href{http://dx.doi.org/10.1088/1751-8113/49/18/185402}{\emph{J. Phys. A}
  {\bf 49} (2016) 185402}, [\href{http://arxiv.org/abs/1503.06083}{{\tt
  1503.06083}}].

\bibitem{Barducci:2017mse}
A.~Barducci, R.~Casalbuoni and J.~Gomis, \emph{{Non-relativistic Spinning
  Particle in a Newton-Cartan Background}},
  \href{http://dx.doi.org/10.1007/JHEP01(2018)002}{\emph{JHEP} {\bf 01} (2018)
  002}, [\href{http://arxiv.org/abs/1710.10970}{{\tt 1710.10970}}].

\bibitem{Kirillov1999}
A.~Kirillov, \emph{{Merits and demerits of the orbit method}},
  \href{http://dx.doi.org/10.1090/S0273-0979-99-00849-6}{\emph{Bulletin of the
  American Mathematical Society} {\bf 36} (1999) 433--488}.

\bibitem{Vogan2000}
D.~A. Vogan~Jr, \emph{{The method of coadjoint orbits for real reductive
  groups}}, vol.~8 of \emph{IAS/Park City Mathematics Series}, pp.~179--238.
\newblock Amer. Math. Soc. Providence, RI, 2000.

\bibitem{Collingwood1993}
D.~H. Collingwood and W.~M. McGovern, \emph{{Nilpotent Orbits in Semisimple Lie
  Algebras: an introduction}}.
\newblock Van Nostrand Reinhold, New York, 1993.

\bibitem{Burgoyne1977}
N.~Burgoyne and R.~Cushman, \emph{{Conjugacy classes in linear groups}},
  \href{http://dx.doi.org/10.1016/0021-8693(77)90186-7}{\emph{Journal of
  algebra} {\bf 44} (1977) 339--362}.

\bibitem{Djokovic1983}
D.~Djokovi{\'c}, J.~Patera, P.~Winternitz and H.~Zassenhaus, \emph{{Normal
  forms of elements of classical real and complex Lie and Jordan algebras}},
  \href{http://dx.doi.org/10.1063/1.525868}{\emph{Journal of Mathematical
  Physics} {\bf 24} (1983) 1363--1374}.

\bibitem{Hanany:2016gbz}
A.~Hanany and R.~Kalveks, \emph{{Quiver Theories for Moduli Spaces of Classical
  Group Nilpotent Orbits}},
  \href{http://dx.doi.org/10.1007/JHEP06(2016)130}{\emph{JHEP} {\bf 06} (2016)
  130}, [\href{http://arxiv.org/abs/1601.04020}{{\tt 1601.04020}}].

\bibitem{Cabrera:2017ucb}
S.~Cabrera, A.~Hanany and Z.~Zhong, \emph{{Nilpotent orbits and the Coulomb
  branch of $T^\sigma (G)$ theories: special orthogonal vs orthogonal gauge
  group factors}}, \href{http://dx.doi.org/10.1007/JHEP11(2017)079}{\emph{JHEP}
  {\bf 11} (2017) 079}, [\href{http://arxiv.org/abs/1707.06941}{{\tt
  1707.06941}}].

\bibitem{Hanany:2017ooe}
A.~Hanany and R.~Kalveks, \emph{{Quiver Theories and Formulae for Nilpotent
  Orbits of Exceptional Algebras}},
  \href{http://dx.doi.org/10.1007/JHEP11(2017)126}{\emph{JHEP} {\bf 11} (2017)
  126}, [\href{http://arxiv.org/abs/1709.05818}{{\tt 1709.05818}}].

\bibitem{Hanany:2018uzt}
A.~Hanany and M.~Sperling, \emph{{Resolutions of nilpotent orbit closures via
  Coulomb branches of 3-dimensional $ \mathcal{N}=4 $ theories}},
  \href{http://dx.doi.org/10.1007/JHEP08(2018)189}{\emph{JHEP} {\bf 08} (2018)
  189}, [\href{http://arxiv.org/abs/1806.01890}{{\tt 1806.01890}}].

\bibitem{Hanany:2018xth}
A.~Hanany and D.~Miketa, \emph{{Nilpotent orbit Coulomb branches of types AD}},
  \href{http://dx.doi.org/10.1007/JHEP02(2019)113}{\emph{JHEP} {\bf 02} (2019)
  113}, [\href{http://arxiv.org/abs/1807.11491}{{\tt 1807.11491}}].

\bibitem{Springer1970}
T.~Springer and R.~Steinberg, \emph{{Conjugacy classes}},  in \emph{Seminar on
  Algebraic Groups and Related Finite Groups: Held at The Institute for
  Advanced Study, Princeton/NJ, 1968/69}, pp.~167--266, 1970.
\newblock \href{http://dx.doi.org/10.1007/BFb0081546}{DOI}.

\bibitem{Gerstenhaber1961}
M.~Gerstenhaber, \emph{{Dominance over the classical groups}},
  \href{http://dx.doi.org/10.2307/1970297}{\emph{Annals of Mathematics} {\bf
  74} (1961) 532--569}.

\bibitem{Knapp2013}
A.~Knapp, \emph{Lie Groups Beyond an Introduction}.
\newblock Birkh{\"a}user Boston, 2013.

\bibitem{Barnich:2017jgw}
G.~Barnich, H.~A. Gonzalez and P.~Salgado-Rebolledo, \emph{{Geometric actions
  for three-dimensional gravity}},
  \href{http://dx.doi.org/10.1088/1361-6382/aa9806}{\emph{Class. Quant. Grav.}
  {\bf 35} (2018) 014003}, [\href{http://arxiv.org/abs/1707.08887}{{\tt
  1707.08887}}].

\bibitem{Ciambelli:2022cfr}
L.~Ciambelli and R.~G. Leigh, \emph{{Universal corner symmetry and the orbit
  method for gravity}},
  \href{http://dx.doi.org/10.1016/j.nuclphysb.2022.116053}{\emph{Nucl. Phys. B}
  {\bf 986} (2023) 116053}, [\href{http://arxiv.org/abs/2207.06441}{{\tt
  2207.06441}}].

\bibitem{Barnich:2022bni}
G.~Barnich, K.~Nguyen and R.~Ruzziconi, \emph{{Geometric action for extended
  Bondi-Metzner-Sachs group in four dimensions}},
  \href{http://dx.doi.org/10.1007/JHEP12(2022)154}{\emph{JHEP} {\bf 12} (2022)
  154}, [\href{http://arxiv.org/abs/2211.07592}{{\tt 2211.07592}}].

\bibitem{Alvarez:1984es}
O.~Alvarez, \emph{{Topological Quantization and Cohomology}},
  \href{http://dx.doi.org/10.1007/BF01212452}{\emph{Commun. Math. Phys.} {\bf
  100} (1985) 279}.

\bibitem{Hanson:1974qy}
A.~J. Hanson and T.~Regge, \emph{{The Relativistic Spherical Top}},
  \href{http://dx.doi.org/10.1016/0003-4916(74)90046-3}{\emph{Annals Phys.}
  {\bf 87} (1974) 498}.

\bibitem{Steinhoff:2015ksa}
J.~Steinhoff, \emph{{Spin gauge symmetry in the action principle for classical
  relativistic particles}},  \href{http://arxiv.org/abs/1501.04951}{{\tt
  1501.04951}}.

\bibitem{Kim:2021rda}
J.-H. Kim, J.-W. Kim and S.~Lee, \emph{{The relativistic spherical top as a
  massive twistor}}, \href{http://dx.doi.org/10.1088/1751-8121/ac11be}{\emph{J.
  Phys. A} {\bf 54} (2021) 335203},
  [\href{http://arxiv.org/abs/2102.07063}{{\tt 2102.07063}}].

\bibitem{Cho:1994gs}
J.-H. Cho, S.~Hyun and J.-K. Kim, \emph{{A Covariant formulation of classical
  spinning particle}},
  \href{http://dx.doi.org/10.1142/S0217732394000599}{\emph{Mod. Phys. Lett. A}
  {\bf 9} (1994) 775--784}, [\href{http://arxiv.org/abs/hep-th/9402012}{{\tt
  hep-th/9402012}}].

\bibitem{Simms:1976dhp}
D.~J. Simms and N.~M.~J. Woodhouse, \emph{{Lectures on Geometric
  Quantization}}.
\newblock Springer Berlin, Heidelberg, 1976,
  \href{http://dx.doi.org/10.1007/3-540-07860-6}{10.1007/3-540-07860-6}.

\bibitem{Bates:1997kc}
S.~Bates and A.~Weinstein, \emph{{Lectures on the geometry of quantization}}.
\newblock American Mathematical Soc., 1997.

\bibitem{Moshayedi:2020spz}
N.~Moshayedi, \emph{{Notes on Geometric Quantization}},
  \href{http://arxiv.org/abs/2010.15419}{{\tt 2010.15419}}.

\bibitem{Wernli:2023pib}
K.~Wernli, \emph{{Six lectures on geometric quantization}},
  \href{http://dx.doi.org/10.22323/1.435.0005}{\emph{PoS} {\bf Modave2022}
  (2023) 005}, [\href{http://arxiv.org/abs/2306.00178}{{\tt 2306.00178}}].

\bibitem{Witten:1987ty}
E.~Witten, \emph{{Coadjoint Orbits of the Virasoro Group}},
  \href{http://dx.doi.org/10.1007/BF01218287}{\emph{Commun. Math. Phys.} {\bf
  114} (1988) 1}.

\bibitem{Dzhordzhadze:1994np}
G.~Dzhordzhadze, L.~O'Raifeartaigh and I.~Tsutsui, \emph{{Quantization of a
  relativistic particle on the SL(2,R) manifold based on Hamiltonian
  reduction}},
  \href{http://dx.doi.org/10.1016/0370-2693(94)90549-5}{\emph{Phys. Lett. B}
  {\bf 336} (1994) 388--394}, [\href{http://arxiv.org/abs/hep-th/9407059}{{\tt
  hep-th/9407059}}].

\bibitem{Ashok:2022thd}
S.~K. Ashok and J.~Troost, \emph{{Path integrals on sl(2, R) orbits}},
  \href{http://dx.doi.org/10.1088/1751-8121/ac802c}{\emph{J. Phys. A} {\bf 55}
  (2022) 335202}, [\href{http://arxiv.org/abs/2204.00232}{{\tt 2204.00232}}].

\bibitem{Enayati:2022hed}
M.~Enayati, J.-P. Gazeau, H.~Pejhan and A.~Wang, \emph{{The de Sitter group and
  its representations: a window on the notion of de Sitterian elementary
  systems}}.
\newblock Springer Cham, 2022.
\newblock \href{http://arxiv.org/abs/2201.11457}{{\tt 2201.11457}}.
\newblock 10.1007/978-3-031-16045-5.

\bibitem{Enayati:2023lld}
M.~Enayati, J.-P. Gazeau, M.~A. del Olmo and H.~Pejhan, \emph{{Anti-de
  Sitterian ''massive'' elementary systems and their Minkowskian and Newtonian
  limits}},  \href{http://arxiv.org/abs/2307.06690}{{\tt 2307.06690}}.

\bibitem{Ortega2013}
J.-P. Ortega and T.~S. Ratiu, \emph{{Momentum maps and Hamiltonian reduction}},
  vol.~222.
\newblock Springer Science \& Business Media, 2013.

\bibitem{Baier2023}
T.~Baier, J.~Hilgert, O.~Kaya, J.~M. Mourão and J.~P. Nunes,
  \emph{{Quantization in fibering polarizations, Mabuchi rays and geometric
  Peter--Weyl theorem}},  \href{http://arxiv.org/abs/2301.10853}{{\tt
  2301.10853}}.

\bibitem{CannasDaSilva1999}
A.~C. Da~Silva and A.~Weinstein, \emph{{Geometric models for noncommutative
  algebras}}, vol.~10.
\newblock American Mathematical Soc., 1999.

\bibitem{Prasad1993}
D.~Prasad, \emph{{Weil representation, Howe duality, and the theta
  correspondence}},  in \emph{Theta functions: from the classical to the
  modern}, vol.~1, pp.~105--127, 1993.

\bibitem{Kudla1996}
S.~Kudla, \emph{{Notes on the local theta correspondence}}, {\emph{{\rm
  unpublished notes available
  \href{http://www.math.toronto.edu/~skudla/castle.pdf}{here}}} (1996) }.

\bibitem{Adams2007}
J.~Adams, \emph{{The theta correspondence over R}},  in \emph{Harmonic
  Analysis, Group Representations, Automorphic Forms, and Invariant Theory},
  pp.~\href{https://pdfs.semanticscholar.org/748b/8e1dd0ce860b37e3c11c7737a409afee09e1.pdf}{1--39},
  2007.

\bibitem{Rawnsley1975}
J.~H. Rawnsley, \emph{{Representations of a semi-direct product by
  quantization}},
  \href{http://dx.doi.org/10.1017/S0305004100051793}{\emph{Mathematical
  Proceedings of the Cambridge Philosophical Society} {\bf 78} (1975)
  345--350}.

\bibitem{Baguis1998}
P.~Baguis, \emph{{Semidirect products and the Pukanszky condition}},
  \href{http://dx.doi.org/10.1016/S0393-0440(97)00028-4}{\emph{Journal of
  Geometry and physics} {\bf 25} (1998) 245--270},
  [\href{http://arxiv.org/abs/dg-ga/9705005}{{\tt dg-ga/9705005}}].

\bibitem{Oblak:2016eij}
B.~Oblak, \emph{{BMS Particles in Three Dimensions}}.
\newblock PhD thesis, U. Brussels, Brussels U., 2016.
\newblock \href{http://arxiv.org/abs/1610.08526}{{\tt 1610.08526}}.
\newblock 10.1007/978-3-319-61878-4.

\bibitem{Arathoon2019}
P.~Arathoon, \emph{{Semidirect Products and Applications to Geometric
  Mechanics}}.
\newblock PhD thesis, University of Manchester, 2019.

\bibitem{Itoh2003}
M.~Itoh, \emph{{Correspondences of the Gelfand invariants in reductive dual
  pairs}}, \href{http://dx.doi.org/10.1017/S1446788700003761}{\emph{Journal of
  the Australian Mathematical Society} {\bf 75} (2003) 263–278}.

\bibitem{Kuzenko:2020ayk}
S.~M. Kuzenko and A.~E. Pindur, \emph{{Massless particles in five and higher
  dimensions}},
  \href{http://dx.doi.org/10.1016/j.physletb.2020.136020}{\emph{Phys. Lett. B}
  {\bf 812} (2021) 136020}, [\href{http://arxiv.org/abs/2010.07124}{{\tt
  2010.07124}}].

\bibitem{Kosinski:2020jmd}
P.~Kosi\'nski and P.~Ma\'slanka, \emph{{Relativistic Symmetries and Hamiltonian
  Formalism}}, \href{http://dx.doi.org/10.3390/sym12111810}{\emph{Symmetry}
  {\bf 12} (2020) 1810}.

\bibitem{Lahlali:2021nrf}
I.~A. Lahlali, N.~Boulanger and A.~Campoleoni, \emph{{Coadjoint Orbits of the
  Poincar\'e Group for Discrete-Spin Particles in Any Dimension}},
  \href{http://dx.doi.org/10.3390/sym13091749}{\emph{Symmetry} {\bf 13} (2021)
  1749}.

\bibitem{Figueroa-OFarrill:2023qty}
J.~Figueroa-O'Farrill, A.~P\'erez and S.~Prohazka, \emph{{Quantum
  Carroll/fracton particles}},  \href{http://arxiv.org/abs/2307.05674}{{\tt
  2307.05674}}.

\bibitem{Mourad:2004fg}
J.~Mourad, \emph{{Continuous spin and tensionless strings}},
  \href{http://arxiv.org/abs/hep-th/0410009}{{\tt hep-th/0410009}}.

\bibitem{Edgren:2005gq}
L.~Edgren, R.~Marnelius and P.~Salomonson, \emph{{Infinite spin particles}},
  \href{http://dx.doi.org/10.1088/1126-6708/2005/05/002}{\emph{JHEP} {\bf 05}
  (2005) 002}, [\href{http://arxiv.org/abs/hep-th/0503136}{{\tt
  hep-th/0503136}}].

\bibitem{Wigner:1963wwt}
E.~P. Wigner, \emph{{Invariant Quantum Mechanical Equations of Motion}},  in
  \emph{{Theoretical Physics}}, (Vienna), pp.~59--82, IAEA, 1963.

\bibitem{Mourad:2005rt}
J.~Mourad, \emph{{Continuous spin particles from a string theory}},
  \href{http://arxiv.org/abs/hep-th/0504118}{{\tt hep-th/0504118}}.

\bibitem{Mourad:2006xk}
J.~Mourad, \emph{{Continuous spin particles from a tensionless string theory}},
  \href{http://dx.doi.org/10.1063/1.2399607}{\emph{AIP Conf. Proc.} {\bf 861}
  (2006) 436--443}.

\bibitem{Bekaert:2005in}
X.~Bekaert and J.~Mourad, \emph{{The Continuous spin limit of higher spin field
  equations}},
  \href{http://dx.doi.org/10.1088/1126-6708/2006/01/115}{\emph{JHEP} {\bf 01}
  (2006) 115}, [\href{http://arxiv.org/abs/hep-th/0509092}{{\tt
  hep-th/0509092}}].

\bibitem{Alkalaev:2017hvj}
K.~B. Alkalaev and M.~A. Grigoriev, \emph{{Continuous spin fields of
  mixed-symmetry type}},
  \href{http://dx.doi.org/10.1007/JHEP03(2018)030}{\emph{JHEP} {\bf 03} (2018)
  030}, [\href{http://arxiv.org/abs/1712.02317}{{\tt 1712.02317}}].

\bibitem{Alkalaev:2018bqe}
K.~Alkalaev, A.~Chekmenev and M.~Grigoriev, \emph{{Unified formulation for
  helicity and continuous spin fermionic fields}},
  \href{http://dx.doi.org/10.1007/JHEP11(2018)050}{\emph{JHEP} {\bf 11} (2018)
  050}, [\href{http://arxiv.org/abs/1808.09385}{{\tt 1808.09385}}].

\bibitem{Fronsdal:1986ui}
C.~Fronsdal, \emph{{Flat space singletons}},
  \href{http://dx.doi.org/10.1103/PhysRevD.35.1262}{\emph{Phys. Rev. D} {\bf
  35} (1987) 1262}.

\bibitem{Ponomarev:2022ryp}
D.~Ponomarev, \emph{{Towards higher-spin holography in flat space}},
  \href{http://dx.doi.org/10.1007/JHEP01(2023)084}{\emph{JHEP} {\bf 01} (2023)
  084}, [\href{http://arxiv.org/abs/2210.04035}{{\tt 2210.04035}}].

\bibitem{Bekaert:2022oeh}
X.~Bekaert, A.~Campoleoni and S.~Pekar, \emph{{Carrollian conformal scalar as
  flat-space singleton}},
  \href{http://dx.doi.org/10.1016/j.physletb.2023.137734}{\emph{Phys. Lett. B}
  {\bf 838} (2023) 137734}, [\href{http://arxiv.org/abs/2211.16498}{{\tt
  2211.16498}}].

\bibitem{Kosinski:2022qrf}
P.~Kosinski and P.~Maslanka, \emph{{Classical and quantum particles from
  nongeneric conformal orbits}},
  \href{http://dx.doi.org/10.1016/j.nuclphysb.2023.116226}{\emph{Nucl. Phys. B}
  {\bf 991} (2023) 116226}, [\href{http://arxiv.org/abs/2207.12756}{{\tt
  2207.12756}}].

\bibitem{Engquist:2005yt}
J.~Engquist and P.~Sundell, \emph{{Brane partons and singleton strings}},
  \href{http://dx.doi.org/10.1016/j.nuclphysb.2006.06.040}{\emph{Nucl. Phys. B}
  {\bf 752} (2006) 206--279}, [\href{http://arxiv.org/abs/hep-th/0508124}{{\tt
  hep-th/0508124}}].

\bibitem{Alkalaev:2009vm}
K.~B. Alkalaev and M.~Grigoriev, \emph{{Unified BRST description of AdS gauge
  fields}},
  \href{http://dx.doi.org/10.1016/j.nuclphysb.2010.04.004}{\emph{Nucl. Phys. B}
  {\bf 835} (2010) 197--220}, [\href{http://arxiv.org/abs/0910.2690}{{\tt
  0910.2690}}].

\bibitem{Alkalaev:2011zv}
K.~Alkalaev and M.~Grigoriev, \emph{{Unified BRST approach to (partially)
  massless and massive AdS fields of arbitrary symmetry type}},
  \href{http://dx.doi.org/10.1016/j.nuclphysb.2011.08.005}{\emph{Nucl. Phys. B}
  {\bf 853} (2011) 663--687}, [\href{http://arxiv.org/abs/1105.6111}{{\tt
  1105.6111}}].

\bibitem{Caron-Huot:2017vep}
S.~Caron-Huot, \emph{{Analyticity in Spin in Conformal Theories}},
  \href{http://dx.doi.org/10.1007/JHEP09(2017)078}{\emph{JHEP} {\bf 09} (2017)
  078}, [\href{http://arxiv.org/abs/1703.00278}{{\tt 1703.00278}}].

\bibitem{Metsaev:2016lhs}
R.~R. Metsaev, \emph{{Continuous spin gauge field in (A)dS space}},
  \href{http://dx.doi.org/10.1016/j.physletb.2017.02.027}{\emph{Phys. Lett. B}
  {\bf 767} (2017) 458--464}, [\href{http://arxiv.org/abs/1610.00657}{{\tt
  1610.00657}}].

\bibitem{Heidenreich:1980xi}
W.~Heidenreich, \emph{{Tensor Products of Positive Energy Representations of
  SO(3,2) and SO(4,2)}}, \href{http://dx.doi.org/10.1063/1.525099}{\emph{J.
  Math. Phys.} {\bf 22} (1981) 1566}.

\bibitem{Girardello:2002pp}
L.~Girardello, M.~Porrati and A.~Zaffaroni, \emph{{3-D interacting CFTs and
  generalized Higgs phenomenon in higher spin theories on AdS}},
  \href{http://dx.doi.org/10.1016/S0370-2693(03)00492-1}{\emph{Phys. Lett. B}
  {\bf 561} (2003) 289--293}, [\href{http://arxiv.org/abs/hep-th/0212181}{{\tt
  hep-th/0212181}}].

\bibitem{Vasiliev:1989re}
M.~A. Vasiliev, \emph{{Higher Spin Algebras and Quantization on the Sphere and
  Hyperboloid}}, \href{http://dx.doi.org/10.1142/S0217751X91000605}{\emph{Int.
  J. Mod. Phys. A} {\bf 6} (1991) 1115--1135}.

\bibitem{Metsaev:2017cuz}
R.~R. Metsaev, \emph{{Cubic interaction vertices for continuous-spin fields and
  arbitrary spin massive fields}},
  \href{http://dx.doi.org/10.1007/JHEP11(2017)197}{\emph{JHEP} {\bf 11} (2017)
  197}, [\href{http://arxiv.org/abs/1709.08596}{{\tt 1709.08596}}].

\bibitem{Metsaev:2017ytk}
R.~R. Metsaev, \emph{{Fermionic continuous spin gauge field in (A)dS space}},
  \href{http://dx.doi.org/10.1016/j.physletb.2017.08.020}{\emph{Phys. Lett. B}
  {\bf 773} (2017) 135--141}, [\href{http://arxiv.org/abs/1703.05780}{{\tt
  1703.05780}}].

\bibitem{Metsaev:2018moa}
R.~R. Metsaev, \emph{{Cubic interaction vertices for massive/massless
  continuous-spin fields and arbitrary spin fields}},
  \href{http://dx.doi.org/10.1007/JHEP12(2018)055}{\emph{JHEP} {\bf 12} (2018)
  055}, [\href{http://arxiv.org/abs/1809.09075}{{\tt 1809.09075}}].

\bibitem{Metsaev:2021zdg}
R.~R. Metsaev, \emph{{Mixed-symmetry continuous-spin fields in flat and AdS
  spaces}}, \href{http://dx.doi.org/10.1016/j.physletb.2021.136497}{\emph{Phys.
  Lett. B} {\bf 820} (2021) 136497},
  [\href{http://arxiv.org/abs/2105.11281}{{\tt 2105.11281}}].

\bibitem{Brylinski1998}
R.~Brylinski, \emph{{Geometric quantization of real minimal nilpotent orbits}},
  \href{http://dx.doi.org/10.1016/S0926-2245(98)00017-5}{\emph{Differential
  Geometry and its Applications} {\bf 9} (1998) 5--58},
  [\href{http://arxiv.org/abs/math/9811033}{{\tt math/9811033}}].

\bibitem{Fronsdal2009}
C.~Fronsdal, \emph{{Deformation Quantization on the Closure of Minimal
  Coadjoint Orbits}},
  \href{http://dx.doi.org/10.1007/s11005-009-0316-5}{\emph{Letters in
  Mathematical Physics} {\bf 88} (2009) 271--320},
  [\href{http://arxiv.org/abs/math/0510580}{{\tt math/0510580}}].

\bibitem{Bekaert:2011js}
X.~Bekaert, \emph{{Singletons and their maximal symmetry algebras}},  in
  \emph{{6th Summer School in Modern Mathematical Physics}}, pp.~71--89, 11,
  2011.
\newblock \href{http://arxiv.org/abs/1111.4554}{{\tt 1111.4554}}.

\bibitem{Metsaev:1995re}
R.~R. Metsaev, \emph{{Massless mixed symmetry bosonic free fields in
  d-dimensional anti-de Sitter space-time}},
  \href{http://dx.doi.org/10.1016/0370-2693(95)00563-Z}{\emph{Phys. Lett. B}
  {\bf 354} (1995) 78--84}.

\bibitem{Metsaev:1998xg}
R.~R. Metsaev, \emph{{Fermionic fields in the d-dimensional anti-de Sitter
  space-time}},
  \href{http://dx.doi.org/10.1016/S0370-2693(97)01446-9}{\emph{Phys. Lett. B}
  {\bf 419} (1998) 49--56}, [\href{http://arxiv.org/abs/hep-th/9802097}{{\tt
  hep-th/9802097}}].

\bibitem{Boulanger:2008kw}
N.~Boulanger, C.~Iazeolla and P.~Sundell, \emph{{Unfolding Mixed-Symmetry
  Fields in AdS and the BMV Conjecture. II. Oscillator Realization}},
  \href{http://dx.doi.org/10.1088/1126-6708/2009/07/014}{\emph{JHEP} {\bf 07}
  (2009) 014}, [\href{http://arxiv.org/abs/0812.4438}{{\tt 0812.4438}}].

\bibitem{Boulanger:2008up}
N.~Boulanger, C.~Iazeolla and P.~Sundell, \emph{{Unfolding Mixed-Symmetry
  Fields in AdS and the BMV Conjecture: I. General Formalism}},
  \href{http://dx.doi.org/10.1088/1126-6708/2009/07/013}{\emph{JHEP} {\bf 07}
  (2009) 013}, [\href{http://arxiv.org/abs/0812.3615}{{\tt 0812.3615}}].

\bibitem{Skvortsov:2009nv}
E.~D. Skvortsov, \emph{{Gauge fields in (A)dS(d) within the unfolded approach:
  algebraic aspects}},
  \href{http://dx.doi.org/10.1007/JHEP01(2010)106}{\emph{JHEP} {\bf 01} (2010)
  106}, [\href{http://arxiv.org/abs/0910.3334}{{\tt 0910.3334}}].

\bibitem{Skvortsov:2009zu}
E.~D. Skvortsov, \emph{{Gauge fields in (A)dS(d) and Connections of its
  symmetry algebra}},
  \href{http://dx.doi.org/10.1088/1751-8113/42/38/385401}{\emph{J. Phys. A}
  {\bf 42} (2009) 385401}, [\href{http://arxiv.org/abs/0904.2919}{{\tt
  0904.2919}}].

\bibitem{Angelopoulos:1980wg}
E.~Angelopoulos, M.~Flato, C.~Fronsdal and D.~Sternheimer, \emph{{Massless
  Particles, Conformal Group and De Sitter Universe}},
  \href{http://dx.doi.org/10.1103/PhysRevD.23.1278}{\emph{Phys. Rev. D} {\bf
  23} (1981) 1278}.

\bibitem{Angelopoulos:1997ij}
E.~Angelopoulos and M.~Laoues, \emph{{Masslessness in n-dimensions}},
  \href{http://dx.doi.org/10.1142/S0129055X98000082}{\emph{Rev. Math. Phys.}
  {\bf 10} (1998) 271--300}, [\href{http://arxiv.org/abs/hep-th/9806100}{{\tt
  hep-th/9806100}}].

\bibitem{Angelopoulos:1999bz}
E.~Angelopoulos and M.~Laoues, \emph{{Singletons on AdS(n)}},
  \href{http://dx.doi.org/10.1007/978-94-015-1276-3_1}{\emph{Math. Phys. Stud.}
  {\bf 21-22} (2000) 3--23}.

\bibitem{Barnich:2014kra}
G.~Barnich and B.~Oblak, \emph{{Notes on the BMS group in three dimensions: I.
  Induced representations}},
  \href{http://dx.doi.org/10.1007/JHEP06(2014)129}{\emph{JHEP} {\bf 06} (2014)
  129}, [\href{http://arxiv.org/abs/1403.5803}{{\tt 1403.5803}}].

\bibitem{Barnich:2015uva}
G.~Barnich and B.~Oblak, \emph{{Notes on the BMS group in three dimensions: II.
  Coadjoint representation}},
  \href{http://dx.doi.org/10.1007/JHEP03(2015)033}{\emph{JHEP} {\bf 03} (2015)
  033}, [\href{http://arxiv.org/abs/1502.00010}{{\tt 1502.00010}}].

\bibitem{Skerritt2019}
P.~Skerritt and C.~Vizman, \emph{{Dual pairs for matrix groups}},
  \href{http://dx.doi.org/10.3934/jgm.2019014}{\emph{Journal of Geometric
  Mechanics} {\bf 11} (2019) 255–275},
  [\href{http://arxiv.org/abs/1805.01519}{{\tt 1805.01519}}].

\bibitem{Kazhdan1978}
D.~Kazhdan, B.~Kostant and S.~Sternberg, \emph{{Hamiltonian group actions and
  dynamical systems of calogero type}},
  \href{http://dx.doi.org/10.1002/cpa.3160310405}{\emph{Communications on Pure
  and Applied Mathematics} {\bf 31} (1978) 481--507}.

\bibitem{Adams1987}
J.~Adams, \emph{{Coadjoint orbits and reductive dual pairs}},
  \href{http://dx.doi.org/10.1016/0001-8708(87)90050-8}{\emph{Advances in
  Mathematics} {\bf 63} (1987) 138--151}.

\bibitem{Przebinda1993}
T.~Przebinda, \emph{{Characters, dual pairs, and unitary representations}},
  \href{http://dx.doi.org/10.1215/S0012-7094-93-06923-2}{\emph{Duke
  Mathematical Journal} {\bf 69} (1993) 547 -- 592}.

\bibitem{Daszkiewicz1997}
A.~Daszkiewicz, W.~Kraśkiewicz and T.~Przebinda, \emph{{Nilpotent Orbits and
  Complex Dual Pairs}},
  \href{http://dx.doi.org/10.1006/jabr.1996.6910}{\emph{Journal of Algebra}
  {\bf 190} (1997) 518--539}.

\bibitem{Pan2010}
S.-Y. Pan, \emph{Orbit correspondences for real reductive dual pairs},
  \href{http://dx.doi.org/10.2140/PJM.2010.248.403}{\emph{Pacific Journal of
  Mathematics} {\bf 248} (2010) 403--427}.

\bibitem{Dwivedi2019}
S.~Dwivedi, J.~Herman, L.~C. Jeffrey, T.~Van~den Hurk et~al.,
  \emph{{Hamiltonian group actions and equivariant cohomology}}.
\newblock Springer, 2019.

\bibitem{Boulanger:2014vya}
N.~Boulanger, D.~Ponomarev, E.~Sezgin and P.~Sundell, \emph{{New unfolded
  higher spin systems in $AdS_3$}},
  \href{http://dx.doi.org/10.1088/0264-9381/32/15/155002}{\emph{Class. Quant.
  Grav.} {\bf 32} (2015) 155002}, [\href{http://arxiv.org/abs/1412.8209}{{\tt
  1412.8209}}].

\end{thebibliography}\endgroup

\end{document}